\documentclass[spanish,12pt,a4,openright,nohyper]{book}
\usepackage{a4wide}
\usepackage{fancyhdr,graphicx}
\usepackage{latexsym,amssymb,epsfig}
\usepackage{amsmath,amssymb}
\usepackage{mathbarr}

\usepackage[spanish]{babel}

\usepackage{graphicx}

\def\slashchar#1{\setbox0=\hbox{$#1$}
   \dimen0=\wd0 \setbox1=\hbox{/} \dimen1=\wd1
   \ifdim\dimen0>\dimen1 \rlap{\hbox to \dimen0{\hfil/\hfil}} #1
   \else  \rlap{\hbox to \dimen1{\hfil$#1$\hfil}} / \fi}

\newcommand{\tr}{{\textrm {tr}}}
\newcommand{\Tr}{{\textrm {Tr}}}
\newcommand{\Det}{{\textrm {Det}}}
\newcommand{\diag}{{\textrm {diag}}}
\newcommand{\SU}{{\textrm {SU}}}

\newcommand{\ga}{{g_A}}
\newcommand{\htr}{{\widehat\tr}}
\newcommand{\Dirac}{{\text {Dirac}}}
\newcommand{\reg}{{\text {reg}}}
\newcommand{\diver}{{\text {div}}}

\newcommand{\MS}{\overline{\text {MS}}}
\newcommand{\PV}{{\text {PV}}}
\newcommand{\YM}{{\text {YM}}}
\newcommand{\NJL}{{\text {NJL}}}
\newcommand{\SQM}{{\text {SQM}}}
\newcommand{\thru}[1]{\mathrel{\mathop{#1\!\!\!/}}}
\newcommand{\thruu}[1]{\mathrel{\mathop{#1\!\!\!\!/}}}
\newcommand{\half}{{\textstyle\frac{1}{2}}}
\newcommand{\cD}{{\mathcal D}}
\newcommand{\cL}{{\mathcal L}}

\newcommand{\bnu}{{\overline \nu}}
\newcommand{\D}{{\widehat D}}
\newcommand{\f}{{f_\pi^*}}
\newcommand{\J}{{\cal J}}
\newcommand{\hb}{b^{T}}

\newcommand{\I}{{\cal I}}
\newcommand{\Nabla}{\overline{\nabla}}
\newcommand{\V}{{\overline V}}

\newcommand{\ochi}{{\overline\chi}}

\newcommand{\A}{{{\mathcal{A}}}}
\newcommand{\F}{{\overline F}}
\newcommand{\FF}{{\widehat F}}
\newcommand{\E}{{\widehat E}}

\newcommand{\be}{\begin{equation}}
\newcommand{\ee}{\end{equation}}
\newcommand{\ba}{\begin{eqnarray}}
\newcommand{\ea}{\end{eqnarray}}

\newcommand{\eq}{\begin{eqnarray}}
\newcommand{\en}{\end{eqnarray}}

\newcommand{\hnu}{{\widehat \nu}}

\begin{document}



\pagenumbering{arabic}
\setcounter{page}{3} \pagestyle{fancy} 

\renewcommand{\chaptermark}[1]{\markboth{\chaptername%
\ \thechapter:\,\ #1}{}}
\renewcommand{\sectionmark}[1]{\markright{\thesection\,\ #1}}

\newpage


\thispagestyle{empty}


\vspace*{1cm}
\begin{center}
{\fontsize{24.pt}{9pt}\selectfont{
{\bf Efectos de Temperatura Finita}\\
\vspace*{0.5cm}
{\bf y Curvatura en QCD }\\
\vspace*{0.5cm}
{\bf y Modelos de Quarks Quirales}\\
}}
\vspace{2.5cm}
{\fontsize{18.pt}{9pt}\selectfont{
Eugenio Meg\'{\i}as Fern\'andez\\
\vspace{0.5cm}
\emph{Departamento de F\'{\i}sica At\'omica, Molecular y Nuclear}}}
\end{center}

\vspace{2.cm}
\begin{figure}[!h]
\begin{center}
\epsfig{file=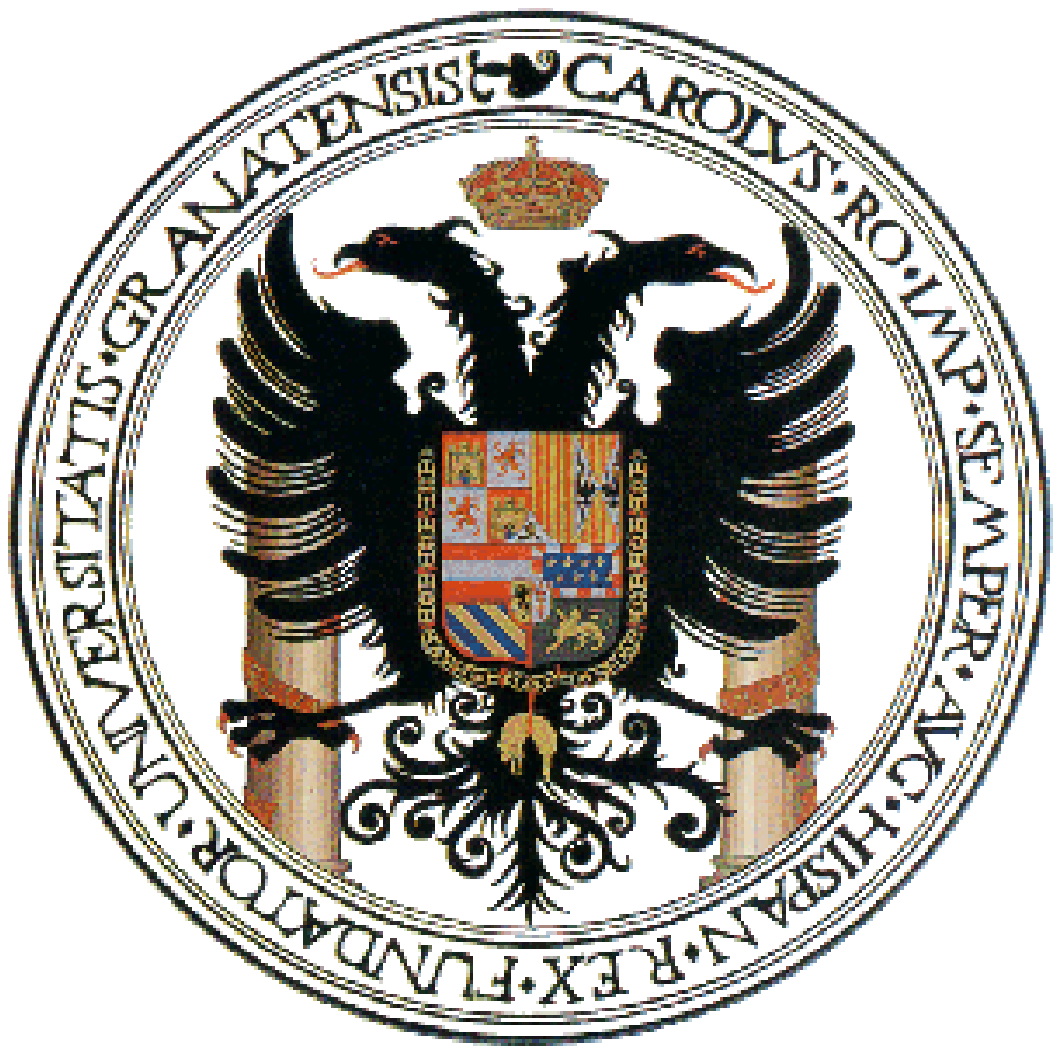,height=4cm}
\end{center}
\end{figure}
\vspace{2.cm}

\begin{center}
{\fontsize{18.pt}{9pt}\selectfont{
\emph{Universidad de Granada}\\
\vspace{0.5cm}
$\cdot$ Abril 2006 $\cdot$}}
\end{center}

\newpage
\thispagestyle{empty}
\mbox{ }
\newpage


\vspace{2cm}
\noindent D. ENRIQUE RUIZ ARRIOLA, Catedr\'atico del Departamento de F\'{\i}sica At\'omica, Molecu-lar y Nuclear y \\
D. LORENZO LUIS SALCEDO MORENO, Profesor titular del Departamento de F\'{\i}sica At\'omica, Molecular y Nuclear,

\vspace{2cm}

{\bf CERTIFICAN:} Que la presente memoria de investigaci\'on, {\it Efectos de Temperatura Finita y Curvatura en QCD y Modelos de Quarks Quirales}, ha sido realizada bajo su direcci\'on en el Departamento de F\'{\i}sica At\'omica, Molecular y Nuclear de la Universidad de Granada, por EUGENIO MEG\'IAS FERN\'ANDEZ, y constituye su Tesis para optar al grado de Doctor en Ciencias F\'{\i}sicas por la Universidad de Granada.
\vspace{1cm}

Y para que as\'{\i} conste, en cumplimiento de la legislaci\'on vigente, presenta ante la Universidad de Granada la referida Tesis.

\vspace{2.5cm}
En Granada, a 27 de abril de 2006.

\vspace{3cm}

Fdo.: Enrique Ruiz Arriola  $\qquad\qquad\qquad$   Fdo.: Lorenzo Luis Salcedo Moreno

\vspace{3cm}

\hspace{4cm} Fdo.: Eugenio Meg\'{\i}as Fern\'andez

\newpage
\thispagestyle{empty}
\mbox{ }
\newpage

{\bf\large AGRADECIMIENTOS    }
\\
\vspace{0.5cm}

     Deseo expresar mi m\'as sincero agradecimiento, en primer lugar a mis dos directores Enrique y Lorenzo Luis, pues se han involucrado por igual en la propuesta y el desarrollo de las diferentes l\'{\i}neas de investigaci\'on que constituyen esta tesis y han sabido aportarme la mejor ciencia que sabe hacer cada uno, que es mucha. 

\vspace{0.3cm}

     Al Departamento de F\'{\i}sica At\'omica, Molecular y Nuclear, por haberme dado la posibilidad de trabajar en \'el, lo que me ha permitido comprobar la enorme calidad cient\'{i}fica y humana de sus miembros.

\vspace{0.3cm}

     A Wojciech Broniowski, por su admirable humanidad. Guardo un grato recuerdo de mi estancia en Cracovia, donde no s\'olo aprend\'{\i} f\'{\i}sica.

\vspace{0.3cm}

     Estoy en deuda con Miguel Angel, mi profesor de f\'{\i}sica en secundaria, por haberme inculcado esa ilusi\'on por la f\'{\i}sica e iniciarme en el camino.
 
\vspace{0.3cm}

     Mis padres Jos\'e Antonio y Aurora han sufrido m\'as directamente mis cambios de humor. Les renocozco su sacrificio, y los admiro por saber dominar los momentos dif\'{\i}ciles y disfrutar de los momentos agradables.

\vspace{0.3cm}

    Finalmente doy las gracias a quien lea total o parcialmente esta tesis, y espero que pueda sacar de ella resultados importantes.

\vspace{0.3cm}

    Este trabajo ha sido parcialmente financiado por la D.G.I. y fondos FEDER con proyecto FIS-2005-00810, la Junta de Andaluc\'{\i}a con proyecto FM-225, EURIDICE con proyecto HPRN-CT-2002-00311 y el Ministerio de Educaci\'on y Ciencia mediante una beca de Postgrado para la Formaci\'on de Profesorado Universitario. Ha sido realizado al amparo del Departamento de F\'{\i}sica At\'omica, Molecular y Nuclear de la Universidad de Granada.


\addtolength{\headheight}{3pt}    
\fancyhead{}
\fancyhead[RE]{\sl\leftmark}
\fancyhead[RO,LE]{\rm\thepage}
\fancyhead[LO]{\sl\rightmark}
\fancyfoot[C,L,E]{}

\tableofcontents      


\chapter{Introducci\'on}
\label{intro}

La extensi\'on de la Teor\'{\i}a de Campos de temperatura cero a temperaturas y densidades finitas es un paso natural que se produjo hace medio siglo~\cite{matsubara,kirzhnits,Pollyakov,Hooftss}. La Teor\'{\i}a de Campos a Temperatura y Densidad Finitas (TCTDF) \cite{das,kapusta,lebellac}, se desarroll\'o a partir de la Teor\'{\i}a Rela\-tivista de Muchos Cuerpos, y constituye una amalgama de Teor\'{\i}a de Campos y Mec\'anica Estad\'{\i}stica. Es aplicable en aquellos problemas de la f\'{\i}sica te\'orica de part\'{\i}culas que tienen caracter\'{\i}sticas de muchos cuerpos. A nivel te\'orico se necesitan formulaciones apropiadas del problema t\'ermico, para el cual se disponen de varios formalismos. Dos ejemplos son el formalismo de Tiempo Imaginario y el de Tiempo Real~\cite{reali}. A pesar de la larga experiencia acumulada en este campo, muchos de los problemas planteados inicialmente a\'un siguen abiertos.  

Muchos son los logros de la TCTDF y se esperan muchos m\'as. Por una parte permite estudiar las teor\'{\i}as ya existentes m\'as all\'a del contexto en el que inicialmente fueron creadas. Esto significa explorar las propiedades de la materia en condiciones extremas, con altas temperaturas y densidades. Un ejemplo de esto es la teor\'{\i}a de QCD \cite{qcdrenormalization}, que se cre\'o como un intento de desarrollar una teor\'{\i}a fundamental de las interacciones fuertes. La TCTDF aplicada a QCD \cite{pisarski} predice que cuando la temperatura y las densidades aumentan, exis\-te una transici\'on a una fase en la que los quarks y gluones est\'an deconfinados (fase de des\-confinamiento del color). TCTDF predice, por tanto, la existencia de un plasma de quarks y gluones que, de hecho, deber\'{\i}a existir en los primeros instantes del universo, de acuerdo con los modelos cosmol\'ogicos actuales. Esto tiene importantes consecuencias en el campo de la astrof\'{\i}sica, ya que la transici\'on de fase podr\'{\i}a haber jugado un papel muy importante en la formaci\'on de materia oscura. Otro campo donde la TCTDF est\'a dando frutos importantes es en el contexto de las colisiones de iones pesados a muy alta energ\'{\i}a. El hecho de que la transici\'on de fase de QCD ocurra a temperaturas no excesivamente altas $T_c \sim 200\,\textrm{MeV}$ hace que estas condiciones se puedan estudiar en el laboratorio. Existen estudios importantes de esta nueva fase de la materia en laboratorios actuales [BNL Relativistic Heavy Ion Collider (RHIC)]~\cite{mclerran_pramana} y es previsible que se contin\'uen posteriormente en futuras instalaciones: Large Hadron Collider (LHC) en el CERN, y Schwerionen-Synchrotron (SIS 200) en el GSI. Finalmente, un tercer lugar donde pueden surgir tales condiciones extremas es en el interior de estrellas de neutrones, donde la densidad es superior a la densidad nuclear.    

Existen distintas t\'ecnicas para estudiar el comportamiento de QCD en funci\'on de la temperatura y la densidad. Estas t\'ecnicas se pueden agrupar en tres categor\'{\i}as diferentes: los m\'etodos perturbativos, los modelos efectivos de QCD en el ret\'{\i}culo y los m\'etodos semicl\'asicos (instantones) \cite{pisarski}. 

\section{Cromodin\'amica Cu\'antica}

La Cromodin\'amica Cu\'antica (QCD, Quantum Chromodynamics) fue desarrollada al comienzo de los a\~nos setenta y responde al intento de mucha gente de crear una teor\'{\i}a fundamental que d\'e cuenta de las interacciones fuertes \cite{ge72,fg73,we73a}. Se trata de una teor\'{\i}a cu\'antica de campos renormalizable. Sus campos fundamentales son espinores de Dirac que describen part\'{\i}culas de esp\'{\i}n~$1/2$, llamados quarks, y campos gauge correspondientes a part\'{\i}culas de esp\'{\i}n~1, llamados gluones. Al contrario que QED (Quantum Electrodynamics) que es una teor\'{\i}a abeliana, QCD es una teor\'{\i}a gauge no abeliana basada en el grupo gauge de color SU($N_c$), de modo que constituye una generalizaci\'on de la teor\'{\i}a de QED para el electromagnetismo. Tanto los quarks como los gluones, que son las part\'{\i}culas intermediarias de la interacci\'on fuerte, llevan asociada una carga, llamada color. Como resultado los gluones pueden interaccionar consigo mismos y con los quarks. QCD viene descrita por el siguiente lagrangiano
\begin{eqnarray}
{\cal L}= -\frac{1}{2g^2} \tr(F_{\mu\nu}^2) + 
\sum_{i=1}^{N_f} \overline{q}_i(\gamma^\mu D_\mu+m_i)q_i \,, \label{eq:Lqcd} \\
D_\mu = \partial_\mu +A_\mu \,, \qquad
F_{\mu\nu} = [D_\mu,D_\nu] \,,
\nonumber
\end{eqnarray}
donde $A_\mu = A_\mu^a T_a$ son los campos de los gluones, $F_{\mu\nu}=F_{\mu \nu}^a T_a$ es el tensor Field Strength de SU($N_c$), $T_a$ son los generadores herm\'{\i}ticos de SU$(N_c)$ y $q_i$ son campos de quarks de varios sabores. La teor\'{\i}a viene parametrizada por una \'unica constante de acoplamiento $g$ y por los par\'ametros~$m_i$ correspondientes a la masa desnuda de los quarks. La evidencia experimental indica que hay tres grados de libertad de color $(N_c=3)$, llamados tradicionalmente rojo, verde y azul, y seis sabores de quarks $(N_f=6)$. Los quarks de tipo up, down y strange son relativamente ligeros, mientras que charm, bottom y top son pesados. 

Gran parte del \'exito de la teor\'{\i}a reside en su habilidad para reproducir  el comportamiento casi sin interacci\'on de los quarks a muy cortas distancias \cite{gw}. Esta propiedad de la teor\'{\i}a, que se conoce como libertad asint\'otica, explica el escalamiento aproximado que se observa en las colisiones profundamente inel\'asticas de leptones con hadrones \cite{gw2,georgi}. QCD tambi\'en parece consistente con mucha de la fenomenolog\'{\i}a existente sobre las interacciones fuertes, como la simetr\'{\i}a quiral aproximada, la noci\'on de confinamiento de color o ciertos modelos de hadrones como el bag o el string.

La teor\'{\i}a de QCD presenta varias simetr\'{\i}as. En primer lugar es
invariante bajo el grupo de simetr\'{\i}a local SU$(N_c)$,  lo cual implica
por ejemplo que la masa de los quarks es independiente de su color.  Cuando la
masa de los quarks es igual a cero, el lagrangiano de QCD~(\ref{eq:Lqcd}) es
invariante bajo el grupo de simetr\'{\i}a global SU$(N_f)_L
\times$SU($N_f)_R$, el cual se suele designar como grupo de simetr\'{\i}a
quiral \cite{cheng-li}. Adem\'as existe una simetr\'{\i}a global U$(1)_B$
relacionada con la conservaci\'on del n\'umero bari\'onico y una simetr\'{\i}a
global axial U$(1)_A$.

Los generadores del \'algebra quiral son conservados y ser\'{\i}a de esperar
que las part\'{\i}culas formaran multipletes degenerados correspondientes a
las representaciones irreducibles de este grupo. Pero no existe evidencia de
que exista esta estructura de multipletes tan amplia, lo cual lleva a la idea
de que la simetr\'{\i}a $\SU(N_f)_L \times \SU(N_f)_R$ est\'a
espont\'aneamente rota. A temperatura cero, o en general a baja temperatura,
el estado fundamental de la teor\'{\i}a rompe espont\'aneamente esta simetr\'{\i}a al grupo~$\SU(N_f)_V$
\begin{equation}
\SU(N_f)_L\times\SU(N_f)_R \xrightarrow{\textrm{RES}} \SU(N_f)_V \,.
\end{equation}
De acuerdo con el teorema de Goldstone esta rotura de la simetr\'{\i}a implica la existencia de $N_f^2-1$ bosones de Goldstone pseudo-escalares sin masa. Para $N_f=2$ estos son los tres piones $\pi^+$, $\pi^-$ y $\pi^0$, y para $N_f=3$ tenemos, adem\'as de los anteriores, los cuatro kaones $K^+$, $K^-$, $K^0$ y $\bar{K}^0$, y el mes\'on $\eta$. La rotura de esta simetr\'{\i}a conduce adem\'as a la aparici\'on de condensados de quarks de la forma $\langle\overline{q}q\rangle \ne 0$. Podemos pensar en $\langle \overline{q}q \rangle$ como en un par\'ametro de orden que caracteriza la rotura de la simetr\'{\i}a quiral. Cuando la temperatura se incrementa por encima de un cierto valor $T_c$, la simetr\'{\i}a se recupera y el condensado de quarks se hace cero.

\section{Simetr\'{\i}a del centro y transici\'on de fase de QCD}

En gluodin\'amica pura, esto es en ausencia de fermiones, la teor\'{\i}a presenta una simetr\'{\i}a global extra asociada al centro~$\mathbb{Z}(N_c)$ del grupo gauge de color~SU($N_c$). En el formalismo de tiempo imaginario, la simetr\'{\i}a $\mathbb{Z}(N_c)$ es generada por la acci\'on de transformaciones gauge locales que son peri\'odicas en la variable temporal, salvo un elemento arbitrario del centro 
\begin{equation}
U(1/T,\vec{x})= z\,U(0,\vec{x})\,, \qquad  z=e^{i2\pi n/N_c}  \,. 
\end{equation}
La transici\'on a la fase de desconfinamiento puede verse como la rotura espont\'anea de la simetr\'{\i}a del centro a temperaturas suficientemente altas. Un par\'ametro de orden natural para la simetr\'{\i}a $\mathbb{Z}(N_c)$ es el valor esperado del loop de Polyakov,\footnote{En esta memoria se har\'a uso en ocasiones de una terminolog\'{\i}a anglosajosa para algunas palabras, y se evitar\'a su traducci\'on con el fin de que el lector pueda identificar estos conceptos en la bibliograf\'{\i}a. 'Loop de Polyakov' puede traducirse como 'bucle de Polyakov'.} que se define como
\begin{equation}
L(T) := \langle {\mathcal P}(\vec{x},T) \rangle  = \left\langle \frac{1}{N_c}\tr_c\, {\mathcal T} \left( 
e^{-\int_0^{1/T} d x_0 A_0 (\vec{x} , x_0) }\right) \right\rangle \,,
\end{equation}
donde $\langle~\rangle$ indica valor esperado en el vac\'{\i}o, $\tr_c$ es la
traza en espacio de color (en representaci\'on fundamental), y ${\mathcal T}$
indica ordenaci\'on a lo largo del camino de integraci\'on. $A_0$ es la
componente temporal del campo glu\'onico (en tiempo eucl\'{\i}deo). Bajo una
transformaci\'on gauge con simetr\'{\i}a del centro, el loop de Polyakov
transforma ${\mathcal P}\rightarrow z {\mathcal P}$, de modo que en la fase en que la teor\'{\i}a presenta la simetr\'{\i}a $\mathbb{Z}(N_c)$ (fase de confinamiento del color), el loop de Polyakov necesariamente vale cero. En la fase de desconfinamiento esta simetr\'{\i}a estar\'a espont\'aneamente rota, y eso vendr\'a caracterizado por un valor no nulo para el loop de Polyakov. C\'alculos recientes muestran que en una teor\'{\i}a glu\'onica pura con $N_c=3$ esta transici\'on ocurre a una temperatura cr\'{\i}tica $T_c \simeq 270\,\textrm{MeV}$ \cite{iwasaki}, y se trata de una transici\'on de primer orden.

F\'{\i}sicamente el promedio t\'ermico del loop de Polyakov en la representaci\'on fundamental determina la energ\'{\i}a libre relativa al vac\'{\i}o de un \'unico quark,
\begin{equation}
e^{-F_q(\vec{x})/T}=\langle {\mathcal P}(\vec{x},T) \rangle \,,   
\end{equation}
y la funci\'on de correlaci\'on de dos loops de Polyakov conduce a la
energ\'{\i}a libre de un par quark-antiquark,
\begin{equation}
e^{-F_{\bar{q}q}(\vec{x}-\vec{y})/T}=\langle {\mathcal P}(\vec{x},T) {\mathcal
P}^\dagger (\vec{y},T) \rangle \,.
\label{eq:corr_polll}
\end{equation}
La renormalizaci\'on del loop de Polyakov es un problema que hoy en d\'{\i}a est\'a abierto~\cite{notes_pis}. Recientemente se ha desarrollado un m\'etodo para renormalizar el loop de Polyakov en el ret\'{\i}culo~\cite{Kaczmarek:2002mc,Kaczmarek:2005ui}, y consiste b\'asicamente en el c\'alculo de la energ\'{\i}a libre a partir de la funci\'on de correlaci\'on de dos loops de Polyakov, ec.~(\ref{eq:corr_polll}). Los datos que se obtienen muestran un comportamiento que difiere claramente del predicho por teor\'{\i}a de perturbaciones~\cite{Gava:1981qd} en la regi\'on cercana a la transici\'on de fase, de modo que los efectos no perturbativos parecen ser dominantes en esta zona de temperaturas.
 


Un punto importante es qu\'e efectos produce la inclusi\'on de fermiones en una teor\'{\i}a gauge pura. En el caso de QCD, cuando se a\~naden quarks en la representaci\'on fundamental, la simetr\'{\i}a del centro~${\mathbb Z}(N_c)$ se rompe expl\'{\i}citamente, y el loop de Polyakov no sirve, en principio, como par\'ametro para caracterizar la transici\'on de desconfinamiento. Una de las consecuencias es la modificaci\'on de las condiciones en que se produce la transici\'on de fase. En concreto, los quarks tienden a suavizar la transici\'on, de tal modo que en la teor\'{\i}a SU(3) se convierte en una transici\'on de fase de segundo orden~\cite{Kaczmarek:2005ui}. 

En cuanto a la simetr\'{\i}a quiral, \'esta se encuentra espont\'aneamente rota a baja temperatura, pero por encima de un cierto valor se recupera. El par\'ametro de orden local en este caso es el condensado de quarks $\langle \overline{q}q \rangle$, que es diferente de cero a baja temperatura, donde la simetr\'{\i}a quiral est\'a rota, y cero por encima de la transici\'on de fase quiral. Por tanto, desde un punto de vista te\'orico la transici\'on de fase de QCD consiste en realidad en dos transiciones de fase distintas, que podemos llamar transici\'on de desconfinamiento de color y transici\'on de restablecimiento de la simetr\'{\i}a quiral. Las simulaciones de QCD en el ret\'{\i}culo sugieren que, cuando se consideran fermiones sin masa, las dos transiciones tienen lugar a la misma temperatura, al menos en el caso de potencial qu\'{\i}mico cero~\cite{Fukushima:2003fm}. En este caso la temperatura de restablecimiento de la simetr\'{\i}a quiral es $T_c\simeq 155-205\,\textrm{MeV}$, donde el valor preciso depende del n\'umero de sabores. Cuando se consideran masas f\'{\i}sicas para los quarks la situaci\'on no est\'a completamente clara. Para valores moderados de la masa, la transici\'on quiral no tiene un par\'ametro de orden bien definido, y no se produce una transici\'on de fase pura sino \'unicamente un cambio r\'apido (crossover).

Obviamente, es de esperar que todos estos fen\'omenos de QCD a temperatura finita sean consistentes con invariancia gauge. La invariancia Lorentz se rompe expl\'{\i}citamente en c\'alculos a temperatura y densidad finitas, debido a que existe un sistema de referencia privilegiado, que es el ba\~no t\'ermico, y que se supone en reposo; no obstante, la invariancia gauge permanece como una simetr\'{\i}a exacta. En c\'alculos concretos en teor\'{\i}a de perturbaciones, la conservaci\'on de la invariancia gauge a temperatura cero se consigue con un n\'umero finito de t\'erminos, sin embargo a temperatura finita es necesario considerar un n\'umero infinito de t\'erminos, lo cual obligar\'{\i}a en un principio a hacer un tratamiento no perturbativo.

\section{Teor\'{\i}as quirales efectivas}

Actualmente los grados de libertad hadr\'onicos se vienen tratando con
teor\'{\i}as quirales efectivas en las cuales un ingrediente b\'asico son los
bosones de Goldstone generados en la rotura espont\'anea de la simetr\'{\i}a
quiral de QCD~\cite{gasser-leutwyler1,chiralanomaly}. La aproximaci\'on por
excelencia es la Teor\'{\i}a Quiral de Perturbaciones
(TQP)~\cite{gasser-leutwyler1,gasser-leutwyler2}. Existen otras aproximaciones
que se basan en la construcci\'on de modelos de quarks quirales como el
modelo sigma \cite{sigmamodel} o el modelo de Nambu--Jona-Lasinio (NJL)
\cite{njlmodel,skyrme,osipov}.

La TQP se fundamenta en la construcci\'on de un lagrangiano efectivo invariante quiral como desarrollo en potencias de los momentos externos de los campos y de la masa de los quarks. Este lagrangiano debe satisfacer ciertos requisitos de simetr\'{\i}a como invariancia gauge, invariancia Lorentz (a temperatura cero), paridad y conjugaci\'on de carga, y se escribe en t\'erminos de constantes de baja energ\'{\i}a que se corresponden con funciones de Green de QCD. Los valores de estas constantes no pueden ser determinados a partir de argumentos de simetr\'{\i}a exclusivamente.  

Los modelos de quarks quirales aspiran, como TQP, a constituir una
aproximaci\'on de la din\'amica de QCD no perturbativa a baja
energ\'{\i}a. Estos modelos hacen uso expl\'{\i}cito de grados de libertad de
quarks. El modelo de Nambu--Jona-Lasinio ha sido muy utilizado en el pasado y
a\'un se sigue utilizando. Las interacciones efectivas de cuatro fermiones del
modelo NJL representan cierta aproximaci\'on a QCD. Sin embargo, desde un
punto de vista te\'orico a\'un no est\'a claro de qu\'e modo estas
interacciones de cuatro quarks surgen de QCD. En el caso de dos sabores uno de
los mecanismos podr\'{\i}a ser las llamadas interacciones de 't~Hooft, que consisten en la interacci\'on de quarks a trav\'es de los modos cero de instantones \cite{diakonovlectures}.

\section{Heat kernel y acci\'on efectiva}

La acci\'on efectiva, una extensi\'on a teor\'{\i}a cu\'antica de campos del
potencial termodin\'amico de mec\'anica estad\'{\i}stica, juega un papel
te\'orico muy importante pues est\'a relacionada con cantidades de inter\'es
f\'{\i}sico. A un loop tiene la forma $c\,\Tr\log ({\mathbf K})$, donde
${\mathbf K}$ es un ope\-rador diferencial que controla las fluctuaciones
cu\'anticas cuadr\'aticas sobre un fondo cl\'asico. Esta magnitud sufre
algunas patolog\'{\i}as matem\'aticas, tales como divergencias ultravioletas y
multivaluaci\'on. Por ello resulta \'util expresar la acci\'on efectiva
mediante la representaci\'on de tiempo propio de Schwinger\footnote{La traza
  funcional de un operador $\hat{A}$ se define
\begin{equation}
\Tr \hat{A} \equiv \int d^D x \, \tr \langle x| \hat{A} |x \rangle \,,
\end{equation}
donde $D$ es la dimensi\'on del espacio-tiempo y $\tr$ indica traza en espacio
interno (color, sabor, Dirac, etc). A lo largo de la tesis haremos uso de esta definici\'on.
}
\begin{equation}
-c\,\Tr \log ({\mathbf K}) 
= c \int_0^\infty \frac{d\tau}{\tau} \,\Tr \,e^{-\tau {\mathbf K}} 
=
c \int_0^\infty \frac{d\tau}{\tau} \int d^D x \, 
\tr \langle x| e^{-\tau {\mathbf K}}    |x\rangle   \,.
\end{equation}
Al contrario que la acci\'on efectiva, el heat kernel (o m\'as concretamente
su elemento de matriz) $\langle x|e^{-\tau {\mathbf K}}|x\rangle$ es univaluado y finito en la regi\'on ultravioleta para valores positivos del par\'ametro de tiempo propio~$\tau$. 

El heat kernel fue introducido por Schwinger \cite{schwinger} en teor\'{\i}a
cu\'antica de campos como una herramienta para regularizar divergencias
ultravioletas de un modo que preserve invariancia gauge. El heat kernel  y su
desarrollo han sido aplicados tambi\'en en el estudio de densidades
espectrales e \'{\i}ndices de operadores de Dirac (${\mathbf D}$)
\cite{gilkey,atiyah} en t\'erminos de operadores de Klein-Gordon $({\mathbf D}^\dagger {\mathbf D})$, para el c\'alculo de la funci\'on $\zeta$ \cite{hawking,zeta}  y anomal\'{\i}as de estos operadores \cite{fujikawa}, para definir la acci\'on efectiva de teor\'{\i}as gauge quirales \cite{ball}, para el efecto Casimir \cite{bordag}, etc. El heat kernel se puede calcular perturbativamente haciendo un desarrollo en potencias del tiempo propio. En la presente memoria va a constituir una herramienta fundamental para el c\'alculo de las diferentes teor\'{\i}as efectivas que vamos a considerar. 

\section{Estructura de la tesis}

Esta tesis est\'a estructurada del siguiente modo:

\vspace{0.4cm}

En el cap\'{\i}tulo~\ref{heat_kernel} se considera el heat kernel a
temperatura cero, y se construye su gene\-ralizaci\'on a temperatura
finita, dentro del formalismo de tiempo imaginario. Con objeto de
conseguir un desarrollo que preserve la invariancia gauge orden por
orden, haremos uso de una generalizaci\'on a temperatura finita del
m\'etodo de los s\'{\i}mbolos~\cite{garcia}, que permite calcular de un modo
sencillo el desarrollo de una funci\'on en t\'erminos de operadores
locales y covariantes gauge. Esto va a conducir a la definici\'on del
loop de Polyakov (sin traza), que es un objeto covariante gauge, y que
aparece de manera natural en el desarrollo. El c\'alculo se hace para
un gauge general y en presencia de campos escalares que pueden ser no
abelianos y no estacionarios.

En el cap\'{\i}tulo~\ref{QCD_efective_action} se considera la teor\'{\i}a gauge SU($N_c$) de QCD, y se calcula su acci\'on efectiva a nivel de un loop en el r\'egimen de temperaturas grandes, haciendo uso del resultado del heat kernel del cap\'{\i}tulo~\ref{heat_kernel}. Se calculan por separado el sector glu\'onico y el sector de quarks, y se hace un estudio de c\'omo los quarks rompen expl\'{\i}citamente la simetr\'{\i}a del centro~${\mathbb Z}(N_c)$. Esta rotura se va a manifestar en que algunos de los m\'{\i}nimos absolutos degenerados que presenta el potencial efectivo de la teor\'{\i}a como funci\'on del loop de Polyakov van a dejar de serlo, y se van a convertir en puntos estacionarios (m\'{\i}nimos o m\'aximos locales).  A temperaturas suficientemente grandes est\'a justificado considerar una teor\'{\i}a efectiva dimensionalmente reducida, pues lo modos de Matsubara no est\'aticos de los campos gauge se hacen muy pesados y desacoplan de la teor\'{\i}a. Dentro del problema de reducci\'on dimensional obtendremos la estructura del lagrangiano dimensionalmente reducido.

En el cap\'{\i}tulo~\ref{condensados} se hace un estudio fundamentado de los datos del loop de Polyakov renormalizado en la fase de desconfinamiento de color, obtenidos en el ret\'{\i}culo. Se estudian las contribuciones no perturbativas existentes, en el marco de un modelo fenomenol\'ogico que las describe como generadas por condensados glu\'onicos invariantes BRST.

En el cap\'{\i}tulo~\ref{quirales_Tfinita} se aborda la problem\'atica que presenta el tratamiento est\'andar de los modelos de quarks quirales a temperatura finita. Discutimos el acoplamiento del loop de Polyakov de color con los quarks, y calculamos el lagrangiano quiral efectivo a bajas energ\'{\i}as, con una predicci\'on para las cons\-tantes de baja energ\'{\i}a. Se estudian asimismo las implicaciones que tiene este modelo, sobre la transiciones de fase quiral y de desconfinamiento de color.

El cap\'{\i}tulo~\ref{tensor_EM_MQQ} est\'a dedicado a estudiar los efectos de curvatura sobre varios modelos de quarks quirales: Quark Constituyente, Georgi-Manohar y Nambu--Jona-Lasinio. En concreto, se estudia
el acoplamiento de la gravedad en estos modelos de un modo que evite la
introducci\'on de nuevos campos aparte de los del caso plano y la
m\'etrica. Se estudia el tensor energ\'{\i}a-impulso a bajas energ\'{\i}as que se obtiene, con valores concretos para las cons\-tantes de baja energ\'{\i}a est\'andar y una predicci\'on para las constantes asociadas a t\'erminos no m\'etricos con contribuci\'on de curvatura.

En el cap\'{\i}tulo~\ref{ae_quiral_SQM} se hace un estudio de la estructura de la acci\'on efectiva del modelo quark espectral acoplado con gravedad. Por una parte se considera la contribuci\'on an\'omala, y por otra la parte no-an\'omala, con una predicci\'on para las constantes de baja energ\'{\i}a. Se estudian los resultados del modelo en el esquema de dominancia vectorial, y se compara con el c\'alculo en el l\'{\i}mite de $N_c$ grande en la aproximaci\'on de una \'unica resonancia.

Por \'ultimo, en el cap\'{\i}tulo~\ref{Conclusiones} se presentan las conclusiones de la memoria.

\chapter{Desarrollo del Heat Kernel}
\label{heat_kernel}

El desarrollo del heat kernel\footnote{Heat kernel puede traducirse como 'N\'ucleo de la ecuaci\'on del calor', pues constituye la soluci\'on a esta conocida ecuaci\'on.} \cite{schwinger,ball} se usa frecuentemente en el contexto de los m\'etodos de integrales de caminos para integrar grados de libertad externos de un modo no perturbativo. El resultado es un desarrollo en los campos que corresponden a aquellos grados de libertad que no han sido integrados. Esto quiere decir que el desarrollo del heat kernel proporciona una teor\'{\i}a de campos efectiva. Los t\'erminos del desarrollo se clasifican de acuerdo con su dimensi\'on.

Nuestro objetivo en este cap\'{\i}tulo consiste en dise\~nar un m\'etodo que permita mantener la invariancia gauge a temperatura finita de forma manifiesta orden por orden en el desarrollo dimensional. Para ello aplicaremos una t\'ecnica conocida como {\it m\'etodo de los s\'{\i}mbolos}, que fue desarrollado a temperatura cero~\cite{metsimb} y extendido posteriormente a temperatura finita~\cite{garcia}. Hay que notar que el tratamiento es inevitablemente complejo pero necesario. 

Como motivaci\'on, estudiaremos el potencial macrocan\'onico de un gas de part\'{\i}culas libres relativistas, donde el loop de Polyakov se reduce a la fugacidad $e^{\beta \mu}$, con $\beta=1/T$ la temperatura inversa y $\mu$ el potencial qu\'{\i}mico. La idea consiste en respetar la propiedad de periodicidad de la exponencial bajo cambios peri\'odicos del potencial qu\'{\i}mico $\mu \rightarrow \mu + i2\pi  T$. Aunque este caso es trivial, ayudar\'a a comprender mejor la idea subyacente del m\'etodo de los s\'{\i}mbolos. 

Este cap\'{\i}tulo est\'a basado en las referencias~\cite{megias,Megias:2003ui}.

\section{Potencial macrocan\'onico de un gas de part\'{\i}culas libres relativistas}
\label{macrocanonico}

Como ilustraci\'on y motivaci\'on del heat kernel, consideraremos el caso de un gas de part\'{\i}culas libres relativistas. Por claridad estudiaremos el caso bos\'onico. La acci\'on eucl\'{\i}dea para esta teor\'{\i}a se escribe
\begin{equation}
S_E[\phi] = \frac{1}{2} \int d^Dx \,\phi(x) (-D_\mu^2+m^2) \phi(x) \,,
\end{equation}
donde $D=d+1$ es la dimensi\'on del espacio-tiempo. Consideramos las siguientes derivadas covariantes:
\begin{equation}
D_0 = \partial_0 - i \mu \,, \qquad
D_i = \partial_i \,.
\label{eq:PM_D0_Di}
\end{equation}
$\mu$ es un potencial qu\'{\i}mico, y el loop de Polyakov correspondiente es~$\Omega = e^{i\beta\mu}$. La funci\'on de partici\'on de esta teor\'{\i}a se calcula f\'acilmente
\begin{equation}
Z = \int {\cal D}\phi \,e^{-S_E[\phi]}  = (\det (-D_\mu^2+m^2))^{-1} \,.
\end{equation}
Usaremos aqu\'{\i} el convenio~$Z=e^{-\Gamma}$, donde $\Gamma$ es la acci\'on efectiva. El potencial macrocan\'onico est\'a relacionado con la acci\'on efectiva a trav\'es de~$\Gamma=\beta \,\Omega_{\rm mc}$. As\'{\i} pues, la acci\'on efectiva se puede calcular a partir del heat kernel del siguiente modo
\begin{equation}
\Gamma=\log\det(-D_\mu^2+m^2) = \Tr\log(-D_\mu^2+m^2)= -\Tr\int_0^\infty\frac{d\tau}{\tau} \langle x|e^{-\tau
(-D_\mu^2+m^2)}|x\rangle \,,
\label{eq:2.17}
\end{equation}
donde hemos hecho uso de la representaci\'on de Schwinger de tiempo propio. $(-D_\mu^2+m^2)$ es un operador de tipo Klein-Gordon, que ser\'a definido en ec.~(\ref{eq:operador_KG}). Si hacemos uso de ec.~(\ref{eq:13}), con la definici\'on de la funci\'on $\varphi_0$ dada en ec.~(\ref{eq:2.15}), sustraemos la parte de temperatura cero (que corresponde a considerar $\varphi_0 \to 1 $), y se realizan las integrales, finalmente llegamos al resultado est\'andar~\cite{kapusta}
\begin{equation}
\Gamma= N\int\frac{d^dxd^dk}{(2\pi)^d}\left[
\log\left(1-e^{-\beta(\omega_k-\mu)}\right)
+\log\left(1-e^{-\beta(\omega_k+\mu)}\right)
\right]\,.
\label{eq:PM_ae}
\end{equation}
$N$ es el n\'umero de especies y $\omega_k=\sqrt{k^2+m^2}$. 
El efecto de introducir otros campos externos puede ser tenido en cuenta mediante los sucesivos \'ordenes del desarrollo del heat kernel (ec.~(\ref{eq:13}) corresponde al primer orden).

\section{M\'etodo de los S\'{\i}mbolos}
\label{metodo_simbolos}

Consideremos un operador gen\'erico
\begin{equation}
\widehat{f} = f(M,D_\mu) \,,
\label{eq:op_general}
\end{equation}
construido con $M$ y $D_\mu$ en un sentido algebraico, esto es, es una combinaci\'on lineal (o serie) de productos de $M$ y $D_\mu$ con coeficientes que son c-n\'umeros. $D_\mu$ es la derivada covariante
\begin{equation}
D_\mu = \partial_\mu + A_\mu(x) \,,
\end{equation}
$A_\mu(x)$ es el campo gauge y $M(x)$ denota una o varias funciones matriciales de $x$ que re\-presentan otros campos externos diferentes de los campos gauge. El m\'etodo de los s\'{\i}mbolos~\cite{garcia,metsimb} permite calcular de un modo sistem\'atico los elementos diagonales del ope\-rador~(\ref{eq:op_general}). 

Consideraremos la siguiente normalizaci\'on para los estados con posici\'on y momento bien definidos
\begin{equation}
\langle x | p\rangle = e^{ipx} \,, \qquad \langle p | p^\prime \rangle = (2\pi)^D \delta(p-p^\prime)  \,,
\end{equation} 
y la relaci\'on de completitud
\begin{equation}
{\mathbf 1}= \int \frac{d^D p }{(2\pi)^D}|p\rangle\langle p| \,.
\label{eq:rel_compl}
\end{equation}
$D$ es la dimensi\'on del espacio-tiempo. Denotaremos por $|0\rangle$ el estado de momento cero, el cual satisface
\begin{equation}
\langle x | 0 \rangle = 1 \,, \qquad 
\widehat{p}_\mu | 0 \rangle = \langle 0 | \widehat{p}_\mu = 0 \,, \qquad
\langle 0 | 0 \rangle = \int d^D x \,.
\end{equation}
En nuestra notaci\'on $p_\mu$ es real, $\int d^D p$ indica integraci\'on est\'andar en ${\mathbb R}^D$ y $\delta(p-p^\prime)$ es la funci\'on delta correspondiente. $p^2$ significa $p_\mu p_\mu$. 
Si consideramos el elemento diagonal $\langle x|f(M,D_\mu) |x\rangle$, se tiene
\begin{eqnarray}
\langle x| f(M,D_\mu)|x\rangle &=& \int \frac{d^D p}{(2\pi)^D} 
\langle x | f(M,D_\mu) |p \rangle \langle p | x \rangle  \nonumber \\
&=&  \int \frac{d^D p}{(2\pi)^D} \langle p | x \rangle\langle x | e^{ip\widehat{x}} e^{-ip\widehat{x}} f(M,D_\mu) e^{ip\widehat{x}} e^{-ip\widehat{x}} |p \rangle \,.
\label{eq:met_simb_demo}
\end{eqnarray}
En la primera igualdad hemos introducido la relaci\'on de completitud (\ref{eq:rel_compl}). Teniendo en cuenta que el operador posici\'on $\widehat{x}$ es el generador de las traslaciones en momentos, tenemos las siguientes transformaciones de semejanza
\begin{equation}
e^{-ip\widehat{x}}D_\mu \,e^{ip\widehat{x}} = D_\mu + ip_\mu \, , \quad  e^{-ip\widehat{x}}M(x) \,e^{ip\widehat{x}} = M(x) \,,
\end{equation}
o en general para $\widehat{f}$, construida en sentido algebraico con $M$ y $D_\mu$,
\begin{equation}
e^{-ip\widehat{x}} f(M,D_\mu) \,e^{ip\widehat{x}} = f(M,D_\mu + ip_\mu)\,. 
\end{equation}
Basta considerar $\langle x | e^{ip\widehat{x}} = e^{ipx}\langle x | $ y $e^{-ip\widehat{x}}|p\rangle=|0\rangle$ en (\ref{eq:met_simb_demo}) para obtener la f\'ormula del m\'etodo de los s\'{\i}mbolos
\begin{equation}
 \langle x|f(M,D_\mu)|x \rangle = \int \frac{d^D { p}}{(2\pi)^D}\langle x|f(M,D_\mu+ip_\mu)|0\rangle \,. 
\label{eq:met_simb}
\end{equation}
Al elemento $\langle x|f(M,D_\mu+ip_\mu)|0\rangle$ se le denomina s\'{\i}mbolo de $\widehat{f}$, y es en realidad una matriz, pues~$M$ y $D_\mu$ son operadores en espacio interno (color, sabor, Dirac, etc ). El problema con (\ref{eq:met_simb}) reside en que la covariancia gauge no se manifiesta de manera expl\'{\i}cita cuando se usa una base en momentos. En efecto, $|0\rangle$ (o m\'as generalmente $|p\rangle$) no es covariante bajo transformaciones gauge locales. Por otra parte, el miembro derecho de la igualdad en ec.~(\ref{eq:met_simb}) es expl\'{\i}citamente invariante bajo transformaciones de tipo boost
\begin{equation}
D_\mu \rightarrow D_\mu + a_\mu \,,
\label{eq:inv_boost_HK}
\end{equation}
donde $a_\mu$ son c-n\'umeros constantes. Esto se debe a que el cambio en $a_\mu$ puede ser compensado mediante un cambio similar en la variable de integraci\'on~$p_\mu$. Esta propiedad es la condici\'on necesaria y suficiente para que exista covariancia gauge, pues implica que en un desarrollo de $f$ en los operadores, $D_\mu$ debe de aparecer s\'olo en el interior de conmutadores.

\section{Desarrollo del Heat Kernel a temperatura cero}
\label{des_hk_temp_cero}

En esta secci\'on aplicaremos el m\'etodo de los s\'{\i}mbolos para el
c\'alculo del heat kernel. Consideramos el operador de
Klein-Gordon\footnote{En este cap\'{\i}tulo haremos uso de una m\'etrica
  eucl\'{\i}dea.}
\begin{equation}
{\mathbf K}=M(x) - D_\mu^2 \,.
\label{eq:operador_KG}
\end{equation} 
El heat kernel se define como el operador $e^{-\tau {\mathbf K}}$. Nosotros
estamos interesados en el c\'alculo del elemento de matriz con puntos
coincidentes $\langle x |e^{-\tau {\mathbf K}} | x\rangle$. A $\tau$ se le denomina par\'ametro de tiempo propio. Este objeto resulta en general dif\'{\i}cil de calcular, y en la pr\'actica interesa estudiar su comportamiento cuando $\tau$ es peque\~no. El heat kernel admite un desarrollo (asint\'otico) en serie de potencias de $\tau$ alrededor de $\tau=0$. Usando la notaci\'on est\'andar
\begin{equation}
\langle x |e^{-\tau {\mathbf K}} | x\rangle = \frac{1}{(4\pi\tau)^{D/2}}\sum_{n=0}^\infty a_n(x) \tau^n \,,
\label{eq:des_HK}
\end{equation}
donde los coeficientes $a_n(x)$ son conocidos como {\it coeficientes de
  Seeley-DeWitt} \cite{dewitt,seeley,Salcedo:2004yh}, y son operadores locales construidos
con una combinaci\'on lineal de productos de $M(x)$ y $D_\mu$. Puesto que el
heat kernel es covariante gauge, la expresi\'on (\ref{eq:des_HK}) debe ser
covariante gauge orden por orden. El heat kernel $e^{-\tau {\mathbf K}}$ no tiene dimensiones si asignamos dimensiones de masa $-2$, $+1$, $+2$ a $\tau$, $D_\mu$ y $M$, respectivamente. Por tanto, el desarrollo en potencias de $\tau$ es equivalente a un contaje de las dimensiones de masa de los operadores locales.

La aplicaci\'on de (\ref{eq:met_simb}) conduce a
\begin{eqnarray}
\langle x|e^{-\tau(M-D_\mu^2)}|x\rangle 
&=& \int \frac{d^D p}{(2\pi)^D} \langle x |e^{-\tau(M-(D_\mu+ip_\mu)^2)} |0\rangle  \nonumber \\
&=& \int \frac{d^D p}{(2\pi)^D} e^{-\tau p^2}\langle x |e^{-\tau(M-D_\mu^2-2ip_\mu D_\mu)} |0\rangle
 \,.
\label{eq:met_simb_HK}
\end{eqnarray}
Notar que $p_\mu$ es un c-n\'umero, de modo que conmuta con todos los operadores. En este punto consideramos el desarrollo de la exponencial. Hasta ${\cal O}(4)$ en dimensiones de masa de los operadores locales se tiene\footnote{Como se ver\'a m\'as adelante, el contaje en $\tau$ es equivalente al contaje en dimensiones de masa \'unicamente despu\'es de integrar en momentos.}
\begin{eqnarray}
&&\langle x| e^{-\tau(M-D_\mu^2)}|x \rangle = 
\int \frac{d^D p}{(2\pi)^D} e^{-\tau p^2} 
\langle x | \Delta_0 + \Delta_1 + \Delta_2 + \Delta_3 + \Delta_4 + \cdots 
|0 \rangle \,,
\label{eq:des_HK_1}
\end{eqnarray}
donde
\begin{eqnarray}
\Delta_0 &=& 1 \,, \nonumber \\
\Delta_1 &=& 2i\tau p_\mu D_\mu \,, \nonumber \\
\Delta_2 &=& -\tau (M-D_\mu^2) - 2\tau^2 p_\mu p_\nu D_\mu D_\nu \,, \nonumber
\\
\Delta_3 &=& -i\tau^2 p_\mu \Big( \{D_\mu, M\} -\{D_\mu,D_\nu^2\}\Big)
-i \frac{4}{3}\tau^3 p_\mu p_\nu p_\alpha  D_\mu D_\nu D_\alpha \,, \nonumber \\
\Delta_4 &=&  \frac{\tau^2}{2}\Big(M^2-\{D_\mu^2,M\}+D_\mu^4\Big) \nonumber \\
&&-\frac{\tau^3}{3} p_\mu p_\nu\Big( \{M,D_\mu D_\nu\}+D_\mu M D_\nu-\{D_\alpha^2,D_\mu D_\nu\} - D_\mu D_\alpha^2 D_\nu\Big) \nonumber \\ 
&&+ \frac{2}{3}\tau^4 p_\mu p_\nu p_\alpha p_\beta D_\mu D_\nu D_\alpha D_\beta \,.  
\label{eq:ordenes_masa}
\end{eqnarray}
Se ha usado la notaci\'on est\'andar para el anticonmutador: $\{A,B\}=AB+BA$. En general, las integrales que aparecen son del tipo
\begin{eqnarray}
\int \frac{d^D{p}}{(2\pi)^D} e^{-\tau p^2}p_{i_1}\cdots p_{i_{2n}} &\equiv& \frac{1}{(4\pi\tau)^{D/2}}\frac{1}{(2\tau)^n}\delta_{i_1 i_2 \cdots i_{2n-1} i_{2n}} \label{eq:int_p}\\
 &=& \frac{1}{(4\pi\tau)^{D/2}}\frac{1}{(2\tau)^{n}}(\delta_{i_1 i_2} \cdots \delta_{i_{2n-1}i_{2n}}+({\rm permutaciones})) \,, \nonumber
\end{eqnarray}
donde $\delta_{i_1i_2 \cdots i_{2n}}$ es el producto sin normalizar y completamente sim\'etrico de $2n$ deltas de Kronecker (es decir, $(2n-1)!!$ t\'erminos). La integral en ec.~(\ref{eq:int_p}) con un n\'umero impar de $p$'s vale cero. Tras integrar en momentos, \'unicamente sobreviven los t\'erminos con dimensi\'on de masa par 
\begin{eqnarray}
\langle x| e^{-\tau(M-D_\mu^2)}|x \rangle &=&
\frac{1}{(4\pi\tau)^{D/2}} 
\Big\langle x \Big| 1-\tau M       \nonumber \\
&+&\tau^2\left(\frac{1}{2}M^2-\frac{2}{3}\{D_\mu^2,M\}
-\frac{1}{6}D_\mu M D_\mu 
+D_\mu^4 + \frac{1}{6} (D_\mu D_\nu)^2 
+ \frac{1}{3} D_\mu D_\nu^2 D_\mu \right) \nonumber \\
&&+ {\cal O}(6) \Big|0 \Big\rangle \,.
\label{eq:des_HK_2}
\end{eqnarray}
Notar que el t\'ermino $2\tau^2p_\mu p_\nu D_\mu D_\nu$ ha cancelado el t\'ermino $\tau D_\mu^2$ en ec.~(\ref{eq:ordenes_masa}), despu\'es de integrar en momentos. Notar que cada orden del desarrollo est\'a formado por un n\'umero finito de t\'erminos. La invariancia del heat kernel bajo la transformaci\'on (\ref{eq:inv_boost_HK}) implica que en ec.~(\ref{eq:des_HK_2}) solamente podr\'an aparecer t\'erminos con derivadas $D_\mu$ dentro de conmutadores. En efecto, el cambio $D_\mu \rightarrow D_\mu + a_\mu$ no tiene efecto cuando $D_\mu$ est\'a dentro de un conmutador, pero da cuenta de las contribuci\'on procedente de t\'erminos con $D_\mu$ fuera de conmutadores. Esto significa que los \'unicos t\'erminos que sobreviven son los multiplicativos en el espacio de posiciones.\footnote{$M(x)$ y $[D_\mu,D_\nu]$ son operadores multiplicativos, mientras que $D_\mu^2$ no lo es. Si $\widehat{h}$ es un operador multiplicativo en espacio de posiciones, $\hat{h}|x\rangle = h(x)|x\rangle$, se tiene
\begin{equation}
\langle x | \hat{h} | 0 \rangle = h(x) \,.
\label{eq:mult_oper}
\end{equation}
}
Como ejemplo, se puede comprobar que
\begin{equation}
\{ D_\mu^2, M\} = [D_\mu,[D_\mu,M]] + 2[D_\mu,M]D_\mu + 2M D_\mu^2 \,.
\end{equation}
Los t\'erminos $2[D_\mu,M]D_\mu $ y $2 M D_\mu^2$ no contribuir\'an en el
desarrollo. El resultado final que se obtiene hasta ${\cal O}(4)$ en dimensiones de masa es
\begin{equation}
\langle x| e^{-\tau(M-D_\mu^2)}|x \rangle =
\frac{1}{(4\pi\tau)^{D/2}} \left( 
1-\tau M 
+\tau^2\left(\frac{1}{2} M^2-\frac{1}{6}M_{\mu \mu}+ 
\frac{1}{12}F_{\mu\nu}^2 \right) 
+{\cal O}(\tau^3)\right)\,.
\label{eq:des_HK_3}
\end{equation}
Al pasar de ec.~(\ref{eq:des_HK_2}) a (\ref{eq:des_HK_3}) hemos quitado~$\langle x |\;|0\rangle$ por la propiedad~(\ref{eq:mult_oper}). En lo sucesivo utilizaremos la siguiente notaci\'on. El tensor de fuerza se define como $F_{\mu\nu}=[D_\mu,D_\nu]$, y del mismo modo el campo el\'ectrico es $E_i=F_{0i}$. Adem\'as, la notaci\'on $\widehat{D}_\mu$ significa la operaci\'on $[D_\mu,\;]$. Por \'ultimo decir que usaremos una notaci\'on con sub\'{\i}ndices del tipo $X_{\mu\nu\alpha}$, lo que significa $\widehat{D}_\mu\widehat{D}_\nu\widehat{D}_\alpha X = [D_\mu,[D_\nu,[D_\alpha,X]]]$. Por ejemplo, $M_{00}=\widehat{D}_0^2 M$, $F_{\alpha\mu\nu}=\widehat{D}_\alpha F_{\mu\nu}$.

Los coeficientes de Seeley-DeWitt est\'an calculados en la literatura. Las expresiones expl\'{\i}citas para los coeficientes $a_{n}(x)$ del desarrollo (\ref{eq:des_HK}) hasta orden $n=3$ son~\cite{ball,vandeVen:1997pf}
\begin{eqnarray}
a_0 &=& 1 \,, \nonumber \\
a_1 &=& -M \,, \nonumber \\
a_2 &=& \frac{1}{2}M^2-\frac{1}{6}M_{\mu\mu}+\frac{1}{12}F_{\mu\nu}^2
\,, \nonumber \\
a_3 &=& -\frac{1}{6}M^3
+\frac{1}{12}\{M,M_{\mu\mu}\}
+\frac{1}{12}M_\mu^2
-\frac{1}{60}M_{\mu\mu\nu\nu}
-\frac{1}{60}[F_{\mu\mu\nu},M_\nu]
-\frac{1}{30}\{M,F_{\mu\nu}^2 \}
\nonumber \\ &&
-\frac{1}{60}F_{\mu\nu}MF_{\mu\nu}
+\frac{1}{45}F_{\mu\nu\alpha}^2
-\frac{1}{30}F_{\mu\nu}F_{\nu\alpha}F_{\alpha\mu}
+\frac{1}{180}F_{\mu\mu\nu}^2
+\frac{1}{60}\{ F_{\mu\nu},F_{\alpha\alpha\mu\nu} \}
 \,.
\label{eq:SeeleyT0}
\end{eqnarray}

El desarrollo del heat kernel se usa frecuentemente para el c\'alculo de la acci\'on efectiva, y en este caso resulta necesario calcular la traza del heat kernel~$\Tr \,e^{-\tau(M-D_\mu^2)}$. A temperatura cero los coeficientes con traza $b_n(x)$ se definen simplemente como
\begin{equation}
\Tr\left(e^{-\tau(M-D_\mu^2)}\right) =
\frac{1}{(4\pi\tau)^{D/2}}\sum_{n=0}^\infty \int d^Dx \,\tr\left(
b_n(x)\right) \tau^n \,.
\label{eq:29}
\end{equation}
Una propiedad importante es que el coeficiente $a_n$ se puede obtener a partir de una variaci\'on en primer orden de~$b_{n+1}$. En efecto, por la propia definici\'on del heat kernel se tiene que 
\begin{equation}
\langle x|e^{-\tau(M-D_\mu^2)}|x \rangle =
 -\frac{1}{\tau}\frac{\delta}{\delta M(x)}\Tr \left(e^{-\tau(M-D_\mu^2)}\right) \, . 
\end{equation}
Si hacemos uso del desarrollo en ambos miembros de la igualdad, a temperatura cero encontramos
\begin{equation}
a_n(x) = -\frac{\delta}{\delta M(x)} \tr \, b_{n+1}(x) \, .
\label{eq:a_var_b_T0}
\end{equation}
Hay cierta libertad en la elecci\'on de los coeficientes~$b_n$. Por supuesto, con tomar $b_n=a_n$ ser\'{\i}a suficiente. No obstante, es conveniente explotar la propiedad c\'{\i}clica de la traza y la integraci\'on por partes con el fin de obtener expresiones m\'as compactas. Haciendo uso de estas dos propiedades, a temperatura cero se encuentra la siguiente forma can\'onica para los coeficientes
\begin{eqnarray}
b_0 &=& 1   \, ,     \nonumber \\ 
b_1 &=& -M  \, ,     \nonumber \\
b_2 &=& \frac{1}{2}M^2+\frac{1}{12}F_{\mu\nu}^2 \, ,   \nonumber \\
b_3 &=& -\frac{1}{6}M^3-\frac{1}{12}M_\mu^2-\frac{1}{12}F_{\mu\nu}M F_{\mu\nu}-\frac{1}{60}F_{\mu\mu\nu}^2+\frac{1}{90}F_{\mu\nu}F_{\nu\alpha}F_{\alpha\mu} \, .\label{eq:trazabT0}
\end{eqnarray}

\section{Desarrollo del Heat Kernel a temperatura finita}
\label{metodo_simbolos_Tfinito}

Es posible extender el m\'etodo de los s\'{\i}mbolos con objeto de realizar c\'alculos a tempera\-tura finita~\cite{garcia}. 

En el formalismo de tiempo imaginario la coordenada temporal est\'a compactificada a un c\'{\i}rculo, de modo que el espacio-tiempo de $D=d+1$ dimensiones tiene topolog\'{\i}a ${\cal M}_{d+1} = S^1 \times {\cal M}_d $. Las funciones de onda para bosones son peri\'odicas en la direcci\'on temporal con per\'{\i}odo $\beta$, la inversa de la temperatura, y antiperi\'odicas para fermiones. Con objeto de que $M$ y $D_\mu$ sean operadores bien definidos en el espacio de Hilbert de las funciones de onda con grados de libertad espacio-temporales e internos, $M(x)$ y $A_\mu(x)$ deben ser funciones peri\'odicas en $x_0$.

En este formalismo usaremos la siguiente normalizaci\'on
\begin{eqnarray}
 \langle x|p\rangle = e^{ipx} \, , \quad
\langle p|p^\prime\rangle = \beta \delta_{p_0 p_0^\prime}(2\pi)^d \delta(\vec{p}-{\vec p}\,{}^\prime) \,.
\end{eqnarray}
La relaci\'on de completitud es
\begin{equation}
\mathbf{1}=\frac{1}{\beta}\sum_{p_0}\int \frac{d^d p}{(2\pi)^d}|p\rangle\langle p| \,.
\label{eq:rel_compl_Tfinita}
\end{equation}
La frecuencia toma los valores de Matsubara $p_0 = 2\pi n/\beta$ para bosones y $p_0=2\pi (n+\frac{1}{2})/\beta$ para fermiones. El m\'etodo de los s\'{\i}mbolos se escribe en este formalismo\footnote{La demostraci\'on de (\ref{eq:met_simb_Tfinita}) es similar a la realizada en la sec.~\ref{metodo_simbolos} para el caso de temperatura cero.}
 \begin{equation}
 \langle x|f(M,D_\mu)|x \rangle = \frac{1}{\beta}\sum_{p_0}\int \frac{d^d p}{(2\pi)^d}\langle x|f(M,D_\mu+ip_\mu)|0\rangle \,.
\label{eq:met_simb_Tfinita}
\end{equation}
Notar que $|0\rangle$ es peri\'odico en la direcci\'on temporal, de modo que la informaci\'on de si estamos trabajando con bosones o fermiones se encuentra ahora contenida en los valores que toma $p_0$.

\subsection{Desarrollo del Heat Kernel: un caso simple}
\label{HKTfinito_simple}

La aplicaci\'on pr\'actica del m\'etodo de los s\'{\i}mbolos a temperatura finita resulta bastante m\'as complicada que a temperatura cero. Con objeto de introducir los conceptos de ma\-nera gradual, vamos a considerar el heat kernel, y estudiaremos su desarrollo en un caso simple. Trataremos el caso en el que no exista potencial vector, el potencial escalar sea indenpendiente de $\vec{x}$, y el t\'ermino de masa sea un c-n\'umero constante:
\begin{equation}
\vec{A}(x)=0\,,\quad A_0=A_0(x_0)\,,\quad M(x)= m^2\,,\quad [m^2,~]=0 \,.
\label{eq:3a}
\end{equation}
El resultado ser\'a el t\'ermino de orden cero de un desarrollo en conmutadores $[D_\mu, \;]$ y $[M, \;]$ del caso general. La aplicaci\'on del m\'etodo de los s\'{\i}mbolos (\ref{eq:met_simb_Tfinita}) conduce a 
\begin{eqnarray}
\langle x| e^{-\tau {\mathbf K}}|x\rangle 
&=&
\frac{1}{\beta}\sum_{p_0}
\int\frac{d^dp}{(2\pi)^d} 
 \langle x|e^{-\tau (m^2+\vec{p}\,{}^2-(D_0+ip_0)^2)}| 0\rangle 
\nonumber \\
&=& 
\frac{e^{-\tau m^2}}{(4\pi\tau)^{d/2}}
\frac{1}{\beta}\sum_{p_0}
 \langle x| e^{\tau (D_0+ip_0)^2}| 0 \rangle \,. 
\label{eq:2a}
\end{eqnarray}
Notar que despu\'es de la transformaci\'on $D_j \rightarrow \partial_j + i p_j$, el operador $D_j=\partial_j$ puede hacerse cero pues actuar\'a sobre $|0\rangle$.

La suma sobre frecuencias de Matsubara implica que el operador
$\frac{1}{\beta}\sum_{p_0} e^{\tau (D_0+ ip_0)^2}$ es una funci\'on
peri\'odica de $D_0$ con periodo $i2\pi/\beta$, y por tanto es una funci\'on
univaluada de $e^{-\beta D_0}$. En efecto, si hacemos uso de la f\'ormula de
Poisson para la sumatoria,\footnote{La f\'ormula de Poisson para la sumatoria
  es:
\begin{equation}
\sum_{n=-\infty}^\infty F ( n ) = \sum_{m=-\infty}^\infty \left\{
\int_{-\infty}^\infty d x F ( x ) e^{ i 2\pi x m } \right\}  \,.
\label{eq:poisson}
\end{equation} 
} se tiene
\begin{equation}
\frac{1}{\beta}\sum_{p_0} e^{\tau (D_0+i p_0)^2} =
\frac{1}{(4\pi\tau)^{1/2}}
\sum_{k\in \mathbb{Z}} (\pm)^k e^{-k\beta D_0} e^{-k^2\beta^2/4\tau}
\label{eq:2.5}
\end{equation}
($\pm$ para bosones y fermiones, respectivamente). En este momento estamos en condiciones de hacer uso de la siguiente identidad operatorial~\cite{garcia}
\begin{equation}
 e^{\beta \partial_0}e^{-\beta D_0} = \Omega(x) \,,
\label{eq:cs1}
\end{equation}
donde $\Omega(x)$ es la l\'{\i}nea de Wilson t\'ermica o loop de Polyakov sin
traza:
\begin{equation}
\Omega(x)= {\mathcal T}\exp\left(-\int_{x_0}^{x_0+\beta}
A_0(x_0^\prime,\vec{x})\,dx_0^\prime\right)
\label{eq:1b}
\end{equation}
[${\mathcal T}$ indica ordenaci\'on temporal.] Si bien es esta secci\'on estamos tratando el caso simple de ec.~(\ref{eq:3a}), la definici\'on (\ref{eq:1b}) es v\'alida para un potencial escalar general $A_0(x)$. El loop de Polyakov surge aqu\'{\i} como la diferencia de fase entre traslaciones temporales covariantes y no covariantes gauge alrededor del tiempo eucl\'{\i}deo compactificado. F\'{\i}sicamente, el loop de Polyakov se puede interpretar como el propagador de part\'{\i}culas pesadas en el fondo del campo gauge. La identidad~(\ref{eq:cs1}) es trivial si uno elije un gauge en el cual $A_0$ es independiente del tiempo (este gauge siempre existe), pues en este caso los operadores $\Omega = e^{-\beta A_0}$, $D_0$, $A_0$ y $\partial_0$ conmutan entre s\'{\i}. Esta identidad es covariante gauge y es v\'alida en cualquier gauge.\footnote{En el ap\'endice~\ref{app:gauge} se hace un estudio detallado de las transformaciones gauge a temperatura finita.}

Un punto importante es que el operador de traslaci\'on en tiempo eucl\'{\i}deo, $e^{\beta \partial_0}$, no tiene otro efecto que producir el cambio $x_0 \rightarrow x_0+\beta$ y esta operaci\'on es la identidad en el espacio de funciones peri\'odicas en que estamos trabajando 
\begin{equation}
e^{\beta\partial_0}= \mathbf{1} \,,
\end{equation}
(incluso en el caso fermi\'onico, ya que despu\'es de aplicar el m\'etodo de los s\'{\i}mbolos las derivadas act\'uan sobre los campos externos y no sobre las funciones de onda de las part\'{\i}culas). Llegamos as\'{\i} al resultado importante de que en este espacio
\begin{equation}
e^{-\beta D_0} = \Omega(x) \,,
\label{eq:D0_Omega} 
\end{equation}
esto es, siempre y cuando el operador diferencial $D_0$ aparezca de manera peri\'odica (con per\'{\i}odo $2\pi i/\beta$), puede ser reemplazado por el operador multiplicativo $-(1/\beta)\log[\Omega(x)]$. La multivaluaci\'on del logaritmo no es efectiva debido a la dependencia peri\'odica. 

Otro punto importante es que $D_0$ (o cualquier funci\'on de $D_0$) act\'ua como un operador covariante gauge sobre los campos externos~$F(x_0,\vec{x})$, y por tanto transforma de acuerdo al grupo de transformaciones gauge locales en el punto $(x_0,\vec{x})$. En particular, el loop de Polyakov ec.~(\ref{eq:1b}), que es tambi\'en covariante gauge, comienza en el instante $x_0$ y no en cero. Esta diferencia ser\'{\i}a irrelevante para el loop de Polyakov con traza, pero no en el contexto de ahora. 

El uso de la regla (\ref{eq:D0_Omega}) en ec.~(\ref{eq:2.5}) conduce a
\begin{equation}
\frac{1}{\beta}\sum_{p_0} e^{\tau (D_0+i p_0)^2} =
\frac{1}{(4\pi\tau)^{1/2}}
\sum_{k\in \mathbb{Z}} (\pm)^k \Omega^k e^{-k^2\beta^2/4\tau} \,.
\label{eq:4}
\end{equation}
En general se tiene
\begin{equation}
\sum_{p_0} f(ip_0 + D_0) = \sum_{p_0}f(ip_0-\frac{1}{\beta}\log(\Omega)) \,,
\label{eq:4b}
\end{equation}
siempre y cuando la sumatoria sea absolutamente convergente, de modo que la suma es una funci\'on peri\'odica de $D_0$. Por futura conveniencia introduciremos el operador $Q$, que se define como
\begin{equation}
Q = ip_0 + D_0 = ip_0-\frac{1}{\beta}\log(\Omega) \,.
\label{eq:3}
\end{equation}
Hay que mencionar que la segunda igualdad se aplica en expresiones de la forma de ec.~(\ref{eq:4b}). Las dos definiciones de $Q$ no son equivalentes en otros contextos (por ejemplo, en $\sum_{p_0}f_1(Q)X f_2(Q)$, a menos que $[D_0,X]=0$.)

El heat kernel en ec.~(\ref{eq:2a}) se puede escribir como
\begin{equation}
\langle x| e^{-\tau K}|x\rangle =
\frac{1}{(4\pi\tau)^{d/2}}
e^{-\tau m^2}
\frac{1}{\beta}\sum_{p_0}
e^{\tau Q^2} = 
\frac{1}{(4\pi\tau)^{(d+1)/2}}e^{-\tau m^2}
\varphi_0(\Omega)\,.
\label{eq:13}
\end{equation}
En la primera igualdad se ha hecho uso de que $\Omega(x)$ es un operador multiplicativo, de modo que es aplicable la ec.~(\ref{eq:mult_oper}). En la segunda igualdad se ha aplicado la definici\'on de las funciones $\varphi_n(\Omega)$, que aparecer\'an con frecuencia en lo sucesivo:
\begin{equation}
\varphi_n(\Omega;\tau/\beta^2) =
\left(4\pi\tau\right)^{1/2}\frac{1}{\beta}\sum_{p_0} \tau^{n/2} Q^n
e^{\tau Q^2} \,,\quad
Q = ip_0-\frac{1}{\beta}\log(\Omega) \,.
\label{eq:2.15}
\end{equation}
Notar que para cada funci\'on existe una versi\'on bos\'onica y otra fermi\'onica, y las dos versiones est\'an relacionadas por el cambio $\Omega\rightarrow-\Omega$. Como se ha indicado, estas funciones dependen s\'olo de la combinaci\'on $\tau/\beta^2$ y son funciones univaluadas de $\Omega$. En el l\'{\i}mite de temperatura cero la suma sobre $p_0$ se transforma en una integral gaussiana
\begin{equation}
\frac{1}{\beta}\sum_{p_0} \xrightarrow{\beta \rightarrow \infty}  \int_{-\infty}^\infty \frac{dp_0}{(2\pi)} \,,
\end{equation}
y se tiene
\begin{equation}
\varphi_n(\Omega;0) = \left\{
\begin{matrix}
(-\frac{1}{2})^{n/2}(n-1)!! & (\text{$n$ par}) \,,
\cr 0 & (\text{$n$ impar})  \,.
\end{matrix}
\right.
\end{equation}
Como se puede ver en la expresi\'on (\ref{eq:4}), para un valor finito de
$\beta$ las correcciones de $\tau$ peque\~no son de orden $e^{-\beta^2/4\tau}$
o menor, y por tanto est\'an exponencialmente suprimidas. La misma supresi\'on
exponencial existe para las correcciones de temperatura peque\~na cuando se considera un valor finito de $\tau$. Ya sea en el l\'{\i}mite de temperatura cero o de tiempo propio cero, \'unicamente queda el modo $k=0$. 

Como motivaci\'on del heat kernel, en la secci\'on~\ref{macrocanonico} se
calcul\'o el potencial macrocan\'onico de un gas de part\'{\i}culas libres
relativistas, que constituye una aplicaci\'on simple de los resultados
obtenidos en esta secci\'on. En vista de ecs.~(\ref{eq:PM_D0_Di}) y
(\ref{eq:PM_ae}), es importante subrayar la relaci\'on entre el potencial
qu\'{\i}mico~$\mu$ y el loop de Polyakov. El potencial qu\'{\i}mico se acopla
al potencial escalar $A_0(x)$ como una constante aditiva. Puesto que es
constante, $\mu$ no contribuye a los operadores locales, ya que $A_0(x)$ s\'olo aparece a trav\'es de la derivada covariante~$\widehat{D}_0$. Notar que si el loop de Polyakov no existiera en las f\'ormulas, $\mu$ no aparecer\'{\i}a en la funci\'on de partici\'on, lo cual obviamente constituye un resultado incorrecto. Asimismo hay que destacar que la dependencia peri\'odica del heat kernel en $\log\Omega$ conduce al hecho bien conocido de que la funci\'on de partici\'on es peri\'odica en $\beta \mu$ con per\'{\i}odo $2\pi i$ (condici\'on de consistencia debido a su acoplamiento con el operador de carga cuantizado). El loop de Polyakov aparece pues, como una generalizaci\'on del factor $e^{\beta\mu}$ para campos gauge no abelianos y no constantes.

\subsection{Coeficientes del desarrollo del Heat Kernel a temperatura finita}
\label{HKTfinito_coef}

En esta secci\'on consideraremos el desarrollo del heat kernel a temperatura finita en el caso totalmente general de campos gauge no abelianos $A_\mu(x)$ y t\'erminos de masa no triviales $M(x)$.

En primer lugar es necesario especificar el contaje del desarrollo. Como vimos en sec.~\ref{des_hk_temp_cero}, a temperatura cero el desarrollo se define en potencias de $\tau$ [despu\'es de extraer el factor geom\'etrico~$(4\pi\tau)^{(d+1)/2}$]. Este contaje en $\tau$ es equivalente a un contaje en las dimensiones de masa de los operadores locales.

A temperatura finita existe una magnitud dimensional adicional, $\beta$, de modo que los dos contajes no van a ser equivalentes y es necesario especificar un desarrollo concreto. Como veremos m\'as adelante un desarrollo estricto del heat kernel en potencias de $\tau$ conducir\'{\i}a al mismo desarrollo asint\'otico que a temperatura cero. Con objeto de extraer correcciones de temperatura finita no triviales ordenaremos nuestro desarrollo de acuerdo con las dimensiones de masa de los operadores locales. Asignaremos dimensiones de masa $0$, $+1$, $+2$ a $\Omega$, $D_\mu$ y $M$, respectivamente. Consideraremos adem\'as un desarrollo en el cual el loop de Polyakov $\Omega(x)$ aparezca a la izquierda en todos los t\'erminos, lo cual es una cuesti\'on de elecci\'on (de manera equivalente, se podr\'{\i}a definir un desarrollo con $\Omega(x)$ a la derecha). Esto es necesario pues el conmutador de $\Omega$ con otros operadores genera conmutadores $[D_0, \;]$ que tienen dimensi\'on~1 en nuestro contaje. Estas especificaciones son suficientes para definir de manera un\'{\i}voca el desarrollo del heat kernel para un grupo gauge gen\'erico, de tal modo que la invariancia gauge sea manifiesta orden por orden.

El desarrollo as\'{\i} definido, en el cual cada t\'ermino contiene funciones arbitrarias del loop de Polyakov pero s\'olo un n\'umero finito de derivadas covariantes (incluyendo derivadas temporales), constituye una extensi\'on natural del desarrollo est\'andar en derivadas covariantes a temperatura cero. Los t\'erminos estar\'an ordenados en potencias de $\tau$ pero con coeficientes que dependen de $\tau/\beta^2$ y $\Omega$:
\begin{equation}
\langle x|e^{-\tau(M-D_\mu^2)}| x\rangle= 
\frac{1}{(4\pi\tau)^{(d+1)/2}}\sum_n
a^T_n(x)\tau^n \,.
\label{eq:18b}
\end{equation}
De la definici\'on se deduce directamente que para una configuraci\'on general el t\'ermino de orden cero es precisamente
\begin{equation}
a_0^T(x) = \varphi_0(\Omega(x);\tau/\beta^2) \,,
\label{eq:13b}
\end{equation}
que fue calculado en la subsecci\'on \ref{HKTfinito_simple}. Esto es debido a que cuando el caso particular~(\ref{eq:3a}) es introducido en el desarrollo general, todos los t\'erminos de orden mayor, con una o m\'as $[D_\mu , \;]$ o $m^2$, se anulan. 


El m\'etodo que vamos a proponer para el c\'alculo del desarrollo del heat kernel a temperatura finita hace uso de los coeficientes de Seeley-DeWitt a temperatura cero. La idea consiste en aplicar la f\'ormula del m\'etodo de los s\'{\i}mbolos~(\ref{eq:met_simb_Tfinita}) en la dimensi\'on temporal \'unicamente, lo cual conduce a
\begin{equation}
\langle x| e^{-\tau(M-D_{\mu}^2)}|x \rangle = \frac{1}{\beta}\sum_{p_0} \langle x_0,{\vec x} | e^{-\tau(M-Q^2-D_i^2)} |0,\vec{x} \rangle \, ,
\qquad
Q = i p_0 + D_0 \,.
\label{eq:kk9}
\end{equation}
Se puede definir el operador de Klein-Gordon efectivo
\begin{equation}
{\mathcal K} = \mathcal Y - D_i^2 \, , \quad \mathcal Y = M - Q^2 \,,
\end{equation}
donde $\mathcal Y$ juega el papel de un t\'ermino de masa no abeliano. Podemos
hacer uso del desarrollo del heat kernel a temperatura cero en $d$ dimensiones
(espaciales) con ese operador efectivo ya que el t\'ermino de masa $\mathcal
Y$,  a pesar de contener derivadas temporales (en~$Q$), no contiene derivadas espaciales, de manera que act\'ua como un operador multiplicativo en el espacio de Hilbert espacial. La aplicaci\'on directa de este argumento dar\'{\i}a lugar al desarrollo
\begin{equation}
\langle x_0,\vec{x} | e^{-\tau(\mathcal Y - D_i^2)} |0,\vec{x} \rangle = \frac{1}{(4\pi\tau)^{d/2}} \sum_{n=0}^{\infty} a_n({\widehat D}_i,\mathcal Y) \tau^n \,,  
\label{eq:desHKmet_alt}   
\end{equation}
donde los coeficientes $a_n({\widehat D}_i,{\mathcal Y})$ son polinomios de dimensi\'on~$2n$ construidos a partir de~$\mathcal Y$ y ${\widehat D}_i = [D_i,\;]$. Los \'ordenes m\'as bajos corresponden a la ec.~(\ref{eq:SeeleyT0}), pero considerando la sustituci\'on del t\'ermino de masa $M$ por el nuevo t\'ermino de masa efectivo $\mathcal Y$, y los \'{\i}ndices s\'olo corren en la dimensi\'on espacial. 

Notamos que para reproducir el primer orden en ec.~(\ref{eq:18b}), $a_0^T(x) = \varphi_0(\Omega(x)) \sim e^{\tau Q^2}$, ser\'{\i}a necesario obtener el desarrollo a todos los \'ordenes en ec.~(\ref{eq:desHKmet_alt}), pues $e^{\tau Q^2}$ no es un polinomio en $Q$. \'Esta es la raz\'on por la cual ec.~(\ref{eq:desHKmet_alt}) introducida en ec.~(\ref{eq:kk9}) no resulta \'util. La manera correcta de proceder ser\'a extraer desde el principio la contribuci\'on~$e^{\tau Q^2}$, lo cual nos llevar\'a a definir un nuevo conjunto de coeficientes polin\'omicos~$\tilde a_n$
\begin{equation}
\langle x_0, \vec{x} |e^{-\tau(M-Q^2-D_i^2)}| 0 , \vec{x} \rangle= 
\frac{1}{(4\pi\tau)^{d/2}}\sum_{n=0}^\infty
e^{\tau Q^2} \tilde a_n(Q^2,M,{\widehat D}_i)\tau^n \,.
\label{eq:18a}
\end{equation}
Consideremos la sustituci\'on de $Q^2$ por $Q^2 + \lambda $ donde $\lambda$ un c-n\'umero constante. Es claro que los coeficientes $\tilde a_n$ no deben cambiar, y por tanto en $\tilde a_n$ el operador $Q^2$ debe aparecer s\'olo dentro de conmutadores de la forma $[Q^2,\; ]$. Para calcular los coeficientes $\tilde a_n$ debemos tener en cuenta la relaci\'on 
\begin{equation}
\sum_{n=0}^{\infty} a_n({\widehat D}_i,\mathcal Y) \tau^n = e^{\tau Q^2}\sum_{n=0}^{\infty}\tilde a_n(Q^2,M,{\widehat D}_i) \tau^n \, .
\label{eq:an_antilde}
\end{equation}
El m\'etodo consiste en partir del desarrollo de la izquierda de la ecuaci\'on~(\ref{eq:an_antilde}) e ir moviendo los operadores $Q^2$ hacia la izquierda haciendo uso de conmutadores $[Q^2,\,\,]$ (por ejemplo $M Q^2 = Q^2 M - [Q^2,M]$). Al final se llega a una situaci\'on en la que existen dos clases de t\'erminos: (i) t\'erminos en que todos los operadores $Q^2$ est\'an dentro de conmutadores y (ii) t\'erminos con factores $Q^2$ no saturados a la izquierda (esto es, con $Q^2$ fuera de conmutadores). Los t\'erminos del tipo (i) se corresponden con el desarrollo $\sum_{n=0}^{\infty}\tilde a_n \tau^n$. Los del tipo (ii) se pueden identificar con el miembro derecho de la ecuaci\'on cuando se realiza un desarrollo de la exponencial $e^{\tau Q^2}$ y se consideran \'ordenes mayores que el primero. Siguiendo esta t\'ecnica, hasta $\tilde a_2$ se tiene
\begin{eqnarray}
\tilde a_0 &=& 1 \, , \nonumber \\
\tilde a_1 &=& -M \, , \nonumber \\
\tilde a_2 &=& \frac{1}{2}M^2-\frac{1}{6}M_{ii}
+\frac{1}{12}F_{ij}^2
+\frac{1}{2}[Q^2,M]+\frac{1}{6}(Q^2)_{ii} \, .
\end{eqnarray}
Una vez que hemos construido por este procedimiento los coeficientes $\tilde a_n$, el siguiente paso consiste en redefinir ec.~(\ref{eq:18a}) como un desarrollo en potencias de $M$, ${\widehat D}_i$ y ${\widehat D}_0$. Para ello debemos expresar $[Q^2,\;]$ que aparece en el desarrollo, en t\'erminos de $[Q,\;] = [D_0,\;] = {\widehat D}_0$. Se usa la siguiente propiedad:
\begin{equation}
[Q^2,X]=Q[Q,X]+[Q,X]Q=2Q[Q,X]-[Q,[Q,X]]=2 Q X_0 - X_{00} \, . 
\end{equation}
Se trata de mover todos los $Q$'s hacia la izquierda, de modo que aparecer\'an operadores ${\widehat D}_0$. Al final los operadores $Q$ fuera de conmutadores quedar\'an todos a la izquierda. Para $\tilde a_2$ se tiene:
\begin{equation}
\tilde a_2 = \frac{1}{2}M^2-\frac{1}{6}M_{ii}+\frac{1}{12}F_{ij}^2-\frac{1}{2}M_{00}+\frac{1}{3}E_i^2 +\frac{1}{6}E_{0ii} + Q\left(M_0-\frac{1}{3}E_{ii}\right) \, .
\end{equation}
Notar que en $\tilde a_2$ existen dos tipos de contribuciones: aquellos t\'erminos con una~$Q$ a la izquierda, y aquellos que no la tienen. En nuestro contaje, estos dos tipos pertenecen a \'ordenes diferentes: dimensi\'on de masa tres y cuatro, respectivamente. Cuando $\tilde a_2$ es introducido en ec.~(\ref{eq:18a}) (queda multiplicado por el factor~$e^{\tau Q^2}$) y despu\'es en ec.~(\ref{eq:kk9}) (suma sobre frecuencias de Matsubara), se obtienen las siguientes contribuciones
\begin{equation}
\tilde a_2 \to \varphi_0(\Omega) \left( \frac{1}{2}M^2  -\frac{1}{6}M_{ii} 
+\frac{1}{12}F_{ij}^2
- \frac{1}{2}M_{00}
+ \frac{1}{3}E^2_i
+ \frac{1}{6}E_{0ii}
\right) \tau^2
+ \varphi_1(\Omega) \left(
 M_0 
- \frac{1}{3}E_{ii}
\right) \tau^{3/2} \,,
\end{equation}
donde se ha hecho uso de la definici\'on de $\varphi_n(\Omega)$, ec.~(\ref{eq:2.15}).

Como vemos cada coeficiente de heat kernel a temperatura cero $a_k$ en ec.~(\ref{eq:desHKmet_alt}) con dimensi\'on de masa $2k$ permite obtener un coeficiente correspondiente $\tilde a_k$. Este coeficiente va a dar contribuci\'on, en general, a varios coeficientes de heat kernel $a_n^T$ (con dimensi\'on de masa $2n$). Las diferentes contribuciones se deben a que pueden existir ciertos factores de $Q$ a la izquierda de cada t\'ermino que no act\'uan como ${\widehat D}_0$, de modo que son adimensionales. Por tanto para un valor de $k$ dado, los valores de $n$ permitidos deben satisfa\-cer $n \le k$,~y la igualdad corresponde a t\'erminos que tienen todos los $Q$'s dentro de conmutadores. Podemos encontrar una cota inferior para $n$ si vemos que el n\'umero m\'aximo de $[Q^2,\,\,]$'s en $\tilde a_k (k \ge 0)$ es $k-1$, y por tanto \'este va a ser el n\'umero m\'aximo de $Q$'s fuera de conmutadores que queden a la izquierda. Esto conduce a la condici\'on $k \le 2n-1$. Adem\'as notemos que un factor $Q^\ell$ va a dar lugar a un coeficiente $\varphi_\ell(\Omega)$ en $a_n^T$. En suma, para el c\'alculo de los coeficientes de heat kernel t\'ermicos vamos a tener el siguiente esquema
\begin{eqnarray}
a_0 &\sim& \tilde a_0 \sim \varphi_0 a_0^T   \nonumber \\
a_1 &\sim& \tilde a_1 \sim \varphi_0 a_1^T   \nonumber \\
a_2 &\sim& \tilde a_2 \sim \varphi_0 a_2^T + \varphi_1 a_{3/2}^T  \nonumber \\
a_3 &\sim& \tilde a_3 \sim \varphi_0 a_3^T + \varphi_1 a_{5/2}^T + \varphi_2 a_2^T \nonumber \\
a_4 &\sim& \tilde a_4 \sim \varphi_0 a_4^T + \varphi_1 a_{7/2}^T + \varphi_2 a_3^T +\varphi_3 a_{5/2}^T \nonumber \\
a_5 &\sim& \tilde a_5 \sim \varphi_0 a_5^T + \varphi_1 a_{9/2}^T + \varphi_2 a_4^T + \varphi_3 a_{7/2}^T + \varphi_4 a_3^T  \nonumber \\
&&\cdots \quad \cdots \quad \cdots \quad \cdots \quad \cdots \quad \cdots \quad \cdots \quad \cdots \quad \cdots \nonumber \\
a_k &\sim& \tilde a_k \sim \varphi_0 a_k^T + \varphi_1 a_{(2k-1)/2}^T + \cdots + \varphi_{k-1}  a_{(k+1)/2}^T 
\end{eqnarray}

Esta mezcla de t\'erminos no ocurre a temperatura cero, no obstante no puede ser evitada a temperatura finita. Vemos que a $Q$ no se le podr\'{\i}a asignar dimensi\'on de masa 1 ya que la suma sobre las frecuencias de Matsubara~$p_0$ no converge para un polinomio en $Q$. Si $p_0$ se cuenta con dimensi\'on cero pero $D_0$ siempre con dimensi\'on 1 la invariancia gauge se perder\'{\i}a. En suma, el hecho de considerar $\Omega$ adimensional y ${\widehat D}_0$ con dimensi\'on 1 es un peque\~no precio que hay que pagar para tener un desarrollo covariante gauge orden por orden.

Del esquema anterior se deduce que para calcular los coeficientes de heat kernel t\'ermicos completos hasta $a_3^T$ debemos buscar contribuciones hasta $a_5$. Como regla general, para $a_n^T$ van a existir contribuciones de $a_k , \, n \le k \le 2n-1$, excepto para $a_0^T$ el cual s\'olo recibe la contribuci\'on trivial de $a_0$. En particular $a_3^T$, aparte de la contribuci\'on que reciba de $a_3$, s\'olo requiere t\'erminos ${\mathcal Y}^n$, con $n = 2, 3, 4$ en $a_4({\widehat D}_i,\mathcal Y)$ y $n = 4, 5$ en $a_5({\widehat D}_i,\mathcal Y)$.

Haciendo uso de este m\'etodo se han calculado los coeficientes de heat kernel t\'ermicos hasta dimensi\'on de masa~6. 
Los resultados son los siguientes:
\begin{eqnarray}
a_0^T &=& \varphi_0 \,, 
\nonumber \\ 
a_{1/2}^T &=& 0 \,, 
\nonumber \\ 
a_1^T &=& -\varphi_0 M  \,, 
\nonumber \\
a_{3/2}^T &=& \varphi_1 \left(M_0-\frac{1}{3}E_{ii}\right)  \,, 
\nonumber  \\
a_2^T &=& 
\varphi_0 \, a_2^{T=0}
+ \frac{1}{6} \overline\varphi_2 (E_i^2+E_{0ii}-2M_{00}) 
 \,,  \nonumber \\
a_{5/2}^T &=& \frac{1}{3}\left(2\varphi_1+\varphi_3\right)M_{000}
+\frac{1}{6}\varphi_1 M_{0ii}-\frac{1}{3}\varphi_1 \left(2M_0 M + M M_0\right)
 \\
&& +\frac{1}{6}\varphi_1 \left ( \{M_i,E_i\}+ \{M,E_{ii}\}  \right)
-\left(\frac{1}{3}\varphi_1 + \frac{1}{5}\varphi_3 \right) E_{00ii}
-\frac{1}{30}\varphi_1 E_{iijj} 
\nonumber 
\end{eqnarray}
\begin{eqnarray}
&& -\left(\frac{5}{6}\varphi_1+\frac{2}{5}\varphi_3\right) E_{0i}E_i
-\left(\frac{1}{2}\varphi_1+\frac{4}{15}\varphi_3\right)E_i E_{0i}
+\frac{1}{30}\varphi_1 [ E_j , F_{iij}] 
\nonumber \\
&& -\varphi_1 \left(\frac{1}{10}F_{0ij}F_{ij}+\frac{1}{15}F_{ij} F_{0ij}\right)
 \,, 
\nonumber  \\
a_3^T &=& \varphi_0 \, a_3^{T=0}
- \left(\frac{1}{4}\overline\varphi_2 - \frac{1}{10}\overline\varphi_4 \right)
M_{0000}
-\frac{1}{60}\overline\varphi_2 \Big( 
 3 M_{00ii}- 15 M_{00}M  - 5 M M_{00} - 15 M_0^2 
\nonumber \\
&& + 4 \{M,E_i^2\} + 2 E_i M E_i
+4 M E_{0ii}+6 E_{0ii}M
+ 4 M_i E_{0i}+ 6 E_{0i}M_i
\nonumber \\
&& 
+ 7 M_0 E_{ii}+ 3 E_{ii}M_0
+6 M_{0i}E_i+ 4 E_i M_{0i}
\Big)
\nonumber \\
&& + \left(\frac{3}{20}\overline\varphi_2 - \frac{1}{15}\overline\varphi_4 \right)
 E_{000ii}
+\frac{1}{60}\overline\varphi_2 E_{0iijj}
+ \left(\frac{1}{2}\overline\varphi_2 - \frac{1}{5}\overline\varphi_4 \right)
E_{00i}E_i
\nonumber \\
&& 
+ \left(\frac{7}{30}\overline\varphi_2 - \frac{1}{10}\overline\varphi_4 \right)
E_i E_{00i}
+ \left(\frac{19}{30}\overline\varphi_2 - \frac{4}{15}\overline\varphi_4 \right)
E_{0i}^2
\nonumber \\
&& +\frac{1}{180}\overline\varphi_2
\Big(
2\{E_i,E_{jji}\}+4\{E_i,E_{ijj}\}+
5E_{ii}^2 + 4E_{ij}^2+4F_{0iij}E_j 
- 2E_j F_{0iij} - 2 E_{0ij}F_{ij}
\nonumber \\
&&-[E_{ij},F_{0ij}]-4E_{0i}F_{jji}
+2F_{jji}E_{0i}+2E_i F_{ij}E_j + 2\{E_i E_j,F_{ij}\}
+7F_{00ij}F_{ij} \nonumber \\
&&+3F_{ij}F_{00ij}+8F_{0ij}^2\Big)
 \,.
\nonumber
\end{eqnarray}
En estas f\'ormulas $a_n^{T=0}$ indican los coeficientes a temperatura cero que aperecen en ec.~(\ref{eq:SeeleyT0}). Por conveniencia hemos introducido las funciones auxiliares
\begin{equation}
\overline\varphi_2= \varphi_0+2\varphi_2 \,,\quad 
\overline\varphi_4= \varphi_0-\frac{4}{3}\varphi_4\,,
\quad \ldots\ldots \quad \, , \quad
\overline\varphi_{2n}= \varphi_0-\frac{(-2)^n}{(2n-1)!!}\varphi_{2n}\,,
\end{equation}
que se anulan en el l\'{\i}mite $\tau/\beta^2=0$. Con nuestro criterio para calcular el desarrollo del heat kernel a temperatura finita conseguimos ordenar las derivadas de manera que las espaciales son las que act\'uan primero y las temporales son las m\'as externas. Esta elecci\'on es \'optima de cara a calcular la traza de los coeficientes $\Tr\,a_n^T(x)$, pues por la propiedad ${\widehat D}_0 \Omega=0$, los t\'erminos de la forma $\varphi_n X_0$ no contribuyen en la traza, como puede verse despu\'es de integrar por partes.

\subsection{Traza de los coeficientes de Heat Kernel}
\label{HKTtraza_coef}

En ec.~(\ref{eq:29}) se definieron los coeficientes de heat kernel con traza a temperatura cero. A temperatura finita podemos definir de manera similar los coeficientes con traza~$b_n^T(x)$
\begin{equation}
\Tr \left( e^{-\tau(M-D_\mu^2)} \right) = \frac{1}{(4\pi\tau)^{(d+1)/2}} \sum_{n=0}^\infty \int_0^{\beta}dx_0 \int d^d{x} \; \tr(b_n^T(x)) \tau^n \, ,
\end{equation}
donde $b_n^T$ presenta una estructura m\'as simple que $a_n^T$. Vamos a elegir una forma can\'onica para estos coeficientes en la cual las funciones de $\Omega$ est\'en situadas a la izquierda de los operadores locales covariantes gauge. Adem\'as de la integraci\'on por partes y propiedad c\'{\i}clica de la traza, deberemos trabajar con conmutadores del tipo $[X,f(\Omega)]\,\,$(en particular ${\widehat D}_\mu f(\Omega)\,$). 

Veamos cuales son las reglas de conmutaci\'on. Consideremos dos operadores cualesquiera $X$ e $Y$, y $f$ una funci\'on gen\'erica. Entonces el conmutador $[X,f(Y)]$ admite el siguiente desarrollo en conmutadores
\begin{eqnarray}
[X,f(Y)] &=& -f^\prime(Y) [Y,X] + \frac{1}{2}f^{\prime\prime}(Y) [Y,[Y,X]] - \frac{1}{3!}f^{(3)}(Y) [Y,[Y,[Y,X]]] + \cdots \nonumber \\
         &=& \sum_{n=1}^{\infty} \frac{(-1)^n}{n!}f^{(n)}(Y)D_Y^n(X) \, ,
\label{eq:kk11}
\end{eqnarray}
donde $D_Y = [Y,\,\,]$. Para probar esto es suficiente con probar que se cumple para funciones del tipo $f(Y)=e^{\lambda Y}$, donde $\lambda$ es un c-n\'umero, ya que el caso general se obtiene por descomposici\'on de Fourier. En este caso, el miembro derecho de (\ref{eq:kk11}) es
\begin{equation}
\sum_{n=1}^{\infty} \frac{(-1)^n}{n!} \lambda^n e^{\lambda Y} D_Y^n(X) = e^{\lambda Y}\left(e^{-\lambda D_Y}-1\right) X = e^{\lambda Y} \left( e^{-\lambda Y} X e^{\lambda Y} - X \right) = [X,e^{\lambda Y}] \, ,
\end{equation}
que coincide con el miembro izquierdo. En esta demostraci\'on hemos hecho uso de la identidad $e^{D_Y}X = e^Y X e^{-Y}$, que es bien conocida.

Particularicemos al caso en que $f$ sea una funci\'on de $\Omega$ (por ejemplo $\varphi_n(\Omega)$). Con $f^{(n)}$ vamos a denotar su derivada $n$-\'esima con respecto a la variable $-\log(\Omega)/\beta$. Entonces de estas f\'ormulas se obtiene

\begin{equation}
 [X,f] = -f^\prime X_0 + \frac{1}{2}f^{\prime\prime} X_{00} - \frac{1}{3!} f^{(3)} X_{000} + \cdots \, .
\label{eq:kk20}
\end{equation}

En el caso de operadores $X = D_\mu $ tendremos 
\begin{eqnarray}
{\widehat D}_0 f &=& 0 \,, \label{eq:kk12}\\
{\widehat D}_i f &=& -f^\prime E_i + \frac{1}{2}f^{\prime\prime} E_{0i} -\frac{1}{3!}f^{(3)}E_{00i} + \cdots \,. \label{eq:kk13}
\end{eqnarray}
La propiedad (\ref{eq:kk12}) se podr\'{\i}a deducir directamente de ${\widehat D}_0 \Omega = [D_0,\Omega] = 0$. Estas f\'ormulas implican que a temperatura finita, al contrario que a temperatura cero, la propiedad c\'{\i}clica de la traza mezcla t\'erminos de \'ordenes diferentes. Esto es debido a que $\widehat{D}_0$ tiene dimensiones de masa, mientras que $\Omega$ es adimensional. As\'{\i}, por ejemplo $\varphi_0(\Omega)$ es de dimensi\'on cero y ${\widehat D}_i$ es de dimensi\'on uno, mientras que ${\widehat D}_i \varphi_0(\Omega)$ contiene t\'erminos de todos los \'ordenes, comenzando con dimensi\'on 2. Para aplicar estas reglas de conmutaci\'on a $a_n^T$ vamos a necesitar adem\'as la relaci\'on
\begin{equation}
\varphi_n^\prime = \sqrt{\tau}(n\varphi_{n-1}+2\varphi_{n+1}) \,,
\end{equation}
que se deduce f\'acilmente a partir de la definici\'on de $\varphi_n$ en ec.~(\ref{eq:2.15}).

La integraci\'on por partes, la propiedad c\'{\i}clica de la traza y estas reglas de conmutaci\'on nos van a permitir escribir expresiones m\'as compactas para los coeficientes $a_n^T$, v\'alidas bajo traza. Hasta dimensi\'on de masa 6 obtenemos  
\begin{eqnarray}
b_0^T =&& \varphi_0 \, ,       \nonumber \\
b_{1/2}^T =&& 0     \, ,       \nonumber \\
b_1^T =&& -\varphi_0 M \, ,    \nonumber \\
b_{3/2}^T =&& 0  \, ,          \nonumber \\
b_2^T =&& \varphi_0 b_2^{T=0} - \frac{1}{6}\overline\varphi_2 E_i^2 \, , \nonumber \\
b_{5/2}^T =&& -\frac{1}{6} \varphi_1 \{M_i,E_i\} \,,  
\label{eq:coef_b}   \\
b_3^T =&& \varphi_0 b_3^{T=0} + \frac{1}{6} \overline\varphi_2 
\left(\frac{1}{2}M_0^2+E_iME_i+\frac{1}{10}E_{ii}^2+\frac{1}{10}F_{0ij}^2-\frac{1}{5}E_iF_{ij}E_j\right) \nonumber \\
&&- \left(\frac{1}{6} \overline\varphi_2-\frac{1}{10} \overline\varphi_4\right) E_{0i}^2  \,. \nonumber 
\end{eqnarray}
Escritos de esta forma, se ve expl\'{\i}citamente que en el l\'{\i}mite de temperatura cero se recupera la simetr\'{\i}a Lorentz. En estas f\'ormulas $b_n^{T=0}$ indican los coeficientes a temperatura cero que aparecen en ec.~(\ref{eq:trazabT0}). El heat kernel es sim\'etrico frente a la transposici\'on de operadores $ ABC \cdots \rightarrow \cdots CBA $, y los $b_n^T$ han sido elegidos de manera que esta simetr\'{\i}a se manifieste en cada orden.

Como hemos dicho, la integraci\'on por partes y la propiedad c\'{\i}clica de la traza hace que exista cierta ambig\"uedad en la expresi\'on de los coeficientes $b_n$ tanto a temperatura cero como a temperatura finita. No obstante a temperatura finita la ambig\"uedad es mayor ya que estas dos propiedades mezclan \'ordenes diferentes. El desarrollo a temperatura finita lo podemos expresar en la forma
\begin{equation}
\Tr\left(e^{-\tau(M-D_\mu^2)}\right) = \frac{1}{(4\pi\tau)^{(d+1)/2}} \sum_{n=0}^\infty B_n^T \tau^n \, , \quad
 B_n^T = \Tr \, b_n^T(x)  \, .
\end{equation}
A temperatura cero el desarrollo se define como un desarrollo en potencias del par\'ametro $\tau$, de modo que $B_n^{T=0}$ no es ambiguo, la ambig\"uedad s\'olo existe en $b_n^{T=0}(x)$. Sin embargo a temperatura finita el desarrollo no est\'a sujeto a un par\'ametro, sino que lo hemos definido como un desarrollo en conmutadores, de modo que existe ambig\"uedad no s\'olo en $b_n^T(x)$ sino tambi\'en en $B_n^T$. En general la elecci\'on concreta de $b_n^T$ va a afectar la forma de los \'ordenes superiores $\,b_{n+1/2}^T, \,b_{n+1}^T, \ldots\,$. Por supuesto, la ambig\"uedad en $B_n^T$ no afecta la suma de la serie, sino que \'unicamente se trata de una reorganizaci\'on de \'esta. Como ejemplo, consideremos que en $b_2^{T=0}$ a\~nadimos el t\'ermino $M_{\mu\mu}$. Nada cambia a temperatura cero, pues ese t\'ermino es un conmutador puro. No obstante, a temperatura finita ese t\'ermino conducir\'{\i}a a la contribuci\'on~$\varphi_0 M_{\mu\mu}$ que no es un conmutador puro, y por tanto va a modificar el funcional~$B_2^T$. De hecho $\varphi_0 M_{\mu\mu}$, que es formalmente de dimensi\'on 4, se puede expresar como una suma de t\'erminos de dimensi\'on 5 y mayores, si hacemos uso de la integraci\'on por partes y de las reglas de conmutaci\'on~(\ref{eq:kk20})-(\ref{eq:kk13}).

El criterio b\'asico que hemos seguido para elegir los coeficientes $b_n^T$ ha consistido en llevarlos de manera recursiva a una forma compacta, comenzando por los de orden inferior. Por ejemplo, bajo traza $a_{3/2}^T$ se puede llevar a una suma de t\'erminos de dimensi\'on 4 o mayor, despu\'es de integrar por partes y aplicar las reglas de conmutaci\'on. Haciendo esto conseguimos $b_{3/2}^T=0$. El siguiente paso consistir\'a en llevar $a_2^T$ (modificado con la contribuci\'on que recibe de $\Tr \, a_{3/2}^T$) a la forma m\'as compacta posible, lo cual en principio producir\'{\i}a contribuciones a $a_{5/2}^T$, y as\'{\i} sucesivamente. Por supuesto, \'esta no es la \'unica posibilidad ya que llevar $b_n^T$ a la forma m\'as simple posible va a implicar en general una mayor complicaci\'on en los \'ordenes superiores. Por ejemplo, se puede ver que es posible ordenar el desarrollo de modo que todos los coeficientes~$b_n^T$ de orden semi-impar se anulen. As\'{\i}, podr\'{\i}amos eliminar $b_{5/2}^T$ con el coste de complicar $b_2^T$.

El an\'alogo de ec.~(\ref{eq:a_var_b_T0}) a temperatura finita va a verse modificado por el hecho de que la variaci\'on de $b_k^T$ contribuye no s\'olo a $a_{k-1}^T$, sino en general a todos los \'ordenes superiores, debido a la propiedad de conmutaci\'on~(\ref{eq:kk20}). Por tanto podemos escribir
\begin{equation}
a_n^T(x) \simeq -\frac{\delta}{\delta M(x)} \sum_{1 \le k \le n+1} B_k^T \tau^{k-n-1} \, , 
\end{equation}
donde el s\'{\i}mbolo $\simeq$ indica que \'unicamente debemos considerar los t\'erminos de dimensi\'on $2n$ en el miembro derecho de la ecuaci\'on. Notar que $k$ puede tomar valores tanto enteros como semi-impares. Hemos comprobado nuestros resultados verificando que esta relaci\'on se cumple para todos los coeficientes.

\section{Conclusiones}

En este cap\'{\i}tulo hemos construido el desarrollo del heat kernel en el contexto de teor\'{\i}a cu\'antica de campos a temperatura finita para espacio-tiempo plano. El desarrollo se ha hecho para un gauge general y en presencia de campos escalares que pueden ser no abelianos y no estacionarios. Se ha puesto un \'enfasis especial en el papel que juega el loop de Polyakov sin traza (o linea de Wilson t\'ermica) para mantener la invariancia gauge expl\'{\i}cita. Esto constituye un problema altamente no trivial, ya que para preservar la invariancia gauge a temperatura finita orden por orden se necesitan infinitos \'ordenes en teor\'{\i}a de perturbaciones. 

Cuando se elige que el ba\~no t\'ermico est\'e en reposo, el loop de Polyakov
es generado por la componente temporal del campo gauge, y \'este se puede
considerar como una generali\-zaci\'on del potencial qu\'{\i}mico para campos gauge no constantes y no abelianos, mediante el factor~$e^{\beta\mu}$. De hecho, hemos aportado argumentos que apoyan esta interpretaci\'on: si el loop de Polyakov no fuera tenido en cuenta, el n\'umero de part\'{\i}culas no podr\'{\i}a ser fijado, lo cual est\'a en contradicci\'on con lo que se espera de los requisitos de la termodin\'amica.

En espacios tiempos curvos, adem\'as del loop de Polyakov de la conexi\'on gauge~$A_\mu$, existe un loop de Polyakov asociado con la conexi\'on de transporte paralelo~$\Gamma_\mu$, con importantes repercusiones en teor\'{\i}a de campos en presencia de campos gravitatorios.

Un ingrediente importante de nuestra t\'ecnica de c\'alculo es que, con objeto
de garantizar la invariancia gauge expl\'{\i}cita, una cierta combinaci\'on
del loop de Polyakov y la temperatura debe tratarse como variable
independiente, $-\log(\Omega)/\beta$. Esto puede hacerse sin necesidad de fijar el gauge.

\chapter{Acci\'on efectiva de QCD a temperatura alta}
\label{QCD_efective_action}

En este cap\'{\i}tulo nos proponemos encontrar un lagrangiano efectivo
de QCD a un loop, incluyendo fermiones sin masa, en la regi\'on de
altas temperaturas. En el c\'alculo de los determinantes funcionales
haremos uso del desarrollo del heat kernel a temperatura finita que
hemos obtenido en el cap\'{\i}tulo~\ref{heat_kernel}. Esto nos permitir\'a calcular el lagrangiano efectivo como un desarrollo en operadores, y aqu\'{\i} obtendremos los \'ordenes m\'as bajos en este desarrollo. 

Existen en la literatura otros m\'etodos equivalentes como el c\'alculo de
diagramas de Feynman a un loop con un n\'umero arbitrario de patas externas
\cite{wirstam}. No obstante suelen ser t\'ecnicamente m\'as complicados y no
dan cuenta autom\'aticamente de invariancia gauge con respecto al campo
externo.

Comenzaremos este cap\'{\i}tulo repasando algunos elementos b\'asicos de la
teor\'{\i}a de Yang-Mills a temperatura finita, para posteriormente entrar de
lleno en el c\'alculo detallado de la acci\'on de QCD a temperatura alta manteniendo la invariancia gauge de manera expl\'{\i}cita. El cap\'{\i}tulo est\'a basado en la referencia~\cite{Megias:2003ui}.

\section{Fundamentos de la Teor\'{\i}a de Yang-Mills a Temperatura Finita}
\label{sec:fundYM}
En esta secci\'on vamos a explicar los fundamentos de la teor\'{\i}a de Yang-Mills a tempe\-ratura finita. Partiremos del hamiltoniano cu\'antico del sistema y deduciremos la funci\'on de partici\'on. 

En una teor\'{\i}a de Yang-Mills el hamiltoniano cu\'antico es
\begin{equation}
H = -\frac{1}{g^2} \int d^3 x \,
\tr \left[ (\partial_0 A_i)^2 + B_i^2  \right]\,, 
\end{equation}
donde $B_i$ es el campo magn\'etico, $B_i = \frac{1}{2}\epsilon_{ijk}F_{jk}$. El espacio de Hilbert est\'a formado por los estados $\{|A_i(x)\rangle\}$. Podemos escribir $e^{-\beta H}$ como $\lim_{N\rightarrow\infty}\left(e^{-\varepsilon H}\right)^N$, $\varepsilon\equiv\beta/N$, y haciendo uso de la relaci\'on de completitud repetidamente se llega a
\begin{equation}
\langle A_i^\prime(\vec{x})|e^{-\beta H}|A_i^{\prime\prime}(\vec{x})\rangle
= \int {\cal D}A_i(x_0,\vec{x})\exp\left\{
\frac{1}{g^2}\int_0^\beta dx_0 \int d^3 x\,
\tr[(\partial_0 A_i)^2+B_i^2]
\right\} \,,
\end{equation}
donde la integral funcional se toma sobre trayectorias  en las que las configuraciones inicial y final est\'an fijas: $A_i(\beta,\vec{x})=A_i^\prime(\vec{x})$ y $A_i(0,\vec{x})=A_i^{\prime\prime}(\vec{x})$. La traza de $e^{-\beta H}$ en el espacio de Hilbert completo es
\begin{eqnarray}
Z_{\YM} &=& \Tr\left(e^{-\beta H}\right)
= \int {\cal D}A_i(\vec{x})
\langle A_i(\vec{x})|e^{-\beta H}|A_i(\vec{x})\rangle
\label{eq:gluon1} \\
&&=\int \mathcal{D}A_i^{(0)}(\vec{x}) \int_{A_i(0,\vec{x})=A_i^{(0)}}^{A_i(\beta,\vec{x})=A_i^{(0)}} \mathcal{D}A_i(x_0,\vec{x}) 
\exp\Big\{\frac{1}{g^2}\int_0^\beta dx_0 \int d^3 x \,\tr\left[ (\partial_0 A_i)^2+B_i^2\right]\Big\} \, .  \nonumber
\end{eqnarray}
Se trata de una integral funcional sobre campos gauge peri\'odicos $A_i(0,\vec{x})=A_i(\beta,\vec{x})$. No obstante, en una teor\'{\i}a gauge hay que sumar, no sobre todos los estados posibles, sino sobre los estados f\'{\i}sicos solamente, esto es, los que satisfacen la ley de Gauss 
\begin{equation}
\vec{D}\cdot \vec{E}(\vec{x})|\psi_{\textrm{fis}}\rangle = 0 
\qquad \forall \,\vec{x} \,,
\label{eq:ymgauss}
\end{equation}
donde $E_i(\vec{x})=\partial_0 A_i(\vec{x})$. Esta relaci\'on expresa la conservaci\'on del flujo el\'ectrico. Para satisfacer (\ref{eq:ymgauss}) basta con que se verifique
\begin{equation}
\exp\left(\int d^3 x \,\tr[\vec{D}\Lambda({\vec{x}})\cdot \vec{E}(\vec{x})]
\right)
|\psi_{\textrm{fis}}\rangle = 
|\psi_{\textrm{fis}}\rangle \,,
\end{equation}
para todo $\Lambda(\vec{x})$ con soporte compacto. $\Omega(U)=\exp(\int
\vec{D}\Lambda\cdot\vec{E})$ es un operador unitario que da lugar a las
transformaciones gauge independientes del tiempo $U=e^\Lambda$. Esto significa que imponer la ley de Gauss es equivalente a exigir que los estados f\'{\i}sicos sean invariantes frente a transformaciones gauge cuyos generadores se anulen en el infinito. Estos estados pueden ser seleccionados introduciendo el proyector $P=\int_{\Lambda(\infty)=0} {\cal D}\Lambda \,\Omega(e^{\Lambda})$ dentro de la integral funcional
\begin{eqnarray}
Z_{\YM} &=& \Tr\left(P e^{-\beta H}\right)
= \int_{\Lambda(\infty)=0} {\cal D}\Lambda(\vec{x}) {\cal D}A_i(\vec{x}) 
\langle A_i^U(\vec{x})|e^{-\beta H}|A_i(\vec{x})\rangle \\
&=&\int_{\Lambda(\infty)=0}{\cal D}\Lambda(\vec{x})
\int_{A_i(\beta,\vec{x})=A_i^U(0,\vec{x})}
{\cal D}A_i(x_0,\vec{x})
\exp\Big\{\frac{1}{g^2}\int_0^\beta dx_0 \int d^3 x \,\tr\left[ (\partial_0 A_i)^2+B_i^2\right]\Big\} \,, \nonumber 
\end{eqnarray}
donde hemos considerado $\langle A_i|\Omega(U)=\langle A_i^U|$. Se trata de una integral funcional sobre campos peri\'odicos salvo transformaci\'on gauge. Con objeto de derivar una expresi\'on que sea estrictamente peri\'odica introducimos el proyector~$P$ m\'as de una vez, lo cual es factible ya que $P$ y $H$ conmutan\begin{eqnarray}
Z_{\YM} &=& \lim_{N\rightarrow\infty} \Tr\left(P e^{-\varepsilon H}\right)^N \\
&& =\int {\cal D}\Lambda(x_0,\vec{x}){\cal D}A_i(x_0,\vec{x})
\exp\Big\{\frac{1}{g^2}\int_0^\beta dx_0 \int d^3 x \,\tr\left[ (\partial_0 A_i-\widehat{D}_i\Lambda)^2+B_i^2\right]\Big\} \,. \nonumber
\end{eqnarray} 
Definiendo el campo $A_0(x_0,\vec{x})=\Lambda(x_0,\vec{x})$, que se anula en $\vec{x}$ infinito, llegamos a
\begin{equation}
Z_{\YM} = \int_{A_\mu(\beta,\vec{x})=A_\mu(0,\vec{x})}
{\cal D}A_\mu(x_0,\vec{x})
\exp\Big\{
\frac{1}{2g^2}\int_0^\beta dx_0 \int d^3 x
\,\tr\left(F_{\mu\nu}^2\right)
\Big\}
=: \int {\cal D}A_\mu(x)e^{-S^E_{\YM}} \,.
\end{equation}  
La ecuaci\'on de movimiento e identidades de Bianchi vienen dadas por
\begin{equation}
{\widehat D}_\mu F_{\mu\nu} = 0 \,, \quad
{\widehat D}_\lambda F_{\mu\nu}
+{\widehat D}_\mu F_{\nu\lambda}
+{\widehat D}_\nu F_{\lambda\mu} = 0 \,.
\end{equation}

En las integrales funcionales existe una condici\'on de periodicidad temporal
en el intervalo $[0,\beta]$ para los campos gauge, que son bos\'onicos.
Adem\'as es necesario integrar sobre todos los valores en los extremos del
intervalo. Si se consideran quarks en la teor\'{\i}a, estos deber\'an
satisfacer condiciones de antiperiodicidad, por ser campos fermi\'onicos. La
funci\'on de partici\'on eucl\'{\i}dea de QCD sin renormalizar se escribe
\begin{equation}
Z_{\rm QCD} = \int_{A_\mu(\beta,\vec{x})=A_\mu(0,\vec{x})} \cD A_\mu(x_0,\vec{x}) 
\int_{q(\beta,\vec{x})=-q(0,\vec{x})}\prod_{\alpha=1}^{N_f}
\cD\overline{q}_\alpha(x_0,\vec{x}) \cD q_\alpha(x_0,\vec{x}) 
\; \exp(-S_E) \,,
\label{eq:func_part_QCD}
\end{equation}
donde la acci\'on eucl\'{\i}dea es
\begin{equation}
S_E=-\frac{1}{2g^2}\int_0^\beta dx_0\int d^3x\,\tr(F_{\mu\nu}^2) 
+\int_0^\beta dx_0\int d^3x \,\sum_{\alpha=1}^{N_f}\overline{q}_\alpha (\thruu{D}+m_\alpha) q_\alpha \,.
\label{eq:42}
\end{equation}
$D_\mu=\partial_\mu+A_\mu$ es la derivada covariante y $A_\mu$ es una matriz antiherm\'{\i}tica de dimensi\'on $N_c$, en la representaci\'on fundamental del \'algebra de Lie del grupo gauge SU($N_c$). $N_f$ es el n\'umero de sabores diferentes de quarks, y~$m_\alpha$ es la masa desnuda de los quarks.

En el tratamiento que haremos para calcular la acci\'on efectiva a un loop, las fluctuaciones cu\'anticas de los campos gauge no van a modificar el sector de los quarks. La contribuci\'on de este sector constituir\'a una correcci\'on a la funci\'on de partici\'on de Yang-Mills, de modo que podremos hacer uso de la siguiente factorizaci\'on
\begin{equation}
Z_{\rm QCD}= Z_q Z_{\YM} \,,
\label{eq:Zq_ZYM}
\end{equation} 
donde $Z_q$ y $Z_{\YM}$ corresponden a la funci\'on de partici\'on del sector fermi\'onico y glu\'onico respectivamente. Esto se justificar\'a en la secci\'on~\ref{sector_gluonico}. Calcularemos cada una de estas contribuciones por separado.

\section{Sector fermi\'onico}
\label{sector_fermionico}

La contribuci\'on de los quarks es m\'as simple que la glu\'onica, de modo que la trataremos en primer lugar para as\'{\i} conseguir una mayor claridad en el desarrollo. Los resultados de esta secci\'on ser\'an v\'alidos para cualquier grupo gauge. En la secci\'on~\ref{sector_gluonico} se particularizar\'an las f\'ormulas para grupos gauge concretos. Consideraremos el caso particular de quarks sin masa~$(m_\alpha=0)$.

La funci\'on de partici\'on sin renormalizar es
\begin{equation}
Z_q[A] = \int_{q(\beta,\vec{x})=-q(0,\vec{x})} \prod_{\alpha=1}^{N_f}\mathcal{D}\overline{q}_\alpha(x_0,\vec{x})\mathcal{D} q_\alpha(x_0,\vec{x}) \exp (-S_q^E)\,,
\end{equation}
con la acci\'on eucl\'{\i}dea 
\begin{equation}
S_q^E= \int_0^\beta dx_0 \int d^3 x \,\sum_{\alpha=1}^{N_f}\overline{q}_\alpha\thruu{D}q_\alpha \, . 
\end{equation}
La integral funcional de los campos de los quarks conduce a
\begin{equation}
Z_q[A] = \Det(\thruu{D}\,)^{N_f} \,,
\end{equation}
y la acci\'on efectiva eucl\'{\i}dea es\footnote{Nuestro convenio para la acci\'on efectiva es $Z=e^{-\Gamma}$.}
\begin{equation}
\Gamma_q^{\rm desn}[A] = -N_f \log \Det(\thruu{D}\,) = - N_f \Tr \log(\thruu{D}\,) \,. 
\end{equation}
Esta expresi\'on es formal debido a la presencia de divergencias ultravioletas. \'Unicamente despu\'es de regularizar y renormalizar estas divergencias se obtiene una acci\'on efectiva finita y bien definida. Existe un gran n\'umero de m\'etodos diferentes para obtener una versi\'on renormalizada, pero un resultado est\'andar de teor\'{\i}a cu\'antica de campos perturbativa es que diferentes definiciones de $\Gamma$ pueden diferir a lo sumo en t\'erminos que son polinomios locales de dimensi\'on can\'onica $d+1$ (donde $d+1$ es la dimensi\'on del espacio-tiempo), construidos con los campos externos y sus derivadas \cite{cheng-li,zuber}. Esto es debido a que todos los diagramas de Feynman son convergentes m\'as all\'a de $d+1$ derivadas en los campos o  en los momentos externos \cite{ramond}. En la pr\'actica vamos a tener que cualquier m\'etodo consistente con la expresi\'on formal de la acci\'on efectiva puede ser usado, puesto que todos ellos van a dar la misma contribuci\'on finita ultravioleta. 

\subsection{Acci\'on efectiva con representaci\'on de Schwinger}
\label{ae_schwinger}

De acuerdo con el tratamiento usual, elevaremos al cuadrado el operador de
Dirac con objeto de obtener un operador de Klein-Gordon. Haciendo uso de la
representaci\'on de Schwinger de tiempo propio podemos escribir la
contribuci\'on del sector fermi\'onico a la acci\'on efectiva de QCD a un loop
como
\begin{equation}
\Gamma_q[A] 
=-\frac{N_f}{2}\Tr \log(\thruu{D}^2)
=\frac{N_f}{2} \int_0^\infty\frac{d\tau}{\tau} \Tr e^{\tau\thruu{D}^2}
=: \int_0^\beta dx_0 \int d^3 x \, {\cal L}_q(x) \, ,
\end{equation}
\begin{equation}
{\cal L}_q(x) = \frac{N_f}{2} \int_0^\infty \frac{d\tau}{\tau} \frac{\mu^{2\epsilon}}{(4\pi\tau)^{D/2}}\sum_n \tau^n \tr(b_{n,q}^T(x)) \, .
\label{eq:Lq_reg_dim}
\end{equation}
Usamos regularizaci\'on dimensional para regular las divergencias ultravioletas en $\tau=0$, con el convenio $D=4-2\epsilon$. El factor $\mu^{2\epsilon}$ restablece la dimensi\'on 4 en masa del lagrangiano efectivo. La traza de Dirac est\'a incluida en $b_{n,q}^T$ y~$\tr$~se refiere a traza en el espacio de color. 
Para aplicar nuestro desarrollo del heat kernel a temperatura finita al c\'alculo de la acci\'on efectiva \'unicamente debemos identificar el operador de Klein-Gordon correspondiente. Usa\-remos el siguiente convenio para las matrices $\gamma_\mu$:
\begin{equation}
\gamma_\mu = \gamma_\mu^\dagger \, , \quad
\{\gamma_\mu,\gamma_\nu\} = 2 \delta_{\mu\nu} \,, \quad
\tr_{\text{Dirac}}(\mathbf{1})=4 \,.
\end{equation}
Se puede escribir
\begin{equation}
-\thruu{D}^2 = -D_\mu^2 - \frac{1}{2}\sigma_{\mu\nu}F_{\mu\nu} \,,
\label{eq:D2_KG}
\end{equation}
donde se ha usado $\gamma_\mu \gamma_\nu = \delta_{\mu\nu} + \sigma_{\mu\nu}$. El operador de ec.~(\ref{eq:D2_KG}) es de tipo Klein-Gordon, y podemos identificar el t\'ermino de masa como $M(x)= -\frac{1}{2}\sigma_{\mu\nu}F_{\mu\nu}$.

\subsection{Traza en espacio de Dirac}
\label{tr_Dirac_space}

El siguiente paso es hacer uso de los coeficientes de heat kernel (\ref{eq:coef_b}) y calcular la traza en el espacio de Dirac. La traza en este espacio muestra que $b_1^T$ y $b_{5/2}^T$ no van a contribuir, lo cual es extensible a todos los t\'erminos del heat kernel con una \'unica $M$. Usamos las siguientes propiedades
\begin{eqnarray}
&&\tr_{\Dirac}(\gamma_{\mu_1}\gamma_{\mu_2}\cdots \gamma_{\mu_{2n+1}}) = 0 \,, \nonumber \\
&&\tr_{\Dirac}(\gamma_\mu\gamma_\nu) =  4 \delta_{\mu\nu}  \,,  \nonumber \\
&&\tr_{\Dirac}(\gamma_\mu\gamma_\nu\gamma_\alpha\gamma_\beta) = 
4(\delta_{\mu\nu}\delta_{\alpha\beta}-\delta_{\mu\alpha}\delta_{\nu\beta}
+\delta_{\mu\beta}\delta_{\nu\alpha})  \, .
\label{eq:trgamma}
\end{eqnarray}
Existe otra propiedad que permite invertir el orden de las matrices $\gamma_\mu$ dentro de la traza
\begin{equation}
\tr_{\Dirac}(\gamma_\mu\gamma_\nu\gamma_\alpha\cdots ) = 
\tr_{\Dirac}(\cdots\gamma_\alpha\gamma_\nu\gamma_\mu)  \, .
\label{eq:trgammainv}
\end{equation}
Hasta dimensi\'on de masa 6 tenemos
\begin{eqnarray}
b_{0,q}^T &=& 4\varphi_0 \,,      \nonumber \\
b_{2,q}^T &=& -\frac{2}{3}\left(\varphi_0 F_{\mu\nu}^2 + \overline\varphi_2 E_i^2\right)\,,   \\
b_{3,q}^T &=& 
\varphi_0  \left(  \frac{32}{45}F_{\mu\nu}F_{\nu\lambda}F_{\lambda\mu} 
+\frac{1}{6}F_{\lambda\mu\nu}^2 - \frac{1}{15}F_{\mu\mu\nu}^2   \right)
+\overline\varphi_2 \left( \frac{1}{15}E_{ii}^2-\frac{1}{10}F_{0ij}^2
-\frac{2}{15}E_iF_{ij}E_j   \right) \nonumber \\
&&+\left( \frac{2}{5}\overline\varphi_4 -\overline\varphi_2  \right) E_{0i}^2
 \,.    \nonumber
\end{eqnarray}
Las funciones $\varphi_n$ corresponden a su versi\'on fermi\'onica, esto es, la suma es sobre las frecuencias de Matsubara $p_0 = 2\pi (n+\frac{1}{2})/\beta$. Los t\'erminos que rompen simetr\'{\i}a Lorentz se han separado expl\'{\i}citamente.

\subsection{Integrales en tiempo propio}
\label{prop_time_integrals}

Como hemos indicado, vamos a hacer uso de la regularizaci\'on dimensional en la integral sobre $\tau$, ec.~(\ref{eq:Lq_reg_dim}). Las integrales van a ser del tipo
\begin{equation}
I_{\ell,n}^\pm(\Omega) := 
\int_0^\infty \frac{d\tau}{\tau} 
(4\pi\mu^2\tau)^\epsilon \tau^{\ell} \varphi_n^\pm(\Omega) \, , \quad
|\Omega|=1 \,,
\label{eq:jk1}
\end{equation}
donde $\varphi_n^\pm$ se refiere a la versi\'on bos\'onica o fermi\'onica, respectivamente. En el sector fermi\'onico $\Omega$ es el loop de Polyakov en la representaci\'on fundamental. A nivel pr\'actico $\Omega$ en realidad va a indicar cada uno de los autovalores del loop de Polyakov. En el ap\'endice \ref{app:integralesQCD} se calculan estas integrales y se discuten algunas de sus propiedades.
Para el sector de los quarks nos va a interesar la versi\'on fermi\'onica de las integrales, y hasta dimensi\'on 6 en masa necesitamos s\'olo valores pares de $n$:
\begin{eqnarray}
I_{\ell,2n}^-(e^{i 2\pi\nu}) 
&=& (-1)^n (4\pi)^\epsilon \left(\frac{\mu\beta}{2\pi}\right)^{2\epsilon}
\left(\frac{\beta}{2\pi}\right)^{2\ell} 
\frac{\Gamma(\ell+n+\epsilon+\frac{1}{2})}
{\Gamma(\frac{1}{2})} \nonumber \\
&&\qquad\qquad\times\Big[ \zeta(1+2\ell+2\epsilon,\half+\nu)
+\zeta(1+2\ell+2\epsilon,\half-\nu) \Big] \,, \nonumber \\
&&\qquad\qquad\qquad\qquad\qquad -\frac{1}{2} < \nu < \frac{1}{2} \,, 
\label{eq:jk2}
\end{eqnarray}
donde hemos hecho uso de la notaci\'on $\Omega=e^{i2\pi\nu}$. $\Gamma(z)$ es la funci\'on Gamma de Euler y $\zeta(z,q)$ la funci\'on de Riemann generalizada~\cite{tablas}. En general las integrales $I_{\ell,n}^\pm(\Omega)$ van a ser funciones univaluadas en $\Omega$, esto es, peri\'odicas en $\nu$ con per\'{\i}odo 1. La f\'ormula~(\ref{eq:jk2}) se ha escrito de manera que sea directamente aplicable en el intervalo $-\frac{1}{2} < \nu < \frac{1}{2}$. Fuera de este intervalo debe considerarse una extensi\'on peri\'odica de la funci\'on. Adem\'as $I_{\ell,2n}^-(\Omega)$ son funciones pares en $\nu$.

Calculemos a continuaci\'on la contribuci\'on al lagrangiano efectivo. El orden cero requiere $I_{-2,0}^-$. Obtenemos
\begin{equation}
I_{-2,0}^- = -\frac{2}{3}\left(\frac{2\pi}{\beta}\right)^4 B_4(\half+\nu) +
{\cal O}(\epsilon) \,,
\end{equation}
donde hemos hecho uso de la relaci\'on $\zeta(1-n,q)=-B_n(q)/n, \, n=1, 2, \dots\,$, y $B_n(q)$ es el polinomio de Bernoulli de orden $n$. Por tanto, tenemos que el potencial efectivo va a ser
\begin{equation}
{\cal L}_{0,q}(x)=\frac{\pi^2 N_f}{\beta^4} \left(\frac{2N_c}{45}-\frac{1}{12}\tr\left[(1-4\nu^2)^2\right]\right)  \, , \quad 
\Omega(x)=e^{i2\pi\nu}\, , \quad 
-\frac{1}{2} < \nu < \frac{1}{2} \,, 
\label{eq:3.11}
\end{equation}
donde $N_c=\tr({\mathbf 1})$ indica el n\'umero de colores. $\tr$ es la traza
en la representaci\'on fundamental del grupo gauge de color, y $\nu$ es la matriz $\,\log(\Omega)/(2\pi i)$ en esta representaci\'on y con valores propios en la rama $|\nu|<1/2$.

Notar que $z=1$ es el \'unico punto singular de la funci\'on $\zeta(z,q)$ (se trata de un polo simple). \'Unicamente las integrales $I_{0,2n}^-$ tienen el polo est\'andar $1/\epsilon$, con lo cual \'este s\'olo va a aparecer en los t\'erminos con dimensi\'on de masa~4, esto es $b_{2,q}^T$. Para estos t\'erminos necesitamos
\begin{eqnarray}
I_{0,0}^- &=& \frac{1}{\epsilon}+\log(4\pi)-\gamma_E+2\log(\mu\beta/4\pi)
-\psi\left(\half+\nu\right)-\psi\left(\half-\nu\right) + {\cal O}(\epsilon) \, ,
\nonumber \\
I_{0,\overline{2}}^- &:=& I_{0,0}^- + 2I_{0,2}^- = -2 + {\cal O}(\epsilon) \, .    
\end{eqnarray}
Las integrales $I_{\ell,\overline{2n}}^\pm$ se definen de forma an\'aloga a $I_{\ell,2n}^\pm$ pero usando $\overline\varphi_{2n}$ en lugar de $\varphi_{2n}$. $\psi(q)$ es la funci\'on digamma, y aqu\'{\i} hemos hecho uso de la relaci\'on
\begin{equation}
\zeta(1+z,q)= \frac{1}{z} - \psi(q) + {\cal O}(z) \,.
\end{equation}

Notar que las funciones $\overline\varphi_{2n}$ se definieron de manera que se anulasen en el l\'{\i}mite $\tau/\beta^2=0$, de modo que las integrales correspondientes van a estar libres de divergencias ultravioletas. 
Los t\'erminos $1/\epsilon+\log(4\pi)-\gamma_E$ que aparecen en $I_{0,0}^-$ son eliminados si adoptamos el esquema de regularizaci\'on $\MS$. Despu\'es de renormalizar, punto que explicaremos en la secci\'on~\ref{renormalizacion_QCD}, tendremos
\begin{equation}
{\cal L}_{2,q}(x) = -\frac{1}{3}\frac{1}{(4\pi)^2}N_f \,\tr 
\left[ \left( 
2\log(\mu\beta/4\pi) - \psi\left(\half+\nu\right) - \psi\left(\half-\nu\right)
\right) F_{\mu\nu}^2
-2E_i^2
\right] \, .
\label{eq:L2qpp}
\end{equation}

Para los t\'erminos con dimensi\'on de masa 6 vamos a necesitar las integrales
\begin{eqnarray}
I_{1,0}^- &=& -\left(\frac{\beta}{4\pi}\right)^2 
\Big(  \psi^{\prime\prime}\left(\half+\nu\right)+ 
        \psi^{\prime\prime}\left(\half-\nu \right)
\Big)  + {\cal O}(\epsilon) \,,   \nonumber \\
I_{1,\overline{2}}^- &=& -2I_{1,0}^- + {\cal O}(\epsilon) \, , \quad
I_{1,\overline{4}}^- = -4I_{1,0}^- + {\cal O}(\epsilon) \, ,
\end{eqnarray}
donde hemos hecho uso de la relaci\'on $\psi^{(n)}(q) = (-1)^{n+1} n! \zeta(n+1,q)\,$. Esto conduce a 
\begin{eqnarray}
{\cal L}_{3,q}(x) &=& -\frac{2}{(4\pi)^4}N_f\beta^2 \tr
\Big[ \Big(\psi^{\prime\prime}\left(\half+\nu\right)
+\psi^{\prime\prime}\left(\half-\nu\right)\Big) \label{eq:L3qpp} \\ 
&& \times \left( \frac{8}{45}F_{\mu\nu}F_{\nu\lambda}F_{\lambda\mu}
+\frac{1}{24}F_{\lambda\mu\nu}^2
-\frac{1}{60}F_{\mu\mu\nu}^2
+\frac{1}{20}F_{0\mu\nu}^2
-\frac{1}{30}E_{ii}^2
+\frac{1}{15}E_{i}F_{ij}E_j
\right)
\Big] \,. \nonumber
\end{eqnarray}


Notar que cada orden del heat kernel est\'a asociado a una potencia en temperatura, ${\cal L}_0\sim T^4$, ${\cal L}_2\sim T^0$, ${\cal L}_3\sim T^{-2}$, lo cual quiere decir que el desarrollo del heat kernel a temperatura finita es esencialmente un desarrollo en potencias de $k^2/T^2$, donde $k$ es el momento glu\'onico t\'{\i}pico. Los t\'erminos de orden $T^2$ est\'an prohibidos ya que no existen operadores invariantes gauge de dimensi\'on 2 disponibles.

\section{Sector glu\'onico}
\label{sector_gluonico}

 A continua\-ci\'on nos vamos a centrar en el t\'ermino de Yang-Mills, para el cual consideraremos espec\'{\i}ficamente el grupo SU($N_c$) (matrices unitarias, una \'unica constante de acoplamiento y matrices del \'algebra de Lie con traza cero en cualquier representaci\'on). 

La funci\'on de partici\'on sin renormalizar es
\begin{equation}
Z_g = \int_{A_\mu(\beta,\vec{x})=A_\mu(0,\vec{x})}
{\cal D}A_\mu(x_0,\vec{x})
\exp(-S^E_{\YM})
\end{equation} 
con la acci\'on eucl\'{\i}dea
\begin{equation}
S^E_{\YM} = -\frac{1}{2g^2} \int_0^\beta dx_0 \int d^3x
\,\tr(F_{\mu\nu}^2) \,.
\label{eq:vb1}
\end{equation}
Se trata de una integral funcional entre configuraciones peri\'odicas.

\subsection{M\'etodo del Campo de Fondo}
\label{Met_campo_fondo}

Para el c\'alculo de la acci\'on efectiva haremos uso del M\'etodo del Campo de Fondo~\cite{metcampofondo,backfieldmethod} que consiste en separar el campo glu\'onico, que por claridad denotaremos aqu\'{\i} como $\overline{A}_\mu$, en un campo cl\'asico $A_\mu$ m\'as una fluctuaci\'on cu\'antica $a_\mu$ en la acci\'on (\ref{eq:vb1}). La fluctuaci\'on es presumiblemente peque\~na. 
\begin{equation}
\overline{A}_\mu(x) = A_\mu(x) + a_\mu(x) \, .  
\end{equation}
Esto va a inducir una separaci\'on en el tensor $F_{\mu\nu}$
\begin{equation}
F_{\mu\nu}[\overline{A}] = F_{\mu\nu}[A] + {\widehat D}_\mu a_\nu
-{\widehat D}_\nu a_\mu + [a_\mu,a_\nu] \, . 
\label{eq:F_met_campo_fondo}
\end{equation}
En ec.~(\ref{eq:F_met_campo_fondo}) la derivada covariante es la asociada al campo cl\'asico, esto es $D_\mu=\partial_\mu + A_\mu$. En nuestra notaci\'on~${\widehat D}_\mu = [D_\mu,\,\,]$. Notar que los campos de los quarks se eligen como una fluctuaci\'on pura, de modo que $a_\mu$ no modifica el sector fermi\'onico a un loop. Esto justifica la factorizaci\'on de ec.~(\ref{eq:Zq_ZYM})

Una transformaci\'on gauge infinitesimal de $\overline{A}_\mu$ con par\'ametro $\Lambda$ puede ser distribuida de muchas maneras sobre los campos $A_\mu$ y $a_\mu$, pero las elecciones m\'as convenientes van a ser la {\it transformaci\'on cu\'antica}
\begin{equation}
\delta A_\mu = 0 \, , \quad  
\delta a_\mu = {\widehat D}_\mu \Lambda + [a_\mu,\Lambda]
\label{eq:ymtranscuantica}
\end{equation}
y la  {\it transformaci\'on del campo de fondo}
\begin{equation}
\delta A_\mu = {\widehat D}_\mu \Lambda \, , \quad  
\delta a_\mu = [a_\mu,\Lambda] \,.
\label{eq:ymtransfondo}
\end{equation}
La clave consiste ahora en a\~nadir a la acci\'on cl\'asica un t\'ermino que fije el gauge (gauge-fixing term) el cual va a romper la invariancia gauge cu\'antica, pero respetar\'a la invariancia gauge del campo cl\'asico de fondo
\begin{equation}
S_{\rm fix} = \frac{1}{\alpha}\int_0^\beta dx_0\int d^3x \,\tr(G^2) \,,
\end{equation}
donde la funci\'on $G$ transforma de modo covariante bajo~(\ref{eq:ymtransfondo}). El t\'ermino de Faddeev-Popov asociado es
\begin{equation}
S_{\rm FP} = 2 \int_0^\beta dx_0 \int d^3x\,\tr
\left(C^\ast\frac{\delta G}{\delta \Lambda}C\right) \, , 
\end{equation}
donde $\delta G/\delta \Lambda$ indica la variaci\'on de $G$ bajo una transformaci\'on gauge cu\'antica. $C$ y $C^\ast$ son los campos ghost y antighost respectivamente, que son objetos que anticonmutan (si bien son peri\'odicos en tiempo eucl\'{\i}deo) y son matrices en la representaci\'on fundamental de su($N_c$). La acci\'on total  $S_{\rm tot}=S^E_{\YM}+S_{\rm fix}+S_{\rm FP}$ aparece en el funcional generador de todas las funciones de Green
\begin{equation}
Z_g[A,J,\eta^\ast,\eta]= 
N \int {\mathcal D}a{\mathcal D}C^\ast{\mathcal D}C 
\,\exp\left( -S_{\rm tot} + J\cdot a + \eta^\ast\cdot C + C^\ast\cdot\eta
\right) \,,
\end{equation}
donde $J$, $\eta$ y $\eta^\ast$ son fuentes y el factor de normalizaci\'on $N$ se elige de modo que se verifique $Z_g[A,0,0,0]=1$. Notar que el campo cl\'asico de fondo no est\'a acoplado con la fuente $J$. El funcional generador para los diagramas conexos viene dado por
\begin{equation}
W_g[A,J,\eta^\ast,\eta] = \log Z_g[A,J,\eta^\ast,\eta] \, .
\end{equation}
Los valores esperados de todos los campos se definen como
\begin{equation}
\tilde{a}=\frac{\delta W_g}{\delta J} \, , \quad
\tilde{C}=\frac{\delta W_g}{\delta \eta^\ast} \, , \quad
\tilde{C}^\ast=\frac{\delta W_g}{\delta \eta} \, ,
\end{equation}
y a partir de ellos, la acci\'on efectiva se define 
\begin{equation}
\Gamma_g[A,\tilde{a},\tilde{C}^\ast,\tilde{C}]
= J\cdot\tilde{a}
+\eta\cdot\tilde{C}^\ast
+\tilde{C}\cdot\eta^\ast
-W_g[A,J,\eta^\ast,\eta] \,.
\end{equation}
Los funcionales $Z_g$ y $W_g$ ser\'an invariantes bajo transformaciones gauge del campo de fondo, ec.~(\ref{eq:ymtransfondo}), si todas las fuentes y campos ghost transforman igual que $a$. Lo mismo se puede decir de la acci\'on efectiva $\Gamma_g$ si uno exige que $\tilde{a}$, $\tilde{C}$ y $\tilde{C}^\ast$ transformen igual que $a$. Una buena elecci\'on es el gauge de Gervais-Neveu generalizado
\begin{equation}
G = {\widehat D}_\mu a_\mu + \lambda a_\mu^2 \, ,
\end{equation}
pero aqu\'{\i} nos vamos a restringir al gauge de Feynman covariante $\alpha=1$, $\lambda=0$, con un t\'ermino de Faddeev-Popov asociado
\begin{equation}
S_{\rm FP} = 2 \int_0^\beta dx_0 \int d^3x \, \tr 
\left( C^\ast {\widehat D}_\mu( {\widehat D}_\mu C 
+[a_\mu,C] )
\right) \, .
\end{equation}
Descompongamos la corriente $J_\mu$ en
\begin{equation}
J_\mu = \mathcal{J}_\mu + j_\mu
\end{equation}
donde definimos
\begin{equation}
j_\mu =\frac{\delta S_{\rm tot}[A,a]}{\delta a_\mu}\Big|_{a=0} \,.
\end{equation}
Podemos escribir la acci\'on total como
\begin{eqnarray}
S_{\rm tot}[A,a] - J \cdot a &=& 
S^E_{\YM}[A] + \frac{1}{2}\int_0^\beta dx_0 \int d^3x \, \tr\left( a_\mu\Delta_{\mu\nu}[A]a_\nu\right) \nonumber \\
&& -\int_0^\beta dx_0 \int d^3x\, \tr\left(C^\ast\Delta[A]C\right)
- S_{\rm int}[A,a]-\mathcal{J}\cdot a  \,,
\end{eqnarray}
con las siguientes definiciones
\begin{eqnarray}
&&\Delta_{\mu\nu}[A] = -\left(
\delta_{\mu\nu}{\widehat D}_\lambda^2 + 2{\widehat F}_{\mu\nu}\right) 
\,, \quad
\Delta[A] = -{\widehat D}_\mu^2 \,, 
\nonumber \\
&&S_{\rm int}[A,a] = \int_0^\beta dx_0\int d^3x \, \tr
\left(  ({\widehat D}_\mu a_\nu) [a_\mu,a_\nu]
+\frac{1}{4}[a_\mu,a_\nu]^2
+ C^\ast{\widehat D}_\mu[a_\mu,C]
\right) \,,
\end{eqnarray}
donde ${\widehat D}_\mu = [D_\mu,\,\,]$ y ${\widehat F}_{\mu\nu}=[F_{\mu\nu},\,\,]$. Notar que la constante de acoplamiento $g$ puede ser absorbida en la normalizaci\'on de los campos.
De este modo obtenemos el nuevo funcional generador
\begin{eqnarray}
&&Z_g[A,\mathcal{J},\eta^\ast,\eta]   \nonumber \\
=&&\int \mathcal{D}a\mathcal{D}C^\ast\mathcal{D}C \, 
\exp\Big[
-S^E_{\YM}[A]-\frac{1}{2}\int_0^\beta dx_0\int d^3x \,a_\mu \Delta_{\mu\nu} a_\nu
+\int_0^\beta dx_0 \int d^3x \,C^\ast \Delta C \nonumber \\
&& \qquad\qquad\qquad\qquad\qquad -S_{\rm int}[A,a]
+\mathcal{J}\cdot a 
+\eta^\ast \cdot C + C^\ast \cdot \eta \,
\Big]   \,,
\label{eq:gluonfuncgen}
\end{eqnarray}
que depende de la nueva corriente $\mathcal{J}$. Notar que el \'unico t\'ermino lineal en el campo gauge $a_\mu$ que aparece en el exponente es el que est\'a acoplado con la corriente $\mathcal{J}$.

Si imponemos ahora las condiciones $\tilde{a} = \tilde{C} = \tilde{C}^\ast = 0$, la acci\'on efectiva se puede escribir como
\begin{equation}
\Gamma_g[A,0,0,0] = -W_g[A,\mathcal{J},\eta^\ast,\eta]\Big|
_{\delta W_g/\delta\mathcal{J}=\delta W_g/\delta\eta^\ast=\delta W_g/\delta \eta=0}
\end{equation}
y es todav\'{\i}a invariante respecto a la transformaci\'on gauge del campo de fondo dada en (\ref{eq:ymtransfondo}). El desarrollo perturbativo de esta acci\'on efectiva solamente contiene diagramas de vac\'{\i}o, que son 1PI.

El desarrollo de la acci\'on efectiva de Yang-Mills en t\'erminos de $\hbar$ viene dado por
\begin{equation}
\Gamma_g[A,0,0,0] = S^E_{\YM}[A] 
+ \frac{1}{2}\Tr\log (-\delta_{\mu\nu}{\widehat D}_\lambda^2 - 2{\widehat F}_{\mu\nu}) 
- \Tr\log (-{\widehat D}_\mu^2)
+{\cal O}(\hbar^2) \,. 
\label{eq:aeg_des}
\end{equation}
El t\'ermino $S^E_{\YM}[A]$ es la contribuci\'on cl\'asica, y los t\'erminos segundo y tercero son las contribuciones cu\'anticas a ${\cal O}(\hbar)$. En (\ref{eq:aeg_des}) se puede ver que el operador de Klein-Gordon sobre los campos cu\'anticos $a_\mu$ (segundo t\'ermino del miembro derecho de la igualdad) act\'ua sobre un espacio interno de dimensi\'on $D\times\widehat{N}_c$, donde en regularizaci\'on dimensional $D=4-2\epsilon$,  que es el n\'umero de polarizaciones del glu\'on (f\'{\i}sicas o no).\footnote{$\widehat{N}_c$ es el n\'umero de generadores del grupo gauge ($N_c^2-1$ en SU($N_c$)) y se corresponde con la dimensi\'on de la representaci\'on adjunta del grupo.} $D$ se corresponde con el \'{\i}ndice de Lorentz $\mu$. Los operadores $\widehat{D}_\mu$ y $\widehat{F}_{\mu\nu}$ act\'uan en la representaci\'on adjunta. Notar que la derivada covariante en este operador de Klein-Gordon es la identidad en el espacio de Lorentz, mientras que el t\'ermino de masa es una matriz en dicho espacio, esto es $(M)_{\mu\nu}=-2{\widehat F}_{\mu\nu}$. Por otra parte, el operador de Klein-Gordon de los campos ghost (tercer t\'ermino del miembro derecho de la igualdad) act\'ua sobre un espacio interno de dimensi\'on $\widehat{N}_c$. La derivada covariante es $D_\mu$ en la representaci\'on adjunta y el t\'ermino de masa es cero.

\subsection{Acci\'on efectiva a un loop}
\label{aeg_1loop}

La acci\'on efectiva total de QCD quiral hasta nivel de un loop queda
\begin{equation}
\Gamma[A] = -\frac{\mu^{-2\epsilon}}{2g_0^2} \int d^4 x \, \tr(F_{\mu\nu}^2)
+\Gamma_q[A] + \Gamma_g[A] \, , 
\end{equation}
donde el primer t\'ermino es la acci\'on de la teor\'{\i}a de Yang-Mills a nivel \'arbol teniendo en cuenta la renormalizaci\'on ($g_0$ es una constante adimensional). El segundo t\'ermino es la contribuci\'on de los quarks, que ha sido calculada en la secci\'on \ref{sector_fermionico}. El tercer t\'ermino corres\-ponde a las contribuciones que surgen despu\'es de integrar los ghosts y las fluctuaciones cu\'anticas de los campos gauge
\begin{equation}
\Gamma_g[A] = \frac{1}{2}\Tr\log (-\delta_{\mu\nu}{\widehat D}_\lambda^2 - 2{\widehat F}_{\mu\nu}) 
- \Tr\log (-{\widehat D}_\mu^2) 
=: \int_0^\beta dx_0 \int d^3x \,{\cal L}_g(x)\,.
\end{equation}
Haciendo uso del desarrollo del heat kernel podemos escribir
\begin{equation}
{\cal L}_g(x) = -\frac{1}{2} \int_0^\infty \frac{d\tau}{\tau} \frac{\mu^{2\epsilon}}{(4\pi\tau)^{D/2}}\sum_n \tau^n \,\widehat{\tr}(b_{n,g}^T(x)) \,.
\label{eq:jk3}    
\end{equation}
La traza sobre el espacio de Lorentz para los gluones est\'a incluida en los coeficientes $b_{n,g}^T$. En esta f\'ormula $\widehat{\tr}$ se refiere a traza en el espacio de color en la representaci\'on adjunta. Se puede comprobar expl\'{\i}citamente en el c\'alculo que el efecto de los ghosts es quitar dos grados de polarizaci\'on del glu\'on, esto es $D \rightarrow D-2$. Debido a la traza de Lorentz todos los t\'erminos con una \'unica $M$ se van a anular, en particular $b_{1,g}^T$ y $b_{5/2,g}^T$ no van a contribuir. Hasta dimensi\'on de masa 6 tenemos
\begin{eqnarray}
b^T_{0,g}&=& (D-2)\varphi_0(\widehat\Omega)  \,, \nonumber \\ 
b^T_{2,g} &=& \left(-2+\frac{D-2}{12}\right)
\varphi_0(\widehat\Omega) {\widehat F}_{\mu\nu}^2
-\frac{D-2}{6} \overline\varphi_2(\widehat\Omega)
{\widehat E}_i^2 \,,
\label{eq:jk4} \\ 
b^T_{3,g} &=& 
\varphi_0(\widehat\Omega)  \left(
\left(\frac{4}{3}+\frac{D-2}{90} \right) 
{\widehat F}_{\mu\nu}{\widehat F}_{\nu\lambda}{\widehat F}_{\lambda\mu}
+\frac{1}{3}{\widehat F}_{\lambda\mu\nu}^2 
-\frac{D-2}{60}{\widehat F}_{\mu\mu\nu}^2
\right)
\nonumber \\
&& 
+ \, 
\frac{1}{6}\overline\varphi_2(\widehat\Omega) \left(
-2{\widehat F}_{0\mu\nu}^2 + 
\frac{D-2}{10}\left( {\widehat E}_{ii}^2+{\widehat F}_{0ij}^2-2{\widehat E}_i {\widehat F}_{ij}{\widehat E}_j \right)\right)
\nonumber \\
&& 
+ \, 
(D-2)\left(\frac{1}{10}\overline\varphi_4(\widehat\Omega) - \frac{1}{6}
\overline\varphi_2(\widehat\Omega) \right) {\widehat E}_{0i}^2
\,.
 \nonumber
\end{eqnarray}
En estas f\'ormulas las funciones $\varphi_n$ corresponden a su versi\'on bos\'onica, esto es, la suma es sobre las frecuencias de Matsubara $p_0 = 2\pi n/\beta$. A diferencia del sector fermi\'onico, el argumento de estas funciones, $\widehat\Omega$, y las derivadas covariantes est\'an en la representaci\'on adjunta. Notar que los t\'erminos con $D-2$ proceden de t\'erminos del heat kernel que no tienen masa. Los t\'erminos que rompen simetr\'{\i}a Lorentz se han separado expl\'{\i}citamente.

Para calcular el lagrangiano efectivo deberemos introducir los coeficientes (\ref{eq:jk4}) en (\ref{eq:jk3}). Las integrales en $\tau$ que aparecen son del tipo (\ref{eq:jk1}). Como sabemos, las versiones bos\'onica y fermi\'onica de las funciones $\varphi_n$ est\'an relacionadas por el cambio $\,\Omega \rightarrow -\Omega\,$, esto es $\,\varphi_n^+(\Omega)=\varphi_n^-(-\Omega)\,$. De aqu\'{\i} tenemos que las integrales $I_{\ell,2n}^+$ van a ser las mismas que $I_{\ell,2n}^-$ de (\ref{eq:jk2}) excepto por el cambio $\nu\rightarrow \widehat\nu-\frac{1}{2}$
\begin{eqnarray}
I^+_{\ell,2n}(e^{i2\pi\widehat\nu})
&=& (-1)^n (4\pi)^\epsilon \left(\frac{\mu\beta}{2\pi}\right)^{2\epsilon}
\left(\frac{\beta}{2\pi}\right)^{2\ell} 
\frac{\Gamma(\ell+n+\epsilon+\frac{1}{2})}
{\Gamma(\frac{1}{2})} \nonumber \\
&&\qquad\qquad\times\Big[ \zeta(1+2\ell+2\epsilon,\widehat\nu)
+\zeta(1+2\ell+2\epsilon,1-\widehat\nu) \Big] \,, \nonumber \\ 
 && \quad\qquad\qquad\qquad\qquad\qquad
      0 < \widehat\nu < 1 \,,
\label{eq:I_2n_gluon}
\end{eqnarray}
donde hemos hecho uso de la notaci\'on~$\widehat\Omega=e^{i2\pi\widehat\nu}$. Esta f\'ormula se ha escrito para que sea v\'alida en el intervalo $0<\widehat\nu<1$. Fuera de este intervalo debe considerarse la extensi\'on peri\'odica de la funci\'on. Las integrales $I_{\ell,2n}^+(e^{i2\pi\widehat{\nu}})$ son pares bajo el cambio $\,\widehat\nu \rightarrow 1-\widehat\nu\,$. 

El orden cero requiere
\begin{equation}
I_{-2,0}^+ = -\frac{1}{3}\left(\frac{2\pi}{\beta}\right)^4 
(B_4(\widehat\nu)+B_4(1-\widehat\nu)) + {\cal O}(\epsilon)  \,.
\end{equation}
El potencial efectivo va a ser
\begin{eqnarray}
\mathcal{L}_{0,g}(x) &=& \frac{\pi^2}{3\beta^4}\widehat{\tr}
(B_4(\widehat\nu)+B_4(1-\widehat\nu))  \nonumber \\
&=& -\frac{\pi^2}{45\beta^4}\widehat{N}_c+
\frac{2\pi^2}{3\beta^4}\,\widehat{\tr}\,[\widehat{\nu}^2(1-\widehat{\nu})^2]
\,,\quad \widehat{\Omega}(x)=e^{i2\pi\widehat{\nu}}
\,,\quad 0 < \widehat{\nu} < 1 \,, 
\label{eq:L0gluonic}
\end{eqnarray}
donde $\widehat{N}_c :=\widehat{\tr}(\mathbf{1})=N_c^2-1$ es el n\'umero de generadores del grupo gauge. $\widehat{\nu}$ es la matriz $\,\log(\widehat{\Omega})/(2\pi i)\,$ con valores propios en la rama $0<\widehat{\nu}<1$. Si se considera~$\widehat{\nu}=0$ en ec.~(\ref{eq:L0gluonic}) se obtiene la presi\'on de un gas ideal de gluones.

Para los t\'erminos con dimensi\'on de masa 4 vamos a necesitar
\begin{eqnarray}
I_{0,0}^+ &=& \frac{1}{\epsilon} + \log(4\pi) -\gamma_E
+2\log(\mu\beta/4\pi) -\psi(\widehat\nu) - \psi(1-\widehat\nu)
+{\cal O}(\epsilon) \,,   \nonumber \\
I_{0,\overline{2}}^+ &:=& I_{0,0}^+ + 2I_{0,2}^+ = -2 + {\cal O}(\epsilon) \,.
\end{eqnarray}
$I_{0,0}^+$ es divergente ultravioleta e $I_{0,\overline{2}}^+$ es finito. Notar que en las $D$'s que aparecen en nuestras expresiones tras hacer la traza de Lorentz tambi\'en existen $\epsilon$'s que hay que tener en cuenta. La parte finita del lagrangiano efectivo, en el esquema~$\MS$, es
\begin{equation}
\mathcal{L}_{2,g}(x)= \frac{1}{(4\pi)^2}\widehat{\tr}
\left[
\frac{11}{12}\left(2\log(\mu\beta/4\pi)+\frac{1}{11}-\psi(\widehat\nu)-\psi(1-\widehat\nu)\right)\widehat{F}_{\mu\nu}^2
-\frac{1}{3}\widehat{E}_i^2
\right] \,.
\label{eq:L2gpp}
\end{equation}
En esta f\'ormula no hemos considerado las contribuciones divergentes. \'Estas ser\'an tratadas en la secci\'on~\ref{renormalizacion_QCD}, donde abordaremos el problema de la renormalizaci\'on.

Para los t\'erminos con dimensi\'on de masa 6 necesitamos las integrales
\begin{eqnarray}
I_{1,0}^+ &=& -\left(\frac{\beta}{4\pi}\right)^2 
\Big(  \psi^{\prime\prime}\left(\widehat\nu\right) 
+ \psi^{\prime\prime}\left(1-\widehat\nu\right)\Big) + {\cal O}(\epsilon) \,,\\
I_{1,\overline{2}}^+ &=& -2I_{1,0}^+ + {\cal O}(\epsilon) \,, \quad
I_{1,\overline{4}}^+ = -4I_{1,0}^+ + {\cal O}(\epsilon) \,. 
\label{eq:jk17}
\end{eqnarray}
Esto conduce a
\begin{eqnarray}
{\cal L}_{3,g}(x) &=&  \frac{1}{2}\frac{\beta^2}{(4\pi)^4 }
\widehat{\tr} \bigg[
\Big(\psi^{\prime\prime}(\widehat{\nu}) 
+ \psi^{\prime\prime}(1-\widehat{\nu}) \Big) \\
&& \times
\left( 
\frac{61}{45}\widehat{F}_{\mu\nu}\widehat{F}_{\nu\lambda}\widehat{F}_{\lambda\mu}
+\frac{1}{3}\widehat{F}_{\lambda\mu\nu}^2
-\frac{1}{30}\widehat{F}_{\mu\mu\nu}^2
+\frac{3}{5}\widehat{F}_{0\mu\nu}^2
-\frac{1}{15}\widehat{E}_{ii}^2
+\frac{2}{15}\widehat{E}_i\widehat{F}_{ij}\widehat{E}_j
\right)
\bigg]
\,.  \nonumber
\end{eqnarray}

\section{Renormalizaci\'on}
\label{renormalizacion_QCD}

Para la renormalizaci\'on deberemos considerar todas las contribuciones divergentes que hemos obtenido, tanto del sector de quarks como del sector glu\'onico. El lagrangiano a nivel \'arbol junto con estas divergencias conduce a
\begin{eqnarray}
&&\mathcal{L}_{\text{\'arbol}}(x) + \mathcal{L}_q^{\diver}(x) + \mathcal{L}_g^{\diver}(x) =   \nonumber \\
&&\qquad\qquad =-\frac{1}{2g_0^2}\tr(F_{\mu\nu}^2)
+\frac{1}{(4\pi)^2}\left(\frac{1}{\epsilon}+\log(4\pi)-\gamma_E\right)
\left(\frac{11}{12}\widehat{\tr}(\widehat{F}_{\mu\nu}^2)-\frac{N_f}{3}\tr(F_{\mu\nu}^2)\right)   \nonumber \\
&&\qquad\qquad = -\frac{1}{2g^2(\mu)}\tr(F_{\mu\nu}^2) \,.
\label{eq:Ltree_div}
\end{eqnarray} 
Hacemos uso de la siguiente identidad
\begin{equation}
\widehat{\tr}\left(\widehat{F}_{\mu\nu}^2\right) =
2\tr(\mathbf 1)\tr\left(F_{\mu\nu}^2\right) - 2\left(\tr\left(F_{\mu\nu}\right)\right)^2 = 2N_c\tr\left(F_{\mu\nu}^2\right) \,, 
\label{eq:glident}
\end{equation}
donde se ha considerado que el grupo gauge es SU($N_c$). De aqu\'{\i} obtenemos en el esquema~$\MS$
\begin{equation}
\frac{1}{g^2(\mu)} = \frac{1}{g_0^2} 
- \beta_0 \left(\frac{1}{\epsilon}+\log(4\pi)-\gamma_E\right) \,,\quad
\beta_0 = \frac{1}{(4\pi)^2}\left(\frac{11}{3}N_c-\frac{2}{3}N_f\right) \,, 
\end{equation}
lo cual garantiza la independencia en escala de ec.~(\ref{eq:Ltree_div}). Notar que nos hemos limitado a renormalizar la constante de acoplamiento. Por invariancia gauge, los campos cl\'asicos $A_\mu$ no necesitan ser renormalizados, si bien para el problema de la reducci\'on dimensional, en general \'esta funciona mejor si los campos se renormalizan de tal modo que todas las contribuciones a $E_i^2$ y $B_i^2$, que proceden de loops no est\'aticos, son canceladas mediante contrat\'erminos \cite{chapmancorto}.

Si consideramos todos los t\'erminos con dimensi\'on de masa 4, (ecs.~(\ref{eq:L2qpp}), (\ref{eq:L2gpp}) y (\ref{eq:Ltree_div})), para ellos el lagrangiano renormalizado de QCD quiral hasta un loop es
\begin{eqnarray}
{\cal L}_{2,{\rm QCD}}(x) &=&
\left(-\frac{1}{2g^2(\mu)} +\beta_0 \log(\mu\beta/4\pi)
+\frac{1}{6}\frac{1}{(4\pi)^2}N_c \right) \,\tr(F_{\mu\nu}^2) 
\nonumber
\\
&& -\frac{11}{12}\frac{1}{(4\pi)^2} \widehat{\tr} \left[
 \big( \psi(\widehat{\nu})+\psi(1-\widehat{\nu}) \big)
\widehat{F}_{\mu\nu}^2
\right]
\nonumber \\
&&
+\frac{1}{3}\frac{1}{(4\pi)^2} N_f \tr \left[
\left(\psi(\half+\nu)+\psi(\half-\nu) \right)F_{\mu\nu}^2
\right]
\nonumber \\
&&
-\frac{2}{3}(N_c-N_f)\frac{1}{(4\pi)^2}  
\tr \left[ E_i^2 \right]
\,,\qquad -\frac{1}{2} < \nu < \frac{1}{2}
\,,\quad 0 < \widehat{\nu} < 1 \,.
\end{eqnarray}

Otra posibilidad es usar el m\'etodo de Pauli-Villars para regular las divergencias ultravioletas \cite{paulivillars}. La regularizaci\'on de Pauli-Villars consiste b\'asicamente en la introducci\'on, en el funcional generador, de nuevos campos $a^\prime_\mu$ y $C^\prime$ que transforman como $a_\mu$ y $C$, pero tienen una masa $M$ que posteriormente se considerar\'a en el l\'{\i}mite $M \rightarrow \infty$. Por conveniencia consideramos que los dos campos tienen la misma masa. La aplicaci\'on de este procedimiento conduce a la siguiente funci\'on de partici\'on regulada
\begin{equation}
Z[A]|_{\reg} = \frac{Z[A]}{Z^\prime[A,M^2]} \,.
\end{equation}
$Z^\prime[A,M^2]$ tiene la misma forma que $Z[A]$ excepto que los t\'erminos
de masa est\'an incluidos. Para un operador gen\'erico ${\mathbf K}$, la expresi\'on regulada para el determinante funcional es
\begin{equation}
\Det(\mathbf{K})|_{\reg} = \frac{\Det(\mathbf{K})}{\Det(\mathbf{K}+M^2)} 
= \exp\left[-\int_0^\infty \frac{d\tau}{\tau} \left(1-e^{-\tau M^2}\right) \Tr \,e^{-\tau \mathbf{K}}\right] \,,
\end{equation}
donde hemos hecho uso de la representaci\'on de tiempo propio de Schwinger. Por tanto la regularizaci\'on de Pauli-Villars corresponde a insertar el factor $(1-e^{-\tau M^2})$ en la integraci\'on en $\tau$. Todas las integrales convergentes (incluidas~$\,I_{0,\overline{2}}^\pm\,$) quedan igual que en regularizaci\'on dimensional (en el l\'{\i}mite~$M \to \infty$), y las integrales que divergen son
\begin{eqnarray}
I_{0,0}^{+,\PV} &=& 2\log(M/\mu) + 2\log(\mu\beta/4\pi)
-\psi(\widehat\nu) - \psi(1-\widehat\nu) + {\cal O}(M^{-1})   \,, \quad
0 < \widehat\nu < 1  \,,  \\
I_{0,0}^{-,\PV} &=& 2\log(M/\mu) + 2\log(\mu\beta/4\pi)
-\psi(\half+\nu) - \psi(\half-\nu)  + {\cal O}(M^{-1})   \,,   \quad
-\frac{1}{2} < \nu < \frac{1}{2} \,. \nonumber     
\end{eqnarray}
El lagrangiano a nivel \'arbol tiene la siguiente constante de acoplamiento desnuda (para cutoff $M$),
\begin{equation}
\frac{1}{g_0^2(M)} = \frac{2}{(4\pi)^2}\left(\frac{11}{3}N_c-\frac{2}{3}N_f\right)\log\frac{M}{\mu}\,.
\end{equation}

El lagrangiano a nivel \'arbol junto con todas las contribuciones divergentes del sector glu\'onico y del sector fermi\'onico conduce a (en este esquema de regularizaci\'on)
\begin{eqnarray}
{\cal L}_{\text{\'arbol}}(x)+{\cal L}_q^{\diver}(x)+{\cal L}_g^{\diver}(x) &=&
-\frac{1}{2g_0^2(M)}\tr(F_{\mu\nu}^2)+\frac{1}{(4\pi)^2}\log(M)
\left(\frac{11}{3}N_c -\frac{2}{3}N_f\right)\tr(F_{\mu\nu}^2)  \nonumber \\
&=& -\frac{1}{2g^2(\mu)}\tr(F_{\mu\nu}^2) \,,
\end{eqnarray}
donde hemos hecho uso de la identidad (\ref{eq:glident}), v\'alida en SU($N_c$). Notar que el t\'ermino divergente ultravioleta $\log(M)$ es cancelado por la constante de acoplamiento desnuda, de modo que al final el cutoff $M$ es reemplazado por el par\'ametro finito $\mu$. Si, como es usual, el par\'ametro $\Lambda$ en cada esquema es definido como la escala $\,\mu=\Lambda\,$ para la cual $1/g^2(\mu)$ se anula, encontramos que los dos esquemas $\MS$ y $\PV$ dan id\'enticos resultados cuando
\begin{equation}
\log\left(\Lambda^2_{\PV}/\Lambda^2_{\MS}\right) 
= \frac{1}{11-\frac{2N_f}{N_c}} \,.
\end{equation}
Si se hace uso de otro esquema de regularizaci\'on, la escala deber\'a modificarse en consecuencia \cite{scales1,scales2}.

\section{Divergencias infrarrojas}
\label{div_infrarrojas}

En el c\'alculo del sector glu\'onico de la acci\'on efectiva existe un problema de divergencias infrarrojas relacionadas con el modo est\'atico de Matsubara. En la representaci\'on adjunta, $N_c-1$ valores propios de $\widehat\Omega$ son necesariamente la unidad, de modo que el valor $\widehat\nu=0$ va a aparecer siempre al tomar la traza adjunta. Notar que para integrales $I_{\ell,n}^+$ con $n \ne 0$, el modo est\'atico no va a contribuir. Sin embargo en $I_{\ell,0}^+$ este modo puede producir divergencias infrarrojas y ultravioletas. En concreto la singularidad de $I_{\ell,n}^+(e^{i2\pi\widehat\nu})$ para valores de $\widehat\nu$ enteros procede de la divergencia infrarroja del modo est\'atico. En regularizaci\'on dimensional la integral $\,I_{\ell,0}^+(1)|_{p_0=0}\,$ se define como cero ya que no tiene una escala natural \cite{renormalizacion}. Tal y como se explica en el ap\'endice~\ref{app:integralesQCD}, las integrales~$I_{\ell,2n}^+$ sin el modo est\'atico vienen dadas por las mismas expresiones que~(\ref{eq:I_2n_gluon}) despu\'es de la sustituci\'on~$\widehat\nu\rightarrow 1+\widehat\nu$ en la primera $\zeta$. En consecuencia, la prescripci\'on va a ser usar las f\'ormulas de ${\cal L}_{2,g}$ y ${\cal L}_{3,g}$ con las sustituciones
\begin{eqnarray}
&& \psi(\hnu)+\psi(1-\hnu)\,\big|_{\hnu=0} \to
\psi(1+\hnu)+\psi(1-\hnu) \,\big|_{\hnu=0}
= -2\gamma_E \,,
\nonumber \\
&& \psi^{\prime\prime}(\hnu)+\psi^{\prime\prime}(1-\hnu)\,\big|_{\hnu=0} \to
\psi^{\prime\prime}(1+\hnu)+\psi^{\prime\prime}(1-\hnu) \,\big|_{\hnu=0}
= -4\zeta(3)\,,
\end{eqnarray}
realizadas \'unicamente en el subespacio~$\widehat\Omega=1$. Esta prescripci\'on preserva la periodicidad de la acci\'on efectiva como funci\'on de $\log(\widehat\Omega)$, de modo que es consistente con la invariancia gauge. 

Un modo alternativo de tratar las divergencias infrarrojas es regul\'andolas a\~nadiendo en las integrales en $\tau$ una funci\'on cutoff $\,e^{-\tau m^2}\,$. Los modos que son finitos en el r\'egimen infrarrojo no se ven afectados en el l\'{\i}mite $m \rightarrow 0$. El modo est\'atico en~$\varphi_0$ da lugar a divergencias que deber\'an ser a\~nadidas al resultado obtenido previamente en regularizaci\'on dimensional. Esto produce la contribuci\'on
\begin{equation}
I_{\ell,0}^+(1)|_{p_0=0} = \frac{\sqrt{4\pi}\Gamma(\half+\ell)}{\beta m^{2\ell+1}} \,.  
\end{equation}
Puesto que las divergencias infrarrojas proceden exclusivamente del modo est\'atico de $\varphi_0$, las relaciones de escala del tipo (\ref{eq:jk17}) para $I^+_{\ell,\overline{2n}}$ no ser\'an v\'alidas. Los t\'erminos divergentes infrarrojos que obtenemos son:
\begin{eqnarray}
{\cal L}_{2,\text{IR}} &=&
\frac{1}{48\pi}\frac{T}{m}\tr\left[ 11 F^2_{\mu\nu\perp}
+2E^2_{i\perp} 
\right] \,,
\nonumber  \\
{\cal L}_{3,\text{IR}} &=&
\frac{1}{240\pi}\frac{T}{m^3}\tr\Big[ 
-\frac{61}{3} F_{\mu\nu\perp}F_{\nu\alpha}F_{\alpha\mu}
+E_{i\perp}F_{ij}E_{j\perp} 
+E_{i}F_{ij\parallel}E_{j}
\label{eq:3.33} \\ &&
-5 F^2_{\mu\nu\lambda\perp}
+\frac{1}{2}F^2_{\mu\mu\nu\perp}
+\frac{9}{2}F^2_{0\mu\nu\perp}
+3 E^2_{0i\perp}
-\frac{1}{2} E^2_{ii\perp}
\Big] \,.
\nonumber
\end{eqnarray}
En nuestra notaci\'on, $F_{\mu\nu\parallel}$ indica la parte de $F_{\mu\nu}$ que conmuta con $\Omega$, y $F_{\mu\nu\perp}$ el resto. Si bien son contribuciones glu\'onicas, se han expresado en la representaci\'on fundamental, que suele ser preferible. En concreto, en el gauge en que $\Omega$ es diagonal, $F_{\mu\nu\parallel}$ es la parte diagonal de $F_{\mu\nu}$. \'Unicamente t\'erminos con al menos una componente perpendicular pueden ser divergentes infrarrojos.

\section{Teor\'{\i}a efectiva dimensionalmente reducida}
\label{reduccion_dimensional}

Desde mediados de la d\'ecada de los noventa la mayor parte del esfuerzo que
se ha dedicado en QCD perturbativa a temperatura alta ha sido en calcular la presi\'on, y s\'olo recientemente se ha obtenido el orden perturbativo m\'as alto posible~\cite{Kajantie:2003ax}, mediante el uso de ideas de reducci\'on dimensional~\cite{Ginsparg:1980ef,Appelquist:1981vg,Nadkarni:1982kb,Braaten:1995jr,Shaposhnikov:1996th}. Estas ideas se basan en el hecho de que a temperatura suficientemente alta el comportamiento de la teor\'{\i}a puede ser descrito, en principio, por una teor\'{\i}a efectiva en tres dimensiones.

Como sabemos, a temperatura finita los campos peri\'odicos se pueden descomponer en modos de Fourier
\begin{equation}
A_\mu(x_0,\vec{x}) = \sum_{n=-\infty}^\infty A_\mu(\omega_n,\vec{x}) 
e^{i\omega_n x_0} \, , \quad \omega_n = \frac{2\pi n}{\beta}  \,.
\end{equation}
Cada modo lleva asociado un propagador de la forma $[{\vec p}\,{}^2 + \omega_n^2]^{-1}$. Para valores de $n$ diferentes de cero, $\omega_n\,$ juega el papel de una masa. En el l\'{\i}mite $T \rightarrow \infty$ todas las frecuencias de Matsubara no nulas son infinitamente grandes. Puesto que las part\'{\i}culas infinitamente pesadas se desacoplan en una teor\'{\i}a de campos a temperatura cero, se puede esperar que ocurra lo mismo para los modos no est\'aticos $n \ne 0$ en teor\'{\i}as de campos a temperatura alta, de modo que una teor\'{\i}a efectiva en tres dimensiones ser\'{\i}a suficiente para explicar el comportamiento.

En la secci\'on~\ref{sector_gluonico} hemos obtenido la acci\'on efectiva haciendo una separaci\'on del campo glu\'onico en un background cl\'asico m\'as una fluctuaci\'on cu\'antica, e integrando \'esta \'ultima a un loop. Se puede adaptar esta t\'ecnica para calcular la acci\'on de la teor\'{\i}a dimensionalmente reducida (que denotaremos en lo sucesivo como ${\cal L}^\prime (\vec{x})$), haciendo lo siguiente:
\begin{itemize}
\item considerar un background estacionario;
\item considerar fluctuaciones puramente no estacionarias.
\end{itemize}
La integraci\'on de los modos fermi\'onicos y los modos glu\'onicos no estacionarios va a dar lugar a una teor\'{\i}a efectiva para los restantes modos estacionarios (independientes del tiempo), que consiste en una teor\'{\i}a gauge SU($N_c$) en tres dimensiones acoplada con un campo escalar $A_0$. A esto se le llama {\it reducci\'on dimensional}
\begin{equation}
\int d^4x \, {\cal L}_{\text{QCD}}(x) \longrightarrow \int d^3x \, {\cal L}^\prime (\vec{x}) \,.
\end{equation}

La segunda condici\'on implica eliminar los modos est\'aticos $n=0$ en todas las sumas de Matsubara. Hay que mencionar que ${\cal L}^\prime(\vec{x})$ no es la acci\'on efectiva de la teor\'{\i}a dimensionalmente reducida, sino que es la acci\'on verdadera (en la aproximaci\'on de un loop), en el sentido de que la integral funcional sobre las configuraciones estacionarias con ${\cal L}^\prime(\vec{x})$ da lugar a la funci\'on de partici\'on.

\subsection{Eliminaci\'on de los modos est\'aticos}
\label{elim_modos_estaticos}

La eliminaci\'on del modo est\'atico s\'olo afecta al sector glu\'onico, y resulta irrelevante en el sector de quarks, de modo que~${\cal L}_q^\prime(\vec{x}) = \beta {\cal L}_q(\vec{x})$.\footnote{Notar que existe un factor $\beta$ extra en ${\cal L}^\prime(\vec{x})$.} El sector glu\'onico a nivel \'arbol tampoco se ve afectado, de modo que para el nivel \'arbol renormalizado se tiene
\begin{equation}
{\cal L}^\prime_{\text{\'arbol}}(\vec{x}) = \beta {\cal L}_{\text{\'arbol}}(\vec{x}) \,.
\end{equation}  

La eliminaci\'on del modo est\'atico en el sector glu\'onico (para $|\widehat\nu|<1$) corresponde a la sustituci\'on $\widehat\nu \rightarrow \widehat\nu+1$ en la primera $\zeta$ de~(\ref{eq:I_2n_gluon}). Esto conduce trivialmente a las siguientes f\'ormulas:
\begin{eqnarray}
{\cal L}^\prime_{0,g}(\vec{x}) &=&
 \frac{2\pi^2}{3} T^3 \,\htr\left[ \hnu^2 
(1+\hnu^2) \right]
\,,\quad \hnu= \log(\widehat\Omega)/(2\pi i)\,,\quad
-1 \le \hnu \le 1\,.
\label{eq:4.1} \\
{\cal L}^\prime_{2,g}(\vec{x}) &=&
\frac{1}{(4\pi)^2T} \htr \Big[
\frac{11}{12} \left( 2\log(\mu/4\pi T) +\frac{1}{11}
-\psi(\hnu)-\psi(-\hnu) \right) \FF_{\mu\nu}^2
-\frac{1}{3} \E_i^2 
\Big]\,,
\label{eq:69a} \\
{\cal L}^\prime_{3,g}(\vec{x}) &=&  
\frac{1}{2}\frac{1}{(4\pi)^4 } \frac{1}{T^3}
\htr \bigg[
\Big(\psi^{\prime\prime}(\hnu) 
+ \psi^{\prime\prime}(-\hnu) \Big)
  \\
&& \times
\left( 
\frac{61}{45}\FF_{\mu\nu}\FF_{\nu\lambda}\FF_{\lambda\mu}
+\frac{1}{3}\FF_{\lambda\mu\nu}^2
-\frac{1}{30}\FF_{\mu\mu\nu}^2
+\frac{3}{5}\FF_{0\mu\nu}^2
-\frac{1}{15}\E_{ii}^2
+\frac{2}{15}\E_i\FF_{ij}\E_j
\right)
\bigg]
\,.
\nonumber
\end{eqnarray}
En el potencial efectivo se ha desechado un t\'ermino que es independiente de $A_0$. Para los t\'erminos con dimensi\'on de masa cuatro y seis se ha hecho uso de la identidad $\psi(1+\hnu)+\psi(1-\hnu) =\psi(\hnu)+\psi(-\hnu) $. En estas f\'ormulas, $\FF_{0\mu\nu}$ indica $[A_0,\FF_{\mu\nu}]$. Notar que la eliminaci\'on del modo est\'atico permite que ${\cal L}^\prime(\vec{x})$ est\'e libre de divergencias infrarrojas.

\subsection{Desarrollo en $A_0$ peque\~no}
\label{des_A0_peq}

Adem\'as de tomar $A_\mu$ estacionario, consideraremos que $A_0$ es peque\~no (en particular $|\widehat\nu|<1$). Esto es v\'alido a temperatura alta, pues en este r\'egimen el potencial efectivo produce una supresi\'on en las configuraciones de $\Omega(\vec{x})$ que est\'an lejos de la unidad. En ausencia de quarks, esto se puede ver como que $\Omega(\vec{x})$ vive cerca de un elemento del centro del grupo gauge (la rotura espont\'anea de la simetr\'{\i}a del centro indica la fase de desconfinamiento). Siempre va a ser posible hacer una transformaci\'on gauge para llevar una configuraci\'on a la regi\'on $|\widehat\nu|<1$. El considerar esta configuraci\'on es importante, pues \'unicamente cuando $A_0$ es peque\~no ($|\widehat\nu|<1$) las fluctuaciones no est\'aticas son las m\'as pesadas.
Si $A_0$ es peque\~no, podremos desarrollar ${\cal L}^\prime (\vec{x})$ en potencias de $A_0$. Notar que en el contaje en dimensiones de masa, el $A_0$ procedente de $\Omega$ tiene dimensi\'on 1 ($\Omega$ no tiene dimensiones de masa). Este nuevo contaje no es compatible con invariancia gauge expl\'{\i}cita bajo transformaciones grandes, aunque s\'{\i} bajo transformaciones estacionarias topol\'ogicamente peque\~nas (pr\'oximas a la identidad).

El potencial efectivo es un polinomio en $A_0$. De ec.~(\ref{eq:3.11}) y (\ref{eq:4.1}), se obtiene
\begin{equation}
{\cal L}^\prime_0(\vec{x})= -\left(\frac{N_c}{3}+\frac{N_f}{6}\right) T 
\langle A_0^2 \rangle
+\frac{1}{4\pi^2T}\langle A_0^2\rangle^2
+\frac{1}{12\pi^2T}(N_c-N_f)\langle A_0^4 \rangle \,.
\label{eq:L0prime}
\end{equation}
En el resto de esta secci\'on usaremos la notaci\'on $\langle X\rangle :=\tr(X)$, para la traza en la representaci\'on fundamental. 

El resultado que se obtiene para los t\'erminos con dimensi\'on de masa cuatro
se puede escribir como\footnote{Mediante un reescalamiento de la constante de
  acoplamiento $g$ y de los campos gauge con factores de renormalizaci\'on
  convenientemente elegidos, se puede conseguir que
  $\cL_{(4)}^\prime(\vec{x})$ presente el mismo aspecto que el nivel \'arbol
  renormalizado a temperatura cero (ec.~(\ref{eq:Ltree_div})). Es necesario
  considerar factores diferentes para la componente espacial y temporal de los
  campos: $g\rightarrow Z_ g^{-1/2} g$,  $A_i\rightarrow Z_M^{1/2}A_i$ y $A_0\rightarrow Z_E^{1/2} A_0$.
\begin{eqnarray}
&&Z_M = Z_g =  1 +2 g^2\beta_0(\log(\mu/4\pi T)+\gamma_E)
-\frac{g^2}{3(4\pi)^2} \left(-N_c+8 N_f\log 2\right)
\,,  \\ 
&&Z_E = Z_M -\frac{2g^2}{3(4\pi)^2}(N_c-N_f)    
\,. 
\end{eqnarray} }
\begin{equation}
\cL_{(4)}^\prime(\vec{x})= 
-\frac{1}{T g_E^2(T)}\langle E_i^2 \rangle
-\frac{1}{T g_M^2(T)}\langle B_i^2 \rangle \,,
\end{equation}
con las siguientes constantes de acoplamiento cromoel\'ectricas y
cromomagn\'eticas 
\begin{eqnarray} 
\frac{1}{ g_E^2(T)} &=& \frac{1}{ g^2(\mu)}
-2\beta_0(\log(\mu/4\pi T)+\gamma_E)
+\frac{1}{3(4\pi)^2}\left(N_c+8N_f\left(\log 2-\frac{1}{4}\right)\right) \,,
\nonumber \\
\frac{1}{g_M^2(T)} &=& \frac{1}{g^2(\mu)}
-2\beta_0(\log(\mu/4\pi T)+\gamma_E) 
+\frac{1}{3(4\pi)^2}\left(-N_c+8N_f\log 2\right) \,.
\label{eq:gE_gM}
\end{eqnarray}
El valor de $g_M^2(T)$ coincide con \cite{Shaposhnikov:1996th} para $N_c=3$. Tambi\'en coincide con \cite{chapmancorto} (con $N_f=0$) si se considera un factor adecuado dependiente de $N_c$ entre la escala $\Lambda$ usada en este art\'{\i}culo y nuestra $\mu$.

Podemos introducir los par\'ametros t\'ermicos $\Lambda$ el\'ectricos y magn\'eticos como \cite{Huang:1994cu}
\begin{equation}
\frac{1}{g_{E,M}^2(T)}  = 2\beta_0 \log(T/\Lambda^T_{E,M}) \,.
\end{equation}
Estos par\'ametros fijan la escala de ambas constantes de acoplamiento a temperatura alta. De ec.~(\ref{eq:gE_gM}) se tiene
\begin{eqnarray}
\log(\Lambda^T_E/\Lambda_{\overline{\text{MS}}}) &=&
\gamma_E -\log(4\pi) - \frac{N_c+8N_f(\log 2-1/4)}{22N_c-4N_f} \,, \nonumber \\
\log(\Lambda^T_M/\Lambda_{\overline{\text{MS}}}) &=&
\gamma_E -\log(4\pi) + \frac{N_c-8N_f\log 2}{22N_c-4N_f} \,.
\end{eqnarray}

Los t\'erminos con dimensi\'on de masa seis proceden de
$\cL_2^\prime(\vec{x})$ (desarrollando la funci\'on digamma hasta orden dos en
$\nu$) y de $\cL_3^\prime(\vec{x})$ (a orden cero). Se obtiene\footnote{Se ha
  hecho uso de las identidades siguientes:
\begin{equation}
{\widehat \tr}\left(\widehat{X}^2\right) = 2N_c\langle X^2\rangle \,,
\quad X\in\text{su($N_c$)}\,,
\label{eq:3.28}
\end{equation}
\begin{equation}
{\widehat \tr}\left(\widehat{X}^2\widehat{Y}^2\right)= 2N_c\langle X^2Y^2\rangle +
2\langle X^2\rangle\langle Y^2\rangle+4\langle XY\rangle^2 \,,\quad X,Y\in\text{su($N_c$)}\,.
\label{eq:4.5}
\end{equation}
}
\begin{eqnarray}
\cL_{(6)}^\prime(\vec{x}) &=&
-\frac{2}{15}\frac{\zeta(3)}{(4\pi)^4T^3}\Bigg[
\big( \frac{2}{3}N_c - \frac{14}{3}N_f \big)
 \langle F_{\mu\nu}F_{\nu\lambda}F_{\lambda\mu} \rangle
+ ( 19 N_c - 28 N_f )  \langle F_{\mu\mu\nu}^2 \rangle
\nonumber \\ &&
+ ( 18 N_c - 21 N_f ) \langle F_{0\mu\nu}^2 \rangle
+ ( 110 N_c - 140 N_f ) \langle A_0^2 F_{\mu\nu}^2 \rangle
- (   2 N_c - 14 N_f ) \langle E_{ii}^2 \rangle
\nonumber \\ &&
+ (   4 N_c - 28 N_f ) \langle E_iF_{ij}E_j \rangle
+ 110 \langle A_0^2 \rangle \langle F_{\mu\nu}^2 \rangle
+ 220 \langle A_0 F_{\mu\nu}\rangle^2
\Bigg]
\,. \label{eq:Lprime6}
\end{eqnarray}
Este resultado coincide con el obtenido en \cite{chapmancorto}, calculado en
el sector glu\'onico y con $N_c$ arbitrario. El lagrangiano de dimensi\'on
seis ha sido calculado asimismo en \cite{wirstam} para el sector de quarks y
en SU(3), en ausencia de campo cromomagn\'etico ($A_i=0$) y eliminando
t\'erminos con m\'as de dos derivadas espaciales (por ejemplo
$E_{ii}^2$). Nuestro c\'alculo reproduce tambi\'en este resultado. En esta misma referencia se hace el c\'alculo para el sector glu\'onico, y tanto nuestro resultado como el de \cite{chapmancorto} se muestran en desacuerdo con \'el.

\section{Resultados en SU(2)}
\label{resultados_su2}

En las secciones precedentes hemos encontrado resultados generales, v\'alidos para cualquier grupo gauge en el sector fermi\'onico (secci\'on~\ref{sector_fermionico}), y para SU($N_c$) en el sector glu\'onico (secci\'on~\ref{sector_gluonico}). Para el c\'alculo de las trazas en espacio de color es necesario particularizar nuestras f\'ormulas a un grupo gauge concreto. Consideraremos aqu\'{\i} espec\'{\i}ficamente el grupo SU(2).

En esta secci\'on s\'olo mostraremos los \'ordenes ${\cal L}_0(x)$ y ${\cal L}_2(x)$. Los resultados completos ${\cal L}_{0,2,3}(x)$ en ambos sectores aparecen en el ap\'endice \ref{app:resultadosSU(2)}.

\subsection{Traza en espacio de color}
\label{traza_color_su2}

Usaremos como base de su(2) las matrices $\vec{t}=\vec{\sigma}/2i$, donde $\vec\sigma$ son las matrices de Pauli. 
\begin{equation}
A_0 = A_0^a t_a = -\frac{i}{2}\vec{\sigma}\cdot\vec{A}_0 \,,  \quad
F_{\mu\nu} = F_{\mu\nu}^a t_a = -\frac{i}{2}\vec{\sigma}\cdot\vec{F}_{\mu\nu}\,, \quad
\dots
\end{equation}
En esta base
\begin{equation}
[t_a,t_b] = \epsilon_{abc}t_c \,, \quad  \tr(t_at_b) = -\frac{1}{2}\delta_{ab} \,.
\end{equation}
En el gauge de Polyakov $A_0$ es independiente del tiempo y diagonal en la representaci\'on fundamental, de modo que resulta especialmente conveniente para nuestro c\'alculo. Para SU(2), en este gauge tenemos
\begin{equation}
A_0 = -\frac{i}{2}\sigma_3\phi \,, \quad \phi = \sqrt{A_0^aA_0^a} \,, 
\end{equation}
de modo que los valores propios del loop de Polyakov en la representaci\'on fundamental $\,\Omega = \exp(-\beta A_0)\,$, son $\,\exp(\pm i\beta\phi/2)\,$. En la representaci\'on adjunta 
\begin{equation}
\widehat{A}_0 = A_0^a \,T_a \,, \quad  (T^a)_{bc}=f_{bac}=-\epsilon_{abc}
\end{equation}
 y de aqu\'{\i} se tiene que los valores propios del loop de Polyakov en la representaci\'on adjunta $\widehat{\Omega}=\exp(-\beta\widehat{A}_0 )$, son $\exp(\pm i\beta\phi)$ y 1.

Para el sector fermi\'onico, tras calcular la traza en el espacio de color, obtenemos
\begin{eqnarray}
{\cal L}_{0,q}(x) &=& \frac{2\pi^2}{3} T^4 N_f \left(\frac{2}{15}-\frac{1}{4}(1-4 \overline \nu^2)^2\right)
\,, \\
 {\cal L}_{2,q}(x) &=& \frac{N_f}{48 \pi^2}\left(2\log\left(\frac{\mu}{4\pi T}\right)-\psi(\half+\overline \nu)-\psi(\half-\overline \nu)-1\right) \vec E_i^2 
\nonumber \\
&& +\frac{N_f }{48 \pi^2}\left(2\log\left(\frac{\mu}{4\pi T}\right)-\psi(\half+\overline \nu)-\psi(\half-\overline \nu)\right)\vec B_i^2
\,,
\end{eqnarray}
donde
\begin{equation}
\overline\nu = \left(\frac{\beta\phi}{4\pi} +\frac{1}{2}\right)\;({\rm mod }\;1)-\frac{1}{2} \,.
\end{equation}
En el sector glu\'onico se obtiene
\begin{eqnarray}
{\cal L}_{0,g}(x) &=& 
\frac{\pi^2}{3}T^4 
\left(-\frac{1}{5}+4\widehat\nu^2 (1-\widehat\nu)^2\right)
\,,  \\
 {\cal L}_{2,g}(x) &=& -  \frac{11}{48 \pi^2}\left(2\log\left(\frac{\mu}{4\pi T}\right)-\frac{1}{11}-\psi(\widehat\nu)-\psi(1-\widehat\nu)\right)  \vec E_{i \parallel}^2  
\nonumber \\
&&- \frac{11}{48 \pi^2}\left(\frac{12}{11}\frac{\pi T}{m}+2\log\left(\frac{\mu}{4\pi T}\right)-\frac{1}{11}+\gamma_E-\frac{1}{2}\psi(\widehat\nu)-\frac{1}{2}\psi(1-\widehat\nu)\right) \vec E_{i \perp}^2
\nonumber \\
&& -  \frac{11}{48 \pi^2}\left(2\log\left(\frac{\mu}{4\pi T}\right)+\frac{1}{11}-\psi(\widehat\nu)-\psi(1-\widehat\nu)\right)  \vec B_{i \parallel}^2  \\
&& -  \frac{11}{48 \pi^2}\left(\frac{\pi T}{m}+2\log\left(\frac{\mu}{4\pi T}\right)+\frac{1}{11}+\gamma_E-\frac{1}{2}\psi(\widehat\nu)-\frac{1}{2}\psi(1-\widehat\nu)\right)  \vec B_{i \perp}^2 
\,,  \nonumber
\end{eqnarray}
donde
\begin{equation}
\widehat\nu = \frac{\beta\phi}{2\pi} \;({\rm mod} \; 1)  \,.
\end{equation}

Hemos hecho uso del esquema $\MS$, y hemos considerado expl\'{\i}citamente un cutoff infrarrojo (ver sec.~\ref{div_infrarrojas}). Nuestros resultados son peri\'odicos en $\phi$. En estas expresiones se ha hecho la separaci\'on de los sectores el\'ectrico y magn\'etico. $B_i = \frac{1}{2}\epsilon_{ijk}F_{jk}\,$ es el campo magn\'etico, y
\begin{equation}
\vec{E}_i = \vec{E}_{i \parallel} + \vec{E}_{i \perp} \,, \quad
\vec{B}_i = \vec{B}_{i \parallel} + \vec{B}_{i \perp} \,,
\end{equation}
es la descomposici\'on de los campos el\'ectrico y magn\'etico en la direcci\'on paralela y perpendicular a $\vec{A}_0$. Esta descomposici\'on es invariante gauge, siempre y cuando se considere que, en un gauge general, la direcci\'on paralela es aquella que venga indicada por el loop de Polyakov (esto es, aquella que conmuta con el loop de Polyakov), y la perpendicular el resto. En la expresi\'on de ${\cal L}_{2,g}(x)$ vemos que las componentes paralelas de los campos est\'an libres de divergencias infrarrojas. Solamente las componentes perpendiculares pueden presentar este tipo de divergencias.

\begin{figure}[t]
\begin{center}
\includegraphics[width=8.5cm]{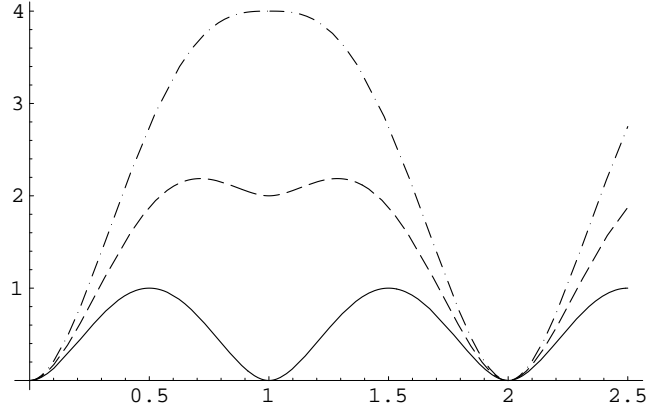}
\end{center}
\caption{Potencial efectivo a un loop para SU(2) como funci\'on de $\beta\phi/2\pi$, en ausencia de fermiones (l\'{\i}nea continua), con un fermi\'on sin masa (rayada) y con dos fermiones (puntos y rayas). Se ha graficado $12\beta^4{\cal L}_0/\pi^2$ y se ha eliminado el t\'ermino constante.}
\label{fig:su2_1}
\end{figure}

En la figura \ref{fig:su2_1} se muestra el comportamiento del potencial efectivo (orden cero del lagrangiano). Observamos que las periodicidades del sector fermi\'onico y del sector glu\'onico se diferencian en un factor 2.

Los lagrangianos efectivos ${\cal L}_{2,q}(x)$ y ${\cal L}_{2,g}(x)$ presentan la siguiente estructura
\begin{eqnarray}
{\cal L}_{2,q}(x) &=& -f_{1,q}(\overline\nu)\vec{E}_{i \perp}^2
-f_{2,q}(\overline\nu)\vec{E}_{i \parallel}^2
-h_{1,q}(\overline\nu)\vec{B}_{i \perp}^2
-h_{2,q}(\overline\nu)\vec{B}_{i \parallel}^2 \,, \\
{\cal L}_{2,g}(x) &=& -f_{1,g}(\widehat\nu)\vec{E}_{i \perp}^2
-f_{2,g}(\widehat\nu)\vec{E}_{i \parallel}^2
-h_{1,g}(\widehat\nu)\vec{B}_{i \perp}^2
-h_{2,g}(\widehat\nu)\vec{B}_{i \parallel}^2 \,.
\label{eq:jkl190}
\end{eqnarray}
Para el sector de quarks se tiene $f_{1,q}=f_{2,q}\equiv f_q$, $h_{1,q}=h_{2,q}\equiv h_q $.
\begin{figure}[t]
\begin{center}
\includegraphics[width=8.5cm]{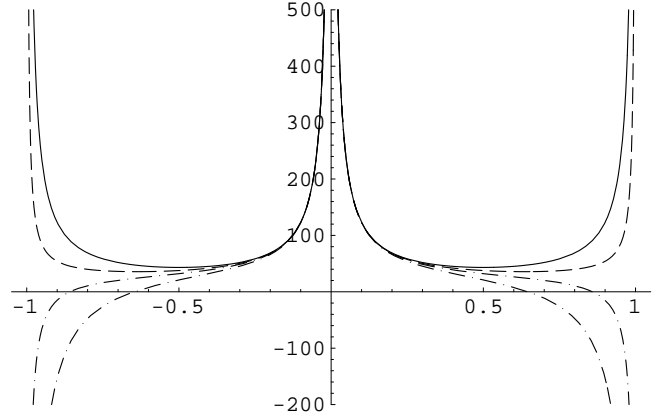}
\end{center}
\caption{Gr\'afico de $96\pi^2(f_{1,q}+f_{1,g})$ como funci\'on de $\beta\phi/2\pi$, en ausencia de fermiones (l\'{\i}nea continua), con dos fermiones sin masa (rayada) y con cuatro y ocho fermiones en orden sucesivo (puntos y rayas). Se han eliminado los t\'erminos constantes.}
\label{fig:su2_2}
\end{figure}

En la figura \ref{fig:su2_2} aparece graficada la funci\'on $f_{1,q}(\overline\nu)+f_{1,g}(\widehat\nu)$. Notar que $f_{1,g}(\widehat\nu)$ es singular en $\nu=0$ lo cual es debido a la contribuci\'on del modo cero. El resto de funciones: $f_{2,q}(\overline\nu)+f_{2,g}(\widehat\nu)\,,\;h_{1,q}(\overline\nu)+h_{1,g}(\widehat\nu) \,\textrm{y}\;h_{2,q}(\overline\nu)+h_{2,g}(\widehat\nu)$; presentan un comportamiento similar.

\subsection{Invariancia gauge del resultado}
\label{inv_gauge_resultado_su2}

Despu\'es de fijar el gauge de Polyakov ($A_0$ diagonal e independiente del
tiempo), a\'un queda una simetr\'{\i}a abeliana residual que consiste en rotaciones gauge arbitrarias independientes del tiempo sobre los generadores de Cartan ($\sigma_3$~en el caso de SU(2)), y de una rotaci\'on gauge dependiente del tiempo (tambi\'en sobre los ejes de Cartan) que va a ser discreta para ser compatible con la periodicidad de $A_i(x_0,\vec{x})$ (ver ap\'endice \ref{app:gauge}). Para una teor\'{\i}a gauge pura SU($2$) esta simetr\'{\i}a residual corresponde a la siguiente transformaci\'on
\begin{equation}
A_\mu \rightarrow U^{-1}\partial_\mu U + U^{-1}A_\mu U \,, \qquad
U(x_0,\vec{x}) = \exp\left[-i\frac{\sigma_3}{2}(\alpha(\vec{x})+x_02\pi n/\beta)\right] \,, 
\label{eq:gltransdiscr}
\end{equation}
donde $n$ es un entero. Notar que no podemos hacer rotaciones sobre un eje que no sea el eje diagonal, ya que esto har\'{\i}a que $A_0$ fuera no diagonal. La dependencia en el tiempo debe ser lineal, ya que en caso contrario se generar\'{\i}a una dependencia temporal en $A_0$. La transformaci\'on gauge (\ref{eq:gltransdiscr}) verifica $U(x_0+\beta,\vec{x})=(-1)^n U(x_0,\vec{x})$. La fase~$(-1)^n$ se debe a la simetr\'{\i}a del centro del grupo gauge, que es ${\mathbb Z}(2)$. En componentes esta transformaci\'on es
\begin{eqnarray} 
A_0^{\prime 3}(\vec{x}) &=& A_0^3(\vec{x}) + 2\pi n/\beta 
\,, \nonumber \\
A_i^{\prime 1}(x_0,\vec{x}) &=& A_i^1\cos\chi  + A_i^2\sin\chi 
\,, \nonumber \\ 
A_i^{\prime 2}(x_0,\vec{x}) &=& -A_i^1\sin\chi + A_i^2\cos\chi 
\,, \nonumber \\
A_i^{\prime 3}(x_0,\vec{x}) &=& A_i^3 + \partial_i \alpha(\vec{x}) \,,
\label{eq:gl190}
\end{eqnarray}
donde $\chi(x_0,\vec{x}) = \alpha(\vec{x}) + x_02\pi n/\beta$. Notar que la primera ecuaci\'on es equivalente a $\widehat\nu^\prime = \widehat\nu + n$. 

En la figura \ref{fig:su2_1} vemos que cuando no hay fermiones los m\'{\i}nimos absolutos del potencial efectivo ocurren para valores enteros de $\beta\phi/2\pi$, y todos ellos son transformaciones gauge de $A_0=0$.  

Se puede comprobar que las combinaciones de campos $\vec{E}^2_{i \parallel}$, $\vec{E}^2_{i \perp}$, $\vec{B}^2_{i \parallel}$, y $\vec{B}^2_{i \perp}$ quedan invariantes bajo la transformaci\'on (\ref{eq:gl190}). Por tanto el sector glu\'onico de la  acci\'on efectiva es invariante gauge.

Al introducir fermiones en la teor\'{\i}a la situaci\'on se modifica ligeramente. En general hay m\'as transformaciones residuales permitidas en una teor\'{\i}a gauge pura SU($N_c$) que en una teor\'{\i}a SU($N_c$) con fermiones. Los fermiones rompen la simetr\'{\i}a del centro del grupo gauge, y la forma m\'as general de la transformaci\'on $U$ en este caso es
\begin{equation}
U(x_0,\vec{x}) = \exp\left[-i\frac{\sigma_3}{2}(\alpha(\vec{x})+x_0 4\pi n/\beta)\right] \,,
\label{eq:gltransdiscr2}
\end{equation} 
que es un subgrupo de la anterior. Esta transformaci\'on produce $\overline\nu^{\prime}=\overline\nu + n$, lo cual respeta la periodicidad de todas las funciones (notar que (\ref{eq:gltransdiscr}) no respeta la periodicidad de las funciones ${\cal L}_{0,q}(\overline\nu),\; f_q(\overline\nu) \;\textrm{y}\; h_q(\overline\nu)$). Como funci\'on de~$x_0$, la transformaci\'on gauge (\ref{eq:gltransdiscr2}) es estrictamente peri\'odica en $[0,\beta]$.

En la figura \ref{fig:su2_1} se observa c\'omo la inclusi\'on de fermiones da lugar a la rotura de la simetr\'{\i}a ${\mathbb Z}(2)$. Esta rotura se manifiesta en que los puntos $\beta\phi/2\pi=2n+1$ dejan de ser m\'{\i}nimos absolutos del potencial efectivo y pasan a ser puntos estacionarios. Los m\'{\i}nimos absolutos $\beta\phi/2\pi=2n$ son transformaciones gauge (\ref{eq:gltransdiscr2}) de $A_0=0$.


\subsection{Comparaci\'on con otros resultados}
\label{com_otros_result_su2}

Podemos comparar nuestros resultados en SU(2) en el sector glu\'onico con los que aparecen en \cite{oswald}, donde se calcula la acci\'on efectiva de una teor\'{\i}a de Yang-Mills SU(2) a altas temperaturas haciendo un desarrollo en derivadas covariantes. Este desarrollo es para configuraciones gauge estacionarias, es a todos los \'ordenes en $A_0$ y se calculan algunos t\'erminos con hasta cuatro derivadas espaciales. Aqu\'{\i} hemos calculado \'unicamente los \'ordenes m\'as bajos en $\widehat{D}_0$, pues as\'{\i} es como est\'a construido el desarrollo del heat kernel, y nuestras configuraciones son generales (no necesariamente est\'aticas). 

El resultado de ref.~\cite{oswald} presenta la estructura de ${\cal L}_{2,g}$ en (\ref{eq:jkl190}). Puesto que el nuestro no es estrictamente un desarrollo en $A_0$ (el loop de Polyakov no se ha desarrollado, con objeto de preservar invariancia gauge), no es posible hacer una comparaci\'on directa con \cite{oswald}. Sin embargo en nuestro tratamiento vemos que si la teor\'{\i}a fuera estacionaria, todos los t\'erminos de la forma $(\widehat{D}_0^nF_{\mu\nu})_{\parallel}$, $n \ge 1$, ser\'{\i}an cero, pues $A_0$ es diagonal. Esto quiere decir que nuestras funciones $f_{2,g}$ y $h_{2,g}$ no reciben ninguna contribuci\'on adicional m\'as all\'a de ${\cal L}_{2,g}$. Estas funciones coinciden con las correspondientes de \cite{oswald}. Por supuesto, el potencial efectivo es correcto a todos los \'ordenes en~$A_0$. 

Por otra parte nuestras funciones $f_{1,g}$ y $h_{1,g}$ contienen divergencias infrarrojas, mientras que en el c\'alculo de \cite{oswald} s\'olo $h_{1,g}$ es divergente. 



\section{Resultados en SU(3)}
\label{resultados_su3}
En esta secci\'on consideraremos espec\'{\i}ficamente el grupo gauge SU(3). Calcularemos el lagrangiano efectivo hasta t\'erminos con dimensi\'on de masa 4.

\subsection{Traza en espacio de color}
\label{traza_color_su3}

Usaremos como base de su(3) las matrices $t_a = \lambda_a/2i $, donde $\lambda_a$, $a=1,\cdots,8$, son las matrices de Gell-Mann
\begin{equation}
A_0 = A_0^a t_a = -\frac{i}{2}\lambda_a A_0^a \,,\quad
F_{\mu\nu} = F_{\mu\nu}^a t_a = -\frac{i}{2}\lambda_a F_{\mu\nu}^a 
\,,\quad \dots
\end{equation}
En esta base
\begin{equation}
[t_a,t_b]=f_{abc}t_c \,, \quad \tr(t_a t_b)=-\frac{1}{2}\delta_{ab} \,.
\end{equation}
Al igual que en la secci\'on \ref{resultados_su2}, elegimos el gauge de Polyakov, de modo que $A_0$ va a ser diagonal en la representaci\'on fundamental,
\begin{equation}
A_0 = -i\frac{\lambda_3}{2}\phi_3 - i\frac{\sqrt{3}}{2}\lambda_8\phi_8 \,.
\end{equation}
Los valores propios del loop de Polyakov en esta representaci\'on son
\begin{equation}
\omega_1 = \exp\left(i\frac{\beta}{2}(\phi_3+\phi_8)\right) \,, \quad
\omega_2 = \exp\left(i\frac{\beta}{2}(-\phi_3+\phi_8)\right)\,, \quad
\omega_3 = \exp\left(-i\beta\phi_8\right) \,,
\end{equation}
y si definimos las magnitudes $\nu_A$ mediante $\omega_A = \exp(i2\pi\nu_A)$, $A=1,2,3$, vamos a poder expresar el sector fermi\'onico del lagrangiano efectivo en t\'erminos de
\begin{equation}
\nu_1 = \frac{\beta}{4\pi}(\phi_3+\phi_8) \,, \quad
\nu_2 = \frac{\beta}{4\pi}(-\phi_3+\phi_8)  \,, \quad
\nu_3 = -\frac{\beta}{2\pi}\phi_8 \,.
\end{equation}
Tras calcular la traza en el espacio de color, obtenemos
\begin{equation}
{\cal L}_{0,q}(x) = -\frac{\pi^2 T^4 N_f}{12}
\left(-\frac{8}{5}+(1-4\overline \nu_1^2)^2+(1-4\overline
\nu_2^2)^2+(1-4\overline \nu_3^2)^2\right) \,,
\end{equation}
\begin{eqnarray}
\cL_{2,q}(x) &=& 
\frac{N_f}{24\pi^2}\left[\log\left(\frac{\mu}{4 \pi T}\right) 
- \frac{1}{2}\right] \vec E_i^2 
+ \frac{N_f}{24\pi^2}\log\left(\frac{\mu}{4 \pi T}\right) \vec B_i^2 
\nonumber \\
&& - \frac{N_f}{12 (4\pi)^2}\left(f^-(\nu_1)+f^-(\nu_2)\right)
\left((F_{\mu\nu}^1)^2+(F_{\mu\nu}^2)^2+(F_{\mu\nu}^3)^2 \right)
\nonumber \\
&& - \frac{N_f}{12 (4\pi)^2}\left(f^-(\nu_1)+f^-(\nu_3)\right)
\left((F_{\mu\nu}^4)^2+(F_{\mu\nu}^5)^2 \right)
\nonumber \\
&& - \frac{N_f}{12 (4\pi)^2}\left(f^-(\nu_2)+f^-(\nu_3)\right)
\left((F_{\mu\nu}^6)^2+(F_{\mu\nu}^7)^2 \right)
\nonumber \\
&& - \frac{N_f}{36 (4\pi)^2}\left(f^-(\nu_1)+f^-(\nu_2)+4 f^-(\nu_3)\right)
(F_{\mu\nu}^8)^2
\nonumber \\
&& - \frac{N_f}{6\sqrt{3}(4\pi)^2}\left(f^-(\nu_1)-f^-(\nu_2)\right)
F_{\mu\nu}^3 F_{\mu\nu}^8 \,,
\label{eq:45a}
\end{eqnarray}
donde
\begin{eqnarray}
   f^-(\nu) &=& \psi(\half+\bnu)+\psi(\half-\bnu)
\,,\qquad
   \bnu = \left(\nu+\frac{1}{2}\right) \;({\rm mod} \;1) -\frac{1}{2}
\,.
\end{eqnarray}

Para el sector glu\'onico debemos calcular la traza en la representaci\'on adjunta. En esta representaci\'on
\begin{equation}
(\widehat{A}_0)_{ab} = (A_0^c T_c)_{ab} = -f_{abc} A_0^c 
= -f_{ab3}\phi_3 -f_{ab8}\sqrt{3}\phi_8 \,,
\end{equation}
donde hemos hecho uso de $(T^c)_{ab}=f_{acb}=-f_{abc}$. De aqu\'{\i} se tiene que los valores propios del loop de Polyakov en la representaci\'on adjunta $\widehat\Omega = \exp(-\beta\widehat{A}_0 )$ son
\begin{equation}
1\,,\quad 1\,,\quad
\exp\left(\pm i\beta\phi_3\right)\,,\quad
\exp\left(\pm i\frac{\beta}{2}(\phi_3+3\phi_8)\right)\,,\quad
\exp\left(\pm i\frac{\beta}{2}(\phi_3-3\phi_8)\right) \,.
\end{equation}
El sector glu\'onico del lagrangiano efectivo se va a poder expresar en t\'erminos de los invariantes
\begin{equation}
\nu_{12} = \frac{\beta}{2\pi}\phi_3 \,, \quad
\nu_{31} = \frac{\beta}{4\pi}(\phi_3+3\phi_8) \,, \quad
\nu_{23} = \frac{\beta}{4\pi}(\phi_3-3\phi_8)\,. 
\end{equation}
Una vez que se calcula la traza en el espacio de color, se obtiene
\begin{eqnarray}
\cL_{0,g}(x) &=& \frac{4}{3}\pi^2 T^4 \left( -\frac{2}{15} + 
\hnu_{12}^2(1-\hnu_{12})^2 
+ \hnu_{31}^2(1-\hnu_{31})^2 
+ \hnu_{23}^2(1-\hnu_{23})^2 
 \right) \,,
\end{eqnarray}
\begin{eqnarray}
\cL_{2,g}(x) &=& 
-\frac{1}{(4\pi)^2}\left(11\log\left(\frac{\mu}{4 \pi T}\right) 
-\frac{1}{2}\right) \vec E_i^2 
-\frac{1}{(4\pi)^2}\left(11\log\left(\frac{\mu}{4 \pi T}\right) 
+\frac{1}{2}\right) \vec B_i^2  \nonumber \\
&& -\frac{T}{4\pi m}\left(\vec E_{i \perp}^2
+\frac{11}{12}\vec B_{i \perp}^2\right)\nonumber \\
&& +\frac{1}{(4\pi)^2}\frac{11}{12}
\left(f^+(0)+f^+(\nu_{12})+\frac{1}{2}f^+(\nu_{31})
+\frac{1}{2}f^+(\nu_{23}) \right)
\left( (F_{\mu\nu}^1)^2+(F_{\mu\nu}^2)^2 \right)
\nonumber \\
&& + \frac{1}{(4\pi)^2}\frac{11}{12}
\left(f^+(0)+\frac{1}{2}f^+(\nu_{12})+f^+(\nu_{31})
+\frac{1}{2}f^+(\nu_{23}) \right)
\left( (F_{\mu\nu}^4)^2+(F_{\mu\nu}^5)^2 \right)
\nonumber \\
&& + \frac{1}{(4\pi)^2} \frac{11}{12}
\left(f^+(0)+\frac{1}{2}f^+(\nu_{12})
+\frac{1}{2}f^+(\nu_{31})+f^+(\nu_{23}) \right)
\left( (F_{\mu\nu}^6)^2+(F_{\mu\nu}^7)^2 \right)
\nonumber \\
&& + \frac{1}{(4\pi)^2} \frac{11}{12}
\left(2 f^+(\nu_{12})+\frac{1}{2}f^+(\nu_{31})
+\frac{1}{2}f^+(\nu_{23}) \right) 
(F_{\mu\nu}^3)^2
\nonumber \\
&& + \frac{1}{(4\pi)^2} \frac{11}{8}
\left(f^+(\nu_{31})+f^+(\nu_{23}) \right)
(F_{\mu\nu}^8)^2
\nonumber \\
&& + \frac{1}{(4\pi)^2}\frac{11}{4\sqrt{3}} 
\left(f^+(\nu_{31})-f^+(\nu_{23})\right)
 F_{\mu\nu}^3 F_{\mu\nu}^8 \,,
\end{eqnarray}
donde
\begin{eqnarray}
   f^+(\nu) &=& \psi(\hnu)+\psi(1-\hnu) \, 
\quad (\nu\not\in\mathbb{Z}) \,, 
\quad    \hnu = \nu \; ({\rm mod} \, 1) \,,
\nonumber \\
   f^+(0) &=& - 2\gamma_E \,. 
\end{eqnarray}
Al igual que hicimos en SU(2), podemos considerar la descomposic\'on de los campos en la direcci\'on paralela y perpendicular a $\vec{A}_0$. La direcci\'on paralela $\vec{F}_{\mu\nu \parallel}$ da cuenta de las componentes 3 y 8. La direcci\'on perpendiculal $\vec{F}_{\mu\nu \perp}$ da cuenta de las componentes 1, 2, 4, 5, 6 y 7. Notar que el subespacio paralelo est\'a libre de divergencias infrarrojas. 

El nivel \'arbol renormalizado es
\begin{equation}
\cL_{\text{\text{\'arbol}}}(x) = \frac{1}{4 g^2(\mu)}\,\vec F_{\mu\nu}^2 \,.
\end{equation}

En las f\'ormulas hasta orden 4 en masa las componentes 1 y 2 juegan el mismo papel. Lo mismo ocurre con las componentes 4 y 5, y con las componentes 6 y 7. Hasta este orden, la estructura que encontramos es de cuatro planos bien definidos: el plano paralelo a $\vec{A}_0$, y tres planos transversales; esto es
\begin{eqnarray}
{\cal L}_{2,q}(x) + {\cal L}_{2,g}(x)  
=&& f_{12}(\phi_3,\phi_8)((E_i^1)^2+(E_i^2)^2)
+f_{45}(\phi_3,\phi_8)((E_i^4)^2+(E_i^5)^2)
\nonumber \\
&+&f_{67}(\phi_3,\phi_8)((E_i^6)^2+(E_i^7)^2) 
+f_{33}(\phi_3,\phi_8)(E_i^3)^2
+f_{88}(\phi_3,\phi_8)(E_i^8)^2
\nonumber \\
&+&f_{38}(\phi_3,\phi_8)(E_i^3 E_i^8)
\nonumber \\
&+&(\textrm{misma estructura para}\;B_i B_i)  \,.
\end{eqnarray}

Se puede comprobar que, eligiendo los generadores del \'algebra del grupo de manera conveniente, la estructura general que se obtiene para SU($N_c$) en nuestro desarrollo hasta orden 4 es de un plano paralelo con $N_c-1$ componentes, y $N_c(N_c-1)/2$ planos transversales, cada uno de ellos formado por dos componentes. Las divergencias infrarrojas solamente van a afectar a estos \'ultimos.

\begin{figure}[ttt]
\begin{center}
\includegraphics[width=8.5cm]{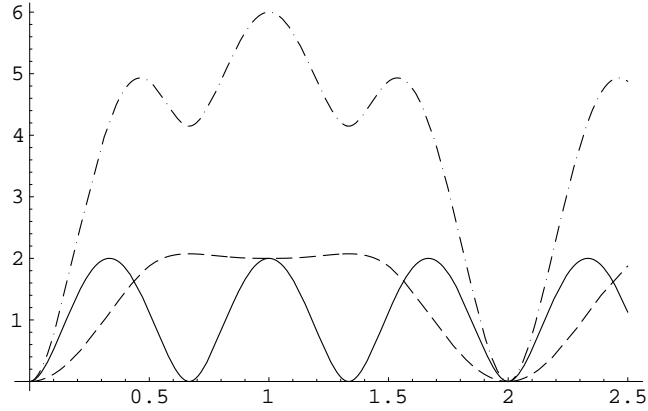}
\end{center}
\caption{Potencial efectivo a un loop para SU(3) como funci\'on de $\rho$. Se considera $\theta=0$ y se muestra el caso en que no hay fermiones (l\'{\i}nea continua), dos fermiones sin masa (puntos y rayas), y fermiones solamente (rayada). Se ha graficado $12\beta^4{\cal L}_0/\pi^2$ y se ha eliminado el t\'ermino constante.}
\label{fig:su3_1}
\end{figure}

\begin{figure}[ttt]
\begin{center}
\includegraphics[width=8.5cm]{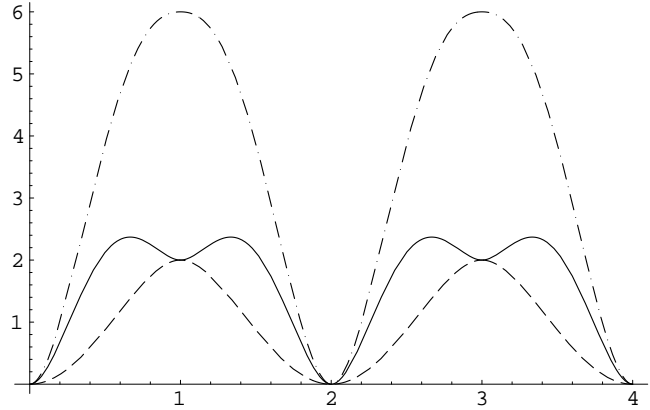}
\end{center}
\caption{Potencial efectivo a un loop para SU(3) como en fig.~{\ref{fig:su3_1}}, pero en la direcci\'on de $\lambda_3$. Se ha graficado como funci\'on de~$\theta$ y se considera $\rho=0$.} 
\label{fig:su3_2}
\end{figure}

\begin{figure}[ttt]
\begin{center}
\includegraphics[width=8.5cm]{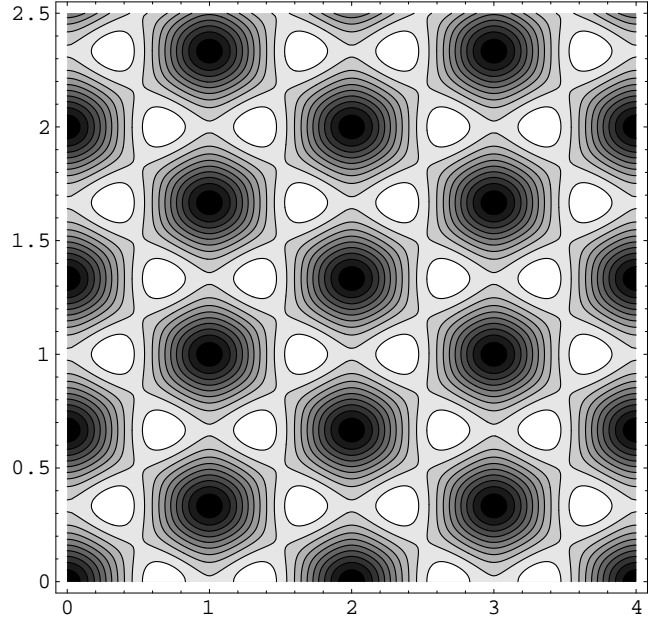}
\end{center}
\caption{Gr\'afico de contorno del potencial efectivo a un loop para SU(3) como funci\'on de $\theta$ (eje horizontal) y $\rho$ (eje vertical) para una teor\'{\i}a gauge pura.}
\label{fig:su3_3}
\end{figure}

\begin{figure}[tbp]
\begin{center}
\includegraphics[width=8.5cm]{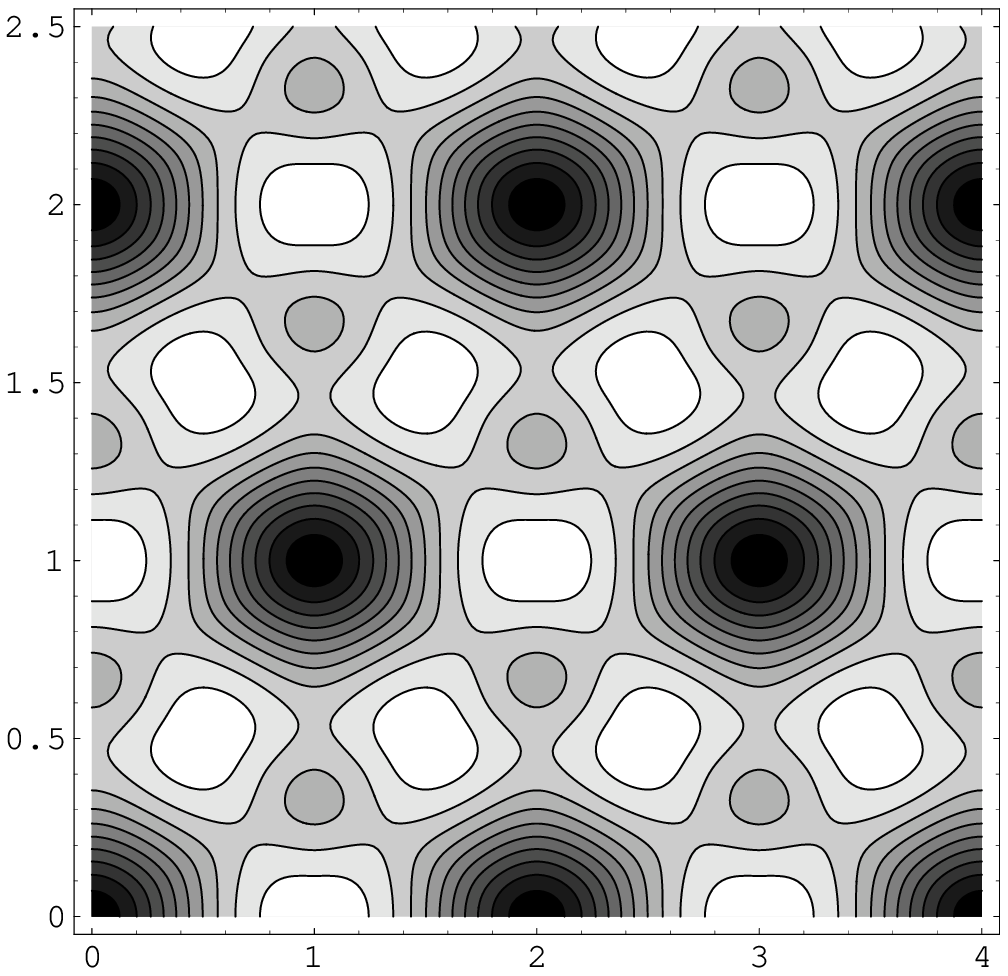}
\end{center}
\caption{Igual que fig.~\ref{fig:su3_3}, pero con dos fermiones sin masa.}
\label{fig:su3_4}
\end{figure}

\subsection{Invariancia gauge del resultado}
\label{inv_gauge_su3}

Con el fin de graficar las funciones definimos las magnitudes $\theta$ y $\rho$ como
\begin{equation}
\theta = \frac{\beta}{2\pi}\phi_3 \,, \qquad
\rho = \frac{\beta}{2\pi}\phi_8 \,.
\end{equation}
En la figura \ref{fig:su3_1} se muestra el potencial efectivo en la direcci\'on de $\lambda_8$. Cuando no hay fermiones los \'unicos m\'{\i}nimos del potencial en esta direcci\'on ocurren en los puntos $\rho=2n/3$ que son justamente transformaciones gauge de $\rho=0$. En la figura \ref{fig:su3_2} aparece el potencial efectivo en la direcci\'on de $\lambda_3$. Los m\'{\i}nimos absolutos son nuevamente transformaciones gauge de $\theta=0$, pero aparecen adem\'as m\'{\i}nimos locales en $\theta=2n+1$. En una gr\'afica del potencial efectivo en dos dimensiones (figura \ref{fig:su3_3}) se puede observar que estos m\'{\i}nimos locales en realidad son cr\'ateres que caen hacia m\'{\i}nimos absolutos en $(\theta,\rho)=(2n+1,1/3)$. En todos estos m\'{\i}nimos la matriz $\Omega=\exp(-\beta A_0)$ tiene los mismos valores propios, de modo que todos ellos son transformaciones gauge de $\Omega=1$.

Como se comenta en el ap\'endice \ref{app:gauge} la introducci\'on de fermiones rompe la simetr\'{\i}a del centro del grupo gauge. La rotura de esta simetr\'{\i}a se manifiesta en la aparici\'on de m\'{\i}nimos locales. En la figura \ref{fig:su3_1} podemos observar que el m\'{\i}nimo absoluto en $\rho=2/3$ para la teor\'{\i}a sin fermiones se transforma, con la inclusi\'on de \'estos, en un m\'{\i}nimo local. Esto es as\'{\i} ya que el m\'{\i}nimo absoluto de la parte gauge del potencial efectivo coincide con el m\'aximo de la parte fermi\'onica. En general, como podemos observar en las figuras \ref{fig:su3_3} y \ref{fig:su3_4}, cada m\'{\i}nimo local de la teor\'{\i}a gauge con fermiones se corresponde exactamente con un m\'{\i}nimo absoluto de la teor\'{\i}a gauge pura.
\begin{figure}[tbp]
\begin{center}
\includegraphics[width=8.5cm]{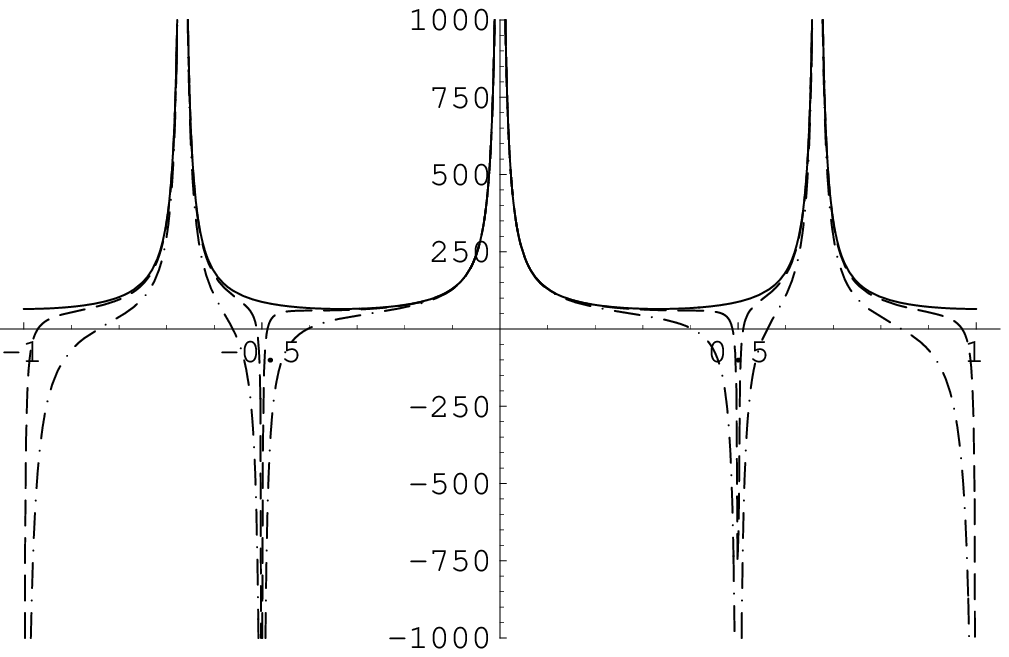}
\end{center}
\caption{Gr\'afico de $-96\pi^2 f_{45}$ como funci\'on de $\rho$. Se considera $\theta=0$ y se muestra el caso en que no hay fermiones (l\'{\i}nea continua), un fermi\'on sin masa (rayada), y ocho fermiones (puntos y rayas). Se han eliminado los t\'erminos constantes y ciertas divergencias que surgen al tomar~$\theta=0$.}
\label{fig:su3_5}
\end{figure}
Con el fin de ilustrar el comportamiento de las funciones que aparecen en ${\cal L}_2$, en la figura \ref{fig:su3_5} se muestra la funci\'on~$f_{45}$ en la direcci\'on de $\lambda_8$. La parte bos\'onica de la funci\'on presenta singularidades en $\rho=2n/3$ lo cual es debido a la contribuci\'on del modo cero. El comportamiento en la direcci\'on de $\lambda_3$ es parecido al de la figura \ref{fig:su2_2}.

\subsection{Comparaci\'on con otros resultados}
\label{result_comp_su3}

En \cite{chapmanlargo}, al igual que en \cite{chapmancorto}, se calcula la acci\'on efectiva a un loop de una teor\'{\i}a de Yang-Mills SU($N_c$) haciendo un desarrollo en derivadas covariantes. El resultado no es covariante gauge pues se considera un desarrollo en potencias de $A_0(\vec{x})$. Como consecuencia de ello el potencial efectivo que se obtiene no es peri\'odico y s\'olo se aproxima al exacto cuando $A_0\rightarrow 0$. De todas formas en \cite{chapmanlargo} se considera el potencial efectivo exacto, que es conocido y ha sido calculado por nosotros, y se hace un estudio en el caso espec\'{\i}fico de una teor\'{\i}a gauge SU(3) incluyendo fermiones. Hemos comprobado que sus resultados coinciden con los nuestros.

\section{Conclusiones}
\label{conclusiones_aeQCD}

En este cap\'{\i}tulo se ha hecho un estudio de la acci\'on efectiva de QCD a un loop a temperatura finita en la regi\'on de inter\'es fenomenol\'ogico correspondiente a la fase de plasma de quarks y gluones. Para tal fin hemos usado la t\'ecnica del heat kernel del cap\'{\i}tulo~\ref{heat_kernel}, que nos ha permitido calcular el determinante fermi\'onico y el determinante bos\'onico corres\-pondiente a fluctuaciones glu\'onicas cu\'anticas en torno a un background cl\'asico (es el conocido como M\'etodo del Campo de Fondo). 

El desarrollo del heat kernel se corresponde en este caso con un desarrollo en derivadas, organizado de un modo muy eficiente. Hemos conseguido reproducir resultados parciales previos, y extenderlos hasta orden $T^{-2}$ incluyendo los efectos del loop de Polyakov, para un grupo gauge general SU($N_c$). Se ha calculado la acci\'on de la teor\'{\i}a efectiva dimensional\-mente reducida hasta ese mismo orden. Finalmente se han particularizado las f\'ormulas para los grupos gauge SU(2) y SU(3), lo cual ha permitido comparar con trabajos previos. 

Un punto de especial relevancia es la invariancia gauge de nuestros resultados.
Hemos estudiado la invariancia frente a la simetr\'{\i}a del centro ${\mathbb Z}(N_c)$ del grupo gauge SU($N_c$) en la teor\'{\i}a sin fermiones, y se ha estudiado expl\'{\i}citamente el mecanismo por el cual los fermiones rompen esta simetr\'{\i}a del centro.

\chapter{Efectos no perturbativos por encima de la transici\'on de fase}
\label{condensados}

\section{Introducci\'on}
\label{introduction_nopert}

El loop de Polyakov juega un papel te\'orico muy importante en QCD a temperatura finita. Representa el propagador de un quark est\'atico test y por tanto es crucial para entender el mecanismo de la transici\'on confinamiento-desconfinamiento. En \cite{Kuti:1980gh,McLerran:1981pb} se encuentra su relaci\'on con la energ\'{\i}a libre de un quark pesado, de tal modo que un valor esperado nulo del loop de Polyakov en quenched QCD indica la fase de confinamiento. La simetr\'{\i}a global ${\mathbb Z}(N_c)$ se encuentra espont\'aneamente rota en la fase de desconfinamiento \cite{Polyakov:1978vu}. El loop de Polyakov constituye un par\'ametro de orden natural para esa transici\'on de fase; bajo transformaciones de gauge peri\'odicas $L$ es un objeto invariante, pero bajo una transformaci\'on de 't Hooft adquiere un factor, que es un elemento del centro del grupo gauge. Diversas teor\'{\i}as efectivas para el loop de Polyakov han sido propuestas en \cite{Pisarski:2000eq}. (Para un an\'alisis detallado ver, por ejemplo, \cite{notes_pis}).  


Al comienzo de los a\~nos ochenta, el c\'alculo perturbativo del loop de Polyakov hasta segundo orden (NLO) fue hecho por Gava y Jengo~\cite{Gava:1981qd}.  Estos resultados muestran que a temperaturas suficientemente grandes el loop de Polyakov renormalizado se aproxima a uno por encima.\footnote{El valor esperado del loop de Polyakov desnudo se anula en el l\'{\i}mite al continuo en cualquier fase.} No se han hecho muchos progresos desde este primer resultado. Actualmente no existen c\'alculos perturbativos del loop de Polyakov m\'as all\'a de NLO. Tal y como se menciona en~\cite{Gava:1981qd}, un c\'alculo directo conducir\'{\i}a a la aparici\'on de un gran n\'umero de diagramas de Feynman debido a las divergencias infrarrojas~\cite{Linde:1980ts}. En este cap\'{\i}tulo discutiremos una aproximaci\'on diferente, relacionada con la t\'ecnica de reducci\'on dimensional.

Desde el punto de vista no perturbativo, el loop de Polyakov desnudo ha sido frecuentemente estudiado en c\'alculos num\'ericos de teor\'{\i}as gauge en el ret\'{\i}culo. No obstante, s\'olo recientemente se ha conseguido una definici\'on conveniente del loop de Polyakov renorma\-lizado. El m\'etodo introducido en ref.~\cite{Kaczmarek:2002mc} para QCD quenched permite calcular el loop de Polyakov a partir del potencial quark-antiquark a temperatura finita, obtenido de la funci\'on de correlaci\'on de dos loops de Polyakov separados. La comparaci\'on con el potencial a temperatura cero para separaciones peque\~nas permite una determinaci\'on muy precisa de la autoenerg\'{\i}a del quark, que debe ser extra\'{\i}da. 

Las temperaturas grandes est\'an relacionadas con regiones cinem\'aticas donde se manifiesta la rotura de la simetr\'{\i}a Lorentz, y  se corresponden con momentos eucl\'{\i}deos grandes para una teor\'{\i}a cu\'antica de campos a temperatura cero. En regularizaci\'on dimensional en el esquema $\MS$ se encuentra que a una temperatura dada~$T$ le corresponde una escala eucl\'{\i}dea $\mu\sim 4\pi T$ \cite{Huang:1994cu}, de modo que para $T_c = 270\;\textrm{MeV}$ se tiene $\mu = 3\;\textrm{GeV}$. En este r\'egimen es de esperar que las ideas del desarrollo en producto de operadores (en ingl\'es Operator Product Expansion, OPE) se puedan aplicar, y m\'as espec\'{\i}ficamente a temperaturas no tan grandes los condensados y las correcciones en potencias de la temperatura deber\'{\i}an de jugar un papel importante. En realidad, siguiendo algunas sugerencias antiguas~\cite{Lavelle:1988eg}, requisitos fenomenol\'ogicos~\cite{Chetyrkin:1998yr}, estudios te\'oricos~\cite{Kondo:2001nq} y an\'alisis en el ret\'{\i}culo~\cite{Boucaud:2001st,RuizArriola:2004en,Boucaud:2005rm} hay actualmente una evidencia creciente de que el condensado invariante BRST de orden m\'as bajo es de dimensi\'on 2. Este condensado es en general no local, pero en el gauge de Landau se convierte en un operador local~$\langle A_{\mu,a}^2\rangle$, donde $A_{\mu,a}$ es el campo del glu\'on. El condensado $\langle A_{0,a}^2\rangle$ tambi\'en aparece como un par\'ametro en el c\'alculo de la presi\'on a temperatura finita~\cite{Kajantie:2000iz}.

El loop de Polyakov est\'a estrechamente relacionado con el valor esperado de
$A_{0,a}^2$ (como veremos, el resultado perturbativo a NLO se puede obtener de
esta manera), de modo que las contribuciones del condensado a esta magnitud tendr\'an un impacto inmediato sobre el loop de Polyakov. En este cap\'{\i}tulo estudiaremos la existencia de contribuciones no perturbativas a este condensado glu\'onico. La situaci\'on es similar a lo que ocurre con el potencial quark-antiquark en QCD a temperatura cero, como funci\'on de la separaci\'on del quark y el antiquark. En esta caso, la teor\'{\i}a de perturbaciones describe bien la regi\'on de cortas distancias, donde la teor\'{\i}a es d\'ebilmente interactuante y el intercambio de un glu\'on produce un potencial tipo Coulomb. A distancias grandes surge el confinamiento y los datos del ret\'{\i}culo sugieren un potencial de tipo lineal~\cite{Bali:2000gf}.

En la secci\'on~\ref{comparison_lattice_data} se analizar\'an los datos en el
ret\'{\i}culo del loop de Polyakov en base a estas ideas.

Este cap\'{\i}tulo est\'a basado en las referencias~\cite{Megias:2005ve,Megias:2007pq}.

\section{Loop de Polyakov perturbativo}
\label{sec:pert_PL}

Con objeto de incluir posteriormente posibles contribuciones provenientes de condensados, trataremos de reproducir en esta secci\'on el resultado perturbativo a orden m\'as bajo para el loop de Polyakov  mendiante la t\'ecnica de reducci\'on dimensional. Adem\'as, esto nos permitir\'a discutir algunas propiedades de las contribuciones perturbativas de orden superior.

\subsection{Resultados perturbativos}

El (valor esperado del) loop de Polyakov se define como
\begin{equation}
L(T)  = \left\langle \frac{1}{N_c}\tr_c\, {\mathcal T} \left( 
e^{i g\int_0^{1/T} d x_0 A_0 (\vec{x} , x_0) }\right) \right\rangle \,,
\label{eq:defPL}
\end{equation}
donde $\langle~\rangle$ indica valor esperado en el vac\'{\i}o, $\tr_c$ es la
traza de color (en representaci\'on fundamental), y ${\mathcal T}$ indica
ordenaci\'on a lo largo del camino de integraci\'on. $A_0$ es la componente
temporal del campo glu\'onico (en tiempo eucl\'{\i}deo). El campo gauge
$A_0(x)$ es un elemento del \'algebra de Lie de SU($N_c$), y puede ser
representado como $A_0=\sum_a T_a A_{0,a}$, donde $T_a$ son los generadores
herm\'{\i}ticos del \'algebra de Lie de SU($N_c$) en la representaci\'on
fundamental. En lo sucesivo consideraremos la normalizaci\'on est\'andar $\tr(T_a T_b)=\delta_{ab}/2$. 

Al ser un operador compuesto, el loop de Polyakov es subceptible de ser
renormalizado. En
refs.~\cite{Polyakov:1980ca,Arefeva:1980zd,Dotsenko:1979wb,Gervais:1979fv} se
estudia la renormalizabilidad del loop de Polyakov en el contexto de
teor\'{\i}a de perturbaciones, donde se muestra el hecho de que se puede
renormalizar perturbativamente, sin mezcla con otros operadores. El c\'alculo
perturbativo de $L(T)$ en gluodin\'amica pura a temperaturas altas fue realizado a comienzos de los a\~nos ochenta por Gava y Jengo~\cite{Gava:1981qd}. Tras incluir efectos de polarizaci\'on de vac\'{\i}o a temperatura finita a trav\'es de la inserci\'on de la masa de Debye, el t\'ermino de orden m\'as bajo resulta ser el ${\cal O}(g^3)$, en lugar del que en un principio cabr\'{\i}a esperar ${\cal O}(g^2)$. Este c\'alculo se hizo en el gauge de Landau hasta NLO (${\cal O}(g^4)$). El resulado es 
\begin{equation}
L(T)= 1+\frac{1}{16 \pi}\frac{N_c^2-1}{N_c}g^2\frac{m_D}{T} +
\frac{N_c^2-1}{32\pi^2}g^4\left(\log\frac{m_D}{2T}+\frac{3}{4}\right)
+{\cal O}(g^5) \,.
\label{eq:Gava_Jengo}
\end{equation}
Este resultado es muy antiguo, y hoy en d\'{\i}a no se dispone de c\'alculos a \'ordenes superiores. La masa de Debye $m_D$ controla el apantallamiento de los modos cromoel\'ectricos en el plasma, y a un loop se escribe~\cite{Nadkarni:1982kb}
\begin{equation}
m_D=gT(N_c/3+N_f/6)^{1/2}\,.
\label{eq:m_debye}
\end{equation}
La dependencia en temperatura de la constante de acoplamiento $g$ se obtiene del an\'alisis est\'andar del grupo de renormalizaci\'on, y es de esperar que (\ref{eq:Gava_Jengo}) constituya una buena aproximaci\'on a temperatura suficientemente alta. Notemos que $L(T)$ se hace mayor que~1, lo cual implica que el loop de Polyakov renormalizado no es una matriz unimodular. 

\subsection{Reducci\'on dimensional}
\label{rd}

En la secci\'on~\ref{reduccion_dimensional} se obtuvo la acci\'on de la
teor\'{\i}a efectiva dimensionalmente reducida de QCD a un loop y en el gauge
de Landau.
Esta teor\'{\i}a queda descrita por la acci\'on tridimensional~$\int d^3x {\cal L}_3(\vec{x})$ \cite{Megias:2003ui,chapmancorto,Shaposhnikov:1996th,Huang:1994cu}, donde
\begin{eqnarray}
T{\cal L}_3(\vec{x}) &=&
m_D^2\tr(A_0^2)
+\frac{g^4(\mu)}{4\pi^2 }(\tr(A_0^2))^2
+\frac{g^4(\mu)}{12\pi^2 }(N_c-N_f)\tr(A_0^4)
\nonumber\\
&& +\frac{g^2(\mu)}{g_E^2(T)}\tr([D_i,A_0]^2)
+\frac{g^2(\mu)}{g_M^2(T)}\frac{1}{2}\tr(F_{ij}^2)
+T\delta {\cal L}_3  \,.
\label{eq:red_dim}
\end{eqnarray}
$g(\mu)$ es la constante de acoplamiento de QCD en el esquema $\overline{\text{MS}}$ (usada tambi\'en en la f\'ormula del loop de Polyakov~(\ref{eq:Gava_Jengo}) y en la masa de Debye~(\ref{eq:m_debye}))
\begin{eqnarray}
\frac{1}{ g^2(\mu)} &=& 
2\beta_0\log(\mu/\Lambda_{\overline{\text{MS}}}) \,,\qquad
\beta_0 = (11N_c/3-2N_f/3)/(4\pi)^2
\label{eq:6}
\end{eqnarray}
y las constantes de acoplamiento cromoel\'ectricas y cromomagn\'eticas vienen dadas por ec.~(\ref{eq:gE_gM}).
El t\'ermino restante $\delta {\cal L}_3$ es no renormalizable y contiene operadores de dimensi\'on~6 o mayores (ver ec.~(\ref{eq:Lprime6})). Adem\'as existen t\'erminos que contribuyen m\'as all\'a de un loop y t\'erminos constantes (indenpendientes de los campos) que son relevantes para el c\'alculo de la presi\'on.

Para obtener el loop de Polyakov a orden m\'as bajo necesitaremos \'unicamente los t\'erminos de masa y de energ\'{\i}a cin\'etica del campo~$A_0$ (t\'erminos primero y cuarto respectivamente en ec.~(\ref{eq:red_dim})). Para simplificar la notaci\'on, en el resto del cap\'{\i}tulo trabajaremos con un campo $A_0$ reescalado
\begin{equation}
 A_0(\vec{x}) = \frac{g(\mu)}{g_E(T)} A_0^{\overline{\textrm{MS}}}(\vec{x})\,,
\end{equation}
donde $A_0^{\overline{\textrm{MS}}}$ es el campo glu\'onico que aparece en f\'ormulas previas. A todos los efectos, el uso de la masa de Debye y la f\'ormula del loop de Polyakov ec.~(\ref{eq:defPL}) que depende del producto de $g A_0$, es equivalente al uso del nuevo campo $A_0$ junto con $g_E(T)$ como constante de acoplamiento. A partir de ahora denotaremos esta constante como $g(T)$ o simplemente $g$,
\begin{eqnarray}
{\cal L}_3(\vec{x}) &=& \frac{m_D^2}{T}\tr(A_0^2)
+\frac{1}{T}\tr([D_i,A_0]^2) +\cdots\,,
\label{eq:L3red}
 \\
\frac{1}{ g^2(T)} &=& 
2\beta_0\log(T/\Lambda_E) \,,
\nonumber
\end{eqnarray}
con
\begin{equation}
\Lambda_E=
\frac{\Lambda_{\overline{\text{MS}}}}{4\pi}
\exp\left(\gamma_E-\frac{N_c+8N_f(\log 2-1/4)}{22N_c-4N_f}\right) \,.
\label{eq:LambdaE}
\end{equation}

En el c\'alculo de la presi\'on de QCD se puede fijar el gauge de cualquier forma para integrar los modos no estacionarios. Por esta raz\'on se suelen utilizar los gauges covariantes, pues los c\'alculos resultan m\'as f\'aciles en estos gauges. Para el loop de Polyakov la situaci\'on es diferente, pues los gauges est\'aticos resultan m\'as convenientes~\cite{Nadkarni:1982kb}. Tal y como se muestra en el ap\'endice~\ref{app:gauge}, un gauge est\'atico es aquel en el que $\partial_0 A_0=0$, y no implica p\'erdida de generalidad ya que este gauge siempre existe. En el gauge est\'atico la ec.~(\ref{eq:defPL}) se escribe
\begin{equation}
L(T) = \frac{1}{N_c}\left\langle \tr\, e^{ig A_0(\vec{x})/T } \right\rangle
\,.
\label{eq:PL_gauge_static}
\end{equation}
Notar que $L$ \'unicamente depende de los modos estacionarios de $A_0$, de modo que si integramos los modos no estacionarios no existir\'a p\'erdida de informaci\'on en el loop de Polyakov. El modo estacionario $A_0(\vec{x})$ coincide con el logaritmo de loop de Polyakov \'unicamente en el gauge est\'atico. Por desgracia, el resultado perturbativo de ${\cal L}_3(\vec{x})$, ec.~(\ref{eq:L3red}), s\'olo se conoce en los gauges covariantes. Por tanto, en un gauge covariante la acci\'on efectiva de los modos estacionarios resulta insuficiente para obtener los valores esperados del loop de Polyakov. El uso del modo estacionario en~(\ref{eq:PL_gauge_static}) equivale a eliminar el operador de ordenaci\'on a lo largo del camino de integraci\'on ${\mathcal{T}}$ en la definici\'on del loop de Polyakov~(\ref{eq:defPL}), dando lugar a una dependencia en el gauge. No obstante, como mostraremos en la subsecci\'on siguiente, la dependencia en el gauge \'unicamente afectar\'a m\'as all\'a de NLO, y seremos capaces de reproducir los dos t\'erminos de~(\ref{eq:Gava_Jengo}) mediante el uso de las f\'ormulas de~\cite{Kajantie:2003ax} para la densidad de energ\'{\i}a de vac\'{\i}o.

Si hacemos un desarrollo en serie de $L(T)$ en ec.~(\ref{eq:PL_gauge_static}), se obtiene
\begin{equation}
L(T)=
1
-\frac{g^2}{2 T^2}\frac{1}{N_c}\langle\tr(A_0^2)\rangle
+\frac{g^4}{24 T^4}\frac{1}{N_c}\langle\tr(A_0^4)\rangle
+\cdots \,.
\label{eq:PL_des_serie}
\end{equation}
En esta f\'ormula hemos usado que $\tr(A_0)$ es cero. Es de esperar que el resto de \'ordenes impares en el campo glu\'onico se anulen debido a la simetr\'{\i}a de conjugaci\'on de QCD, $A_\mu(x)\to-A^T_\mu(x)$. La contribuci\'on de orden m\'as bajo $\langle\tr(A_0^2)\rangle$ tiene dimensiones de masa al cuadrado, de modo que esta contribuci\'on no existir\'{\i}a en un c\'alculo a temperatura cero. A temperatura finita debe de escalar como $T^2$ (m\'odulo correcciones radiativas con una d\'ebil dependencia en $T$, que incluyen el running de la constante de acoplamiento y dimensiones an\'omalas).

Sea $D_{00}(\vec{k})\delta_{ab}$ la componente temporal del propagador en espacio de momentos para los campos gauge normalizados can\'onicamente $T^{-1/2}A_{0,a}(\vec{x})$. Integrando el propagador obtenemos el valor esperado de los campos
\begin{equation}
\langle A_{0,a}^2\rangle
=(N_c^2-1)T\int\frac{d^3k}{(2\pi)^3}D_{00}(\vec{k}) \,.
\label{eq:val_esp_A0}
\end{equation}
A orden m\'as bajo en teor\'{\i}a de perturbaciones, el propagador se escribe
\begin{equation}
D^{\text{Pert}}_{00}(\vec{k})=
\frac{1}{\vec{k}^2+m_D^2} \,.
\label{eq:prop_pert}
\end{equation}
Si introducimos~(\ref{eq:prop_pert}) en (\ref{eq:val_esp_A0}) obtenemos la contribuci\'on perturbativa de orden m\'as bajo para el condensado glu\'onico de dimensi\'on dos (hacemos uso de las reglas de regularizaci\'on dimensional)
\begin{equation}
\langle A_{0,a}^2\rangle^{\text{Pert}} = -(N_c^2-1)\frac{T m_D}{4\pi} \,.
\end{equation}
Este resultado introducido en ec.~(\ref{eq:PL_des_serie}) (y usando que $\tr(A_0^2)=A_{0,a}^2/2$) reproduce el valor perturbativo de $L(T)$ hasta orden ${\cal O}(g^3)$.

\begin{figure}[tbp]
\begin{center}
\epsfig{figure=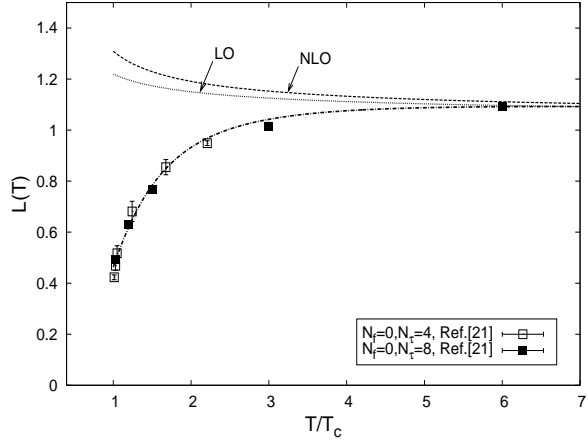,height=6cm,width=8cm}
\end{center}
\caption{Dependencia en temperatura del loop de Polyakov renormalizado en
 gluo\-din\'amica ($N_c=3$). Los datos en el ret\'{\i}culo son de ref.~\cite{Kaczmarek:2002mc}. Para comparar, se muestran los resultados perturbativos LO y NLO de ec.~(\ref{eq:Gava_Jengo}). La curva es un ajuste del par\'ametro $b$ en ec.~(\ref{eq:adjust_pert}) con los datos del ret\'{\i}culo.}
\label{fig:PL1}
\end{figure}
En la figura~\ref{fig:PL1} se compara el valor perturbativo de $L(T)$ en
ec.~(\ref{eq:Gava_Jengo}) con datos del ret\'{\i}culo obtenidos recientemente
en gluodin\'amica pura y $N_c=3$~\cite{Kaczmarek:2002mc}. Podemos observar que
en la regi\'on de temperatura alta, $T$ pr\'oximo a $6T_c$, los valores del ret\'{\i}culo para $L(T)$ son mayores que 1, tal y como predice el c\'alculo perturbativo. Adem\'as el valor num\'erico en esta regi\'on es consistente con teor\'{\i}a de perturbaciones. Este acuerdo desaparece r\'apidamente a medida que nos aproximamos a la temperatura cr\'{\i}tica: los datos del ret\'{\i}culo decrecen hasta producir una transici\'on de fase (en este caso de primer orden), mientras que la curva perturbativa crece ligeramente. Como es de esperar, el resultado perturbativo es lentamente variable con la temperatura, pues esta variaci\'on procede de correcciones radiativas logar\'{\i}tmicas.

\subsection{Resultados perturbativos a \'ordenes superiores}
\label{rpos}

En la secci\'on~(\ref{rd}) discutimos la t\'ecnica de reducci\'on dimensional y llegamos a obtener el valor perturbativo de $L(T)$ a orden m\'as bajo en teor\'{\i}a de perturbaciones ${\cal O}(g^3)$. En este apartado vamos a hacer una discusi\'on de las contribuciones de \'ordenes superiores al loop de Polyakov. 

El lagrangiano renormalizable tridimensional tiene la siguiente estructura
\begin{eqnarray}
{\cal L}_3^{\text{ren}}(\vec{x})&=&
\frac{1}{2}\tr(F_{ij}^2)+\tr([D_i,\A_0]^2)+m_3^2\tr(\A_0^2)
+\lambda_1(\tr(\A_0^2))^2+\lambda_2\tr(\A_0^4) \,, \nonumber \\
D_i &=& \partial_i-ig_3\A_i \,.
\end{eqnarray}
con $\A_\mu\sim T^{-1/2}A_\mu$, $m_3\sim gT$, $g_3 \sim T^{1/2}g$, y
$\lambda_1\sim\lambda_2\sim g^4T$. Para $N_c=2$ y $N_c=3$ el t\'ermino $\lambda_2$ es redundante y podemos considerar $\lambda_2=0$. 

La densidad de energ\'{\i}a de vac\'{\i}o de esta teor\'{\i}a, $\epsilon(g_3,m_3,\lambda_1)$, ha sido calculada hasta cua\-tro loops en~\cite{Kajantie:2003ax}, con $g_3$, $m_3$ y $\lambda_1$ como par\'ametros independientes. Esto permite calcular los condensados $\langle A_0^2 \rangle$ y $\langle A_0^4\rangle$ tomando derivadas de $\epsilon$ con respecto a $m_3^2$ y $\lambda_1$ respectivamente, lo cual va a permitir obtener sucesivos \'ordenes perturbativos del loop de Polyakov mediante ec.~(\ref{eq:PL_des_serie}).

La estructura general de la densidad de energ\'{\i}a de vac\'{\i}o es~\cite{Kajantie:2003ax}
\begin{equation}
\epsilon(g_3,m_3,\lambda_1)=\sum_{\ell\ge 1}
\sum_{k=0}^{\ell-1} f_{\ell k}\,m_3^{4-\ell} g_3^{2k}\lambda_1^{\ell-k-1} \,,
\label{eq:energia_vacio}
\end{equation}
donde $\ell$ indica el n\'umero de loops y los coeficientes $f_{\ell k}$ dependen logar\'{\i}tmicamente de $m_3$. Para las magnitudes que aparecen en~(\ref{eq:PL_des_serie}) se tiene
\begin{eqnarray}
\frac{g^2}{T^2}\langle\tr(A_0^2)\rangle
&\sim& 
\frac{g^2}{T}\frac{\partial \epsilon(g_3,m_3,\lambda_1)}{\partial m_3^2}
\sim
\sum_{\ell\ge 1}\sum_{n=\ell+2}^{3\ell}g^n \,,
\nonumber \\
\frac{g^4}{T^4}\langle\tr(A_0^4)\rangle
&\sim& 
\frac{g^4}{T^2}\frac{\partial \epsilon(g_3,m_3,\lambda_1)}{\partial \lambda_1}
\sim
\sum_{\ell\ge 2}\sum_{n=\ell+4}^{3\ell}g^n  \,.
\label{eq:der_evacio}
\end{eqnarray}
Teniendo en cuenta que en ref.~\cite{Kajantie:2003ax} se calcula la densidad de energ\'{\i}a de vac\'{\i}o hasta 4 loops, la primera contribuci\'on a $L(T)$ que no se tendr\'{\i}a en cuenta ser\'{\i}a ${\cal O}(g^7)$, correspondiente a $\ell=5$ en el t\'ermino~$\langle \tr{(A_0^2)}\rangle$. La contribuci\'on de orden m\'as bajo de~$\langle \tr{(A_0^4)}\rangle$ a 5 loops es ${\cal O}(g^9)$, y la primera contribuci\'on de~$\langle \tr{(A_0^6)}\rangle$, no disponible en el c\'alculo, comenzar\'{\i}a en~${\cal O}(g^9)$ a 3 loops. Esto quiere decir que en principio, con el resultado de~\cite{Kajantie:2003ax} se podr\'{\i}a extender el resultado perturbativo de~$L(T)$ hasta~${\cal O}(g^6)$. Desafortunadamente las relaciones que conectan los par\'ametros de la teor\'{\i}a dimensionalmente reducida $m_3$, $g_3$ y $\lambda_1$ con los correspondientes de QCD en cuatro dimensiones solamente se conocen en gauges covariantes, para los cuales la relaci\'on~(\ref{eq:PL_gauge_static}) no se cumple. En particular, la raz\'on~$g(\mu)/g_E(T)$ tiene una dependencia en el gauge que comienza en~${\cal O}(g^2)$ para las contribuciones de dos loops, lo cual dar\'{\i}a lugar a una dependencia en el gauge a~${\cal O}(g^5)$ en $L(T)$.

Deber\'{\i}amos estudiar asimismo la contribuci\'on de los t\'erminos no renormalizables~$\delta {\cal L}_3$. Los t\'erminos de orden m\'as bajo de ese tipo son~\cite{Megias:2003ui,chapmancorto}
\begin{equation}
\delta{\cal L}_3=
\frac{g^2}{T^2}\tr([D_i,F_{\mu\nu}]^2)
+ \frac{g^3}{T^{3/2}}\tr(F_{\mu\nu}^3)
+ \frac{g^4}{T}\tr(A_0^2 F_{\mu\nu}^2) \,.
\end{equation}
Si tenemos en cuenta la relaci\'on efectiva~$D_i \sim gT$, el primer t\'ermino corresponde a una correcci\'on~${\cal O}(g^4)$ en la energ\'{\i}a cin\'etica, de modo que comenzar\'a a contribuir a~${\cal O}(g^7)$ como una correcci\'on del LO en~$L(T)$. Los otros t\'erminos son de orden superior.

Teniendo en cuenta las relaciones~(\ref{eq:der_evacio}) y la ecuaci\'on~(\ref{eq:PL_des_serie}), reproducimos el t\'ermino de orden~${\cal O}(g^4)$ que aparece en el resultado de Gava y Jengo, ec.~(\ref{eq:Gava_Jengo}). Encontramos asimismo la siguiente contribuci\'on de orden ${\cal O}(g^5)$ en~$L(T)$
\begin{eqnarray}
{\cal O}(g^5)&=&\frac{(N_c^2-1) g^4 T m_D}{384\pi^3}\bigg[ -\frac{m_D^2}{(gT)^2}(9Q+3c_m+4N_f+2N_c(6\zeta-7))\nonumber \\
&& \qquad\qquad +\frac{N_c^2}{4}(89+4\pi^2-44\log{2})\bigg] \,,
\end{eqnarray}
donde 
\begin{eqnarray}
Q= \frac{22}{3}N_c \log \frac{\mu}{\mu_T} - \frac{4}{3}N_f \log \frac{4\mu}{\mu_T} \,, \qquad
c_m = \frac{10N_c^2+2N_f^2+9N_f/N_c}{6N_c+3N_f} \,,
\end{eqnarray}
y $\mu_T=4\pi e^{-\gamma_E}T$ es la escala t\'ermica est\'andar que surge en la reducci\'on dimensional perturbativa en el esquema $\overline{\textrm{MS}}$. $m_D$ viene dada por ec.~(\ref{eq:m_debye}). $\zeta$ es un par\'ametro que depende del gauge.

Si bien los t\'erminos~${\cal O}(g^5)+{\cal O}(g^6)$ tienen una dependencia en el gauge, num\'ericamente se observa que no producen una contribuci\'on sustancial a $L(T)$, pues son cualitativamente y cuantitativamente similares a los obtenidos en~\cite{Gava:1981qd}. Nuevamente la naturaleza radiativa de estos t\'erminos perturbativos produce una dependencia logar\'{\i}tmica en temperatura que es muy plana. 

Encontramos que teor\'{\i}a de perturbaciones resulta ser incapaz de explicar el comportamiento que se observa en el ret\'{\i}culo del loop de Polyakov en el r\'egimen~$T_c <T<6T_c$ (ver figura~\ref{fig:PL1}), y este hecho refuerza la necesidad de incluir en el c\'alculo efectos no perturbativos.

\subsection{Ansatz gaussiano}

Con objeto de simplificar el tratamiento, consideraremos que en la fase de desconfinamiento el campo $A_0(\vec{x})$ se encuentra suficientemente bien descrito por una distribuci\'on gaussiana. En este caso, todos los valores esperados conexos de $A_0$ m\'as all\'a de $\langle A_0^2\rangle$ se anulan, y haciendo uso del desarrollo est\'andar en cumulantes, se encuentra
\begin{equation}
 L  = \exp\left[-\frac{g^2\langle A_{0,a}^2\rangle}{4N_cT^2}
 \right]
\label{eq:PL_gaussiano}
\end{equation}
de modo que\footnote{Esta f\'ormula es v\'alida tambi\'en para la teor\'{\i}a unquenched, puesto que hasta este orden $N_f$ \'unicamente aparece a trav\'es de la masa de Debye.}
\begin{equation}
\langle A_{0,a}^2\rangle^{\text{Pert}}=
-\frac{N_c^2-1}{4 \pi} m_D T 
-\frac{N_c(N_c^2-1)}{8\pi^2}g^2T^2
\left(\log\frac{m_D}{2T}+\frac{3}{4}\right)+{\cal
O}(g^3) \,.
\label{eq:A0_pert}
\end{equation}
De~(\ref{eq:der_evacio}) se observa que la contribuci\'on a~$L(T)$ proveniente de~$\langle A_0^4\rangle$ comienza en~${\cal O}(g^6)$, de modo que el ansatz gaussiano ser\'a v\'alido hasta orden~${\cal O}(g^5)$ a temperatura suficientemente alta, donde la teor\'{\i}a se convierte en d\'ebilmente interactuante debido a la propiedad de libertad asint\'otica. Es exacto en el l\'{\i}mite de $N_c$ grande ya que los valores esperados conexos de \'ordenes mayores se encuentran suprimidos por potencias de $1/N_c$. $A_{0,a}^2$ escala como $N_c^2-1$, de modo que $L$ tiene un l\'{\i}mite bien definido para $N_c\rightarrow\infty$, con la prescripci\'on est\'andar de mantener fijo $g^2 N_c$.

Los c\'alculos en el ret\'{\i}culo muestran una distribuci\'on gaussiana para
el loop de Polyakov \cite{Engels:1988if}. El ansatz gaussiano es equivalente a
desarrollar la exponencial en ec.~(\ref{eq:PL_gauge_static}), promediar sobre
grados de libertad de color y finalmente hacer uso de la hip\'otesis de
saturaci\'on de vac\'{\i}o ($\langle A_0^{2k} \rangle =(2 k-1) !!  \langle
A_0^{2} \rangle^k$), usada habitualmente en las reglas de suma de QCD a
temperatura cero. En este contexto el loop de Wilson fue discutido en
ref.~\cite{Shifman:1980ui} mediante el uso del condensado glu\'onico
est\'andar de dimensi\'on 4, dando como resultado un t\'ermino proporcional al
cuadrado del \'area del contorno para contornos peque\~nos. El problema fue
discutido nuevamente en ref.~\cite{Kondo:2002xn} en el contexto de condensados
de dimensi\'on 2, dando lugar a una ley proporcional al \'area. Esto se
muestra de acuerdo con la observaci\'on de ref.~\cite{Chetyrkin:1998yr} de que
los condensados de dimensi\'on 2 podr\'{\i}an considerarse de manera efectiva como masas glu\'onicas taqui\'onicas, lo cual proporciona el comportamiento a cortas distancias de las fuerzas que son confinantes a distancias grandes.

\section{Contribuciones no perturbativas en el loop de Polyakov}
\label{non_pert_contr}

En la secci\'on~\ref{sec:pert_PL} de este cap\'{\i}tulo hemos hecho un estudio de las contribuciones perturbativas para el loop de Polyakov, y encontramos que teor\'{\i}a de perturbaciones reproduce \'unicamente los datos del ret\'{\i}culo a temperaturas suficientemente altas $(T \sim 6T_c)$. Este hecho aparece ilustrado en la figura~\ref{fig:PL1}. 

Nuestra motivaci\'on para dar cuenta de las contribuciones no perturbativas puede entenderse bien si se muestra la analog\'{\i}a que existe con el potencial quark-antiquark a temperatura cero en QCD quenched. Este potencial se puede obtener a partir de la funci\'on de correlaci\'on de dos l\'{\i}neas de Wilson. El r\'egimen perturbativo del potencial $V_{\bar{q}q}(r)$ es el correspondiente a separaciones peque\~nas, donde el potencial es aproximadamente coulombiano. Para separaciones del orden de $1/\Lambda_{\textrm{QCD}}$ (no existe otra escala en gluodin\'amica) surge un t\'ermino lineal confinante que comienza a ser dominante~\cite{Bali:2000gf}. Estas dos contribuciones del potencial evolucionan bajo el grupo de renormalizaci\'on siguiendo una ley logar\'{\i}tmica. Por tanto, m\'odulo correcciones radiativas, $rV_{\bar{q}q}(r)$ est\'a formado por una parte perturbativa que es constante y por un t\'ermino del tipo $\Lambda_{\textrm{QCD}}\,r^2$ que es no perturbativo. De manera an\'aloga, a temperaturas grandes podemos considerar el comportamiento de la magnitud adimensional $\langle
\tr(A_0^2)\rangle/T^2$, que tambi\'en est\'a directamente relacionada con la funci\'on de correlaci\'on de dos l\'{\i}neas de Wilson t\'ermicas. El an\'alogo de la escala $r$ en el caso anterior es aqu\'{\i} la escala $1/T$, y por suspuesto para $T$ grande la magnitud $\langle\tr(A_0^2)\rangle/T^2$ es perturbativa y plana (m\'odulo una dependencia logar\'{\i}tmica). A temperaturas no tan grandes habr\'{\i}a que considerar la posibilidad de que surjan t\'erminos no perturbativos en potencias del tipo $\Lambda_{\textrm{QCD}}^2/T^2$.



Con objeto de dar cuenta de contribuciones no perturbativas provenientes de condensados glu\'onicos, consideraremos en el propagador $D_{00}(\vec{k})$ nuevos t\'erminos fenomenol\'ogicos con par\'ametros dimensionales positivos. En concreto
\begin{equation}
D_{00}(\vec{k})=D^{\text{Pert}}_{00}(\vec{k})+D^{\text{No Pert}}_{00}(\vec{k}) \,,
\label{eq:prop_p_np}
\end{equation}
con el t\'ermino no perturbativo
\begin{equation}
D^{\text{No Pert}}_{00}(\vec{k})= \frac{m_G^2}{(\vec{k}^2+m_D^2)^2} \,.
\label{eq:prop_np}
\end{equation}
Este ansatz es equivalente al que se realiza a temperatura cero en presencia de condensados \cite{Lavelle:1988eg,Chetyrkin:1998yr}. Si introducimos ec.~(\ref{eq:prop_np}) en ec.~(\ref{eq:val_esp_A0}), podemos ver que este t\'ermino nuevo genera una contribuci\'on no perturbativa para el condensado:
\begin{equation}
\langle A_{0,a}^2\rangle^{\text{No Pert}} = \frac{(N_c^2-1)T m^2_G}{8\pi m_D} \,.
\end{equation}
Si suponemos que el par\'ametro $m_G$ es independiente de la temperatura (salvo correcciones radiativas), el condensado ser\'a asimismo T-independiente (m\'odulo esas mismas correcciones radiativas). En t\'erminos del condensado, la contribuci\'on no perturbativa al propagador se escribe
\begin{equation}
D^{\text{No Pert}}_{00}(\vec{k})= 
\frac{8\pi}{N_c^2-1}\frac{m_D}{T}
\frac{\langle A_{0,a}^2\rangle^{\text{No Pert}}}
{(\vec{k}^2+m_D^2)^2} \,.
\end{equation}
Notar que un condensado positivo~$\langle A_{0,a}^2\rangle^{\text{No Pert}}$ indica lo que ser\'{\i}a una masa glu\'onica taqui\'onica~$-m_G^2$, al igual que en ref.~\cite{Chetyrkin:1998yr}.

Si hacemos uso del ansatz gaussiano, ec.~(\ref{eq:PL_gaussiano}), y sumamos las contribuciones perturbativa y no perturbativa de $\langle A_{0,a}^2\rangle$, se obtiene
\begin{eqnarray} 
-2\log L = 
\frac{ g^2 \langle A_{0,a}^2 \rangle^{\text{Pert}}
 }{ 2N_c  T^2 } 
+
\frac{ g^2\langle A_{0,a}^2 \rangle^{\text{No Pert}}
 }{ 2N_c  T^2 } \,.
 \label{eq:log_L}
\end{eqnarray}
El hecho de que $\langle A_{0,a}^2\rangle^{\text{Pert}}$ escale como $T^2$ mientras que $\langle A_{0,a}^2\rangle^{\text{No Pert}}$ sea independiente de la temperatura (m\'odulo correcciones radiativas), sugiere que la f\'ormula anterior se pueda reescribir de la siguiente forma
\begin{eqnarray}
- 2 \log L = a + b \left(\frac{T_c}T \right)^2 \,,  
\label{eq:adjust}
\end{eqnarray}
donde se espera que los par\'ametros $a$ y $b$ tengan una dependencia d\'ebil en temperatura. Esta f\'ormula muestra que la contribuci\'on no perturbativa da lugar a una dependencia en temperatura que sigue una ley de potencia, la cual no est\'a presente en los c\'alculos perturbativos.

\section{Comparaci\'on con datos del ret\'{\i}culo}
\label{comparison_lattice_data}

Recientemente se han desarrollado diferentes m\'etodos para renormalizar el loop de Polyakov en el ret\'{\i}culo. Por supuesto, estos c\'alculos son completamente no perturbativos. Uno de los procedimientos de renormalizaci\'on se basa en el c\'alculo de funciones de corre\-laci\'on singlete y octete a temperatura finita de una pareja de quark y antiquark pesados~\cite{Kaczmarek:2002mc,Kaczmarek:2005ui}
\begin{eqnarray}
 e^{-F_1(\vec{x},T)/T +C(T)}&=& \frac{1}{N_c}\langle \Tr \,\Omega^{\text{desn}}(\vec{x}) \,\Omega^{\dagger \text{desn}}(0)\rangle  \,, \label{corr_loop_Polyakov}  \\
 e^{-F_8(\vec{x},T)/T + C(T)} &=&  \frac{1}{N_c^2-1}\langle \Tr \, \Omega^{\text{desn}}(\vec{x}) \,\Tr \,\Omega^{\dagger \text{desn}}(0)\rangle - \frac{1}{N_c(N_c^2-1)} \langle \Tr \,\Omega^{\text{desn}}(\vec{x}) \,\Omega^{\dagger \text{desn}}(0) \rangle \,. \nonumber
\end{eqnarray}
En estas f\'ormulas $\Omega^\text{desn}(\vec{x})$ indica el operador loop de
Polyakov desnudo (sin renormalizar) localizado en el punto $\vec{x}$. Los dos
loops de Polyakov se renormalizan mediante la extracci\'on de la
autoenerg\'{\i}a del quark (que es dependiente de $T$, pero independiente de
la separaci\'on), de tal modo que se reproduzca a peque\~nas distancias el
potencial quark-antiquark est\'andar a temperatura cero. El valor esperado del
loop de Polyakov se obtiene considerando en las f\'ormulas anteriores el
l\'{\i}mite de separaci\'on grande. Si $\Omega_{\cal R}(\vec{x})$ denota el loop de Polyakov renormalizado en el punto $\vec{x}$,  
\begin{equation}
\frac{1}{N_c}\langle \Tr \, \Omega_{\cal R}(\vec{x}) \, \Omega_{\cal R}^\dagger(0)\rangle = 
\frac{1}{N_c} e^{-C(T)} \langle \Tr \, \Omega^{\text{desn}}(\vec{x}) \, \Omega^{\dagger \text{desn}}(0)\rangle = e^{-F_1(r,T)/T}
\underset{{r\to\infty}}{\longrightarrow} L^2(T) \,.
\label{eq:PL_correlacion}
\end{equation}
Tal y como muestran los autores de~\cite{Kaczmarek:2002mc}, existe una ambig\"uedad en su procedimiento, que corresponde a a\~nadir una constante al potencial quark-antiquark a temperatura cero. Esta ambig\"uedad se traduce en una ambig\"uedad aditiva en $F_1(r,T)$ en ec.~(\ref{eq:PL_correlacion}), lo cual conducir\'{\i}a a un t\'ermino del tipo $1/T$ en $\log(L(T))$. Para eliminar esta ambig\"uedad los autores han adoptado la prescripci\'on de Cornell, que consiste en elegir $v_1=0$ en $V_{\bar{q}q}(r)\sim v_0/r+v_1+v_2 r$.

\subsection{Resultados en gluodin\'amica}
\label{comparison_lattice_data_nf0}

\begin{figure}[tbp]
\begin{center}
\epsfig{figure=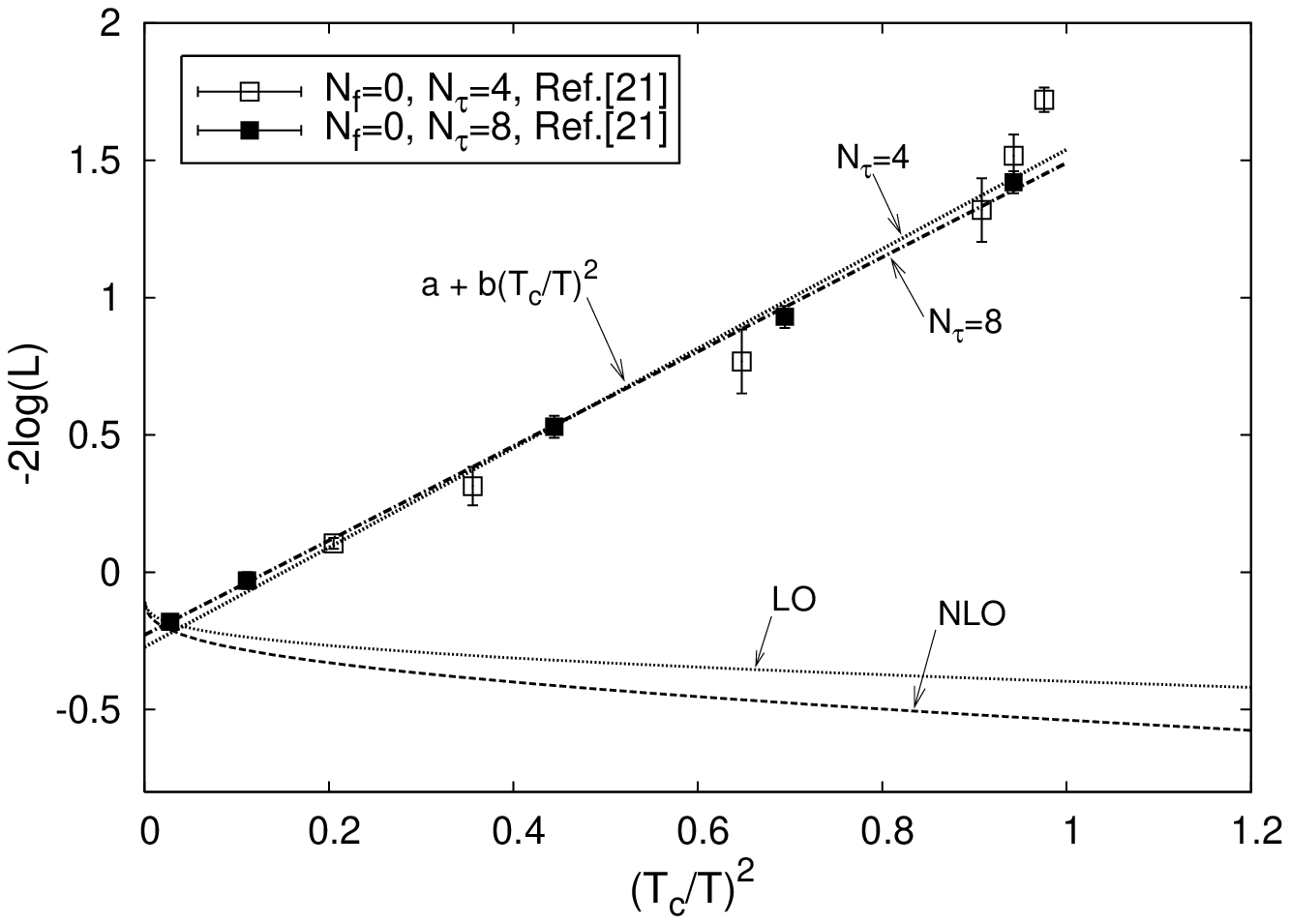,height=6.5cm,width=8.5cm}
\end{center}
\caption{Logaritmo del loop de Polyakov renormalizado en gluodin\'amica $(N_c=3)$ frente al cuadrado de la inversa de la temperatura en unidades de la temperatura de transici\'on de fase. Los datos del ret\'{\i}culo son de ref.~\cite{Kaczmarek:2002mc}. En los ajustes se usa ec.~(\ref{eq:adjust}) con $a$ y $b$ como par\'ametros libres, y datos del ret\'{\i}culo por encima de $1.03\,T_c$ para $N_\tau=4$ y $N_\tau=8$. Para comparar, se muestran los resultados perturbativos LO y NLO para $N_f=0$.}
\label{fig:PL2}
\end{figure}
En ref.~\cite{Kaczmarek:2002mc} se hace un estudio del loop de Polyakov renormalizado, siguiendo el m\'etodo se\~nalado anteriormente, para gluodin\'amica pura y $N_c=3$. Motivado por el resultado de nuestro modelo, ec.~(\ref{eq:adjust}), en la figura~\ref{fig:PL2} mostramos los datos en el ret\'{\i}culo de $-2\log L(T)$ frente a $(T_c/T)^2$. Se observa que los datos presentan un comportamiento pr\'acticamente lineal. Este patr\'on es claramente diferente del que predice teor\'{\i}a de perturbaciones, que es mucho m\'as plano, y muestra de manera inequ\'{\i}voca la existencia de la correcci\'on en potencias de temperatura t\'{\i}pica de un condensado de dimensi\'on 2.

Si identificamos (\ref{eq:adjust}) con (\ref{eq:log_L}) obtenemos las siguientes relaciones:
\begin{eqnarray}
a &=& 
-\frac{1}{8 \pi}\frac{N_c^2-1}{N_c}g^2\frac{m_D}{T} -
\frac{N_c^2-1}{16\pi^2}g^4\left(\log\frac{m_D}{2T}+\frac{3}{4}\right)
+{\cal O}(g^5) 
\,,  \label{eq:apert}\\
g^2 \langle A_{0,a}^2 \rangle^{\text{No Pert}} &=& 2 N_c T_c^2 b\,.
\label{eq:cpert}
\end{eqnarray}
Haremos un primer ajuste de los datos del ret\'{\i}culo considerando para $a$ el valor que predice teor\'{\i}a de perturbaciones a NLO~(\ref{eq:apert}), y dejando $b$ como par\'ametro libre
\begin{eqnarray}
-2 \log{L} =a^{\text{NLO}}+ b \left(\frac{T_c}{T}\right)^2 \,.
 \label{eq:adjust_pert}
\end{eqnarray}
El resultado se muestra en la tabla~\ref{tab:adjust_nf0_aNLO}.

\begin{table}[htb]
\begin{center}
\begin{tabular}{|c|c|c|c|}
\hline
$N_\tau$ & $b$ & $g^2 \langle A_{0,a}^2 \rangle^{\text{No Pert}}\;(\text{GeV})^2$ & $\chi^2/{\rm DOF}$\\
\hline
4 & 2.20(6) & $(0.98(2))^2$ & 0.75 \\
8 & 2.14(4) & $(0.97(1))^2$ & 1.43 \\
\hline
\end{tabular}
\end{center}
\caption{\label{tab:adjust_nf0_aNLO} Resultado del ajuste con ec.~(\ref{eq:adjust_pert}) de los datos en el ret\'{\i}culo del loop de Polyakov renormalizado en gluodin\'amica~\cite{Kaczmarek:2002mc}. Se han incluido datos por encima de $1.03\,T_c$. El valor del condensado se ha obtenido a partir de $b$ y la ecuaci\'on~(\ref{eq:cpert}).}
\end{table}

En el ajuste hemos incluido datos del ret\'{\i}culo para temperaturas por encima de $1.03\,T_c$. Hacemos uso de $T_c/\Lambda_{\overline{\text{MS}}}=1.14(4)$~\cite{Bali:2000gf,Beinlich:1997ia}, y $T_c=270(2)\;{\rm MeV}$~\cite{Beinlich:1997ia}. En el resto de esta secci\'on usaremos la constante de acoplamiento que se obtiene de la funci\'on beta hasta tres loops y $\Lambda_E$ de ec.~(\ref{eq:LambdaE}) como par\'ametro de escala. Si suponemos que la diferencia entre los dos resultados del ret\'{\i}culo ($N_\tau=4$ y $N_\tau=8$) es debida \'unicamente a efectos de cutoff finito, y consideramos que el efecto principal va como $1/N_\tau$, encontramos como estimaci\'on para $g^2 \langle A_{0,a}^2\rangle^{\text{No Pert}}$ en el l\'{\i}mite del continuo $(0.95(4) \, {\rm GeV} )^2$.

Hemos considerado tambi\'en un segundo ajuste de los datos del ret\'{\i}culo considerando $a$ y $b$ par\'ametros libres. El resultado se muestra en la tabla~\ref{tab:adjust_nf0_acte}.

\begin{table}[htb]
\begin{center}
\begin{tabular}{|c|c|c|c|c|}
\hline
$N_\tau$ & $a$ & $b$ & $g^2 \langle A_{0,a}^2 \rangle^{\text{No Pert}}\;(\text{GeV})^2$ & $\chi^2/{\rm DOF}$\\
\hline
4 & -0.27(5) & 1.81(13) & $(0.89(3))^2$ & 1.07 \\
8 & -0.23(1) & 1.72(5)  & $(0.87(2))^2$ & 0.45 \\
\hline
\end{tabular}
\end{center}
\caption{\label{tab:adjust_nf0_acte} Igual que tabla~\ref{tab:adjust_nf0_aNLO}, con $a$ y $b$ como par\'ametros libres.}
\end{table}

Los valores de $\chi^2 /{\rm DOF}$ son ligeramente mejores que los correspondientes al ajuste con $a^{\text{NLO}}$, y los valores del condensado son un poco m\'as peque\~nos que antes. La correspondiente estimaci\'on del l\'{\i}mite del continuo es~$g^2 \langle A_{0,a}^2 \rangle^{\text{No Pert}} = (0.84(6) \, {\rm GeV} )^2$. 

La identificaci\'on de $a$ con el resultado perturbativo debe de funcionar mejor a tempera\-turas grandes. De ec.~(\ref{eq:apert}) se obtiene para la temperatura m\'as alta $6\,T_c$
\begin{eqnarray}
a^{\text{NLO}} = -0.22(1)\qquad (T= 6\,T_c) \,,
\end{eqnarray}
lo cual muestra un acuerdo razonable con los valores ajustados. Notar que las correcciones no perturbativas en potencias de $T$ contribuyen poco a esta temperatura $(\sim 20 \% )$. Se puede concluir que el resultado perturbativo NLO evoluciona a temperaturas peque\~nas m\'as r\'apidamente de lo que sugiere el ajuste. Ser\'{\i}a interesante tener en cuenta correcciones logar\'{\i}tmicas al valor del condensado y quiz\'as ciertas correcciones de dimensi\'on an\'omala para \'este. Sin embargo, los datos actuales del ret\'{\i}culo no permiten una extracci\'on limpia de esos detalles. 

En un intento por determinar una posible correcci\'on de tipo $1/T^4$, hemos considerado en ec.~(\ref{eq:adjust}) el t\'ermino extra $c(T_c/T)^4$. El resultado del ajuste de los datos del ret\'{\i}culo para $N_\tau=8$ se muestra en la tabla~\ref{tab:adjust_nf0_c}.

\begin{table}[htb]
\begin{center}
\begin{tabular}{|c|c|c|c|c|}
\hline
$N_\tau$ & $a$ & $b$ & $c$ & $\chi^2/{\rm DOF}$ \\
\hline
8 & $a^{\rm NLO}$ & 2.18(20) & -0.04$\pm$0.24 & 1.89 \\
8 & -0.22(2) & 1.61(24)  & 0.13$\pm$0.28 & 0.42 \\
\hline
\end{tabular}
\end{center}
\caption{\label{tab:adjust_nf0_c} Resultado del ajuste de los datos del ret\'{\i}culo del loop de Polyakov renorma\-lizado en gluodin\'amica~\cite{Kaczmarek:2002mc}, con ec.~(\ref{eq:adjust}) y un t\'ermino extra $c(T_c/T)^4$. En la primera fila se han tomado $b$ y $c$ como par\'ametros libres, y se considera para $a$ el valor perturbativo a NLO, ec.~(\ref{eq:apert}). En la segunda fila se toman $a$, $b$ y $c$ como par\'ametros libres.}
\end{table}

El valor de $c$ es compatible con cero en los dos casos, y los errores se superponen con los valores centrales de $a$ y $b$ $(N_\tau=8)$, en tab.~\ref{tab:adjust_nf0_aNLO} y tab.~\ref{tab:adjust_nf0_acte} respectivamente. Es necesario disponer de datos m\'as precisos con objeto de identificar posibles contribuciones de condensados de dimensi\'on 4.

Un ajuste de los datos excluye por completo la existencia de un t\'ermino del tipo~$1/T$ en $\log(L(T))$. Este t\'ermino no tiene base te\'orica, pues no existe un condensado de dimensi\'on uno. La ausencia de este t\'ermino en los datos se debe a que los autores han adoptado la prescripci\'on de Cornell para el potencial quark-antiquark.

\subsection{Resultados unquenched}
\label{comparison_lattice_data_nf2}

El loop de Polyakov renormalizado ha sido calculado tambi\'en en ref.~\cite{Kaczmarek:2005ui} en el caso unquenched para QCD con dos sabores, siguiendo el m\'etodo explicado al comienzo de la secci\'on~\ref{comparison_lattice_data}. 
\begin{figure}[tbp]
\begin{center}
\epsfig{figure=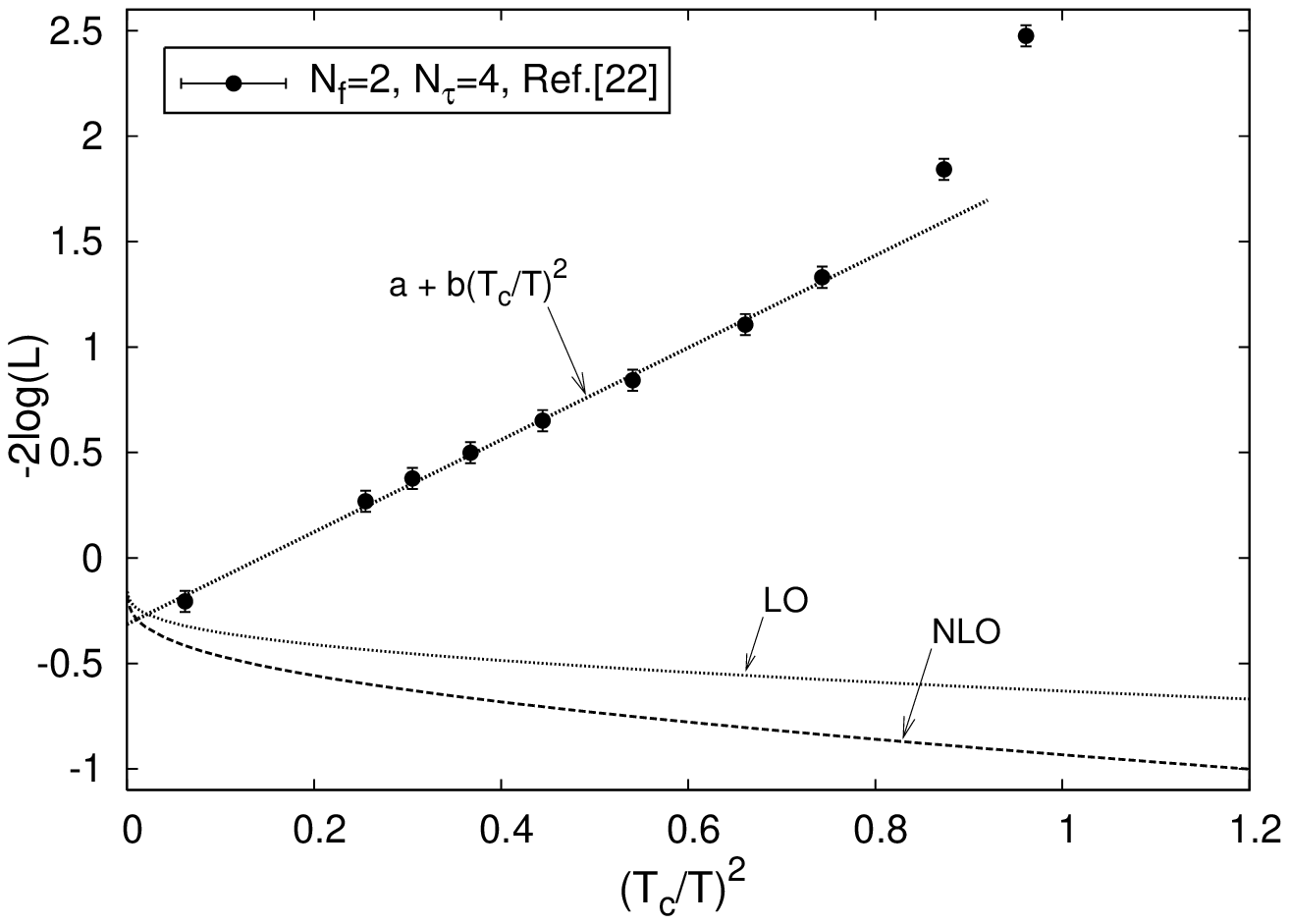,height=6.5cm,width=8.5cm}
\end{center}
\caption{Logaritmo del loop de Polyakov renormalizado en QCD unquenched con dos sabores frente al cuadrado de la inversa de la temperatura en unidades de la temperatura de transici\'on de fase. Los datos del ret\'{\i}culo son de ref.~\cite{Kaczmarek:2005ui}. En los ajustes se usa ec.~(\ref{eq:adjust}) con $a$ y $b$ como par\'ametros libres, y datos del ret\'{\i}culo por encima de $1.15\,T_c$ para $N_\tau=4$. Para comparar, se muestran los resultados perturbativos LO y NLO para $N_f=2$.}
\label{fig:PL3}
\end{figure}
En la figura~\ref{fig:PL3} mostramos estos datos para $N_\tau=4$. En este caso, los datos siguen un comportamiento pr\'acticamente lineal para temperaturas por encima de $1.15\,T_c$. Cerca de la temperatura de transici\'on los datos comienzan a salirse del patr\'on de la ec.~(\ref{eq:adjust}), lo cual es se\~nal de que se hace necesaria una descripci\'on m\'as rica a medida que nos aproximamos a la transici\'on de fase.

En la tabla \ref{tab:adjust_nf2} se muestran los resultados del ajuste de los datos del ret\'{\i}culo para $T>1.15\,T_c$. 

\begin{table}[htb]
\begin{center}
\begin{tabular}{|c|c|c|c|c|}
\hline
$N_\tau$ & $a$ & $b$ & $g^2 \langle A_{0,a}^2 \rangle^{\text{No Pert}}\;(\text{GeV})^2$ & $\chi^2/{\rm DOF}$ \\
\hline
4 & $a^{\rm NLO}$ & 2.99(12) & $(0.86(2))^2$ & 1.87 \\
4 & -0.31(6) & 2.19(13)  & $(0.73(3))^2$ & 0.25 \\
\hline
\end{tabular}
\end{center}
\caption{\label{tab:adjust_nf2} Resultado del ajuste con ec.~(\ref{eq:adjust}) de los datos en el ret\'{\i}culo del loop de Polyakov renormalizado en QCD con dos sabores~\cite{Kaczmarek:2005ui}. Se han incluido datos por encima de $1.15\, T_c$. En la primera fila se ha tomado $b$ como par\'ametro libre, y se considera para $a$ el valor perturbativo a NLO, ec.~(\ref{eq:apert}). En la segunda fila se toman $a$ y $b$ como par\'ametros libres.}
\end{table}

Hemos usado $T_c/\Lambda_{\overline{\rm MS}}=0.77(9)$, con $T_c=202(4)\;{\rm
MeV}$~\cite{Karsch:2000ps} y $\Lambda_{\overline{\rm MS}}=261(31)\;{\rm MeV}$~\cite{Gockeler:2005rv}. En el ajuste hemos considerado el mismo peso para todos los puntos, y el valor de $\chi^2$ corresponde a un error representativo de $\pm 0.05$ en $2\log (L(T))$ (similar al caso quenched).

Al igual que en el caso quenched, el valor de $a$ es consistente con el valor perturbativo a temperatura grande
\begin{eqnarray}
a^{\text{NLO}} = -0.35(2)\qquad (T= 6\,T_c) \,.
\end{eqnarray}

La p\'erdida del patr\'on lineal para temperaturas por debajo de $1.15\,T_c$ no se explica convenientemente si consideramos nuevos condensados de dimensi\'on mayor. En efecto, hemos sido incapaces de extraer de los datos un condensado de dimensi\'on 4. En la tabla~\ref{tab:adjust_nf2_c} se muestra el resultado del ajuste para $T>1.0\,T_c$ al considerar en ec.~(\ref{eq:adjust_pert}) el t\'ermino extra $c(T_c/T)^4$. 

\begin{table}[htb]
\begin{center}
\begin{tabular}{|c|c|c|c|c|}
\hline
$N_\tau$ & $a$ & $b$ & $c$ & $\chi^2/{\rm DOF}$ \\
\hline
4 & $a^{\rm NLO}$ & 2.44(21) & 1.07(19) & 12.8 \\
\hline
\end{tabular}
\end{center}
\caption{\label{tab:adjust_nf2_c} Ajuste de los datos en el ret\'{\i}culo del loop de Polyakov renormalizado en QCD con dos sabores~\cite{Kaczmarek:2005ui}, con ec.~(\ref{eq:adjust}) y un t\'ermino extra $c(T_c/T)^4$. Se han incluido datos por encima de $1.0\, T_c$.}
\end{table}

El ajuste no es bueno, y la gran correlaci\'on que encontramos entre $b$ y $c$ hace que no se pueda extraer informaci\'on fiable de este nuevo par\'ametro.

\subsection{Otros resultados quenched}
\label{comparison_lattice_data_nf0_pisarski}

Recientemente ha aparecido en la literatura un m\'etodo alternativo para
renormalizar el loop de Polyakov en el ret\'{\i}culo. En
ref.~\cite{Dumitru:2003hp} los autores consideran loops de Polyakov aislados
en gluodin\'amica pura, y hacen una renormalizaci\'on multiplicativa mediante
la extracci\'on de la autoenerg\'{\i}a del quark. Si ${\mathcal P}_{\cal
  R}(\vec{x})$ denota el loop de Polyakov renormalizado en una
representaci\'on irreducible arbitraria ${\cal R}$ en el punto $\vec{x}$, se
tiene\footnote{En nuestra notaci\'on ${\mathcal P}_{\cal
    R}(\vec{x})=\frac{1}{N_c}\Tr \,\Omega_{\cal R}(\vec{x})\,.$}
\begin{equation}
\langle {\mathcal P}_{\cal R}(\vec{x}) \rangle = 
\frac{1}{{\cal Z}_{\cal R}} \langle {\mathcal P}^{\rm desn}(\vec{x})\rangle \,, \qquad
{\cal Z}_{\cal R} = \exp\left( -\frac{m_{\cal R}^{\rm div}}{T} \right) \,,
\end{equation}
donde se ha dividido por una constante de renormalizaci\'on apropiada ${\cal
  Z}_{\cal R}$. ${\mathcal P}^{\rm den}(\vec{x})$ indica el operador
  loop de Polyakov desnudo. \'Este es un tipo est\'andar de renormalizaci\'on
  de masa, si bien aqu\'{\i} se debe tener en cuenta que, puesto que la l\'{\i}nea de Wilson es un operador no local, la constante de renormalizaci\'on depender\'a de la longitud del camino: en general, para un camino de longitud $\ell$ se tiene ${\cal Z}_{\cal R}=\exp(-m_{\cal R}^{\rm div}\ell)$.

El problema principal reside en c\'omo determinar las masas divergentes de un modo no perturbativo. En un espacio-tiempo de cuatro dimensiones la masa divergente para un quark test $m_{\cal R}^{\rm div}$ es lineal con el cutoff ultravioleta, el cual es proporcional al inverso del espaciado del ret\'{\i}culo, $a$, esto es:
\begin{equation}
m_{\cal R}^{\rm div} \sim \frac{1}{a} \,.
\end{equation}
Los autores consideran diferentes ret\'{\i}culos, todos a la misma temperatura f\'{\i}sica $T$, pero con diferentes valores del espaciado $a$. Puesto que el n\'umero de puntos en la direcci\'on temporal $N_\tau = 1/(aT)$ es diferente en estos ret\'{\i}culos, obtienen la masa divergente $a m_{\cal R}^{\rm div}$ mediante comparaci\'on de los valores del loop de Polyakov desnudo en los diversos ret\'{\i}culos.

\begin{figure}[tbp]
\begin{center}
\epsfig{figure=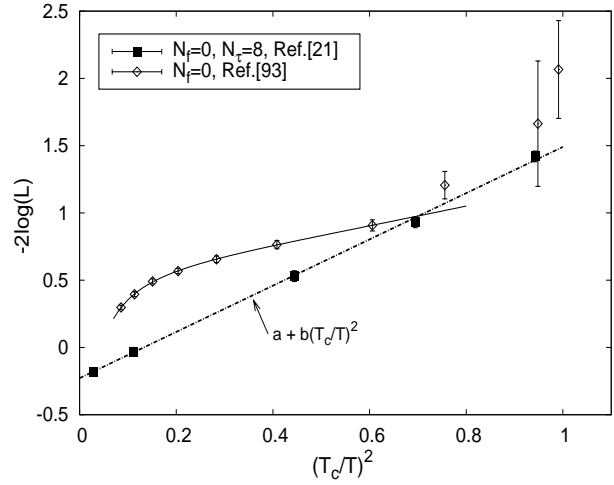,height=6.5cm,width=8.5cm}
\end{center}
\caption{Logaritmo de loop de Polyakov renormalizado en gluodin\'amica $(N_c=3)$ frente al cuadrado de la inversa de la temperatura en unidades de la temperatura de transici\'on de fase. Los datos del ret\'{\i}culo son de refs.~\cite{Kaczmarek:2002mc} y \cite{Dumitru:2003hp}. Los ajustes usan ec.~(\ref{eq:adjust}) con $a$ y $b$ como par\'ametros libres para \cite{Kaczmarek:2002mc}, y ec.~(\ref{eq:adjust_pis}) con $a$ como par\'ametro libre para \cite{Dumitru:2003hp}. }
\label{fig:PL4}
\end{figure}

Siguiendo este m\'etodo, los autores de \cite{Dumitru:2003hp} calculan el loop de Polyakov renormalizado en varias representaciones de SU(3). Nuestro inter\'es se centra en la representaci\'on fundamental, y cuando comparamos con los datos de \cite{Kaczmarek:2002mc} encontramos que ambos resultados difieren cualitativamente, principalmente para temperaturas por encima de $1.3\,T_c$. En la figura~\ref{fig:PL4} se muestran los dos conjuntos de datos.

El origen de la discrepancia entre ambos resultados no est\'a del todo claro, aunque los autores de \cite{Dumitru:2003hp} no excluyen la posibilidad de que se deba a efectos del espaciado finito del ret\'{\i}culo, que no hayan sido tenidos en cuenta de manera conveniente. 

Existen varias razones para pensar que los resultados de \cite{Kaczmarek:2002mc} son m\'as fiables. Por una parte este m\'etodo resulta t\'ecnicamente m\'as simple y susceptible de ser comprobado. Los autores pueden comprobar que a cortas distancias los dos loops de Polyakov reproducen de una manera muy precisa el potencial quark-antiquark a temperatura cero como funci\'on de $r$ para todas las temperaturas. El contacto entre el potencial a temperatura cero y el correspondiente a temperatura finita es casi total hasta una separaci\'on $r(T)$, relacionada con la masa de Debye, lo cual permite una determinaci\'on muy precisa del contrat\'ermino $C(T)$ de ec.~(\ref{eq:PL_correlacion}). Adem\'as, el c\'alculo est\'a hecho para dos tama\~nos diferentes del ret\'{\i}culo, $N_\tau=4$ y $N_\tau=8$ (tambi\'en $N_\tau=16$ en \cite{Zantow:2003uh}), y los resultados muestran una dependencia muy peque\~na en el cutoff, lo cual significa que el l\'{\i}mite del continuo ha sido alcanzado.

El m\'etodo de ref.~\cite{Dumitru:2003hp} es t\'ecnicamente m\'as complicado, pues necesita comparar tama\~nos diferentes del ret\'{\i}culo a la misma temperatura f\'{\i}sica $T$. La extracci\'on del contrat\'ermino es asimismo m\'as compleja, pues el an\'alogo de $C(T)$ en ec.~(\ref{eq:PL_correlacion}) se escribe como una serie en potencias de $T$ con coeficientes que deben de ser ajustados con los datos del loop de Polyakov desnudo. Por otra parte, desde el punto de vista del modelo que proponemos en nuestro trabajo, esperamos que las correcciones no perturbativas sean despreciables a las temperaturas m\'as altas de los dos datos del ret\'{\i}culo, pero \'unicamente \cite{Kaczmarek:2002mc} parece ser consistente con teor\'{\i}a de perturbaciones \cite{Gava:1981qd} a esas temperaturas. 

El m\'etodo de \cite{Dumitru:2003hp} renormaliza el logaritmo del loop de Polyakov siguiendo este esquema\footnote{Nos vamos a limitar a analizar los datos correspondientes al loop de Polyakov en representaci\'on fundamental. En sec.~\ref{transicion_fase_modelo} se discute el comportamiento del loop de Polyakov adjunto obtenido en el contexto de modelos de quarks quirales a temperatura finita.}
\begin{equation}
-\log L^{\rm desn}(T) =
f^{\text{div}}N_\tau+f^{\text{ren}}+f^{\text{lat}}N_\tau^{-1} \,,
\end{equation}
donde
\begin{equation}
L^{\rm desn}(T) = \frac{1}{N_c}\langle \Tr \, \Omega^{\rm desn}(\vec{x})\rangle \,, \qquad
L(T) = \frac{1}{N_c}\langle \Tr \, \Omega_{\cal R}(\vec{x})\rangle = e^{-f^{\rm ren}} \,.
\end{equation}
Podemos especular con esta f\'ormula suponiendo que los t\'erminos que dependen del cutoff no han sido extra\'{\i}dos completamente en los datos, o bien que despu\'es de haber sido extra\'{\i}dos permanezcan t\'erminos del mismo tipo a los extra\'{\i}dos. En concreto, consideraremos el siguiente patr\'on de ajuste
\begin{equation}
- 2 \log L = a + b \left(\frac{T_c}{T} \right)^2  +
\delta a_{-1}\frac{T_c}{ T}+ \delta a + \delta a_1\frac{T}{T_c} \,.
\label{eq:adjust_pis}
\end{equation}
En la tabla \ref{tab:adjust_nf0_pis} se muestran los resultados del ajuste de los datos del ret\'{\i}culo (figura 8 de ref.~\cite{Dumitru:2003hp}) para el loop de Polyakov en la representaci\'on fundamental, en el r\'egimen $1.3\,T_c <T < 3.5\,T_c$.

\begin{table}[htb]
\begin{center}
\begin{tabular}{|c|c|c|c|c|c|c|}
\hline
$a$ & $\delta a$ & $a+\delta a$ & $b$ & $\delta a_{-1}$ & $\delta a_1$ & $\chi^2/{\rm DOF}$ \\
\hline
$a^{\rm NLO}$ & 1.8 $\pm$ 1.8 & $-$ & 1.4 $\pm$ 2.6 & -1.0$\pm$3.8 & -0.29$\pm$0.26 & 0.0349 \\
\hline
$-$  & $-$ & 1.6 $\pm$ 1.8 & 1.3 $\pm$ 2.6 & -1.4 $\pm$ 3.8 & -0.28 $\pm$ 0.26 & 0.0350 \\
\hline
\end{tabular}
\end{center}
\caption{\label{tab:adjust_nf0_pis} Resultado del ajuste con ec.~(\ref{eq:adjust_pis}) de los datos en el ret\'{\i}culo del loop de Polyakov renormalizado en gluodin\'amica~\cite{Dumitru:2003hp}. En la primera fila se ha tomado para $a$ el valor $a^{\rm NLO}$ de ec.~(\ref{eq:apert}), y en la segunda se ha considerado $a$ como par\'ametro libre.}
\end{table}

Un hecho alentador es que el valor del condensado parece ser compatible con el obtenido en la secci\'on \ref{comparison_lattice_data_nf0} a partir de los datos de ref.~\cite{Kaczmarek:2002mc}. No obstante, esta especulaci\'on no es totalmente concluyente y ser\'{\i}a deseable un acuerdo entre los resultados de ambos grupos antes de sacar nuevas consecuencias.

\subsection{Relaci\'on con otras determinaciones del condensado}
\label{cond_T0}

Si bien nuestra determinaci\'on del condensado se ha hecho en el gauge est\'atico y a temperatura finita, resulta tentador comparar con condensados a temperatura cero $g^2 \langle A_{\mu,a}^2 \rangle$, calculados en la literatura en quenched QCD y en el gauge de Landau. En la tabla \ref{tab:PL_cond_T0} se muestran algunos valores de este condensado obtenidos recientemente por diferentes procedimientos. El acuerdo entre ellos es aceptable.

\begin{table}[htb]
\begin{center}
\begin{tabular}{|c|c|}
\hline
Referencia & $g^2 \langle A_{\mu,a}^2 \rangle\;(\text{GeV})^2$\\
\hline
Del propagador del glu\'on \cite{Boucaud:2001st}   & $(2.4 \pm 0.6)^2$ \\
Del v\'ertice sim\'etrico de tres gluones \cite{Boucaud:2001st} &   $(3.6 \pm 1.2)^2$ \\
De la cola del propagador del quark \cite{RuizArriola:2004en}  & $(2.1 \pm 0.1)^2$      \\
De la cola del propagador del quark \cite{Boucaud:2005rm}   & $(3.0-3.4 )^2$ \\
\hline
\end{tabular}
\end{center}
\caption{\label{tab:PL_cond_T0} Valores del condensado $g^2 \langle A_{\mu,a}^2 \rangle$ a temperatura cero, en el gauge de Landau en quenched QCD.}
\end{table}

A temperatura cero todas las componentes de Lorentz contribuyen de igual forma, lo cual sugiere un factor de conversi\'on 4 al pasar de $g^2 \langle A_{\mu,a}^2 \rangle$ a $g^2 \langle A_{0,a}^2 \rangle$. Sin embargo, de acuerdo con ref.~\cite{Lavelle:1988eg}, en el gauge de Landau el condensado total escala como $D-1$, donde $D$ es la dimensi\'on del espacio eucl\'{\i}deo, lo cual sugiere un factor de conversi\'on 3. En cualquier caso, si tenemos en cuenta tanto las incertidumbres de los datos del ret\'{\i}culo como las te\'oricas, el acuerdo es significativo, pues estamos comparando resultados a temperaturas y gauges diferentes.

Podemos comparar asimismo nuestro resultado para el condensado glu\'onico con c\'alculos realizados a temperatura finita basados en el estudio de contribuciones no perturbativas de la presi\'on en gluodin\'amica pura~\cite{Kajantie:2000iz,Kajantie:2002pu}. Estos resultados conducen a
\begin{equation}
g^2 \langle A_{0,a}^2\rangle^{\text{No Pert}}= (0.93(7)\; \text{GeV})^2 \,,
\end{equation} 
en el gauge de Landau.\footnote{Este valor ha sido obtenido a partir de los datos del ret\'{\i}culo de la figura 2 de ref.~\cite{Kajantie:2000iz}, y tambi\'en de la figura 1 de ref.~\cite{Kajantie:2002pu}, en la regi\'on de temperaturas usada en nuestros ajustes de la secci\'on~\ref{comparison_lattice_data_nf0}.} Todos estos an\'alisis muestran un esquema coherente en su conjunto.

\section{Energ\'{\i}a libre de un quark pesado}
\label{free_energy}

El potencial quark-antiquark a temperatura finita se puede obtener a partir de la funci\'on de correlaci\'on de dos loops de Polyakov separados. Como sabemos, si se toma el l\'{\i}mite de separaci\'on grande se obtiene el valor esperado del loop de Polyakov, ec.~(\ref{eq:PL_correlacion}). En el l\'{\i}mite de separaci\'on peque\~na los efectos t\'ermicos son despreciables, y este potencial coincide con el potencial quark-antiquark a temperatura cero.

Hasta ahora hemos aplicado nuestro modelo fenomenol\'ogico de ecs.~(\ref{eq:prop_p_np})-(\ref{eq:prop_np}) para dar cuenta de las correcciones no perturbativas en el loop de Polyakov. En esta secci\'on aplicaremos este modelo para describir los datos del ret\'{\i}culo de la energ\'{\i}a libre de un quark pesado.

\subsection{Contribuciones no perturbativas en la energ\'{\i}a libre}
\label{free_energy_1}

El potencial quark-antiquark puede relacionarse con la amplitud de scattering correspondiente al intercambio de un \'unico glu\'on. En el l\'{\i}mite no relativista, para la energ\'{\i}a libre en el canal singlete se tiene
\begin{equation}
F_1(\vec{x},T) = -\frac{N_c^2-1}{2N_c} g^2 \int \frac{d^3 k}{(2\pi)^3} e^{i\vec{k} \cdot \vec{x} } D_{00}(\vec{k}) \,.
\label{eq:def_free_energy}
\end{equation}

Podemos estudiar contribuciones no perturbativas en la energ\'{\i}a libre aplicando el modelo que desarrollamos en la secci\'on~\ref{non_pert_contr}. Si sustituimos (\ref{eq:prop_p_np}) en (\ref{eq:def_free_energy}) obtenemos adem\'as de las contribuciones perturbativas a LO $({\cal O}(g^2))$ y NLO $({\cal O}(g^3))$, nuevas contribuciones no perturbativas\footnote{Hacemos uso de las reglas de regularizaci\'on dimensional y consideramos $g$ independiente de~$\vec{k}$.}
\begin{equation}
F_1(r,T) = -\frac{N_c^2-1}{2N_c} \left(
\frac{g^2}{4 \pi r} + \frac{1}{N_c^2-1}\frac{g^2 \langle A_{0,a}^2 \rangle^{\text{No Pert}}}{T}
\right) e^{-m_D r}
-\frac{N_c^2-1}{2N_c}\frac{g^2 m_D}{4\pi} + \frac{g^2 \langle A_{0,a}^2\rangle^{\text{No Pert}}}{2N_c T} \,.
\label{eq:free_energy}
\end{equation}

Si consideramos el l\'{\i}mite $r\rightarrow\infty$ en (\ref{eq:free_energy}), se obtiene esencialmente el logaritmo del loop de Polyakov
\begin{equation}
F_\infty(T) \equiv F_1(r\rightarrow\infty,T) = -2 T \log L(T) =  
-\frac{N_c^2-1}{2N_c}\frac{g^2 m_D}{4\pi} + \frac{g^2 \langle A_{0,a}^2\rangle^{\text{No Pert}}}{2N_c T} +{\cal O}(g^4)\,.
\label{eq:F_infinito}
\end{equation}
Esta expresi\'on coincide con ec.~(\ref{eq:adjust}), teniendo en cuenta ec.~(\ref{eq:cpert}) para $b$ y ec.~(\ref{eq:apert}) hasta ${\cal O}(g^3)$ para $a$.

En el l\'{\i}mite de temperatura cero, para lo cual consideramos $m_D r \rightarrow 0$ en (\ref{eq:free_energy}), se tiene
\begin{equation}
F_{1}(r,T) \stackrel{T \rightarrow 0} \sim 
-\frac{N_c^2-1}{2N_c}\frac{g^2}{4\pi r} 
+ \sigma \, r  \equiv V_{\bar{q}q}(r) \,,
\label{eq:t_zero}
\end{equation}
donde
\begin{equation}
\sigma = \left(\frac{N_c}{3}+\frac{N_f}{6}\right)^{1/2}\frac{g^3 \langle A_{0,a}^2 \rangle_{T=0}}{2N_c} \,.
\label{eq:string_tension_condensado}
\end{equation}
En esta expresi\'on, $\langle A_{0,a}^2 \rangle_{T=0}$ denota el condensado a
temperatura cero. En este l\'{\i}mite se llega obviamente a la expresi\'on del
potencial quark-antiquark a temperatura cero \cite{Necco:2001xg}. El t\'ermino
de Coulomb es el resultado perturbativo est\'andar a LO, mientras que el
segundo t\'ermino es una contribuci\'on lineal no perturbativa bien conocida
en la literatura. Ec.~(\ref{eq:t_zero}) con $g=\pi/2$ corresponde al modelo de cuerda bos\'onica, y reproduce los datos del ret\'{\i}culo para $V_{\bar{q}q}(r)$ en el rango $0.75 \,{\rm GeV}^{-1}\le r \le 4\, {\rm GeV}^{-1}$ con un error del~$1\%$ \cite{Necco:2001xg}. Nuestro modelo predice un valor concreto para la tensi\'on de la cuerda~$\sigma$.

Como vemos, el modelo predice para la energ\'{\i}a libre unos comportamientos asint\'oticos totalmente coherentes con la fenomenolog\'{\i}a conocida. Esto refuerza nuestra suposici\'on de existencia de contribuciones no perturbativas dadas por condensados glu\'onicos.

\subsection{Comparaci\'on con datos del ret\'{\i}culo}
\label{free_energy_2}

\begin{figure}[tbp]
\begin{center}
\epsfig{figure=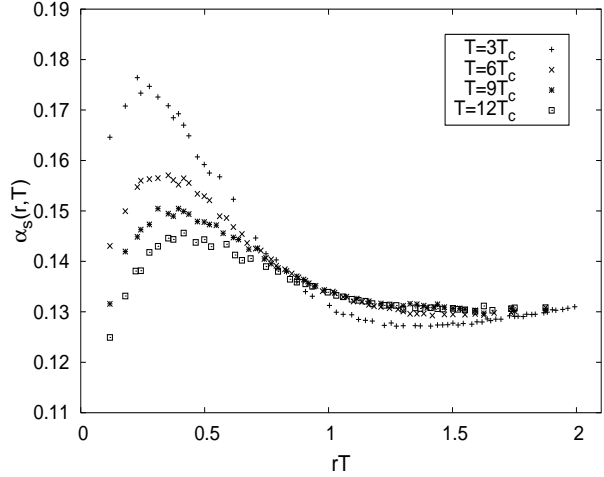,height=6.5cm,width=8.5cm}
\end{center}
\caption{Constante de acoplamiento $\alpha_s$ frente a $rT$ en gluodin\'amica pura $(N_c=3)$, para diferentes valores de $T$. Datos obtenidos a partir del ajuste de ec.~(\ref{eq:free_energy}) con los datos del ret\'{\i}culo de la figura 5 de ref.~\cite{Kaczmarek:2004gv}.}
\label{fig:PL5}
\end{figure}

Podemos comparar nuestro resultado, ec.~(\ref{eq:free_energy}), con datos del ret\'{\i}culo existentes para la energ\'{\i}a libre. Puesto que conocemos el valor del condensado $g^2 \langle A_{0,a}^2\rangle^{\text{No Pert}}$, esto nos va a permitir obtener la dependencia en $r$ y $T$ de la constante de acoplamiento $\alpha_s \equiv g^2/4\pi$. En la figura~\ref{fig:PL5} se muestra el valor de $\alpha_s$ frente a $rT$ para diferentes valores de la temperatura. Las curvas se han obtenido tras ajustar ec.~(\ref{eq:free_energy}) con los datos de ref.~\cite{Kaczmarek:2004gv} (figura 5) para gluodin\'amica $(N_c=3)$. Como valor de $g^2 \langle A_{0,a}^2\rangle^{\text{No Pert}}$ consideramos el de la tabla~\ref{tab:adjust_nf0_acte} con $N_\tau=8$.

Se observa un comportamiento suave para $\alpha_s$ y los valores son relativamente peque\~nos, lo cual contrasta con an\'alisis recientes en el ret\'{\i}culo a temperatura finita~\cite{Kaczmarek:2005ui,Kaczmarek:2004gv}. Estos autores tienen en cuenta los efectos no perturbativos que observan en los datos del ret\'{\i}culo de la energ\'{\i}a libre mediante el uso de dos constantes: $\alpha_s$ y $\overline\alpha_s$; y esta \'ultima se diferencia del valor perturbativo por un factor multiplicativo:
\begin{equation}
\overline\alpha_s(r,T) = \lambda \,\alpha_s^{\rm Pert} (r,T) \,, \qquad \lambda>1 \,.
\end{equation}  
Esto no tiene justificaci\'on te\'orica, y se trata en realidad de un esquema de an\'alisis dema\-siado forzado, pues la constante $\lambda$ no es tal, sino que tiene una dependencia en temperatura, de tal modo que vale 1 en el l\'{\i}mite $T\rightarrow\infty$.\footnote{El ajuste que se considera en ref.~\cite{Kaczmarek:2005ui,Kaczmarek:2004gv} se hace en base a la f\'ormula
\begin{equation}
F_{\text{fit}}(r,T) = -\frac{4\alpha(T)}{3r} \exp\left(-\sqrt{4\pi\overline\alpha(T)}\,rT\right) + b(T) \,,
\end{equation}
donde $\alpha(T)$ y $\overline\alpha(T)$ se usan como dos par\'ametros de ajuste independientes. Esta f\'ormula \'unicamente les permite ajustar el comportamiento de $F_1(r,T)$ a grandes distancias, en contraste con ec.~(\ref{eq:free_energy}), que reproduce correctamente tambi\'en el comportamiento a $r$ peque\~no, que viene dado por $V_{\bar{q}q}(r)$.
}
Los valores que obtienen para las~$\alpha_s$'s son excesivamente grandes. Por el contrario, al considerar nuestro modelo, el ajuste de los datos del ret\'{\i}culo de la energ\'{\i}a libre resulta m\'as natural. Notar que el comportamiento $r \to \infty$ de $\alpha_s(r,T)$ que se observa en fig.~\ref{fig:PL5} es consistente con el hecho de que nuestro mejor ajuste de los datos del loop de Polyakov renormalizado sea con~$a={\rm constante}$.

\subsection{Analog\'{\i}a entre el loop de Polyakov y el potencial quark-antiquark a temperatura cero}
\label{free_energy_3}

Al comparar (\ref{eq:F_infinito}) con (\ref{eq:t_zero}) se observa que las expresiones son similares desde un punto de vista formal, con la identificaci\'on~$r \leftrightarrow 1/m_D$. Si consideramos que no existe dependencia en $r$ y $T$ para la constante de acoplamiento $g$ y el condensado $\langle A_{0,a}^2\rangle$, de ec.~(\ref{eq:F_infinito}) a LO y de ec.~(\ref{eq:t_zero}) se deduce la siguiente propiedad
\begin{equation}
F_\infty (T) = V_{\bar{q}q}(r)\bigg|_{r=1/m_D} \,.
\label{eq:duality}
\end{equation}
\begin{figure*}[tb]
\begin{center}
\epsfig{figure=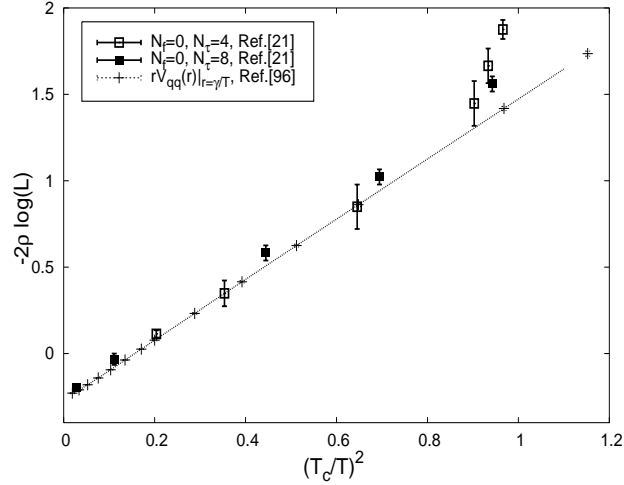,height=6.5cm,width=8.5cm}
\end{center}
\caption{ Logaritmo del loop de Polyakov renormalizado en gluodin\'amica
  $(N_c=3)$, reescalado con $\rho$, ec.~(\ref{eq:duality_2_alternativa}), frente al cuadrado de la inversa de la temperatura en unidades de la temperatura de transici\'on de fase. Los cuadrados negros y blancos corres\-pon\-den a datos del ret\'{\i}culo para el loop de Polyakov de ref.~\cite{Kaczmarek:2002mc}. Las cruces corres\-pon\-den a datos en el ret\'{\i}culo del potencial quark-antiquark a temperatura cero, $r V_{\bar{q}q}(r)$, de ref.~\cite{Necco:2001xg}, y modificados con el cambio $r=\gamma/T$. La l\'{\i}nea continua representa el modelo de cuerda bos\'onica que reproduce muy bien los datos del ret\'{\i}culo para $r V_{\bar{q}q}$(r) en la regi\'on~$0.75\,{\rm GeV}^{-1} \le r \le 4\,{\rm GeV}^{-1}$. Con el cambio~$r=\gamma/T$, esta regi\'on corresponde a~$0.06 \le (T_c/T)^2 \le 1.6 $.}
\label{fig:PL6}
\end{figure*}

Notar que (\ref{eq:duality}) es v\'alida s\'olo a LO en teor\'{\i}a de perturbaciones. Con objeto de comprobar num\'ericamente esta propiedad debemos tener en cuenta los diferentes comportamientos asint\'oticos de $\alpha_s$. Usaremos la siguiente notaci\'on:
\begin{equation}
\alpha_s(r) \equiv \alpha_s(r,T=0) \,, \qquad
\alpha_s(T) \equiv \alpha_s(r\rightarrow\infty,T) \,.
\end{equation}
La propiedad (\ref{eq:duality}) se escribir\'a ahora\footnote{Esta propiedad
  tambi\'en se puede expresar como
\begin{equation} 
\rho F_\infty(T)/T = r V_{\bar{q}q}(r)\bigg|_{r=\gamma/T} \,, \qquad
\rho=\gamma B \,, \label{eq:duality_2_alternativa}
\end{equation}
donde $B$ y $\gamma$ est\'an definidos en ec.~(\ref{eq:duality_factores}). 
}
\begin{equation}
B F_\infty (T) = V_{\bar{q}q}(r)\bigg|_{r=\gamma/T} \,,
\label{eq:duality_2}
\end{equation}
donde
\begin{equation}
B=\left(\frac{\alpha_s(r)}{\alpha_s(T)}\right)^{3/4}  \,, \qquad
\gamma=\frac{1}{\sqrt{4\pi(N_c/3+N_f/6)}}\left(\frac{\alpha_s(r)}{\alpha_s(T)^3}\right)^{1/4} \,.
\label{eq:duality_factores}
\end{equation}
La propiedad (\ref{eq:duality_2}), con los valores de los par\'ametros $B$ y $\gamma$ dados en ec.~(\ref{eq:duality_factores}), se ha deducido suponiendo que se cumple
\begin{equation}
\alpha_s(T) \langle A_{0,a}^2 \rangle_T^{\text{No Pert}} = \alpha_s(r) \langle A_{0,a}^2\rangle_{T=0}  \,.
\label{eq:alfa_s_suposicion}
\end{equation}
El miembro izquierdo de la igualdad $\alpha_s(T)\langle A_{0,a}^2 \rangle^{\text{No Pert}}$ ha sido ajustado en la secci\'on~\ref{comparison_lattice_data_nf0}. El valor de~$\alpha_s(r) \langle A_{0,a}^2\rangle_{T=0}$ puede obtenerse a partir del valor conocido para la tensi\'on de la cuerda, $\sigma = (0.42 \,{\rm GeV})^2$, y la ecuaci\'on~(\ref{eq:string_tension_condensado}). Num\'ericamente encontramos que ec.~(\ref{eq:alfa_s_suposicion}) es correcta con un error del $9\%$.
En la figura~\ref{fig:PL6} se muestran los datos del ret\'{\i}culo en gluodin\'amica para $-2\rho\log L$ frente a $(T_c/T)^2$ (ref.~\cite{Kaczmarek:2002mc}), y se comparan con el potencial quark-antiquark a temperatura cero $r V_{\bar{q}q}(r)$~\cite{Necco:2001xg} despu\'es de haber considerado el cambio de variable que se especifica en ec.~(\ref{eq:duality_2_alternativa}). Se observa un acuerdo excelente. Esta dualidad sugiere la existencia de una profunda analog\'{\i}a entre el potencial quark-antiquark a temperatura cero y el loop de Polyakov.

\section{Conclusiones}

Tres son los resultados importantes de este cap\'{\i}tulo. Por una parte, tras analizar de manera conveniente los datos en el ret\'{\i}culo del loop de Polyakov renormalizado por encima de la transici\'on de fase de QCD, encontramos la contribuci\'on inequ\'{\i}voca de un condensado de dimensi\'on 2 no perturbativo. Estas contribuciones no han sido consideradas hasta ahora en el contexto del loop de Polyakov, pero de hecho son dominantes en la regi\'on cercana a la transici\'on de fase y permiten describir los datos de \cite{Kaczmarek:2002mc} en la fase de desconfinamiento hasta $1.03\, T_c$ para gluodin\'amica y de \cite{Kaczmarek:2005ui} hasta $1.15\,T_c$ para dos sabores. 

En segundo lugar, hemos sugerido identificar este condensado con el
condensado glu\'onico de dimensi\'on 2 invariante BRST. El valor
num\'erico de $g^2\langle A_{0,a}^2\rangle^{\text{No Pert}}$ que
obtenemos a partir del loop de Polyakov es totalmente consistente con
el valor que se deduce de la presi\'on en
gluodin\'amica~\cite{Kajantie:2000iz,Kajantie:2002pu}. Adem\'as, aun
habiendo definido el condensado en un gauge est\'atico, su valor es
significativamente pr\'oximo al valor de $g^2 \langle A_{\mu,a}^2\rangle /
4$, obtenido a temperatura cero y en el gauge de Landau.

En tercer lugar, a la luz de estos resultados hemos encontrado una analog\'{\i}a entre el potencial quark-antiquark a temperatura cero y el loop de Polyakov, la cual se manifiesta en la relaci\'on que predice nuestro modelo entre la tensi\'on de la cuerda y la pendiente del loop de Polyakov.

\chapter{Modelos de Quarks Quirales a Temperatura Finita}
\label{quirales_Tfinita}

En este cap\'{\i}tulo estudiaremos algunos modelos de quarks quirales en el contexto de temperatura finita. Haciendo uso de nuestra t\'ecnica del heat kernel del cap\'{\i}tulo~\ref{heat_kernel}, obtendremos el acoplamiento m\'{\i}nimo entre el loop de Polyakov glu\'onico y los quarks, lo cual solucionar\'a algunas inconsistencias presentes en el tratamiento est\'andar de estos modelos a temperatura finita a nivel de un loop de quarks.

En primer lugar se estudiar\'an algunas propiedades de las transformaciones gauge a temperatura finita, lo cual nos llevar\'a a considerar la simetr\'{\i}a del centro como aquella que es generada por la acci\'on de transformaciones gauge locales que son peri\'odicas en la variable temporal salvo un elementro arbitrario del centro del grupo gauge. Para m\'as detalles sobre este punto, ver ap\'endice~\ref{app:gauge}. 

Posteriormente introduciremos dos modelos: modelo de Nambu--Jona-Lasinio y modelo quark espectral. Con ellos ilustraremos la problem\'atica del tratamiento est\'andar a temperatura finita que se viene haciendo en los modelos de quarks quirales, y definiremos un modelo quark quiral con acoplamiento del loop de Polyakov que permitir\'a compatibilizar los resultados con los conocidos de Teor\'{\i}a Quiral de Perturbaciones. Calcularemos el lagrangiano quiral efectivo en estos modelos hasta ${\cal O}(p^4)$ en un desarrollo en momentos externos, y se estudiar\'a la estructura que presenta este lagrangiano a temperatura finita.

Se har\'a un estudio de algunas correcciones de orden mayor, tales como correcciones m\'as all\'a de un loop de quarks, correcciones glu\'onicas y correcciones locales en el loop de Polyakov. Finalmente se calcular\'an dos observables de inter\'es: condensado quiral y valor esperado del loop de Polyakov; para lo cual se har\'a un tratamiento unquenched, y se estudiar\'a el mecanismo de rotura de la simetr\'{\i}a del centro que conduce a la transici\'on de fase de QCD.

El cap\'{\i}tulo est\'a basado en las referencias~\cite{Megias:2004hj,Megias:2006bn}.

\section{Transformaciones gauge grandes}
\label{large_gauge_transformations}

En el ap\'endice~\ref{app:gauge} discutimos las transformaciones gauge en el contexto de la Teor\'{\i}a Cu\'antica de Campos a temperatura finita. Al comienzo de este cap\'{\i}tulo vamos a hacer un repaso de las principales propiedades de estas transformaciones, y la importancia que tienen para el estudio de los procesos de desconfinamiento de color en QCD. Esta secci\'on podr\'{\i}a haberse incluido igualmente en el cap\'{\i}tulo~\ref{QCD_efective_action}, pero hemos preferido ponerla aqu\'{\i} para que el cap\'{\i}tulo quede autoconsistente, pues como veremos los modelos de quarks quirales nos van a permitir una descripci\'on de la transici\'on de fase de QCD.

\subsection{Transformaciones gauge a temperatura finita}
\label{gauge_tfinita}

En el formalismo de Tiempo Imaginario el espacio-tiempo es un cilindro topol\'ogico, de tal modo que el tiempo imaginario eucl\'{\i}deo est\'a compactificado y las integrales funcionales se eval\'uan bajo la condici\'on de que los campos sean peri\'odicos para bosones y antiperi\'odicos para fermiones en el intervalo temporal $[0,\beta]$, donde $\beta=1/T$. En un principio, s\'olo estar\'{\i}an permitidas las transformaciones gauge peri\'odicas
\begin{equation}
g(x_0,\vec{x}) = g(x_0+\beta,\vec{x}) \,,
\label{eq:transf_gauge_period}
\end{equation}
pues los campos de los quarks y los bosones son estables frente a este tipo de
transformaciones. Un ejemplo de tal transformaci\'on para el grupo gauge
SU($N_c$), en el gauge de Polyakov, $\partial_0 A_0 = 0$, con $A_0$ una matriz
diagonal $N_c \times N_c$ de traza cero, es 
\begin{equation}
g(x_0) = e^{i2\pi x_0 \Lambda/\beta} \,,
\end{equation} 
donde $\Lambda$ es una matriz diagonal de enteros, de traza cero, en el espacio de color, esto es $\Lambda_{ij} = n_i \delta_{ij} \,, n_i \in \mathbb{Z}\,, \sum_{j=1}^{N_c} n_j = 0$. Esta transformaci\'on no puede estar pr\'oxima a la identidad, y en este sentido se considera una {\it transformaci\'on gauge grande}. Bajo ella, el campo $A_0$ transforma
\begin{equation}
A_0 \rightarrow A_0 + \frac{2\pi}{\beta} \Lambda \,,
\end{equation}
de modo que en este gauge, la invariancia gauge se manifiesta en la periodicidad del campo glu\'onico $A_0$. El problema de teor\'{\i}a de perturbaciones radica en que esta invariancia a temperatura finita se rompe expl\'{\i}citamente si se hace un desarrollo perturbativo de $A_0$, ya que el desarrollo de una funci\'on peri\'odica da lugar a un polinomio, que no es peri\'odico.

Esta problem\'atica de la invariancia gauge a temperatura finita conduce a la necesidad de tratar el campo $A_0$ de una manera no perturbativa, y a tales efectos se considera el loop de Polyakov (o l\'{\i}nea de Wilson sin traza) como grado de libertad independiente $\Omega(x)$. Transforma de manera covariante en $x$ bajo una transformaci\'on gauge peri\'odica
\begin{equation}
\Omega(x) \rightarrow g^{-1}(x) \Omega(x) g(x)\,,
\end{equation}
y en el gauge de Polyakov, $\Omega(x)= e^{i\beta A_0(\vec{x})}$, es invariante gauge.

\subsection{Simetr\'{\i}a del centro}

En gluodin\'amica pura a temperatura finita la condici\'on (\ref{eq:transf_gauge_period}) resulta en realidad demasiado restrictiva, y es posible considerar transformaciones gauge aperi\'odicas
\begin{equation}
g(x_0+\beta,\vec{x}) = z\, g(x_0,\vec{x}) \,, \qquad z^{N_c}=1\,. 
\label{eq:g_tr_c} 
\end{equation}
z es un elemento de $\mathbb{Z}(N_c)$, que es el centro del grupo gauge SU($N_c$), esto es $z=e^{i2\pi n /N_c}\,, n \in \mathbb{Z}(N_c)$. Un ejemplo de esa transformaci\'on, en el gauge de Polyakov, es
\begin{equation}
g(x_0) = e^{i2\pi x_0 \Lambda/N_c\beta} \,,
\label{eq:transf_gauge_centro}
\end{equation}
donde $z=e^{i2\pi/N_c}$. El campo $A_0$ y el loop de Polyakov transforman bajo (\ref{eq:transf_gauge_centro}) como
\begin{equation}
A_0 \rightarrow A_0 + \frac{2\pi}{N_c\beta}\Lambda \,, \qquad
\Omega \rightarrow z \Omega \,.
\label{eq:trans_g_centro}
\end{equation}
$\Omega$ transforma como la representaci\'on fundamental del grupo $\mathbb{Z}(N_c)$. F\'{\i}sicamente el promedio t\'ermico del loop de Polyakov (con traza) en la representaci\'on fundamental determina la energ\'{\i}a libre relativa al vac\'{\i}o de un \'unico quark,
\begin{equation}
e^{-\beta F_q(x)} = \frac{1}{N_c}\langle \tr_c \,\Omega (x)\rangle \,. 
\end{equation}
De ec.~(\ref{eq:trans_g_centro}) se deduce (por invariancia gauge) que
\begin{equation}
\langle \tr_c \,\Omega (x)\rangle = z \,\langle \tr_c \,\Omega (x)\rangle \,,
\end{equation}
y por tanto $\langle \tr_c \,\Omega (x)\rangle=0$ en la fase en que la simetr\'{\i}a del centro se preserva (fase de confinamiento). De manera m\'as general, se obtiene
\begin{equation}
\langle \tr_c \,\Omega^n (x) \rangle = 0  \qquad \textrm{para} \qquad  n \ne m N_c \,, \qquad m\in\mathbb{Z} \,.
\end{equation}
La simetr\'{\i}a del centro est\'a espont\'aneamente rota por encima de una cierta temperatura ($T_D \approx 270\;\textrm{MeV}$ para $N_c=3$), lo cual indica una fase de desconfinamiento. En esta fase $\langle \tr_c \,\Omega (x)\rangle$ puede tomar valores diferentes de cero.

\subsection{Rotura de la simetr\'{\i}a del centro por fermiones}

Las funciones de onda de los fermiones deben satisfacer condiciones antiperi\'odicas en la direcci\'on temporal, esto es $q(\beta,\vec{x}) = -q(0,\vec{x})$, de modo que bajo una transformaci\'on del tipo (\ref{eq:g_tr_c})
\begin{equation}
q(\beta,\vec{x}) \rightarrow g(\beta,\vec{x}) q(\beta,\vec{x}) = -z g(0,\vec{x}) q(0,\vec{x}) \,,
\label{eq:quarks_s_c}
\end{equation}
en lugar de $-g(0,\vec{x})q(0,\vec{x})$. Notar que $\overline{q}(n\beta)q(0)\rightarrow z^{-n} \overline{q}(n\beta)q(0)$, lo cual implica que en la fase confinante (con simetr\'{\i}a del centro)
\begin{equation}
\langle \overline{q}(n\beta)q(0)\rangle = 0 \qquad \textrm{para} \qquad
n\ne m N_c \,, \qquad m \in \mathbb{Z} \,.
\label{eq:regla_selec1}
\end{equation}
Esto genera una regla de selecci\'on en gluodin\'amica pura. 

Los fermiones no son estables bajo las transformaciones del tipo
(\ref{eq:g_tr_c}) (\'unicamente lo son bajo transformaciones peri\'odicas), de
modo que rompen expl\'{\i}citamente la simetr\'{\i}a del centro. Esto
significa que la regla de selecci\'on~(\ref{eq:regla_selec1}) no se realiza en
la pr\'actica.  No obstante, esta regla ser\'a importante en el contexto de
modelos de quarks quirales en el l\'{\i}mite de $N_c$ grande, tal y como veremos m\'as adelante.

\section{Modelos de Quarks Quirales}
\label{modelos_quark_quirales}

En esta secci\'on explicaremos dos modelos de quarks quirales de especial relevancia: el modelo de Nambu--Jona-Lasinio~\cite{njlmodel} y el modelo quark espectral~\cite{spectralqm}.

\subsection{Modelo Quark de Nambu--Jona-Lasinio}
\label{sec:njl1}

El lagrangiano eucl\'{\i}deo del modelo de Nambu--Jona-Lasinio generalizado es
\begin{equation}
{\cal L}_{\NJL}= \overline{q}(\thru{\partial}+\hat{m}_0)q 
+\frac{1}{2a_s^2}\sum_{a=0}^{N_f^2-1}((\overline{q}\lambda_a q)^2
+(\overline{q}\lambda_a i\gamma_5 q)^2)
+\frac{1}{2a_v^2}\sum_{a=0}^{N_f^2-1}((\overline{q}\lambda_a \gamma_\mu q)^2+
(\overline{q}\lambda_a \gamma_\mu \gamma_5 q)^2) \,,
\label{eq:lag_njl}
\end{equation}
donde $q=(u,d,s,\ldots)$ representa el campo de los quarks con $N_c$ colores y $N_f$ sabores. Las $\lambda$'s son las matrices de Gell-Mann del grupo U($N_f$) y $\hat{m}_0=\diag(m_u,m_d,m_s,\ldots)$ es la matriz de masa de los quarks. $1/a_s^2$ y $1/a_v^2$ son las constantes de acoplamiento. Este lagrangiano es invariante bajo simetr\'{\i}a global de color SU($N_c$).

El funcional generador en presencia de campos externos bos\'onicos $(s,p,v,a)$ y fermi\'onicos $(\eta,\overline\eta)$ es
\begin{equation}
Z_{\NJL}[s,p,v,a,\eta,\overline\eta] = \int {\cal D}\overline{q} {\cal D}q
\exp\Big[
-\int d^4 x ({\cal L}_{\NJL}+\overline{q}(\thru{v}+\thru{a}\gamma_5+s+i\gamma_5 p)q 
+\overline\eta q + \overline{q}\eta)
\Big] \,.
\end{equation}
Los s\'{\i}mbolos $s$, $p$, $v_\mu$ y $a_\mu$ indican campos externos (en espacio de sabor) de tipo escalar, pseudoescalar, vector y axial, respectivamente. En funci\'on de los generadores del grupo de sabor U($N_f$), estos se escriben
\begin{equation}
s = \sum_{a=0}^{N_f^2-1} s_a \frac{\lambda_a}{2} \,, \qquad \cdots 
\end{equation}
La acci\'on del modelo puede ser bosonizada mediante la introducci\'on de campos bos\'onicos auxiliares, lo cual va a transformar la interacci\'on local de cuatro puntos en un acoplamiento tipo Yukawa \cite{eguchi}. El nuevo funcional generador es
\begin{eqnarray}
Z_{\NJL}[s,p,v,a,\eta,\overline\eta] &=& 
\int {\cal D}\overline{q} {\cal D}q{\cal D}S {\cal D}P {\cal D}V {\cal D}A
\exp\Biggl[
-\int d^4 x \Big(
\overline{q}(\thru{\partial}+\thru{\cal V}+\thru{\cal A}\gamma_5+{\cal S}+i\gamma_5{\cal P})q
\nonumber \\
&&\quad\qquad+\frac{a_s^2}{4}\tr((S-\hat{m}_0)^2+P^2)-\frac{a_v^2}{4}\tr(V_\mu^2+A_\mu^2)
+\overline\eta q +\overline{q}\eta\Big) 
\Biggl] \,,
\label{eq:njlfuncgen1} 
\end{eqnarray}
donde hemos escrito en notaci\'on corta ${\cal S}=s+S$, ${\cal P}=p+P$, ${\cal V}=v+V$, ${\cal A}=a+A$. En esta f\'ormula $(S,P,V,A)$ representan campos bos\'onicos din\'amicos internos de tipo escalar, pseudoescalar, vector y axial respectivamente. Los campos ${\cal S}(x)$, ${\cal P}(x)$, ${\cal V}_\mu (x)$ y ${\cal A}_\mu (x)$ son matrices en espacio interno (que se entiende como espacio de sabor), son la identidad en espacio de Dirac y operadores multiplicativos en el espacio~$x$. ${\cal S}(x)$ es herm\'{\i}tico y  ${\cal P}(x)$, ${\cal V}_\mu (x)$ y ${\cal A}_\mu (x)$ son antiherm\'{\i}ticos. En la secci\'on~\ref{coupling_polyakov} extende\-remos los campos~${\cal A}_\mu (x)$ y ${\cal V}_\mu (x)$ para que sean matrices no triviales en espacio de color, lo que nos permitir\'a acoplar el loop de Polyakov de color en el modelo. Por conveniencia en nuestro desarrollo hemos incluido la rotura expl\'{\i}cita de la simetr\'{\i}a quiral (proporcional a $\hat{m}_0$) en el t\'ermino bos\'onico local. Podemos integrar formalmente sobre fermiones, lo cual conduce a 
\begin{eqnarray}
Z_{\NJL}[s,p,v,a,\eta,\overline\eta] &=& 
\int {\cal D}S {\cal D}P {\cal D}V {\cal D}A\;
\Det({\bf D})^{N_c}\exp(\langle \overline\eta|{\bf D}^{-1}|\eta\rangle) \label{eq:njlfuncgen2} \\
&& \exp\Biggl[
-\int d^4 x \Big(\frac{a_s^2}{4}\tr((S-\hat{m}_0)^2+P^2)-\frac{a_v^2}{4}\tr(V_\mu^2+A_\mu^2)
\Big)
\Biggl]\nonumber 
\end{eqnarray}
donde
\begin{equation}
{\bf D} = \thru\partial + \thru{\cal V} + \thru{\cal A}\gamma_5
+{\cal S} + i\gamma_5{\cal P}\,
\label{eq:njlopdirac}
\end{equation}
es un operador de Dirac. Este operador se puede escribir en la forma
\begin{equation}
{\bf D}=\thruu{D}_{\cal V}+\thru{\cal A}\gamma_5 + M U^{\gamma_5} \,, 
\label{eq:bnm1}
\end{equation}
donde $D_\mu^{\cal V}=\partial_\mu + {\cal V}_\mu$ es la derivada covariante vector, $M$ es la masa constituyente de los quarks, y $U$ es una matriz en espacio de sabor que representa los octetes pseudoescalares de los mesones en la representaci\'on no lineal. Para tres sabores, $N_f=3$, se escribe $U=e^{i\sqrt{2}\Phi/f_\pi}$, con
\begin{eqnarray} 
        \Phi = \left( \begin{matrix} 
        \frac{1}{\sqrt{2}} \pi^0 +
        \frac{1}{\sqrt{6}} \eta & \pi^+ & K^+  \cr  \pi^- & -
        \frac{1}{\sqrt{2}} \pi^0 + \frac{1}{\sqrt{6}} \eta & K^0  \cr 
        K^- & \bar{K}^0 & - \frac{2}{\sqrt{6}} \eta 
        \end{matrix}
        \right) .
\end{eqnarray}
$f_\pi$ es la constante de desintegraci\'on d\'ebil del pi\'on en el l\'{\i}mite quiral. 

En lo que sigue consideraremos la acci\'on efectiva a nivel de un loop de quarks y a nivel \'arbol para los mesones. En este caso
\begin{equation}
\Gamma_{\NJL} = \Gamma_q[{\bf D}] + \Gamma_m \,,
\label{eq:njlacefectE}
\end{equation}
donde
\begin{eqnarray}
\Gamma_q[{\bf D}] &=& -N_c \Tr \log({\bf D}) \,, \label{eq:njlacefectq} \\
\Gamma_m &=& \int d^4 x \left\{\frac{a_s^2}{4}\tr(S^2+P^2)
-\frac{a_s^2}{2}\tr(\hat{m}_0 S)+\frac{a_s^2}{4}\tr(\hat{m}_0^2)
-\frac{a_v^2}{4}\tr(V_\mu^2+A_\mu^2)
\right\}
\label{eq:njlacefect}
\end{eqnarray}
En adelante nos vamos a referir al t\'ermino $\Gamma_q[{\bf D}]$ como la contribuci\'on de los quarks a un loop. \'Este ser\'a el t\'ermino que calculemos como aplicaci\'on de nuestro desarrollo del heat kernel a temperatura finita.  

La contribuci\'on de los quarks a la acci\'on efectiva se puede separar en una parte $\gamma_5$-par y otra $\gamma_5$-impar, correspondiente a procesos de paridad normal y anormal, respectivamente. En espacio eucl\'{\i}deo, la primera corresponde a la parte real de la acci\'on efectiva, y la segunda a la parte imaginaria. Introduciremos el operador
\begin{eqnarray}
{\bf D}_5 [ {\cal S}, {\cal P}, {\cal V}, {\cal A} ] = \gamma_5 {\bf D} [ {\cal S},-{\cal P},{\cal V},-{\cal A}] \gamma_5 \,,
\label{eq:defD5}
\end{eqnarray}
que en espacio eucl\'{\i}deo se corresponde con el herm\'{\i}tico conjugado ${\bf D}^\dagger$. La contribuci\'on de paridad normal es cuadr\'aticamente divergente y puede ser regularizada de un modo invariante gauge quiral mediante el esquema de Pauli-Villars~\cite{paulivillars}
\begin{equation} 
\Gamma_q^+[{\bf D}] = -{N_c \over 2} {\rm Tr} \sum_i c_i \log({\bf D}_5 {\bf D} +
\Lambda_i^2)  \,,
\label{eq:njlacefectreal1}
\end{equation}
donde los reguladores de Pauli-Villars satisfacen $c_0=1$, $\Lambda_0=0$ y $\sum_i c_i=0$, $\sum_i c_i \Lambda_i^2 = 0$, lo cual permitir\'a hacer finitas las divergencias logar\'{\i}tmicas y las cuadr\'aticas, respectivamente. Haciendo uso de la representaci\'on de Schwinger de tiempo propio, esta contribuci\'on se escribe
\begin{equation}
\Gamma_q^+[{\bf D}] = \frac{N_c}{2}\int_0^\infty \frac{d\tau}{\tau}
\,\phi(\tau) \,\Tr \,e^{-\tau {\bf D}_5 {\bf D} }  \,, 
\label{eq:njlacefectreal}
\end{equation}
donde 
\begin{equation}
\phi(\tau) = \sum_i c_i \,e^{-\tau \Lambda_i^2} \,.
\end{equation}
Las funciones de Green se pueden obtener a partir de (\ref{eq:njlacefectE}) derivando respecto a los campos medios mes\'onicos. De particular inter\'es es la funci\'on a un punto. Si en (\ref{eq:njlacefectE}) consideramos solamente la parte real de la contribuci\'on de los quarks a un loop, esto es (\ref{eq:njlacefectreal1}), esta acci\'on presenta un punto estacionario invariante traslacional en $(S,P)=(\Pi,0)$, $(V_\mu,A_\mu)=(0,0)$
\begin{equation}
\frac{\delta \Gamma_{\NJL}^+[S]}{\delta S(x)}\Big|_{S(x)=\Pi}
= \frac{a_s^2}{2}\tr(\Pi-\hat{m}_0)
-\frac{N_c}{2} \,\Tr \left(({\bf D}_5 {\bf D})^{-1}\frac{\delta ({\bf D}_5 {\bf D)}}{\delta S(x)}\right)_{S(x)=\Pi}
= 0 \,.
\label{eq:njlcondgap}
\end{equation}
El punto estacionario $\Pi$ se identifica con el valor esperado en el vac\'{\i}o del campo $S$ en la aproximaci\'on de un loop de quarks. Introduciendo la acci\'on efectiva regularizada (\ref{eq:njlacefectreal}) en (\ref{eq:njlcondgap}) obtenemos la siguiente ecuaci\'on para $\Pi$
\begin{equation} 
a_s^2 (\Pi-\hat{m}_0) -8 N_c\Pi \,g(\Pi)=0 \,,
\label{eq:njlgap}
\end{equation}
donde
\begin{equation}
g(\Pi)=\int \frac{d^4 p}{(2\pi)^4} 
\int_0^\infty d\tau \, \phi(\tau) \,e^{-\tau(p^2+\Pi^2)} \,.
\end{equation}
En adelante nos referiremos a (\ref{eq:njlgap}) como ecuaci\'on del gap pues esta ecuaci\'on determina el gap de energ\'{\i}a $2\Pi$ entre los estados de quarks con energ\'{\i}a positiva y negativa. $\Pi$ juega el papel de la masa constituyente de los quarks. 

El condensado de quarks $\langle\overline{q}q\rangle$ viene dado por $\langle\overline{q}q\rangle=\delta\Gamma_{\NJL}^+/\delta\hat{m}_0$. De (\ref{eq:njlacefect}) se obtiene inmediatamente
\begin{equation}
\langle\overline{q}q\rangle = -\frac{a_s^2}{2}\tr(\Pi-\hat{m}_0) \,. 
\end{equation}

\subsection{Modelo Quark Espectral}
\label{sec:mqe1}

El Modelo Quark Espectral, desarrollado recientemente por E. Ruiz Arriola y W. Broniowski \cite{spectralqm}, es aplicable a f\'{\i}sica hadr\'onica en el rango de baja energ\'{\i}a. La novedad reside en el uso de una regularizaci\'on espectral basada en la introducci\'on a nivel formal de la representaci\'on de Lehmann~\cite{zuber} del propagador del quark. Esta regularizaci\'on permite resolver de una manera simple las identidades de Ward-Takahashi quiral y electromagn\'etica mediante el uso de la llamada {\it prescripci\'on gauge} \cite{gaugetechnique}. Consideraremos el modelo a nivel de un loop fermi\'onico y en el l\'{\i}mite quiral en que la masa de los quarks es cero.   

En esta secci\'on vamos a seguir la referencia \cite{spectralqm}. El punto de partida es el propagador del quark, que en espacio de momentos se define
\begin{equation} 
S(p) = \int d^4 x e^{-px} \langle 0| T\{q(x)\overline{q}(0)\}|0 \rangle 
\,.
\end{equation}
Consideraremos una representaci\'on espectral para el propagador
\begin{equation}
S(p) = \int_{\cal C} d\omega \frac{\rho(\omega)}{\thru{p}-\omega} \,,
\label{eq:prop_q_sqm_re}
\end{equation}
donde $\rho(\omega)$ es la funci\'on espectral y ${\cal C}$ indica un contorno de integraci\'on en el plano complejo $\omega$ elegido de un modo conveniente. Este propagador puede ser parametrizado en la forma est\'andar
\begin{equation}
S(p)=A(p)\thru{p}+B(p)=Z(p)\frac{\thru{p}+M(p)}{p^2-M^2(p)} \,,
\end{equation}
donde
\begin{eqnarray}
A(p)=\int_{\cal C} d\omega \frac{\rho(\omega)}{p^2-\omega^2} \,,
\qquad
B(p)= \int_{\cal C} d\omega \frac{\rho(\omega)\omega}{p^2-\omega^2} \,. 
\label{eq:ab_sqm}
\end{eqnarray}
La masa y factor de renormalizaci\'on vienen dados por
\begin{equation}
M(p)= \frac{B(p)}{A(p)} \,,
\qquad
Z(p)=(p^2-M^2(p))A(p) \,, \label{eq:mz_sqm}
\end{equation}
respectivamente. Notar que si $\rho(\omega)=\rho(-\omega)$ tendr\'{\i}amos $M(p)=0$ y no existir\'{\i}a rotura espont\'anea de la simetr\'{\i}a quiral. Por tanto es de esperar que $\rho(\omega)$ no sea una funci\'on par en general. La funci\'on espectral debe ser tal que proporcione valores finitos para los observables hadr\'onicos. Esto dar\'a lugar a una serie de condiciones que deben cumplir los momentos y los momentos logar\'{\i}tmicos de $\rho(\omega)$,
\begin{equation}
\rho_n = \int_{\cal C} d\omega \omega^n \rho(\omega) \,,
\qquad
\rho_n^\prime = \int_{\cal C} d\omega \log(\omega^2/\mu^2) \omega^n \rho(\omega) \,,
\qquad 
n \in {\mathbb Z} \,.
\end{equation}
Aqu\'{\i} $\mu$ es una cierta escala. Notar que por normalizaci\'on $\rho_0=1$. Como ejemplo conside\-remos el condensado de quarks (por el momento trabajaremos a temperatura cero)
\begin{equation}
\langle\overline{q}q\rangle = -N_c \int_{\cal C} d\omega \rho(\omega)
\int \frac{d^4 p}{(2\pi)^4} \tr_{\Dirac} \frac{1}{\thru{p}-\omega} \,.
\end{equation}
Tras tomar la traza en el espacio de Dirac, la integral es cuadr\'aticamente divergente. Un modo de regularizarla es haciendo uso de un cutoff tridimensional con la siguiente sustituci\'on
\begin{equation}
\int d^4 p \longrightarrow 4\pi \int d p_0 \int_0^\Lambda dp \;p^2 \,,
\qquad p=|\vec{p}| \,.
\end{equation}
Con esta regularizaci\'on obtenemos
\begin{equation}
\langle\overline{q}q\rangle =
-\frac{N_c}{4\pi^2} \int_{\cal C} d\omega\, \omega \rho(\omega)
\left[ 2\Lambda^2 + \omega^2\log\left(\frac{\omega^2}{4\Lambda^2}\right)
+\omega^2
\right] \,.
\end{equation}
Puesto que el resultado debe ser finito en el l\'{\i}mite
$\Lambda\rightarrow\infty$, es necesario imponer las condiciones $\rho_1=0$ y
$\rho_3=0$, lo cual conduce a
$\langle\overline{q}q\rangle=-N_c\rho_3^\prime/(4\pi^2)$. El c\'alculo de
otros observables va a dar lugar a condiciones adicionales. En general todos los observables van a ser proporcionales a los momentos inversos y a los momentos logar\'{\i}tmicos, y para que sean finitos se debe cumplir $\rho_n=0$, $n>0$. El modelo espectral no se ha desarrollado m\'as all\'a de un loop.

La prescripci\'on gauge fue usada en el pasado en la obtenci\'on de soluciones de las ecuaciones de Schwinger-Dyson. Haciendo uso de ella se pueden resolver en este modelo las identidades de Ward-Takahashi. Sin embargo en situaciones en las que las l\'{\i}neas de propagadores de los quarks est\'an cerradas es m\'as conveniente el formalismo de la acci\'on efectiva. Consideraremos, como en el modelo de Nambu--Jona-Lasinio, acoplamientos escalar, pseudo-escalar, vector y axial. El acoplamiento quark-pi\'on debe satisfacer la relaci\'on de Goldberger-Treiman~\cite{cheng-li}. Con estas premisas, la acci\'on efectiva de este modelo a nivel de un loop de quarks se puede escribir como
\begin{equation}
\Gamma_{\SQM} = -N_c \int d^4 x \int_{\cal C} d\omega \rho(\omega) 
\tr \log\left(
\thruu{D}_V + \thruu{A}\gamma_5 + \omega U^{\gamma_5}
\right) \,,   
\label{eq:ae_sqm_Tfin}
\end{equation}
donde $D_\mu^V = \partial_\mu + V_\mu$ es la derivada covariante vector. 
En el modelo NJL, $M$ jugaba el papel de la masa constituyente de los
quarks. En el modelo espectral $M$ se convierte en la variable de
integraci\'on $\omega$ de la funci\'on espectral. La diferencia esencial con
el modelo NJL, y en general con todos los modelos de quarks quirales, es que aqu\'{\i} no consideramos un cutoff que separe el r\'egimen de baja energ\'{\i}a, donde se supone que el modelo funciona, y el r\'egimen de alta energ\'{\i}a.  

En el cap\'{\i}tulo \ref{ae_quiral_SQM} se har\'a un estudio m\'as extenso del modelo espectral considerando un espacio-tiempo curvo, y se introducir\'a el {\it esquema de dominancia del mes\'on vectorial}, que constituye una realizaci\'on simple del modelo y proporciona una forma expl\'{\i}cita para la funci\'on espectral.

\section{Problem\'atica de los modelos de quarks quirales a temperatura finita}
\label{problematica_modelosqq}

El tratamiento est\'andar de los Modelos de Quarks Quirales a Temperatura Finita presenta algunas inconsistencias. Por una parte, en el c\'alculo de observables aparecen involucrados estados excitados con cualquier n\'umero de quarks, y esto ocurre incluso para temperaturas bajas. Sorprendentemente, durante mucho tiempo no ha habido demasiada preocupaci\'on por parte de los autores en resolver este problema, y normalmente lo han atribuido a fallos del propio modelo, tales como falta de confinamiento.

\subsection{Tratamiento est\'andar a temperatura finita}
\label{tratamiento_estandar}

El tratamiento est\'andar consiste en pasar de las f\'ormulas con $T=0$ hasta otras f\'ormulas para $T\ne 0$, mediante la aplicaci\'on de la regla
\begin{eqnarray}
\int \frac{d k_0}{2\pi}  
F( k_0 , \vec k) \to i T
\sum_{n=-\infty}^\infty  F( i w_n ,
\vec k) \,,
\label{eq:regla_estandar}
\end{eqnarray}
donde $F$ puede representar el propagador de un quark, en espacio de momentos. $\omega_n$ son las frecuencias de Matsubara fermi\'onicas, $\omega_n=2\pi T (n+1/2)$. Si aplicamos esta regla en el condensado quiral, a temperatura finita y a un loop se tiene
\begin{equation}
\langle \overline{q}q \rangle = -iN_c \sum_{n=-\infty}^{\infty} (-1)^n \tr_{\textrm{Dirac}} S(x)|_{x_0=in\beta}
= 4MT \tr_c \sum_{\omega_n} 
\int \frac{d^3k}{(2\pi)^3} \frac{1}{\omega_n^2+\vec{k}^2+M^2} \,.
\end{equation}
Despu\'es de hacer la integraci\'on en momentos, y aplicar la f\'ormula de
Poisson para la sumatoria, ec.~(\ref{eq:poisson}), queda
\begin{equation}
\langle \overline{q}q \rangle_T = \langle \overline{q}q \rangle_{T=0} -2\frac{N_cM^2 T}{\pi^2} \sum_{n=1}^\infty \frac{(-1)^n}{n} K_1(nM/T)  \nonumber 
\end{equation}
\begin{equation}
\qquad\stackrel{{\rm T} \; \text{peque\~no}} \sim  \langle \overline{q} q \rangle_{T=0} - \frac{N_c}{2} \sum_{n=1}^\infty  (-1)^n \left( \frac{2MT}{n\pi} \right)^{3/2} e^{-nM/T} \,,
\label{eq:cond_pois}
\end{equation}
donde se ha hecho uso del comportamiento asint\'otico de la funci\'on de Bessel $K_n(z)$ para el r\'egimen de temperatura peque\~na $K_n(z)\sim e^{-z}\sqrt{\pi/2z}$.

\subsection{Generaci\'on de estados multi-quarks}
\label{multi_quarks}

Ec.~(\ref{eq:cond_pois}) se puede interpretar en t\'erminos del propagador del quark en espacio de coordenadas
\begin{equation}
S(x) = \int \frac{d^4k}{(2\pi)^4} \frac{e^{-i k \cdot x}}{\thru{k} - M} =  
(i\thru\partial + M) \frac{M^2}{4\pi^2 i } \frac{K_1(\sqrt{-M^2 x^2})}{\sqrt{-M^2 x^2}} \,.
\label{eq:S(x)}
\end{equation}
El comportamiento de (\ref{eq:S(x)}) a temperatura peque\~na es
\begin{equation}
S(\vec{x},i\beta) \stackrel{{\rm T} \; \text{peque\~no}}\sim e^{-M/T} \,,
\end{equation}
lo cual representa la supresi\'on exponencial a temperatura peque\~na correspondiente al propagador de un \'unico quark. Si nos fijamos en ec.~(\ref{eq:cond_pois}), esto significa que el condensado de quarks se puede escribir en t\'erminos de factores de Boltzmann estad\'{\i}sticos con masa $M_n=nM$. Esto constituye un problema, pues significa que el ba\~no t\'ermico est\'a formado por quarks constituyentes libres, sin ning\'un confinamiento de color.\footnote{Este c\'alculo se puede extender a cualquier observable que sea singlete de color en el l\'{\i}mite de tempe\-ratura cero, y el resultado general que se obtiene es que los c\'alculos en modelos de quarks a temperatura finita en la aproximaci\'on de un loop van a generar todos los estados posibles de quarks, esto es
\begin{equation}
{\cal O}^T = {\cal O}^{T=0} + {\cal O}_q e^{-M/T} + {\cal O}_{qq} e^{-2M/T} + \cdots \,.
\end{equation}
Notar que, si bien el t\'ermino ${\cal O}_q$ corresponde al estado de un quark aislado, el siguiente t\'ermino ${\cal O}_{qq}$ tiene que ser un estado diquark $qq$, correspondiente a un \'unico quark que se propaga dando dos vueltas alrededor del cilindro t\'ermico. Este t\'ermino no puede ser un estado mes\'onico $\overline{q}q$, puesto que a un loop este estado viene de la l\'{\i}nea de un quark que primero sube y despu\'es baja en tiempo imaginario. En este caso el camino no da ninguna vuelta alrededor del cilindro t\'ermico, y por tanto su contribuci\'on est\'a ya incluida en el t\'ermino de temperatura cero ${\cal O}^{T=0}$.
}

El condensado de quarks a temperatura finita no es invariante gauge (en el sentido de transformaciones gauge grandes). En efecto, del ejemplo del condensado se tiene
\begin{equation}
\langle \overline{q} q \rangle_T = \sum_{n=-\infty}^\infty (-1)^n 
\langle \overline{q}(x_0) q(0)\rangle |_{x_0=in\beta} \,,
\label{eq:qq_sum}
\end{equation}
o sea, el condensado a temperatura finita se puede escribir como una suma coherente de condensados de quarks no locales a temperatura cero. Notar que la contribuci\'on de temperatura cero corresponde al t\'ermino $n=0$ en la sumatoria. Bajo una transformaci\'on gauge de tipo central se tiene
\begin{eqnarray}
\langle \bar q q \rangle_T &\to & \sum_{n=-\infty}^\infty (-z)^n \langle
\bar q(x_0 ) q (0) \rangle \Big|_{x_0=i  n \beta }\,.
\label{eq:sum_coher}
\end{eqnarray} 
Esto significa que (\ref{eq:qq_sum}) no es invariante gauge, y el condensado
se puede descomponer en una suma de representaciones irreducibles con una
trialidad dada~$n$, lo cual genera estados con cualquier n\'umero de
quarks~$\sim e^{-n\beta M}$.

Este problema se puede evitar imponiendo a mano que el condensado sea invariante gauge. Esto se har\'{\i}a eliminando de la suma en~(\ref{eq:sum_coher}) los t\'erminos que no tienen trialidad cero, esto es
\begin{eqnarray}
\langle \bar q q \rangle_T \Big|_{\rm singlete} &= &
\sum_{n=-\infty}^\infty (-1)^n \langle \bar q(x_0 ) q (0) \rangle
\Big|_{x_0=i N_c n \beta } \,.
\end{eqnarray} 
Esta f\'ormula genera como primera correcci\'on un t\'ermino bari\'onico $~N_c
\,e^{-N_c\beta M}$. El factor $N_c$ es generado por el loop de quarks.

\subsection{Conflicto con Teor\'{\i}a Quiral de Perturbaciones}
\label{eq:conflicto_CHPT}

Aparte del problema de la generaci\'on de estados multi-quarks que no preservan triali\-dad, surge otra problem\'atica cuando comparamos nuestros resultados con los de Teor\'{\i}a Quiral de Perturbaciones a temperatura finita. En el l\'{\i}mite quiral, esto es para $m_\pi \ll 2 \pi T \ll 4 \pi f_\pi $, las correcciones t\'ermica de orden m\'as bajo al condensado de quarks (por ejemplo, para $N_f=2$), vienen dadas por
\begin{eqnarray}
\langle \bar q q \rangle_T \Big|_{\rm TQP} &= & \langle \bar q q
\rangle_{T=0} \left( 1- \frac{T^2 } {8 f_\pi^2} - \frac{T^4 } {384 f_\pi^4}
+ \cdots \right) \,.
\label{eq:TQP_fpi}
\end{eqnarray} 
Puesto que $f_\pi \sim \sqrt{N_c}$, las correcciones de temperatura finita est\'an suprimidas en $N_c$ en relaci\'on a la contribuci\'on de temperatura cero. Este hecho contradice el resultado de ec.~(\ref{eq:qq_sum}), pues de ah\'{\i} se obtiene que todas las correcciones t\'ermicas son del mismo orden en un contaje en $N_c$.

El resultado de TQP, ec.~(\ref{eq:TQP_fpi}), se ha obtenido considerando loops pi\'onicos, los cuales son dominantes para $T \ll M$. El problema reside en que incluso sin loops pi\'onicos los modelos de quarks quirales predicen una transici\'on de fase quiral en torno a $T_c \sim 170\;\textrm{MeV}$, lo cual concuerda bien, aunque de manera injustificada, con los resultados en el ret\'{\i}culo.

\section{Acoplamiento del loop de Polyakov en los Mode\-los de Quarks Quirales}
\label{coupling_polyakov}

A temperatura cero es posible preservar la invariancia gauge mediante el acoplamiento de los gluones con el modelo. Dentro del esp\'{\i}ritu del modelo, estos grados de libertad deber\'{\i}an tratarse de un modo perturbativo, pues los quarks constituyentes llevan cierta informaci\'on sobre efectos glu\'onicos no perturbativos.

A temperatura finita la situaci\'on es diferente pues, como hemos dicho ya, un tratamiento perturbativo de la componente cero del campo glu\'onico romper\'{\i}a expl\'{\i}citamente la invariancia gauge. Por tanto, tiene sentido considerar aqu\'{\i} el loop de Polyakov glu\'onico y su acoplamiento con los modelos quirales. K. Fukushima~\cite{Fukushima:2003fw} sugiere este acoplamiento en virtud de la analog\'{\i}a que existe entre el loop de Polyakov y el potencial qu\'{\i}mico (ver sec.~\ref{macrocanonico}). El tratamiento que vamos a considerar nosotros parte del uso del heat kernel a temperatura finita (cap\'{\i}tulo~\ref{heat_kernel}). Nuestra aproximaci\'on es similar a la de Fukushima, excepto por el hecho de que consideraremos un loop de Polyakov local $\Omega(\vec{x})$ sujeto a fluctuaciones cu\'anticas. Un tratamiento de campo medio no permitir\'{\i}a tener en cuenta estas fluctuaciones, y al final del cap\'{\i}tulo veremos que \'estas pueden ser importantes para que los resultados del modelo se muestren compatibles con estudios recientes en el ret\'{\i}culo.

\subsection{Acoplamiento m\'{\i}nimo del loop de Polyakov}
\label{acoplamiento_minimo}

En los modelos de quarks quirales debemos considerar quarks con grados de libertad de sabor y de color. A partir de ahora consideraremos el operador de Dirac, ec.~(\ref{eq:bnm1}), como un operador no trivial en espacio de color. Lo podemos escribir de la siguiente forma:
\begin{equation}
{\bf D}=\thruu{D}_{\cal V}+\thru{\cal A}^f\gamma_5 + M U^{\gamma_5} \,, 
\end{equation}
donde ${\cal D}_\mu^{\cal V} = \partial_\mu + {\cal V}_\mu^f + g {\cal
  V}_\mu^c \delta_{\mu 0}$ es la derivada covariante vector. ${\cal V}_\mu^f$ y ${\cal A}_\mu^f$ son matrices antiherm\'{\i}ticas en espacio de sabor y la identidad en el espacio de color. ${\cal V}_\mu^c$ es la identidad en espacio de sabor y matriz antiherm\'{\i}tica en espacio de color.  Los acoplamientos gauge de sabor dar\'an lugar a loops de Polyakov con quiralidades right y left.\footnote{El {\it loop de Polyakov quiral de sabor} se define 
\begin{eqnarray}
\Omega_f(x_0,\vec{x}) &=& {\mathcal T} \exp\left(
-\int_{x_0}^{x_0+\beta} dx_0^\prime \,({\cal V}_0^f(x_0^\prime,\vec{x})
+\gamma_5 {\cal A}_0^f(x_0^\prime,\vec{x})) 
\right) \,.
\end{eqnarray}
$\Omega_f$ es una matriz en espacio de sabor, y la identidad en espacio de color. En t\'erminos de campos right y left se escribe como $\Omega_f = \Omega_R P_R+ \Omega_L P_L$, donde  
\begin{eqnarray}
\Omega_{R,L}(x_0,\vec{x}) &=& {\mathcal T} \exp\left(
-\int_{x_0}^{x_0+\beta} dx_0^\prime \,({\cal V}_0^f(x_0^\prime,\vec{x})
\pm {\cal A}_0^f(x_0^\prime,\vec{x})) 
\right)
\,,
\end{eqnarray}
y $P_{R,L}=\half (1\pm\gamma_5)$. Notar que la simetr\'{\i}a gauge grande en espacio de sabor a temperatura finita precisa del uso del {\it loop de Polyakov quiral}.
} 
Los acoplamientos gauge de color dar\'an lugar al loop de Polyakov con grados de libertad de color, 
\begin{equation}
\Omega_c(x_0,\vec{x}) = {\mathcal T} \, \exp \left( 
- g \int_{x_0}^{x_0+\beta} dx_0^\prime \,
{\cal V}_0^c(x_0^\prime,\vec{x})
\right) \,.
\end{equation}
$\Omega_c$ es una matriz en espacio de color, y la identidad en espacio de sabor. En esta tesis \'unicamente nos vamos a preocupar del loop de Polyakov de color, que denotaremos como lo venimos haciendo hasta ahora, $\Omega$, de modo que el loop de Polyakov de sabor lo consideraremos igual a la identidad.  

Si nos fijamos en ec.~(\ref{eq:njlacefectreal}), podemos hacer uso del desarrollo del heat kernel a temperatura finita (cap\'{\i}tulo \ref{heat_kernel}) para obtener el lagrangiano efectivo como un desarrollo en derivadas covariantes. El lagrangiano va a tener la forma
\begin{equation}
{\cal L}(x) = \sum_n \tr [ f_n(\Omega(x)) {\cal O}_n(x)] \,,
\end{equation}
donde $\tr$ es la traza sobre todos los grados de libertad internos, $n$ etiqueta todos los operadores locales covariantes gauge~${\cal O}_n$ (esto es, que contienen derivadas covariantes), y $f_n(\Omega(x))$ son funciones dependientes de la temperatura y del loop de Polyakov. Estas funciones reemplazan los coeficientes num\'ericos presentes en el caso de temperatura cero.

En estos c\'alculos, el loop de Polyakov aparece m\'{\i}nimamente acoplado a trav\'es de las frecuencias de Matsubara fermi\'onicas modificadas\footnote{En nuestro tratamiento, $\widehat\omega_n$ es el \'unico sitio donde aparece la dependencia expl\'{\i}cita en los grados de libertad de color, de modo que se puede pensar en $\widehat\nu$ como el conjunto de sus autovalores.}
\begin{equation}
\widehat\omega_n = 2\pi T (n+1/2+\widehat\nu) \,, \qquad
\widehat\nu = (2\pi i)^{-1} \log \Omega \,.
\end{equation} 
En nuestra notaci\'on $\Omega = e^{i 2\pi\widehat\nu}$, donde
$\widehat\nu(\vec{x})=i g {\cal V}_0(\vec{x})/(2\pi T)$. El efecto de este
cambio en las frecuencias de Matsubara da lugar a la siguiente regla para
pasar a las f\'ormulas con $T \ne 0$
\begin{equation}
\tilde{F}(x;x) \rightarrow \sum_{n=-\infty}^\infty
(-\Omega(\vec{x}))^n \tilde{F}(\vec{x},x_0+in\beta;\vec{x},x_0) \,.
\label{eq:F_wpl}
\end{equation}
$F(x;x)$ es el propagador fermi\'onico a temperatura finita que comienza y acaba en el mismo punto. En ec.~(\ref{eq:F_wpl}) aparece el factor $(-\Omega(\vec{x}))^n$, en lugar del factor $(-1)^n$ que se obtiene de la regla est\'andar, ec.~(\ref{eq:regla_estandar}), despu\'es de usar la f\'ormula de Poisson para la sumatoria, ec.~(\ref{eq:poisson}), y considerar la transformada de Fourier.

\begin{figure}[tbc]
\begin{center}
\epsfig{figure=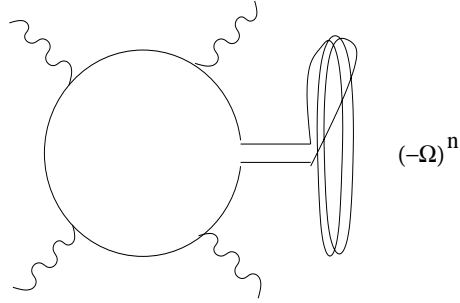,height=4cm,width=6cm}
\end{center}
\caption{Diagrama t\'{\i}pico de un loop de quarks con una l\'{\i}nea de Wilson no trivial. Para $n$ vueltas alrededor de la direcci\'on temporal compactificada U(1), surge un factor topol\'ogico~$\Omega^n$ adem\'as del factor estad\'{\i}stico de Fermi-Dirac~$(-1)^n$. Las l\'{\i}neas onduladas son campos externos. La contribuci\'on total del diagrama se obtiene sumando sobre todas las vueltas y calculando la traza en espacio de color.}
\label{fig:loop}
\end{figure}

La interpretaci\'on de ec.~(\ref{eq:F_wpl}) se puede visualizar en fig.~\ref{fig:loop}. En un loop de quarks a temperatura finita con un n\'umero arbitrario de campos externos y con una l\'{\i}nea de Wilson no trivial, cada vez que los quarks dan una vuelta alrededor de la direcci\'on temporal compatificada, estos adquieren una fase $(-1)$ debido a la estad\'{\i}stica de Fermi-Dirac, y un factor no abeliano de Aharonov-Bohm\footnote{\'Esta es una fase de tipo el\'ectrico, diferente a la fase magn\'etica est\'andar. No obstante, el nombre es apropiado puesto que la fase el\'ectrica fue discutida por primera vez en el art\'{\i}culo original AB.} $\Omega$. La contribuci\'on total del diagrama se obtiene sumando sobre todas las vueltas y calculando la traza en espacio de color.

\subsection{Promedio sobre el grupo}
\label{promedio_grupo}

En la secci\'on~\ref{acoplamiento_minimo} se ha considerado el acoplamiento m\'{\i}nimo del loop de Polyakov con el modelo quark quiral, que consiste simplemente en hacer la sustituci\'on
\begin{equation}
\partial_0 \rightarrow \partial_0 + g {\cal V}_0^c \,,
\label{eq:ac_min}
\end{equation}
en el operador de Dirac, ec.~(\ref{eq:bnm1}). El modelo quark quiral acoplado con el loop de Polyakov se obtiene considerando el acoplamiento m\'{\i}nimo de ec.~(\ref{eq:ac_min}), y una integraci\'on del campo glu\'onico ${\cal V}_0$ de un modo que preserve invariancia gauge. Esto va a generar una funci\'on de partici\'on de la forma
\begin{eqnarray}
Z = \int {\cal D}U {\cal D}\Omega  \; e^{-\Gamma_G [\Omega]} e^{-\Gamma_Q [ U , \Omega ]} \,, 
\label{eq:Z_pnjl} 
\end{eqnarray} 
donde ${\cal D}U$ es la medida de Haar del grupo quiral de sabor SU$(N_f)_L
\times$SU$(N_f)_R$, y ${\cal D}\Omega$ la medida de Haar del grupo de color
SU($N_c$). $\Gamma_G$ es la acci\'on efectiva glu\'onica y $\Gamma_Q$
corresponde a la acci\'on efectiva de los quarks. Ec.~(\ref{eq:Z_pnjl}) es una
expresi\'on gen\'erica, v\'alida tanto para el modelo NJL como para el modelo
espectral, siempre y cuando se considere la correspondiente acci\'on efectiva
de los quarks: ec.~(\ref{eq:njlacefectE}) en el primer caso y
ec.~(\ref{eq:ae_sqm_Tfin}) para el segundo.

Si no se tuviera en cuenta la medida de Haar de color, y se considerara ${\cal V}_0^c=0$ y $\Omega=1$, se obtendr\'{\i}a la forma original del modelo quark quiral, donde existe una relaci\'on uno a uno entre el desarrollo en loops y el desarrollo en $N_c$ grande, tanto a temperatura cero como a temperatura finita. De manera equivalente se podr\'{\i}a considerar una aproximaci\'on de punto de silla y sus correcciones. En presencia del loop de Polyakov tal correspondencia no existe, de modo que consideraremos un desarrollo en loops de quarks, esto es, una aproximaci\'on de punto de silla para el campo bos\'onico $U$, y mantendremos la integraci\'on en el loop de Polyakov (constante) $\Omega$. En el trabajo de \cite{Fukushima:2003fw} se hace uso de la aproximaci\'on de punto de silla para $\Omega$.

La integraci\'on del loop de Polyakov $\Omega$ debe realizarse de acuerdo con la din\'amica de QCD. Esto implica un promedio sobre el loop de Polyakov local con cierto peso norma\-lizado $\rho(\Omega;\vec{x})\, {\cal D}\Omega$. Aqu\'{\i} $\rho(\Omega;\vec{x})$ es la distribuci\'on de probabilidad (independiente de la temperatura) de $\Omega(\vec{x})$ en el grupo gauge. Para una funci\'on general $f(\Omega)$, se tiene\footnote{$f(\Omega)$ se entiende como una funci\'on ordinaria $f(z)$ evaluada en $z=\Omega$.}
\begin{eqnarray}
\left\langle \frac{1}{N_c} \tr_c f(\Omega) \right\rangle &=&
\int_{{\rm SU}(N_c)}\!\!\! {\cal D}\Omega\,\rho(\Omega)  \frac{1}{N_c}
\sum_{j=1}^{N_c}f(e^{i\phi_j})
= \int_{-\pi}^{\pi}\frac{d\phi}{2\pi}\hat\rho(\phi) f(e^{i\phi}) \,,
\label{eq:int_f}
\end{eqnarray}
donde $e^{i\phi_j}$, $j=1,\ldots,N_c$ son los valores propios de $\Omega$ y
\begin{eqnarray}
\hat\rho(\phi) &:=& 
\int_{{\rm SU}(N_c)}\!\!\! {\cal D}\Omega \, \rho(\Omega) \frac{1}{N_c}
\sum_{j=1}^{N_c}2\pi\delta(\phi-\phi_j) \,.
\end{eqnarray}

A temperatura suficientemente peque\~na, la distribuci\'on del loop de Polyakov
se encuentra muy cercana a la medida de Haar de SU($N_c$).\footnote{Esto se justificar\'a en sec.~\ref{correcciones_gluonicas}.} En este caso la funci\'on $\hat\rho(\phi)$ es simplemente
\begin{equation}
\hat\rho(\phi) = 1-\frac{2(-1)^{N_c}}{N_c} \cos \left(N_c\phi\right) \,.
\label{eq:sigma_haar}
\end{equation}
Introduciendo ec.~(\ref{eq:sigma_haar}) en ec.~(\ref{eq:int_f}) se obtienen f\'acilmente las siguientes f\'ormulas para el promedio sobre la medida de Haar de SU($N_c$)
\begin{eqnarray}
\langle\tr_c(-\Omega)^n\rangle_{{\rm SU}(N_c)}  = 
\left\{\begin{matrix} 
N_c\,, & n=0 \label{eq:p1biss}\\
-1  \,, & n=\pm N_c \label{eq:p2biss} \\ 
  0 \,, & {\textrm{otro} \; \textrm{caso}} \label{eq:p3biss}\\
\end{matrix}
 \right. \,.
\label{eq:p123}
\end{eqnarray}

\subsection{Soluci\'on de la problem\'atica}
\label{solucion_problematica}

Si aplicamos este formalismo al condensado de quarks, nuestro modelo conduce a\footnote{
La f\'ormula (\ref{eq:qq_wpl}) es la an\'aloga a ec.~(\ref{eq:qq_sum}), pero considerando la fase no abeliana $\Omega$ y el promedio sobre el grupo.} 
\begin{equation}
\langle \overline{q}q \rangle_T = \sum_{n=-\infty}^\infty
\frac{1}{N_c} \langle \tr_c (-\Omega)^n \rangle \langle
\overline{q}(x_0)q(0)\rangle|_{x_0=in\beta} \,.
\label{eq:qq_wpl}
\end{equation}
Si tenemos en cuenta ec.~(\ref{eq:p123}), observamos que en nuestro modelo el
loop de Polyakov no s\'olo permite eliminar los t\'erminos que rompen
trialidad, sino que las contribuciones t\'ermicas est\'an suprimidas en $N_c$
en relaci\'on al valor de temperatura cero, tal y como se espera de TQP. Esto
resuelve la problem\'atica que discutimos en la
secci\'on~\ref{problematica_modelosqq}.

El condensado de quarks a temperatura finita, a un loop de quarks es
\begin{equation}
\langle \overline{q}q\rangle_T = 
\langle \overline{q}q\rangle_{T=0} + 
\frac{2M^2 T}{\pi^2N_c}K_1(N_cM/T) + \cdots \stackrel{{\rm T}\; \text{peque\~no}}\sim  \langle \overline{q}q\rangle_{T=0} 
+4\left(\frac{MT}{2\pi N_c} \right)^{3/2} e^{-N_cM/T} \,.
\label{eq:qq_wpl_aver2}
\end{equation}
Los puntos indican efectos glu\'onicos o del mar de quarks de orden superior. Notar que debido a la supresi\'on exponencial, las correcciones t\'ermicas de \'orden m\'as bajo a nivel de un loop de quarks comienzan s\'olo a temperaturas cercanas a la transici\'on de fase de desconfinamiento. Hemos denominado a este efecto el {\it enfriamiento de Polyakov}~\cite{Megias:2004hj,Megias:2006bn}, ya que es generado por el promedio de los loops de Polyakov sobre el grupo. Esto significa que en la aproximaci\'on quenched, no se debe de esperar ning\'un efecto t\'ermico importante sobre los observables de los quarks por debajo de la transici\'on de fase, y el cambio m\'as grande deber\'{\i}a de provenir de loops de bosones pseudoescalares a bajas temperaturas. Esto es justo lo que se espera de TQP. Veremos m\'as adelante c\'omo estas propiedades se modifican en presencia del determinante fermi\'onico.

\section{Lagrangiano Quiral a Temperatura Finita}
\label{lagrangiano_quiral_Tfinita}

La estructura de QCD a bajas energ\'{\i}as se puede describir muy bien en
teor\'{\i}a quiral de perturbaciones. El desarrollo quiral corresponde a un
desarrollo en potencias de los momentos externos de los campos. Los campos
pseudoescalares $U$ son de orden ${\cal O}(p^0)$, los campos vector ${\cal V}_\mu$, axial ${\cal A}_\mu$ y cualquier derivada $\partial_\mu$ son de orden ${\cal O}(p)$. Los campos externos escalar ${\cal S}$, pseudoescalar ${\cal P}$ y la matriz de masa de los quarks $\widehat{m}_0$ son de orden~${\cal O}(p^2)$.

Como se muestra en trabajos previos a temperatura cero~\cite{Espriu:1989ff,Bijnens:1992uz,njlarriola,Megias:2004uj,Megias:2005fj}, los modelos de quarks quirales permiten entender de un modo cuantitativo y microsc\'opico la estructura del lagrangiano efectivo a bajas energ\'{\i}as que se deduce de TQP para los mesones pseudoescalares a {\it nivel \'arbol}. En concreto, proporcionan valores num\'ericos para las contribuciones de orden m\'as bajo en $N_c$ de las constantes de baja energ\'{\i}a.

En esta secci\'on vamos a extender los resultados de temperatura cero a
temperatura finita, y consideraremos la influencia del loop de
Polyakov. Siguiendo el m\'etodo desarrollado en el
cap\'{\i}tulo~\ref{heat_kernel}, y que ya aplicamos en el
cap\'{\i}tulo~\ref{QCD_efective_action} para el c\'alculo de la acci\'on
efectiva de QCD en el r\'egimen de temperatura alta, se puede escribir la estructura del lagrangiano efectivo a baja energ\'{\i}a para los mesones pseudoescalares a temperatura finita a nivel \'arbol, mediante un desarrollo de tipo heat kernel para los modelos de quarks quirales a nivel de un loop. En TQP a temperatura finita se considera en general que las constantes de baja energ\'{\i}a son {\it independientes de la temperatura}. \'Esta es una suposici\'on bastante razonable, ya que la aplicabilidad de TQP se basa en la existencia de un gap de masa entre los bosones de Goldstone y el resto del espectro hadr\'onico. Para mesones no extra\~nos el gap viene dado por la masa del mes\'on~$\rho$, $M_V$, de modo que es de esperar que la dependencia en temperatura de las constantes de baja energ\'{\i}a sea del orden de~$e^{-M_V/T}$. En un modelo quark quiral, los mesones pseudoescalares son part\'{\i}culas compuestas de quarks constituyentes con una masa~$M$, y los efectos t\'erminos tambi\'en deber\'{\i}an de influir en su estructura microsc\'opica. El c\'alculo que realizaremos en esta secci\'on va a permitir analizar esto de una manera cuantitativa.

\subsection{Estructura del lagrangiano}

El c\'alculo del lagrangiano quiral efectivo a temperatura finita en los modelos de quarks quirales se limita, desde un punta de vista t\'ecnico, al c\'alculo
de trazas en espacio de Dirac y en espacio de sabor. En el ap\'endice~\ref{sec:hhkk} se hace en detalle. Mostraremos aqu\'{\i} el resultado final.

El lagrangiano efectivo a baja energ\'{\i}a escrito en la notaci\'on de Gasser-Leutwyler~\cite{gasser-leutwyler2} y en espacio eucl\'{\i}deo se escribe
\begin{eqnarray}
\cL_q^{* (0)} &=& \frac{2N_f}{(4\pi)^2}\langle\tr_c\J_{-2}(\Lambda,M, \widehat\nu)\rangle
\,,  \label{eq:el_lq0} \\
\cL_q^{* (2)} &=& \frac{\f^2}{4}\tr_f\left( \D_\mu U^\dagger\D_\mu
U -(\chi^\dagger U +\chi U^\dagger) \right) \,,
\label{eq:el_lq2} \\
\cL_q^{* (4)} &=& -L^*_1(\tr_f(\D_\mu U^\dagger \D^\mu U))^2
-L^*_2\tr_f(\D_\mu U^\dagger \D_\nu U)\tr_f(\D^\mu U^\dagger \D^\nu U)
\nonumber  \\
&&-L^*_3\tr_f(\D_\mu U^\dagger \D^\mu U \D_\nu U^\dagger
\D^\nu U) -\overline{L}^*_3 \tr_f(\D_0 U^\dagger \D^0 U \D_\mu U^\dagger
\D^\mu U) \nonumber \\
&&+L^*_4\tr_f(\D_\mu U^\dagger \D^\mu
U)\tr_f(\chi^\dagger U + \chi U^\dagger) \nonumber \\
&&+L^*_5\tr_f(\D_\mu U^\dagger \D^\mu U(\chi^\dagger U +
U^\dagger\chi)) +\overline L^*_5\tr_f(\D_0 U^\dagger \D^0
U(\chi^\dagger U + U^\dagger\chi)) \nonumber \\
&&+\overline L_5^{* \prime}
\tr_f(\D_0\D^0 U^\dagger \chi + \D_0\D^0U\chi^\dagger)
-L^*_6(\tr_f(\chi^\dagger U +\chi U^\dagger))^2
-L^*_7(\tr_f(\chi^\dagger U -\chi U^\dagger))^2 \nonumber \\
&&+\overline L^{* \prime} \tr_f(U^\dagger \D_0\D^0 U - U\D_0\D^0
U^\dagger) \tr_f(\chi^\dagger U - \chi U^\dagger) \nonumber \\
&&-L^*_8\tr_f(\chi^\dagger U\chi^\dagger U +\chi U^\dagger \ochi
U^\dagger) \nonumber \\ &&-L^*_9\tr_f(F^R_{\mu\nu}\D^\mu U^\dagger
\D^\nu U +F^L_{\mu\nu}\D^\mu U \D^\nu U^\dagger) \nonumber \\
&&-\overline L^*_9 \tr_f(E_i^R(\D^0 U^\dagger \D^i U -\D^i U^\dagger \D^0
U)+E_i^L(\D^0 U \D^i U^\dagger -\D^i U \D^0 U^\dagger)) \nonumber \\
&& -\overline L_9^{* \prime}\tr_f(\D_0E_i^R U^\dagger \D^i U +\D_0E_i^L U \D^i
U^\dagger) \nonumber \\
&&+L^*_{10}\tr_f(U^\dagger F_{\mu\nu}^L U
F^{\mu\nu R}) \nonumber \\
&&+H^*_1\tr_f((F_{\mu\nu}^R)^2+(F_{\mu\nu}^L)^2) +\overline
H^*_1 \tr_f((E_i^R)^2+(E_i^L)^2)- H^*_2\tr_f(\chi^\dagger \chi) \,.
\label{eq:lag_4}
\end{eqnarray}
$\tr_f$ es la traza en espacio de sabor. Las derivadas covariantes quirales son
\begin{eqnarray} 
\D_\mu U &=& D_\mu^L U-U D_\mu^R =
\partial_\mu U+l_\mu U - U  r_\mu, \label{eq:der_cov_quiral_U}\\ 
 F_{\mu\nu}^R &=& [ D_\mu^R, D_\nu^R] = 
\partial_\mu r_\nu -\partial_\nu r_\mu
+ [ r_\mu , r_\nu ], \nonumber \\
 F_{\mu\nu}^L &=& [ D_\mu^L, D_\nu^L] = 
\partial_\mu l_\nu -\partial_\nu l_\mu
+ [ l_\mu , l_\nu ], \nonumber  
\end{eqnarray} 
donde $r_\mu = {\cal V}_\mu + {\cal A}_\mu$, y $l_\mu = {\cal V}_\mu - {\cal
  A}_\mu$. $\langle \ldots \rangle$ indica promedio sobre el grupo gauge de
  color SU($N_c$). La estructura de este lagrangiano resulta bastante interesante. Por una parte existen t\'erminos que se pueden escribir como los del lagrangiano a temperatura cero, pero con acoplamientos efectivos dependientes de la temperatura. Adem\'as de estos, existen nuevos t\'erminos que rompen invariancia Lorentz. Curiosamente, en el lagrangiano aparecen menos t\'erminos del segundo tipo de los que en un principio se podr\'{\i}a pensar en base a las simetr\'{\i}as conocidas. Todav\'{\i}a no entendemos del todo este hecho, que parece sugerir la existencia de alguna simetr\'{\i}a accidental. Si bien sospechamos que esta simetr\'{\i}a existe s\'olo a un loop, ser\'{\i}a interesante encontrarla expl\'{\i}citamente.

\subsection{LEC para el modelo de Nambu--Jona-Lasinio}
\label{LEC_NJL}

Si bien \'esta es la estructura general que se ha encontrado para los modelos
de quarks quirales, los valores de los coeficientes de baja energ\'{\i}a
dependen del modelo en particular. Mostramos aqu\'{\i} los valores de las
constantes de baja energ\'{\i}a (LEC) obtenidas para el modelo de
Nambu--Jona-Lasinio. Para evitar complicaciones con el loop de Polyakov, hemos
considerado el modelo NJL sin integraci\'on de los campos de esp\'{\i}n~1
(vector y axial).\footnote{En el cap\'{\i}tulo~\ref{tensor_EM_MQQ} se
  calcular\'a la acci\'on efectiva del modelo NJL generalizado a temperatura
  cero con integraci\'on en estos campos.}
\begin{eqnarray}
&&\cL_q^{* (0)} = \frac{2N_f}{(4\pi)^2}\langle \tr_c \J_{-2}\rangle \,,
\quad \f^2 = \frac{M^2}{4\pi^2}\langle \tr_c\J_{0}\rangle \,, 
\quad \f^2 B^*_0 = \frac{M}{4\pi^2}\langle\tr_c\J_{-1}\rangle \,, \nonumber \\
&&L^*_1 = \frac{M^4}{24(4\pi)^2}\langle\tr_c\J_{2}\rangle \,, 
\qquad L^*_2 = 2L^*_1 \,,  
\qquad L^*_3 = -8L^*_1 +\frac{1}{2}L^*_9 \,,  \nonumber \\
&&\overline{L}^*_3 =-\frac{M^2}{6(4\pi)^2}\langle\tr_c\overline\J_{1}\rangle \,,
\qquad L^*_4 = 0 \,,
\qquad L^*_5 =\frac{M}{2B^*_0}\left(\frac{\f^2}{4M^2}-3L^*_9\right) \,, \nonumber \\
&&\overline{L}^*_5 = \frac{1}{2}\overline{L}^*_3 \,, 
\qquad \overline{L}_5^{* \prime} = \frac{1}{2}\overline{L}^*_3 \,, 
\qquad L^*_6 = 0\,,
\qquad L^*_7 = \frac{1}{8N_f} \left(-\frac{\f^2}{2B^*_0M}+L^*_9\right)
\,, \nonumber \\
&&\overline{L}^{* \prime} = -\frac{1}{4N_f}\overline{L}^*_3 \,,
\qquad L^*_8 = \frac{1}{16B^*_0}\left(\frac{1}{M}-\frac{1}{B^*_0}\right)\f^2
-\frac{1}{8}L^*_9 \,, \label{CBE_NJL} \\
&&L^*_9 =
\frac{M^2}{3(4\pi)^2}\langle\tr_c\J_{1}\rangle \,, 
\qquad \overline{L}^*_9 = -\overline{L}^*_3 \,, 
\qquad \overline{L}_9^{* \prime} = -\overline{L}^*_3 \,,
\qquad L^*_{10} = -\frac{1}{2}L^*_9 \,, \nonumber \\
&&H^*_1 = -\frac{\f^2}{24M^2} +\frac{1}{4}L^*_9 \,,
\qquad \overline{H}^*_1 = -\frac{1}{6(4\pi)^2}\langle\tr_c\overline\J_{0}\rangle \,,
\qquad H^*_2 = -\frac{\f^2}{8B^{*2}_0}+\frac{1}{4}L^*_9\,, \nonumber
\end{eqnarray}
donde las integrales $\J_l$ est\'an definidas en
ecs.~(\ref{eq:Jl1})-(\ref{eq:Jl5}). Los coeficientes de Gasser-Leutwyler
est\'andar se pueden expresar en t\'erminos de $f_\pi^{* 2}, B^*_0, L^*_1$ y
$L^*_9$, o de manera equi\-valente, en t\'erminos de las integrales
$\langle\tr_c\J_{-1}\rangle, \langle\tr_c\J_{0}\rangle,
\langle\tr_c\J_{1}\rangle$ y $\langle\tr_c\J_{2}\rangle$. Notar que todos los
t\'erminos que rompen la simetr\'{\i}a Lorentz, excepto $\overline{H}_1^*$,
son proporcionales.

Si el loop de Polyakov $\Omega$ se considera igual a la unidad, las expresiones (\ref{CBE_NJL}) siguen siendo v\'alidas, salvo por el hecho de que el promedio en el grupo y la traza de color deben sustituirse por un factor~$N_c$.

\subsection{LEC para el Modelo Quark Espectral}
\label{CBE_SQM}

En este modelo se debe hacer un promedio sobre la masa constituyente de los quarks con una funci\'on espectral~$\rho(\omega)$ que act\'ua como peso (ver sec.~\ref{sec:mqe1}). Notar que $M$ no s\'olo aparece como argumento de las integrales $\J_l$, sino que tambi\'en aparece en forma de factores multiplicativos. Esto dar\'a lugar a un n\'umero mayor de funciones independientes en comparaci\'on con el modelo NJL.
\begin{eqnarray}
\cL_q^{* (0)} &=& \frac{2N_f}{(4\pi)^2}\langle\tr_c\J_{-2}\rangle \,,
\quad \f^2 = \frac{1}{4\pi^2}\langle\omega^2\tr_c\J_{0}\rangle \,, 
\quad \f^2B^*_0 = \frac{1}{4\pi^2}\langle
\omega\tr_c\J_{-1}\rangle \,, \nonumber \\ 
L^*_1 &=&\frac{1}{24(4\pi)^2}\langle \omega^4 \tr_c\J_{2}\rangle \,, 
\quad L^*_9 = \frac{1}{3(4\pi)^2}\langle \omega^2\tr_c\J_{1}\rangle \,, 
\quad \overline{L}^*_3 = -\frac{1}{6(4\pi)^2}\langle \omega^2
\tr_c\overline\J_{1}\rangle \,, \nonumber\\ 
L^*_5 &=& \frac{1}{2(4\pi)^2B^*_0}(\langle\omega\tr_c\J_{0}\rangle
-\langle\omega^3\tr_c\J_{1}\rangle)\,, \nonumber \\ 
L^*_7 &=&
\frac{1}{2(4\pi)^2N_f}\left(-\frac{1}{2B^*_0}\langle\omega\tr_c\J_{0}\rangle+4\pi^2L^*_9\right)
\,, \nonumber \\ L^*_8 &=&
\frac{1}{4(4\pi)^2B^*_0}\langle\omega\tr_c\J_{0}\rangle
-\frac{\f^2}{16B_0^{* 2}} -\frac{1}{8}L^*_9 \,, \nonumber \\ 
H^*_1 &=&
-\frac{1}{6(4\pi)^2}\langle\tr_c\J_{0}\rangle + \frac{1}{4}L^*_9 \,, \nonumber \\ \overline{H}^*_1 &=&
-\frac{1}{6(4\pi)^2}\langle\tr_c\overline\J_{0}\rangle \,, \nonumber \\ 
H^*_2 &=& \frac{1}{2(4\pi)^2B^*_0}\left(\frac{1}{B^*_0}\langle\tr_c\J_{-1}\rangle-\langle
\omega\tr_c\J_{0}\rangle\right)-\frac{\f^2}{8B^{*2}_0}+\frac{1}{4}L^*_9 \,. \label{eq:CBE_SQM}
\end{eqnarray}
Para simplificar la notaci\'on, con $\langle \ldots \rangle$ indicamos tanto
el promedio sobre el loop de Polyakov como el promedio espectral~$\int_{\cal C} d\omega \rho(\omega) \dots$. El resto de coeficientes satisfacen las mismas relaciones geom\'etricas que se obtuvieron para el modelo NJL. En ambos modelos se obtiene la relaci\'on 
\begin{equation}
L^*_7 =-\frac{1}{N_f}\left(\frac{\f^2}{16B^{*2}_0}+L^*_8\right) \,.
\end{equation}
Podemos calcular expl\'{\i}citamente las integrales haciendo uso del esquema de dominancia vectorial de la funci\'on espectral~$\rho(\omega)$ (ver sec.~\ref{MDM_results} y ref.~\cite{spectralqm}). Despu\'es de calcular el promedio en el grupo SU($N_c$), se obtiene
\begin{eqnarray}
\langle\tr_c\J_{-2}\rangle &=& -\frac{N_c}{2}\rho_4^\prime
-\frac{2M_V^4}{3x_V^4}\left(48+24x_V+6x_V^2+x_V^3\right)e^{-x_V/2} \,,
\nonumber \\ \langle\tr_c\J_{-1}\rangle &=& N_c
\rho_2^\prime-\frac{2M_V^2}{3x_V^2}\left(12+6x_V
+x_V^2\right)e^{-x_V/2} \,, \nonumber \\
\langle\omega\tr_c\J_{-1}\rangle &=&
\rho_3^\prime\left(N_c-2e^{-x_S/2}\right)\,, \nonumber \\
\langle\tr_c\J_{0}\rangle &=& -N_c(\rho_0+\rho_0^\prime) +
2\gamma_E-4\log(4)+4\log(x_V)-2\psi(5/2) \nonumber \\
&&\qquad-\frac{x_V^5}{1800} \; {}_1 F_2\left[\{\frac{5}{2}\},\{\frac{7}{2},\frac{7}{2}\},\left(\frac{x_V}{4}\right)^2\right]
\nonumber \\ &&\qquad-\frac{x_V^2}{12} \;
{}_2 F_3\left[\{1,1\},\{-\frac{1}{2},2,2\},\left(\frac{x_V}{4}\right)^2\right]
\,, \nonumber 
\end{eqnarray}
\begin{eqnarray}
 \langle\omega\tr_c\J_{0}\rangle &=& -N_c\rho_1^\prime
-\frac{2\rho_3^\prime}{M_S^2}(2+x_S)e^{-x_S/2} \,, \nonumber \\
\langle\omega^2\tr_c\J_{0}\rangle &=&
-N_c\rho_2^\prime-\frac{M_V^2}{6}(2+x_V)e^{-x_V/2} \,, \nonumber \\
\langle\omega^2\tr_c\J_{1}\rangle &=& N_c\rho_0
-\frac{1}{6}\left(12+6x_V+x_V^2\right)e^{-x_V/2} \,, \nonumber \\
\langle\omega^3\tr_c\J_{1}\rangle &=&-\frac{\rho_3^\prime
  x_S^2}{2M_S^2}e^{-x_S/2} \,, \nonumber \\
 \langle\omega^4\tr_c\J_{2}\rangle
&=& N_c\rho_0 -\frac{1}{24}\left(48+24x_V +
6x_V^2+x_V^3\right)e^{-x_V/2} \,, \nonumber \\
\langle\tr_c\overline\J_{0}\rangle &=&
-\frac{1}{3}\left(12+6x_V+x_V^2\right)e^{-x_V/2} \,, \nonumber \\
\langle\omega^2\tr_c\overline\J_{1}\rangle &=&
-\frac{x_V^2}{12}(2+x_V) e^{-x_V/2} \,, \label{eq:integrales_SQM}
\end{eqnarray}
con la notaci\'on
\begin{equation}
x_V := N_c\beta M_V \,,
\qquad\quad
x_S := N_c\beta M_S \,,
\end{equation}
donde $M_V$ es la masa del mes\'on vectorial (masa del $\rho$), y $M_S$ es la
masa del escalar. ${}_p F_q[a_1,\dots , a_p;b_1,\dots,b_q;z]$ son las
funciones hipergeom\'etricas generalizadas \cite{tablas}.

\section{Correcciones de orden superior}
\label{correcciones_orden_mayor}

En las secciones~\ref{coupling_polyakov} y \ref{lagrangiano_quiral_Tfinita} hemos considerado los modelos de quarks quirales a nivel de un loop de quarks. Esto corresponde a la aproximaci\'on quenched dentro del modelo. Asimismo se ha hecho uso de que a temperaturas suficientemente peque\~nas basta con considerar el promedio sobre el grupo gauge de color SU($N_c$). En esta secci\'on discutiremos algunas consecuencias importantes que se obtienen al ir m\'as all\'a de estas aproximaciones.

\subsection{M\'as all\'a de un loop de quarks}
\label{mas_alla_1loop}

El ir m\'as all\'a de la aproximaci\'on de un loop de quarks puede conducir a c\'alculos bastante tediosos (ver refs.~\cite{broniowski,Oertel:2000jp} para c\'alculos expl\'{\i}citos del modelo NJL est\'andar sin loop de Polyakov). Aqu\'{\i} no nos vamos a preocupar de hacer un c\'alculo expl\'{\i}cito, no obstante se pueden deducir algunas consecuencias importantes basadas en ciertas reglas de contaje en~$N_c$ a temperatura finita.

\begin{figure*}[tbc]
\begin{center}
\epsfig{figure=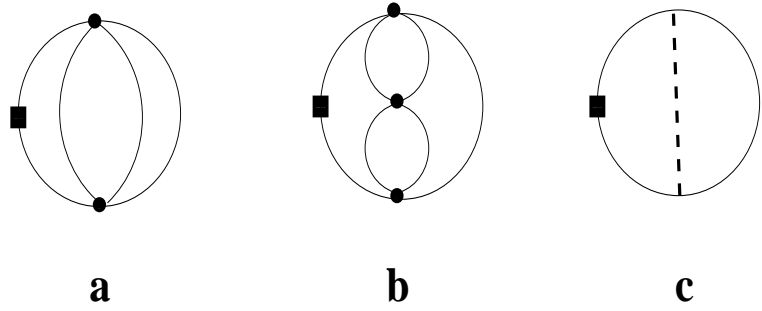,height=4cm,width=10cm}
\end{center}
\caption{Diagrama t\'{\i}pico m\'as all\'a de un loop para el operador del condensado de quarks $\overline{q}q$. Las l\'{\i}neas de los quarks con momentos independientes pueden dar $n$ vueltas alrededor del tiempo eucl\'{\i}deo compactificado, dando lugar al factor de Fermi-Polyakov~$(-\Omega)^n$. La conservaci\'on de trialidad solamente permite que las l\'{\i}neas internas de quark-antiquark den una \'unica vuelta y en sentidos opuestos, lo cual genera una supresi\'on exponencial~$e^{-2M\beta}$ para el diagrama a). Una supresi\'on similar ocurre para el diagrama b) si las vueltas del quark-antiquark ocurren en cada una de las burbujas. El diagrama c) se corresponde con una suma de todos los estados intermedios con los mismos n\'umeros cu\'anticos, y puede interpretarse como la l\'{\i}nea de un mes\'on.}
\label{fig:3loop}
\end{figure*}

Consideremos, por ejemplo, el diagrama a tres loops de la figura~\ref{fig:3loop}, que contribuye al condensado quiral en el modelo NJL en t\'erminos de los propagadores de los quarks. La contribuci\'on de este diagrama se escribe\footnote{Por simplicidad, escribimos \'unicamente las frecuencias de Matsubara.}
\begin{eqnarray} 
{\rm Fig}.(2a) &=& \sum_{w^{(1)}, w^{(2)} , w^{(3)}} S ( w^{(1)} ) \otimes S(
w^{(1)})  \otimes  S( w^{(2)}) \otimes S( w^{(3)}) \otimes
S( w^{(1)}+ w^{(3)}- w^{(2)}) \,.\nonumber  
\end{eqnarray} 
Haciendo uso de la f\'ormula de Poisson para la sumatoria, ec.~(\ref{eq:poisson}), y yendo a espacio eucl\'{\i}deo se tiene
\begin{eqnarray} 
{\rm Fig}.(2a) &=& \sum_{n_1,n_2,n_3} \langle \Omega^{n_1+n_2+n_3}
\rangle \int_{-\infty}^\infty d \tau_1 d \tau_3 \;S (\tau_1 ) \otimes S( -\tau_1 - \tau_3 + n_1 \beta + n_3 \beta ) \otimes \nonumber \\
&&\otimes S( -\tau_3 + n_2 \beta +n_3 \beta ) \otimes S(
\tau_3 ) \otimes S( \tau_3 - n_3 \beta ) \nonumber \\ 
&\stackrel{\beta \to \infty}  \sim &
\sum_{n_1,n_2,n_3} \langle \Omega^{n_1+n_2+n_3} \rangle e^{-\beta M
(|n_1|+|n_2|+|n_3|)} \,.
\end{eqnarray} 
La conservaci\'on de trialidad para este diagrama implica, $n_1+n_2+n_3=k
N_c$, y el valor m\'{\i}nimo del exponente se consigue con $n_1=n_2=n_3=0$,
que es la contribuci\'on de tempe\-ra\-tura cero. La primera correcci\'on
t\'ermica a temperatura peque\~na viene dada por $n_1=0$, $n_2=-n_3=1$, de
modo que el diagrama a 3 loops de fig.(2a) se encuentra suprimido en un
factor~$\sim e^{-2\beta M}$, en comparaci\'on con la supresi\'on de un loop de
quarks~$\sim e^{-N_c\beta M}$. Una supresi\'on t\'ermica similar se obtiene si
introducimos la suma est\'andar sobre burbujas, que puede acoplarse a los
n\'umeros cu\'anticos de los mesones transformando el argumento del exponente
en $2 M \rightarrow M_{\overline{q}q}$. Obviamente, esta contribuci\'on
resulta m\'as importante para el pi\'on m\'as ligero. En realidad, el diagrama quark-mes\'on de la fig.(2b) es similar al diagrama bosonizado de dos loops que se muestra en fig.(2c). Para este diagrama bosonizado los argumentos previos resultan m\'as simples, ya que el n\'umero de loops es igual al n\'umero de propagadores de quarks. El operador de polarizaci\'on del pi\'on, proporcional al propagador del pi\'on, se puede tomar a temperatura cero, ya que la supresi\'on m\'as importante viene de las l\'{\i}neas de quarks que no est\'an acopladas a los n\'umeros cu\'anticos del pi\'on.

Para un diagrama bosonizado con $L$ loops de quarks, tenemos que considerar L gene\-ra\-lizaciones de las correcciones a nivel de un loop de quarks, ec.~(\ref{eq:F_wpl}). El an\'alisis es m\'as simple en espacio de coordenadas. En lugar del n\'umero total de propagadores de quarks, consideramos la suma de Poisson de L propagadores. Esto se puede hacer mediante la f\'ormula
\begin{eqnarray}
\sum_{n,m=-\infty}^\infty \int_0^\beta d x_4 F( x_4 + n \beta + m \beta ) = \sum_{n=-\infty}^\infty \int_{-\infty}^\infty 
d x_4 F( x_4 + n \beta ) \,.
\end{eqnarray} 
Esto significa que es posible eliminar tantas sumas de Poisson como integrales en coordenadas aparecen en las expresiones. Haciendo uso de $L=I-(V-1)$ y $4V=E+2I$ tenemos\footnote{$L$ es el n\'umero de loops de quarks, $V$ el n\'umero de v\'ertices, $I$ el n\'umero de l\'{\i}neas de quarks y $E$ el n\'umero de patas externas.}
\begin{eqnarray}
{\prod}_{i=0}^L \int d^4 z_i G^{2L} \sum_{n_1, \dots , n_L}
{\prod}_{i=1}^L (-\Omega)^{n_i} S ( \vec x_i , t_i + {\rm i} n_i \beta
) \,.
\end{eqnarray}
En realidad, esta regla no depende de la forma precisa de la interacci\'on de
los quarks. A bajas temperaturas, cada l\'{\i}nea de quark con un \'{\i}ndice
de Poisson independiente genera una supresi\'on dada por una masa
constituyente de quark. Por tanto, la contribuci\'on a un observable se puede descomponer esquem\'aticamente del siguiente modo
\begin{eqnarray} 
{\cal O}^T = \sum_L \sum_{n_1 , \dots , n_L } {\cal O}_{n_1 \dots n_L
} \langle \Omega^{n_1+ \dots n_L } \rangle \, e^{- M \beta ( |n_1| +
\dots + |n_L| ) } \,.
\label{eq:loopkk}
\end{eqnarray} 
La conservaci\'on de trialidad de la medida $\Omega \rightarrow z \Omega$ a este nivel conduce a
\begin{eqnarray} 
n_1 + \dots + n_L = k N_c 
\label{eq:n-ality}
\end{eqnarray} 
con $k=0,1,2,\dots$. El t\'ermino dominante en el desarrollo de
ec.~(\ref{eq:loopkk}) es aquel para el que $n_1= \ldots = n_L=0$ con un
n\'umero arbitrario de loops de quarks $L$, y se corresponde con la
contribuci\'on de temperatura cero. Adem\'as, se ve que para $L=1$
\'unicamente se tienen contribuciones de $n_1=k N_c$, lo cual da lugar a
correcciones~$e^{-N_c M \beta}$, que permiten reproducir los resultados de las
secciones~\ref{coupling_polyakov} y \ref{lagrangiano_quiral_Tfinita}. A partir
de la ec.~(\ref{eq:loopkk}) podemos ver c\'omo se organiza el desarrollo
t\'ermico para temperaturas bajas. Las contribuciones t\'ermicas m\'as
importantes vienen de minimizar~$\sum_{i=1}^L |n_i|$, sujeto al requerimiento
de conservaci\'on de tria\-li\-dad, ec.~(\ref{eq:n-ality}). A temperatura
finita y para $N_c \ge 3$ se tiene que la primera correcci\'on t\'ermica viene
dada por $L=2$ y $n_1=-n_2=1$ con $n_3 = \ldots = n_L=0$, lo cual da el factor
$e^{-2\beta M}$ y se corresponde con un estado mes\'onico $\overline{q}q$.
Esta contribuci\'on est\'a suprimida por un factor $1/N_c$ en relaci\'on con
la contribuci\'on de temperatura cero. Para $N_c=3$ el siguien\-te t\'ermino
en el desarrollo corresponder\'{\i}a a $L \ge 3$ y $n_1=n_2=n_3=1$, lo cual da
lugar a una supresi\'on t\'ermica~$e^{-N_c \beta M}$. Para~$N_c \ge 5$ se
tendr\'{\i}a $L \ge 4$ con $n_1=-n_2=n_3=n_4=1$ y $n_5= \ldots = n_L=0$. Si
consideramos el caso $N_c=3$ se tiene\footnote{En el caso en que no se
  considerara la existencia del loop de Polyakov, se tendr\'{\i}a~$Z_{ q^{N_q}
    (\bar q q)^{N_M} } \sim \frac1{N_c^{N_M}} e^{-(2 N_M +N_q ) M / T}$, de
  modo que las contribuciones de orden m\'as bajo corresponder\'{\i}an a
  estados de un quark.}
\begin{eqnarray} 
Z_{\bar q q } &\sim&   \frac1{N_c} e^{-2 M / T} \,, \\ 
Z_{qqq} & \sim&  e^{-N_c M / T} \,, \\
Z_{qqq \bar q q } &\sim&   \frac1{N_c} e^{-(2+N_c) M / T} \,, \\ 
&& \dots \\ 
Z_{(\bar q q)^{N_M} (qqq)^{N_B} } &\sim& \frac1{N_c^{N_M}} e^{-(2 N_M
+N_B N_c) M / T} \,.
\end{eqnarray} 
Obviamente, para $N_c=3$ la contribuci\'on del loop mes\'onico es m\'as dominante que la del loop bari\'onico. Los argumentos previos se han hecho sin tener en cuenta el efecto de confinamiento de los quarks, de modo que en realidad deber\'{\i}amos considerar la masa f\'{\i}sica del mes\'on~$m$, y en este caso se tendr\'{\i}a
 \begin{eqnarray} 
{\cal O}^T = {\cal O}^{T=0} + \sum_{m} {\cal O}_{m} \frac1{N_c}\,e^{-m/T}  
+ \sum_{B} {\cal O}_B \,e^{-M_B/T } + \cdots \,.  
\end{eqnarray} 
As\'{\i} es como funciona la dualidad quark-hadr\'on en los modelos de quarks
quirales a tempera\-tura finita. Como vemos, las contribuciones de los loops
pi\'onicos son las m\'as importantes, incluso si se tiene en cuenta que
est\'an suprimidas en~$1/N_c$. La siguiente contribuci\'on al observable total
a temperatura finita viene dada por los estados mes\'onicos sucesivos. En su conjunto, esto es lo que se espera como consecuencia de la inclusi\'on del loop de Polyakov en los modelos de quarks quirales, teniendo en cuenta la proyecci\'on sobre el sector singlete de color invariante gauge.

En definitiva, a temperatura finita se tiene una supresi\'on est\'andar~$\frac{1}{N_c} \, e^{-2M/T}$ proveniente de loops mes\'onicos y una supresi\'on~$e^{-N_c M/T}$ de loops bari\'onicos. Obviamente, las contribuciones m\'as importantes para $N_c$ grande o $T$ peque\~no  son las debidas a loops mes\'onicos.   

La discusi\'on anterior est\'a centrada en observables que contienen quarks. Para el valor esperado del loop de Polyakov, por ejemplo, se tiene 
\begin{equation} 
\sum_L \sum_{n_1 , \ldots , n_L } {\cal O}_{n_1 \ldots n_L
} \langle \Omega^{1+n_1+ \cdots + n_L } \rangle e^{- M \beta ( |n_1| +
\cdots + |n_L| ) }
\label{eq:loop1}
\end{equation}
y
\begin{eqnarray} 
1+n_1 + \cdots + n_L = k N_c \,.
\label{eq:n-ality1}
\end{eqnarray}
La contribuci\'on t\'ermica de orden m\'as bajo (no existe contribuci\'on de temperatura cero) es $n_1=-1$, $n_2=\ldots=n_L=0$, que se corresponde con un \'unico loop de antiquark que apantalla la carga del loop de Polyakov test. Este t\'ermino escala como $e^{-M/T}$. Al contrario que para observables con quarks como el condensado quiral, este comportamiento no se ve afectado por loops pi\'onicos. En sec.~\ref{mas_alla_quenched} obtendremos expresiones expl\'{\i}citas para estos observables en el modelo NJL.

\subsection{Correcciones glu\'onicas}
\label{correcciones_gluonicas}

Hasta ahora hemos considerado simplemente una integraci\'on sobre la medida del grupo gauge. Desafortunadamente, no conocemos ning\'un argumento general por el cual tenga que existir una supresi\'on exponencial de los grados de libertad glu\'onicos a temperaturas bajas, y por tanto dejando la medida de Haar como \'unico vestigio de los gluones. No obstante, los resultados basados en desarrollos con acoplamientos grandes~\cite{Gross:1983pk,Polonyi:1982wz} y en la aproximaci\'on de gluones masivos a un loop~\cite{Meisinger:2001cq,Meisinger:2003id} proporcionan esta supresi\'on, y de hecho los resultados recientes en el ret\'{\i}culo confirman una sorprendente universalidad en todas las representaciones de los grupos, y favorece el mecanismo dominante del promedio simple sobre el grupo~\cite{Dumitru:2003hp}.

De manera m\'as espec\'{\i}fica, de los datos del ret\'{\i}culo~\cite{Dumitru:2003hp} y de la medida del grupo se encuentra que
\begin{equation}
\langle |\tr_c \,\Omega|^2\rangle = 1\,,
\end{equation}
en la fase de confinamiento, o de manera equivalente $\langle
\widehat\tr_c\,\widehat\Omega \rangle = 0$, para la representaci\'on adjunta.
Notar que en la aproximaci\'on de campo medio~\cite{Fukushima:2003fw} $\langle
|\tr_c \,\Omega|^2\rangle$ se anula, debido a la ausencia de fluctuaciones.

El potencial glu\'onico a orden m\'as bajo que se deduce del desarrollo con acoplamientos grandes viene dado por~\cite{Gross:1983pk,Polonyi:1982wz}
\begin{equation}
\Gamma_G[\Omega] = V_{\text{glue}}[ \Omega ]\cdot a^3/T=-2(d-1)\,\mathrm{e}^{-\sigma a/T} \bigl|\tr_c \Omega \bigr|^2  \,,
\label{eq:V_glue_Fuk}
\end{equation}
para $N_c=3$ con la tensi\'on de la cuerda $\sigma=(425\,\text{MeV})^2$. A nivel de campo medio~$V_{\rm glue }[\Omega]$ da lugar a una transici\'on de fase de primer orden con el acoplamiento cr\'{\i}tico~$2(d-1)\mathrm{e}^{-\sigma a/T_{D}}=0.5153$. Se puede fijar la temperatura de transici\'on a su valor emp\'{\i}rico~$T_{D}=270\,\text{MeV}$ mediante la elecci\'on~$a^{-1}=272\,\text{MeV}$~\cite{Fukushima:2003fw}. La masa correspondiente es~$m_G =\sigma a = 664\,{\rm MeV}$. A temperaturas peque\~nas se puede desarrollar la exponencial en potencias de la acci\'on glu\'onica
\begin{equation}
 e^{-\Gamma_G[\Omega]} = 1 - \Gamma_G[\Omega] + \frac{1}{2} \Gamma_G[\Omega]^2 + \cdots  \,,
\end{equation}
lo que genera una supresi\'on exponencial del tipo~$ e^{- m_G /T }$. Esto da lugar a la siguiente f\'ormula de masas para el argumento de Boltzmann en la exponencial
\begin{eqnarray}
{\cal M} = n N_c M_q + m M_{\bar q q } + l m_G  \,,
\end{eqnarray} 
que muestra claramente que las contribuciones t\'ermicas de orden m\'as bajo a temperaturas bajas vienen dadas nuevamente por los loops t\'ermicos pi\'onicos, lo cual corresponde a tomar $n = l = 0$ y $m=1$, pues $N_c M_q \gg m_G \gg M_{\overline{q}q}=m_\pi$. Notar que num\'ericamente, incluso la contribuci\'on de dos loops pi\'onicos resultar\'{\i}a m\'as importante que las correcciones glu\'onicas.

En una serie de trabajos recientes~\cite{Meisinger:2001cq,Meisinger:2003id} se ha obtenido la ecuaci\'on de estado para un gas de gluones masivos con una masa dependiente de temperatura en presencia del loop de Polyakov, lo cual permite reproducir los datos del ret\'{\i}culo de manera bastante precisa por encima de la transici\'on de fase. La densidad de energ\'{\i}a de vac\'{\i}o se escribe
\begin{equation}
 V_{\text{glue}}[ \Omega ]= T \int \frac{d^3 k}{(2\pi)^3} \widehat \tr_{c} \ln \left[ 1 - e^{-\beta \omega_k} \widehat \Omega \right] \,,
\end{equation}
donde $ \omega_k = \sqrt{k^2 + m_G^2 } $, con $m_G$ la masa del glu\'on. La dependencia en temperatura que se considera en estos trabajos es~$m_G(T) = T g(T)\sqrt{2}$, que en la transici\'on de fase ($T=T_D$) toma el valor~$m_G(T_D)= 1.2-1.3\,T_D$. Si se toma un valor constante para la masa del glu\'on por debajo de la transici\'on de fase, a bajas temperaturas se obtiene
\begin{equation}
 V_{\text{glue}}[ \Omega ]=- T \sum_{n=1}^\infty \frac1{n}
\left(|\tr_c \,\Omega^{n}|^2-1\right)\, \int \frac{d^3 k}{(2\pi)^3}
e^{-n \beta \omega_k} \,,
\end{equation}
donde se ha hecho uso de la identidad
\begin{eqnarray}
\widehat \tr_c \, \widehat \Omega^n = |\tr_c \,\Omega^{n}|^2 -1 \,.
\label{eq:Ladj_Lfund_1}
\end{eqnarray} 
Haciendo uso de la representaci\'on asint\'otica de las funciones de Bessel, se obtiene una supresi\'on similar a la que se encuentra en el l\'{\i}mite de acoplamientos grandes.

\subsection{Correcciones locales en el loop de Polyakov}
\label{correcciones_locales}

Vamos a considerar aqu\'{\i} un tratamiento preliminar de las correcciones
locales en el loop de Polyakov. Hasta ahora se ha considerado un
campo~$\Omega$ constante en el espacio. De manera general, el loop de Polyakov
depende tanto del tiempo eucl\'{\i}deo como de las coordenadas espaciales. En
el gauge de Polyakov la dependencia en tiempo eucl\'{\i}deo es simple, pero
a\'un queda una dependencia en coordenadas que es desconocida. En tal caso, las reglas anteriores deben ser modificadas, ya que las inserciones del loop de Polyakov llevar\'an un momento, y el resultado depende de su ordenamiento. Si seguimos considerando, como hasta ahora, que el loop de Polyakov es la \'unica fuente de color en el problema, nos vamos a encontrar con funciones de correlaci\'on de loops de Polyakov. En la fase de confinamiento es de esperar una descomposici\'on basada en la existencia de propiedades de agrupamiento para cada par de variables. Por ejemplo, se tiene
\begin{eqnarray}
\langle \tr_c \Omega (\vec x_1 , \beta ) \,\tr_c\Omega^{-1} (\vec x_2, \beta )
\rangle \simeq e^{- \beta \sigma |\vec x_1 - \vec x_ 2 |} \,. 
\end{eqnarray} 
Por tanto, valores muy diferentes en la coordenada espacial est\'an suprimidos, de modo que tiene sentido considerar una aproximaci\'on local dentro de la longitud de correlaci\'on, y desarrollar las funciones de correlaci\'on en gradientes dentro de esta regi\'on. En una primera aproximaci\'on, esto se corresponde con la sustituci\'on del volumen cuatridimensional por un dominio de correlaci\'on, mediante la regla
\begin{equation}
V = \frac{1}{T}  \int d^3x \longrightarrow \frac{1}{T} \int d^3 x \, e^{-\sigma r/T } 
= \frac{8\pi T^2}{\sigma^3} \,.
\label{eq:regla_V_VT}
\end{equation}
 En el lagrangiano quiral a bajas energ\'{\i}as, que se obtiene desarrollando la acci\'on efectiva en derivadas de los campos mes\'onicos, aparecen tambi\'en gradientes del loop de Polyakov. Este hecho se comenta en ref.~\cite{Megias:2006bn}. En realidad, puesto que estamos acoplando el loop de Polyakov de manera efectiva como un potencial qu\'{\i}mico de color dependiente de $x$, nuestra aproximaci\'on es similar a una generalizaci\'on no abeliana de la aproximaci\'on de densidad local en teor\'{\i}a de muchos cuerpos de f\'{\i}sica nuclear y materia condensada, dentro del esp\'{\i}ritu de la teor\'{\i}a del funcional de la densidad.

\subsection{Resultados m\'as all\'a de la aproximaci\'on quenched}
\label{mas_alla_quenched}

En esta secci\'on nos proponemos ir m\'as all\'a de la aproximaci\'on quenched
en el c\'alculo de algunos observables concretos, y para ello deberemos tener en cuenta la contribuci\'on del determinante fermi\'onico. El modelo quark quiral completo con acoplamiento del loop de Polyakov viene dado por ec.~(\ref{eq:Z_pnjl}). La contribuci\'on de los quarks a la funci\'on de partici\'on del modelo NJL se escribe como
\begin{equation}
Z_Q[U,\Omega] := e^{-\Gamma_Q[U,\Omega]} = \Det ({\bf D}) \, \exp \left( -\frac{a_s^2}{4}\tr_f\int d^4 x  \,(M-\hat{m_0})^2 \right) \,,
\end{equation} 
que se obtiene a partir de ecs.~(\ref{eq:njlacefectq})-(\ref{eq:njlacefect}) donde se ha aplicado la ecuaci\'on del gap, ec.~(\ref{eq:njlcondgap}). En la secci\'on \ref{lagrangiano_quiral_Tfinita} se calcul\'o del determinante fermi\'onico en presencia de un loop de Polyakov (lentamente variable), como un desarrollo en momentos externos de los campos
\begin{equation}
\Det({\bf D}) = e^{-\int d^4 x \,{\cal L}^*_q (x)} = \exp\left(-\int d^4 x
  \,({\cal L}_q^{* (0)} (x) + {\cal L}_q^{* (2)}(x) + {\cal L}_q^{* (4)}(x) + \cdots )\right) \,.
\end{equation}
De acuerdo con la discusi\'on de la secci\'on~\ref{correcciones_locales}, la
aproximaci\'on de loop de Polyakov lentamente variable tiene sentido en una
regi\'on donde existen correlaciones fuertes entre loops de Polyakov. Para
nuestros prop\'ositos, bastar\'a con considerar aqu\'{\i} la contribuci\'on de
vac\'{\i}o ${\cal L}_q^{* (0)}$. En el modelo de NJL esta contribuci\'on se escribe
\begin{eqnarray}
{\cal L}_q^{* (0)}(x) &=& -\frac{N_c N_f}{(4\pi)^2} \sum_i c_i (\Lambda_i^2+M^2)^2 \log (\Lambda_i^2+M^2)  \nonumber \\
&&\qquad +\frac{N_f}{\pi^2}\left(M T\right)^2 \sum_{n=1}^\infty (-1)^n \frac{K_2(n M/T)}{n^2} 
\left( \tr_c \Omega^n(x) + \tr_c \Omega^{-n}(x) \right) \,, \nonumber \\
&=& {\cal L}_q^{(0)}(T=0) + {\cal L}_q^{(0)} (\Omega(x), T)  \,,  
\label{eq:Lq0_aZ}
\end{eqnarray}
que se obtiene a partir de ec.~(\ref{eq:el_lq0}) y ec.~(\ref{eq:Jl4}). El lagrangiano se ha escrito separando dos contribuciones: temperatura cero y temperatura finita. Esta \'ultima contiene el loop de Polyakov. Notar que en este punto a\'un no hemos considerado la integraci\'on en el grupo gauge SU($N_c$), de modo que no escribimos los corchetes $\langle \ldots \rangle$ como hicimos en la secci\'on~\ref{lagrangiano_quiral_Tfinita}. En ec.~(\ref{eq:Z_pnjl}) la integraci\'on en ${\cal D}U$ la hemos realizado a nivel cl\'asico, mediante el uso de las ecuaciones cl\'asicas de movimiento del campo~$U$, ec.~(\ref{eq:ecm}), (para detalles, ver ap\'endice~\ref{sec:hhkk}). La funci\'on de partici\'on se puede escribir
\begin{equation}
Z = \int {\cal D}\Omega \, e^{-\Gamma_G[\Omega]} \exp \left(-\int d^4 x \left\{ \frac{a_s^2}{4} \tr_f(M-\hat{m}_0)^2 + {\cal L}_q^{(0)}(T=0) + {\cal L}_q^{(0)}(\Omega(x),T)\right\} \right) \,.  \nonumber \\
\label{eq:Z_unquenched}
\end{equation}
El valor esperado del loop de Polyakov se escribe
\begin{equation}
L = \frac{1}{N_c} \langle \tr_c \Omega\rangle = \frac{1}{N_c Z} \int {\cal D}\Omega \, e^{-\Gamma_G[\Omega]} e^{-\Gamma_Q[\Omega]} \,\tr_c \Omega (x) \,,
\label{eq:L_esp_Z}
\end{equation} 
donde no indicamos dependencia de~$\Gamma_Q$ en $U$, pues nos limitamos a
considerar~${\cal L}_q^{* (0)}$ que no tiene dependencia en los campos mes\'onicos. El c\'alculo de ec.~(\ref{eq:L_esp_Z}) puede hacerse anal\'iticamente en el l\'{\i}mite de temperatura peque\~na. En este r\'egimen pueden despreciarse las correcciones glu\'onicas~$e^{-\Gamma_G[\Omega]}$ (ver secci\'on \ref{correcciones_gluonicas}), de modo que en el promedio sobre el grupo solamente contribuir\'a la medida de Haar~${\cal D}\Omega$. Cuando $T$ es suficientemente peque\~no, se puede considerar el desarrollo del t\'ermino ${\cal L}_q^{(0)}(\Omega,T)$ en la exponencial de ec.~(\ref{eq:Z_unquenched}). A primer orden en este desarrollo aparecen las siguientes funciones de correlaci\'on entre loops de Polyakov\footnote{Si se considera un loop de Polyakov independiente de $x$, se tiene la siguiente f\'ormula de integraci\'on sobre el grupo SU($N_c$)
\begin{equation}
\int {\cal D}\Omega\; \Omega_{ij} \Omega_{kl}^* = \frac{1}{N_c}\delta_{ik}\delta_{jl} \,,
\end{equation}
que conduce trivialmente a
\begin{equation}
\int {\cal D}\Omega\; \tr_c\Omega \,\tr_c \Omega^{-1} = 1 \,.
\end{equation}
Al considerar correcciones locales, se tiene
\begin{equation}
\int {\cal D}\Omega\; \tr_c\Omega(\vec{x}) \; \tr_c \Omega^{-1}(\vec{y}) = e^{-\sigma|\vec{x}-\vec{y}|/T} \,.
\end{equation}
}
\begin{equation}
\int d^4x \int {\cal D}\Omega \;\tr_c \Omega(\vec{x}) \, \tr_c \Omega (\vec{y}) = 0 \,, \label{eq:corrLL}
\end{equation}
\begin{equation}
\int d^4x \int {\cal D}\Omega \;\tr_c \Omega(\vec{x}) \, \tr_c \Omega^{-1}(\vec{y}) 
= \int d^4x  \; e^{-\sigma |\vec{x}-\vec{y}|/T} = \frac{8\pi T^2}{\sigma^3} \,.
\label{eq:corrLLm}
\end{equation}
La primera expresi\'on es cero por conservaci\'on de trialidad. La segunda expresi\'on constituye la regla que mencionamos en ec.~(\ref{eq:regla_V_VT}), que permite sustituir el cuadrivolumen infinito $\int d^4x$, por un volumen efectivo que especifica un dominio de correlaci\'on~$8\pi T^2/\sigma^3$. Con todo esto se llega finalmente al siguiente resultado en el modelo NJL
\begin{equation}
L(T) \stackrel{{\rm T}  \; \text{peque\~no}} \sim \frac{4N_f}{N_c\sigma^3}\sqrt{\frac{2M^3 T^9}{\pi}} e^{-M/T} \,.
\label{eq:L_low_T}
\end{equation}
Notar que la trialidad no se preserva, debido a la presencia de quarks
din\'amicos, y la escala relevante es la masa constituyente de los quarks.
Gracias a esta supresi\'on exponencial, est\'a justificado usar de manera
efectiva el loop de Polyakov como un par\'ametro de orden para la
simetr\'{\i}a del centro incluso en el caso unquenched. En realidad, nuestro
an\'alisis sugiere que un c\'alculo del loop de Polyakov en QCD completo
podr\'{\i}a constituir un m\'etodo para extraer una masa constituyente de los
quarks invariante gauge.  En cualquier caso, ser\'{\i}a deseable disponer de
datos en el ret\'{\i}culo del loop de Polyakov para temperaturas bajas, $T\le
50\, \text{MeV}$, con objeto de hacer un an\'alisis preciso.


Para el condensado de quarks, hacemos uso de 
\begin{equation}
\langle \overline{q}q \rangle_T = -f_\pi^{* 2} B^*_0 = -\frac{M}{4\pi^2} \langle \tr_c \J_{-1} \rangle\,,
\end{equation}
que obtuvimos en ec.~(\ref{CBE_NJL}). Al tener en cuenta la contribuci\'on del determinante fermi\'onico, se tiene
\begin{equation}
\langle \overline{q}q \rangle_T = -\frac{M}{4\pi^2}\frac{1}{Z} \int {\cal D}\Omega \, e^{-\Gamma_G[\Omega]} e^{-\Gamma_Q[\Omega]} \,\tr_c \J_{-1}(M,\Omega) \,.
\label{eq:qq_esp_Z}
\end{equation}
La expresi\'on de $\J_{-1}$ viene dada en ec.~(\ref{eq:Jl3}). A partir de aqu\'{\i}, el procedimiento para hallar el comportamiento de $\langle \overline{q}q \rangle_T$ a baja temperatura es id\'entico al caso del valor esperado del loop de Polyakov. En el r\'egimen de $T$ peque\~no, nuevamente $e^{-\Gamma_G[\Omega]}$ se puede despreciar, y podemos desarrollar el t\'ermino ${\cal L}_q^{(0)}(\Omega,T)$ que aparece en $e^{-\Gamma_Q[\Omega]}$. Teniendo en cuenta las integrales (\ref{eq:corrLL})-(\ref{eq:corrLLm}), se llega a
\begin{equation}
\langle \overline{q}q\rangle_T 
\stackrel{{\rm T}  \; \text{peque\~no}} \sim 
\langle \overline{q}q\rangle_{T=0} 
+ \frac{8N_f}{\pi^2} \frac{M^3 T^6}{\sigma^3} e^{-2M/T} \,.
\label{eq:qq_low_T}
\end{equation}
En el modelo quark espectral se obtiene el resultado de ec.~(\ref{eq:qq_low_T}), con la sustituci\'on $2M\rightarrow M_V$ (la masa del mes\'on $\rho$), y un factor multiplicativo ligeramente diferente.

Como vemos, en el c\'alculo unquenched el enfriamiento de Polyakov persiste,
aunque es un poco menos efectivo que en el c\'alculo quenched. Este mismo
an\'alisis se puede hacer para otros observables, por ejemplo las constantes
de baja energ\'{\i}a del lagrangiano efectivo quiral tienen un
comportamiento~$ L_i^T - L_i^{T=0} \stackrel{{\rm T}\; \text{peque\~no}}\sim
e^{- M_V /T} $ \cite{Megias:2006bn}.

Finalmente, ser\'{\i}a necesario incluir m\'as loops de quarks, o equivalentemente excitaciones mes\'onicas. Esto dar\'{\i}a exactamente el resultado de TQP con piones sin masa dominando en la regi\'on de temperaturas peque\~nas. Por tanto, vemos que cuando el loop de Polyakov se acopla de manera conveniente a los modelos de quarks quirales, se obtiene una explicaci\'on natural de los resultados encontrados hace tiempo en modelos puramente hadr\'onicos.

\section{Implicaciones sobre la transici\'on de fase de QCD}
\label{transicion_fase_modelo}

En la secci\'on~\ref{correcciones_orden_mayor} se hizo un estudio anal\'{\i}tico del comportamiento a baja temperatura del loop de Polyakov y del condensado quiral en QCD unquenched. Resultar\'{\i}a interesante estudiar el comportamiento que predice nuestro modelo para estos observables en la regi\'on de la transici\'on de fase, y para ello deberemos integrar num\'ericamente las ecuaciones~(\ref{eq:L_esp_Z}) y (\ref{eq:qq_esp_Z}). A diferencia de nuestro tratamiento, en ref.~\cite{Fukushima:2003fw} se hace un estudio en la aproximaci\'on de campo medio, en el cual la probabilidad de encontrar un loop de Polyakov dado es una funci\'on delta. La integral en el grupo permite tener en cuenta una dispersi\'on de esa probabilidad debido a efectos cu\'anticos.

Para $N_c=3$ el loop de Polyakov contiene dos variables independientes. En el gauge de Polyakov, $\partial_0 A_0 =0$, se puede parametrizar como una matriz diagonal del siguiente modo
\begin{equation}
\Omega = \diag (e^{i\phi_1}, e^{i\phi_2}, e^{-i(\phi_1+\phi_2)}) \,.
\end{equation}
Con esta parametrizaci\'on, podemos calcular la funci\'on de partici\'on como
\begin{equation}
Z = \int {\cal D}\Omega \,e^{-\Gamma_G[\Omega]} e^{-\Gamma_Q[\Omega]} =
\int_{-\pi}^{\pi} \frac{d\phi_1}{2\pi} \frac{d\phi_2}{2\pi} \,\rho_G(\phi_1,\phi_2) \rho_Q(\phi_1,\phi_2) \,,
\end{equation}
donde
\begin{equation}
{\cal D}\Omega\, e^{-\Gamma_G[\Omega]} = \frac{d\phi_1}{2\pi} \frac{d\phi_2}{2\pi}\,\rho_G(\phi_1,\phi_2) \,,
\qquad e^{-\Gamma_Q[\Omega]} = \rho_Q(\phi_1,\phi_2) \,.
\end{equation}
En $\Gamma_Q[\Omega]$ no indicamos dependencia en los campos mes\'onicos~$U$, pues al igual que en sec.~\ref{mas_alla_quenched} nos limitaremos a considerar la contribuci\'on de vac\'{\i}o ${\cal L}_q^{(0)}$ del lagrangiano quiral, ec.~(\ref{eq:Lq0_aZ}). Para una funci\'on general $f(\Omega)$, se tiene
\begin{eqnarray}
\langle \tr_c f(\Omega) \rangle &=& \frac{1}{Z} \int_{-\pi}^{\pi} \frac{d\phi_1}{2\pi} \frac{d\phi_2}{2\pi} \, \rho_G(\phi_1,\phi_2) \rho_Q(\phi_1,\phi_2) (f(e^{i\phi_1}) + f(e^{i\phi_2})+f(e^{-i(\phi_1+\phi_2)})) \nonumber \\
&=& \frac{1}{Z}\int_{-\pi}^\pi \frac{d\phi_1}{2\pi} \, \hat\rho(\phi_1) f(e^{i\phi_1}) \,,
\label{eq:trftrans}
\end{eqnarray}
donde
\begin{equation}
\hat\rho(\phi_1) = 3 \int_{-\pi}^{\pi} \frac{d\phi_2}{2\pi} \, \rho_G(\phi_1,\phi_2) \rho_Q(\phi_1,\phi_2) \,.
\label{eq:rogtrans}
\end{equation}
Por invariancia gauge, tanto la medida de Haar ${\cal D}\Omega$, como las correcciones glu\'onicas $e^{-\Gamma_G[\Omega]}$, las contribuciones fermi\'onicas~$e^{-\Gamma_Q[\Omega]}$, y $\tr_c f(\Omega)$ son invariantes frente al intercambio de los autovalores del loop de Polyakov. Esto permite expresar $\langle \tr_c f(\Omega)\rangle$ como una integral en un \'unico par\'ametro, tal y como se expresa en ec.~(\ref{eq:trftrans}), con la funci\'on peso adecuada, ec.~(\ref{eq:rogtrans}).

En nuestro tratamiento consideramos la integraci\'on sobre el grupo SU(3) y una minimi\-zaci\'on con respecto a $M$, lo cual se corresponde con ec.~(\ref{eq:njlcondgap}). Esto \'ultimo permite calcular la dependencia en temperatura de la masa constituyente, y de ah\'{\i} obtener el condensado quiral
\begin{equation}
\langle \overline{q}q \rangle_T = -\frac{a_s^2}{2}\tr_f (M(T)-\hat{m}_0) \,.
\end{equation} 
Puesto que la constante de acoplamiento de cuatro quarks $a_s$ parametriza
informaci\'on sobre los gluones, deber\'{\i}a de tener una dependencia
en~$\Omega$. No obstante, $a_s$ incorpora informaci\'on sobre todos los grados
de libertad glu\'onicos, de modo que no deber\'{\i}a de verse muy afectado por
la contribuci\'on de $\Omega$, donde $\Omega$ viene dado \'unicamente por la componente temporal de los gluones.

\begin{figure}[ttt]
\begin{center}
\epsfig{figure=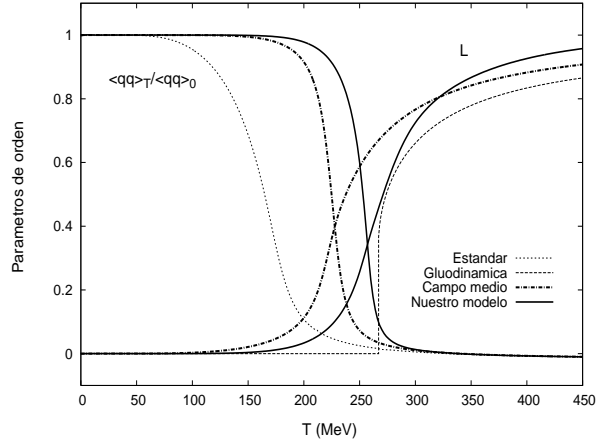,height=6cm,width=8cm}
\end{center}
\caption{Dependencia en temperatura del condensado quiral~$\langle
  \bar q q \rangle $ en unidades relativas, y del valor esperado del loop de
  Polyakov~$ L= \langle \tr_c \Omega \rangle /N_c $. El resultado est\'andar
  de $\langle \overline{q}q \rangle_T$ se corresponde con el modelo NJL sin
  acoplamiento con el loop de Polyakov. Se compara tambi\'en por una parte con
  la aproximaci\'on de campo medio de ref.~\cite{Fukushima:2003fw}, donde el
  loop de Polyakov es cl\'asico y est\'a acoplado con los quarks, y por otra
  con nuestro modelo basado en la integraci\'on sobre el grupo de color SU(3)
  y considerando correcciones locales en el loop de Polyakov. Se muestra
  asimismo el comportamiento de $L$ en gluodin\'amica dentro del esquema de
  desarrollo con acoplamientos grandes, ec.~(\ref{eq:V_glue_Fuk}). Se ha
  considerado~$N_f=2$.}
\label{fig:phase_transition}
\end{figure}

En fig.~\ref{fig:phase_transition} se muestra el comportamiento del condensado quiral $\langle \overline{q}q\rangle_T$ y el valor esperado del loop de Polyakov $L=\langle \tr_c\Omega\rangle/N_c$ en diferentes tratamientos del modelo NJL. Se compara la predicci\'on est\'andar del modelo NJL, con el c\'alculo en aproximaci\'on de campo medio de ref.~\cite{Fukushima:2003fw}, que corresponde a minimizar la energ\'{\i}a de vac\'{\i}o como funci\'on de la masa constituyente $M$ y del valor esperado del loop de Polyakov $L$. Comparamos asimismo con el resultado que obtenemos al considerar una integraci\'on en el loop de Polyakov~$\Omega$ con correcciones locales. En la figura se muestra adem\'as el comportamiento de $L$ que se obtiene en gluodin\'amica, con el modelo de ec.~(\ref{eq:V_glue_Fuk}) en su tratamiento de campo medio, lo cual conduce a una transici\'on de fase de primer orden en $T_D=270\,{\rm MeV}$. En nuestros c\'alculos estamos considerando el modelo quark quiral con dos sabores~$N_f=2$, y para la masa desnuda de los quarks $\hat{m}_0=\diag(m_u, m_d)$ consideramos el l\'{\i}mite en que hay simetr\'{\i}a de isosp\'{\i}n, $m_u=m_d \equiv m_q$. En los tres modelos hemos tomado $m_q=5.5\,{\rm MeV}$, y $a_s^2=76.2\cdot 10^{-3}\,{\rm GeV}^2$. La integraci\'on en momentos est\'a regulada por un cut-off~$\Lambda_{\rm PV}=828\,{\rm MeV}$ con regularizaci\'on de Pauli-Villars. Este valor es el que se necesita para reproducir el valor experimental de la constante de desintegraci\'on d\'ebil del pi\'on~$f_\pi=93.2\,{\rm MeV}$, con la masa constituyente~$M=300\,{\rm MeV}$. Para la tensi\'on de la cuerda consideramos su valor a temperatura cero~$\sigma=(425\,{\rm MeV})^2$. Este par\'ametro aparece cuando se calculan funciones de correlaci\'on de loops de Polyakov (por ejemplo, ec.~(\ref{eq:corrLLm})).

\begin{figure}[ttt]
\begin{center}
\epsfig{figure=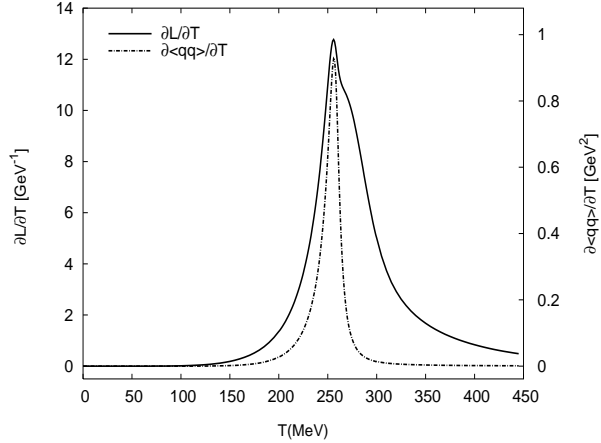,height=6cm,width=8cm}
\end{center}
\caption{Dependencia en temperatura de $\partial \langle \bar
q q \rangle /\partial T$ y~$ \partial L/\partial T$, obtenida con el modelo
NJL basado en la integraci\'on sobre el grupo de color SU(3) y considerando
correcciones locales en el loop de Polyakov. Se ha tomado~$N_f=2$.}
\label{fig:susceptibilidad}
\end{figure}

El efecto neto de la integraci\'on sobre el grupo de color SU(3) consiste en un desplazamiento de la temperatura de transici\'on quiral a valores mayores, respecto a las tempera\-turas que se obtienen en los tratamientos est\'andar y de campo medio. Por tanto, el modelo basado en la integraci\'on sobre el grupo de color proporciona un enfriamiento efectivo, no s\'olo en el r\'egimen de temperaturas peque\~nas (ver secciones~\ref{solucion_problematica} y \ref{mas_alla_quenched}), sino tambi\'en en el r\'egimen de la transici\'on de fase. Como se ve en fig.~\ref{fig:phase_transition}, el acoplamiento del modelo quark quiral con gluodin\'amica modifica la transici\'on de fase de primer orden de gluodin\'amica en una transici\'on de fase de segundo orden. Un estudio de la susceptibilidad de los pa\-r\'a\-me\-tros de orden quiral $\langle \overline{q}q\rangle$ y de desconfinamiento $L$, permite ver que con nuestro modelo ambas transiciones de fase (quiral y de desconfinamiento) se producen simult\'aneamente: $T_\chi=T_D=256(1)\,{\rm MeV}$; (ver fig.~\ref{fig:susceptibilidad}).

\begin{figure}[ttt]
\begin{center}
\epsfig{figure=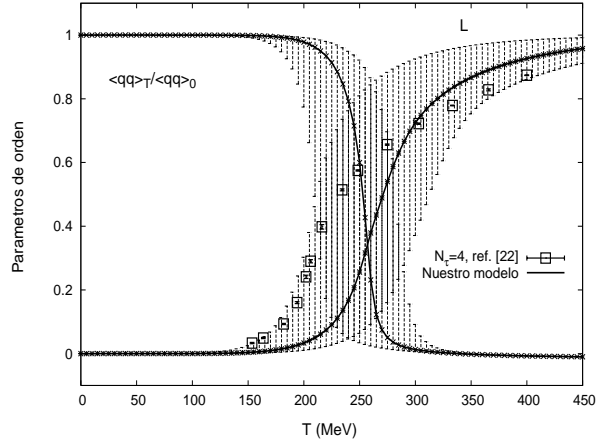,height=6cm,width=8cm}
\end{center}
\caption{Dependencia en temperatura del condensado quiral~$\langle
\bar q q \rangle $ y del valor esperado del loop de Polyakov~$ L=
\langle \tr_c \Omega \rangle /N_c $, obtenido con el modelo NJL basado en la integraci\'on sobre el grupo de color SU(3) y
considerando correcciones locales en el loop de Polyakov. Se ha tomado~ $N_f=2$. Las bandas de error corresponden a una incertidumbre en la tensi\'on de la cuerda $\sigma = 0.181 \pm 0.085 \, {\rm GeV}^2$. Se compara
con los datos del ret\'{\i}culo para QCD con 2 sabores, obtenidos
en~\cite{Kaczmarek:2005ui}.}
\label{fig:qq_Lfund_lattice}
\end{figure}

En fig.~\ref{fig:qq_Lfund_lattice} se compara el comportamiento del loop de Polyakov obtenido en nuestro modelo, con c\'alculos en el ret\'{\i}culo para QCD unquenched $(N_f=2)$ en la zona de transici\'on de fase. Estos datos se han calculado en un ret\'{\i}culo de tama\~no~$16^3 \times 4$, con $m_q/T=0.4$~\cite{Kaczmarek:2005ui}. Se muestra asimismo el comportamiento del condensado quiral. Hemos comprobado que una dependencia en temperatura de la tensi\'on de la cuerda~$\sigma$ permite compatibilizar los resultados de nuestro modelo con los obtenidos en el ret\'{\i}culo. Esto conduce a un rango de incertidumbre en la tensi\'on de la cuerda, $\sigma = 0.181 \pm 0.085 \,{\rm GeV}^2$, que da cuenta en cierto sentido de la incertidumbre existente en el modelo. En fig.~\ref{fig:qq_Lfund_lattice} la banda de error asociada a esta incertidumbre conduce a una temperatura de transici\'on de $T_\chi=T_D= 255 \pm 50 \, {\rm MeV}$.
Si se ignoraran en el modelo las correcciones glu\'onicas dadas por ec.~(\ref{eq:V_glue_Fuk}), no existir\'{\i}a un efecto apreciable por debajo de la transici\'on de fase, si bien \'esta aumentar\'{\i}a en~$30\,{\rm MeV}$, un valor que se encuentra dentro de nuestra estimaci\'on del error.
 
\begin{figure*}[tbc]
\begin{center}
\epsfig{figure=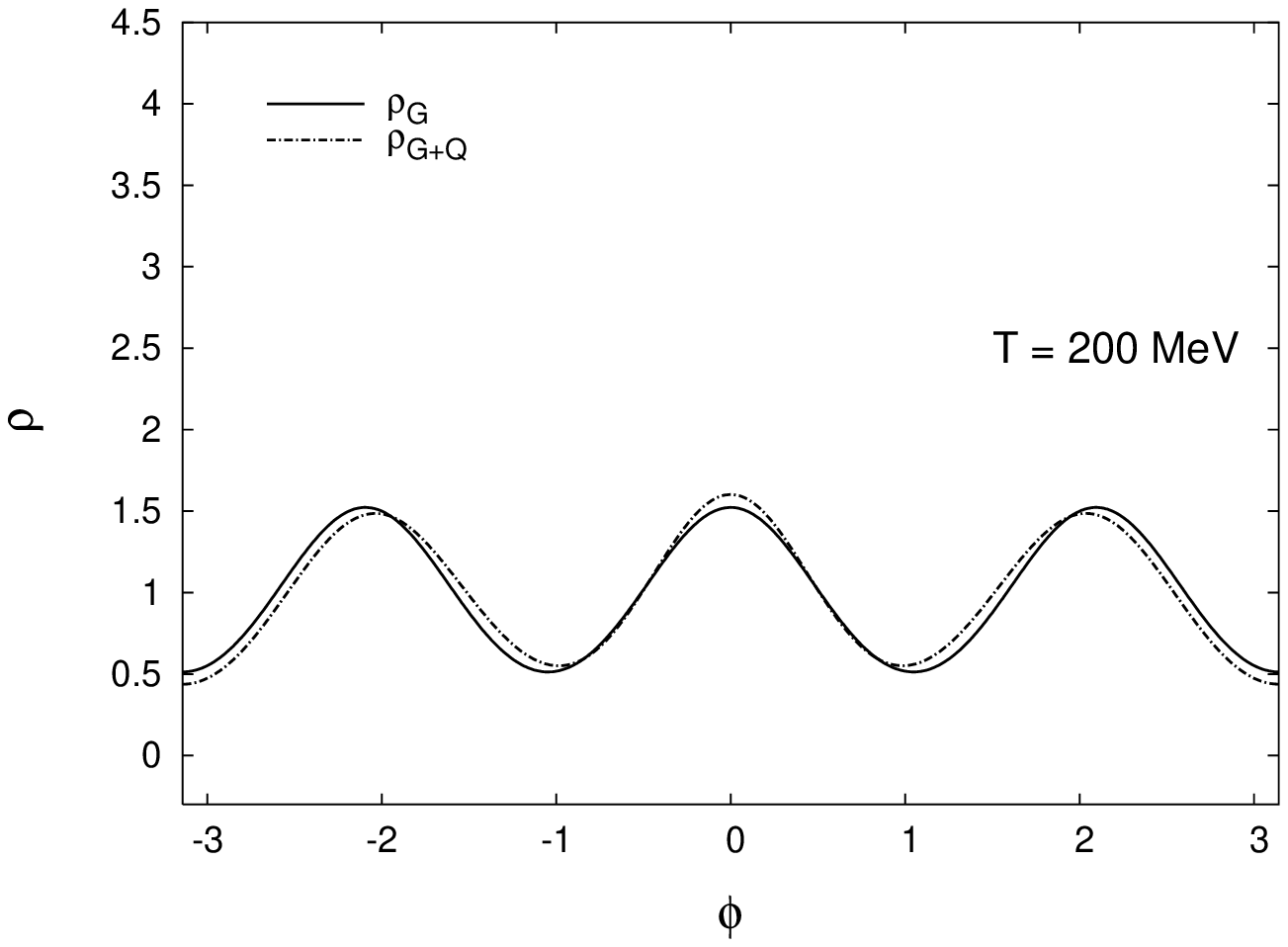,height=5cm,width=5cm}
\epsfig{figure=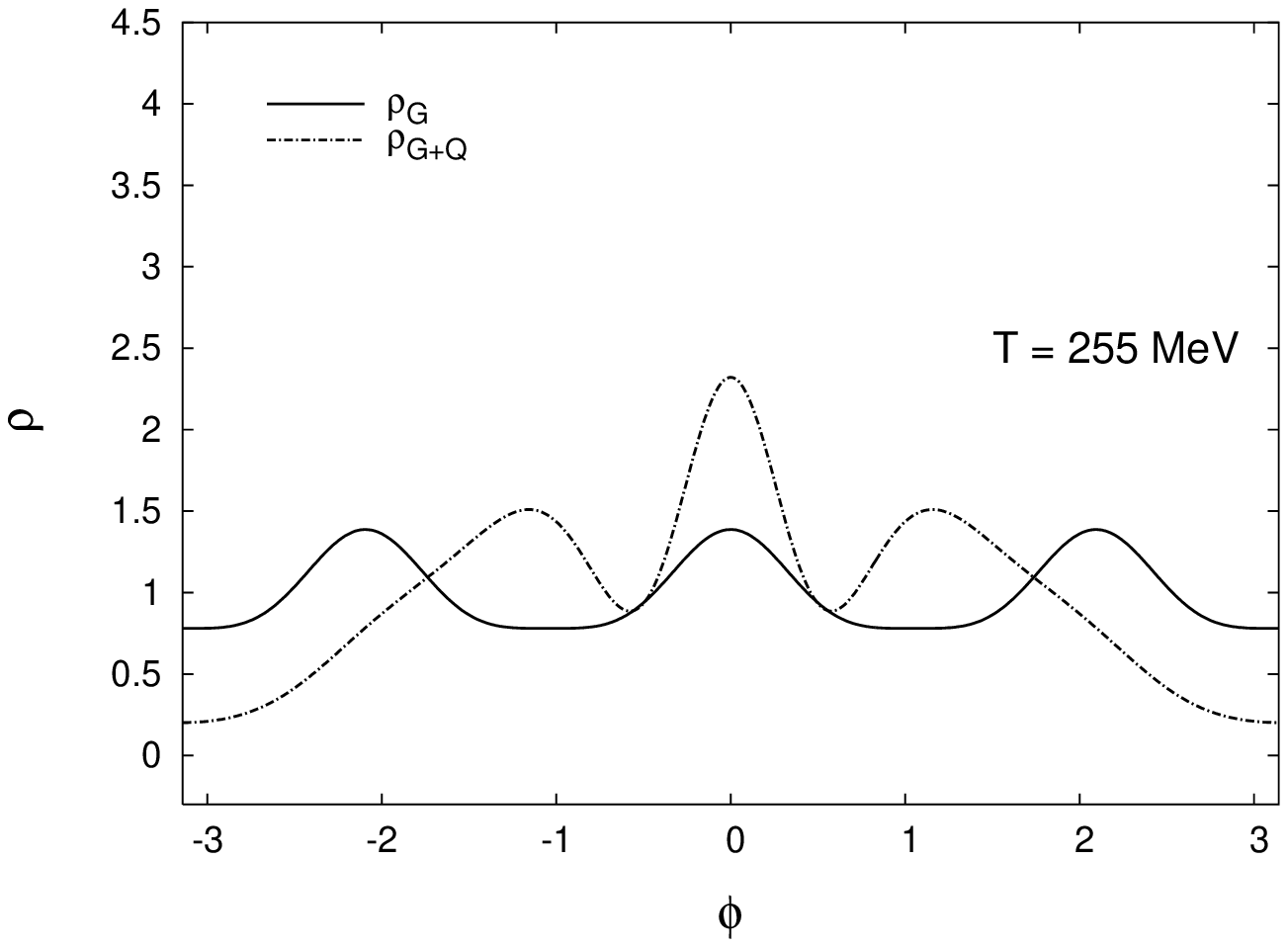,height=5cm,width=5cm}
\epsfig{figure=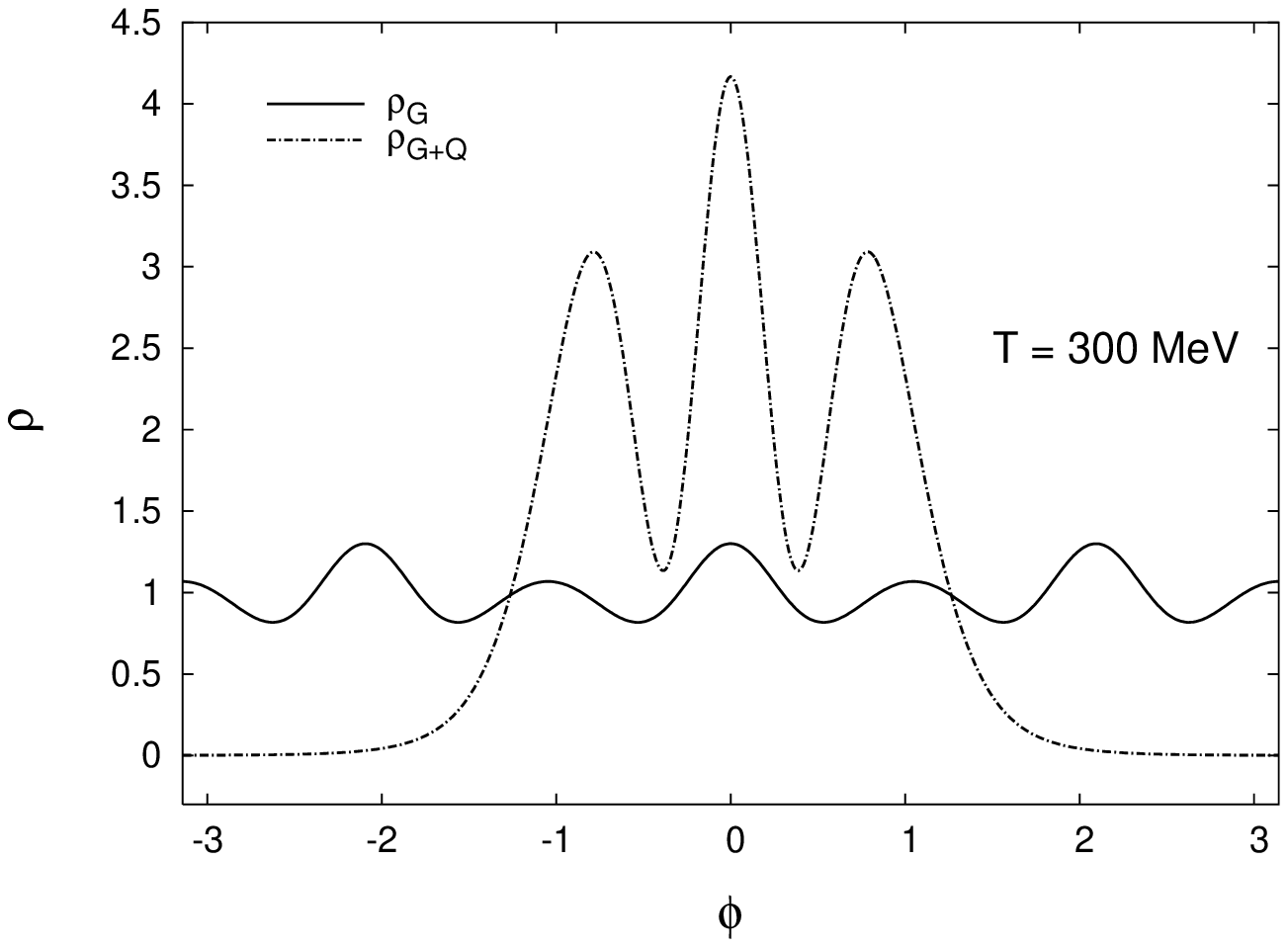,height=5cm,width=5cm} 
\end{center}
\caption{Dependencia en temperatura de la distribuci\'on del loop de Polyakov $\hat\rho(\phi)$, ec.~(\ref{eq:rogtrans}). $\hat\rho_G$ corresponde a la distribuci\'on en gluodin\'amica (sin contribuci\'on de quarks) procedente de la medida de Haar junto con el esquema del desarrollo en acoplamientos grandes a orden m\'as bajo, ec.~(\ref{eq:V_glue_Fuk}), y $\hat\rho_{G+Q}$  incluye contribuciones de quarks de acuerdo con el modelo NJL. Se toma $N_f=2$. Se consideran tres temperaturas: $T=200, 255, 300\,{\rm MeV}$; por debajo de la transici\'on de fase, en la transici\'on y por encima, respectivamente.}
\label{fig:polyakov_prob} 
\end{figure*}

Con objeto de comprender el mecanismo de rotura de la simetr\'{\i}a del centro en nuestro modelo, podemos estudiar c\'omo evoluciona la distribuci\'on~$\hat\rho(\phi)$, ec.~(\ref{eq:rogtrans}), a trav\'es de la transici\'on de fase, y observar expl\'{\i}citamente los efectos generados por las contribuciones fermi\'onicas~$e^{-\Gamma_Q[\Omega]}$. En fig.~\ref{fig:polyakov_prob} se muestra esta evoluci\'on. Por debajo de la transici\'on de fase la funci\'on de distribuci\'on $\hat\rho(\phi)$ presenta tres m\'{\i}nimos en valores de $\phi$ equidistantes, tal y como exige la simetr\'{\i}a del centro~${\mathbb Z}(3)$. En este caso el determinante fermi\'onico no produce una modificaci\'on importante. Cuando la transici\'on de fase tiene lugar, aparece una concentraci\'on interesante de \'angulos en la regi\'on cercana al origen $\phi=0$, debida a los quarks, lo que genera una fuerte rotura de la simetr\'{\i}a del centro. A medida que la temperatura aumenta, la distribuci\'on del loop de Polyakov tiende a ser m\'as picuda en torno a $\phi=0$, y este pico domina la integral en $\phi$. Notar que la distribuci\'on~$\hat\rho_G(\phi)$ en gluodin\'amica no presenta rotura expl\'{\i}cita de la simetr\'{\i}a del centro para ning\'un valor de $T$, de modo que el \'unico mecanismo posible en este caso es la rotura espont\'anea.

\begin{figure}[ttt]
\begin{center}
\epsfig{figure=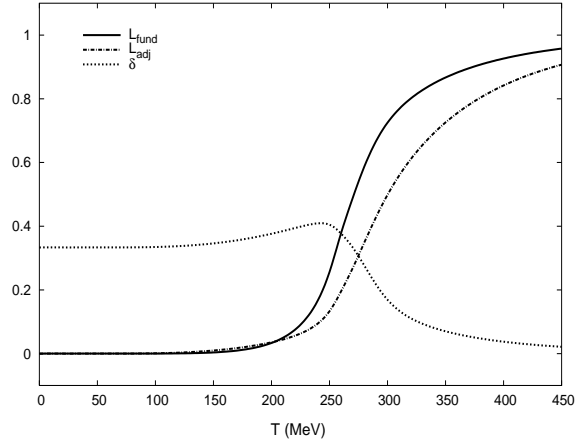,height=6cm,width=8cm}
\end{center}
\caption{Dependencia en temperatura del valor esperado del loop de Polyakov en representaci\'on fundamental~$\langle \tr_c \Omega \rangle /N_c $ y en representaci\'on adjunta $\langle \widehat {\rm tr}_c \, \widehat \Omega \rangle/(N_c^2-1)$, y fluctuaci\'on total del loop de Polyakov $\delta$. Resultados obtenidos en el modelo NJL con integraci\'on en el grupo de color SU(3). Se considera $N_f=2$.}
\label{fig:L_adj_fund_fluc}
\end{figure}

Nuestro modelo permite calcular el valor esperado del loop de Polyakov en otras repre\-sentaciones. En fig.~\ref{fig:L_adj_fund_fluc} se muestra el comportamiento del valor esperado del loop de Polyakov en representaci\'on adjunta, $\langle\widehat\tr_c \widehat\Omega \rangle/(N_c^2-1)$. Para ello hemos hecho uso de la identidad (\ref{eq:Ladj_Lfund_1}) con $n=1$. De acuerdo con los datos del ret\'{\i}culo obtenidos con el modelo matricial de ref.~\cite{Dumitru:2003hp}, el valor esperado se anula por debajo de la transici\'on de fase. Notar que este hecho no se cumple en el tratamiento de campo medio, para el cual se obtiene el valor~$-1/(N_c^2-1)$ (de ec.~(\ref{eq:Ladj_Lfund_1})). El considerar la integraci\'on sobre el grupo conduce a unos resultados acordes con lo que se espera de los estudios en el ret\'{\i}culo. En fig.~\ref{fig:L_adj_fund_fluc} se muestra tambi\'en el comportamiento del loop de Polyakov en representaci\'on fundamental, y la fluctuaci\'on total del loop de Polyakov, que definimos como
\begin{equation}
\delta \equiv \frac{1}{N_c}\sqrt{\langle \tr_c\Omega \,\tr_c \Omega^{-1}\rangle - \langle \tr_c \Omega \rangle^2} 
= \frac{1}{N_c}\sqrt{1+ \langle \widehat\tr_c \widehat\Omega\rangle - 
\langle  \tr_c \Omega\rangle^2 } \,.
\label{eq:fluctuation_L}
\end{equation}
$\delta$ da cuenta de manera conjunta de las fluctuaciones en la parte real e
imaginaria de $\Omega$. Esta fluctuaci\'on tiende a cero a temperaturas
grandes, lo cual es compatible con el hecho de que la
distribuci\'on~$\hat\rho(\phi)$ se hace muy picuda en torno a $\phi=0$ en el
r\'egimen de $T$ grande. 

\section{Conclusiones}
\label{conclusiones_modelos_quirales_T}

En este cap\'{\i}tulo hemos estudiado c\'omo la introducci\'on del loop de Polyakov permite resolver los problemas que presentan los modelos de quarks quirales a temperatura finita en su tratamiento est\'andar. Con objeto de preservar la invariancia gauge expl\'{\i}cita a tempera\-tura finita es necesario mantener de un modo no perturbativo ciertos grados de libertad glu\'onicos. En la pr\'actica, y en gauges particulares tales como el gauge de Polyakov, esto se corresponde con tratar la componente cero del campo del glu\'on como un potencial qu\'{\i}mico dependiente del color en el propagador del quark. Esto da lugar a una fuente de color que va a generar todos los estados posibles de quarks, los cuales pueden no ser singletes de color (incluso a bajas temperaturas, en la fase de confinamiento de color). Para evitar este problema, es necesario proyectar sobre los estados f\'{\i}sicos que son singletes de color, lo cual se consigue de un modo elegante haciendo la integral funcional sobre el campo ${\cal V}_0^c$ de un modo que se preserve la invariancia gauge. 

De este an\'alisis en la aproximaci\'on quenched y a nivel de un loop de quarks, encontramos que existe una supresi\'on de los efectos t\'ermicos en los observables hadr\'onicos por debajo de la transici\'on de fase, que surge de la conservaci\'on de la trialidad en una fase en que la simetr\'{\i}a quiral est\'a espont\'aneamente rota. A este efecto lo hemos denominado {\it enfriamiento de Polyakov} de las excitaciones de los quarks. En particular, la transici\'on de fase quiral no puede ocurrir antes que la transici\'on de desconfinamiento del color. En esta situaci\'on, el mayor cambio a bajas temperaturas en los observables tales como el condensado de quarks debe de provenir de los loops de pseudoescalares, y quiz\'as a temperaturas intermedias de resonancias mes\'onicas de orden mayor. Esto es precisamente lo que se espera de TPQ o de las aproximaciones unitarias con inclusi\'on efectiva de estos loops en las resonancias.

Nuestros argumentos muestran tambi\'en c\'omo, debido al enfriamiento de Polyakov, los modelos de quarks quirales se muestran de acuerdo con las suposiciones te\'oricas de TQP a temperatura finita. Para ver c\'omo se materializa esto en la pr\'actica hemos calculado el lagrangiano quiral a temperatura finita a nivel de un loop de quarks y a nivel \'arbol para los mesones. El lagrangiano resultante se puede descomponer en una parte con la misma estructura que a temperatura cero, pero con constantes de baja energ\'{\i}a dependientes de la temperatura, y otra parte con nuevos t\'erminos que rompen la invariancia Lorentz, que surgen como consecuencia de que el ba\~no t\'ermico est\'a en reposo. En cualquier caso, los efectos t\'ermicos en las constantes de baja energ\'{\i}a a este nivel de aproximaci\'on muestran el enfriamiento de Polyakov. En otras palabras, por debajo de la transici\'on de fase cualquier dependencia en temperatura sobre las constantes de baja energ\'{\i}a a nivel \'arbol puede ser despreciada. \'Esta es precisamente la suposici\'on inicial de TQP.

En el cap\'{\i}tulo hemos analizado algunas consecuencias que se obtienen al considerar el tratamiento de los modelos de quarks quirales acoplados con el loop de Polyakov, m\'as all\'a de un loop de quarks. Como consecuencia de la integraci\'on en el grupo gauge de color SU($N_c$), encontramos que para observables que contienen quarks las contribuciones m\'as importantes a temperaturas peque\~nas proceden de loops mes\'onicos, con una supresi\'on est\'andar a bajas temperaturas de~$\frac{1}{N_c} e^{-2M/T}$. Los loops bari\'onicos producen contribuciones m\'as peque\~nas~$e^{-N_c M/T}$. Un an\'alisis de las correcciones glu\'onicas permite ver que \'estas tienden a contribuir de manera apreciable \'unicamente por encima de la transici\'on de fase. 

Hemos estudiado c\'omo se modifican los resultados al considerar la introducci\'on del determinante de quarks en el c\'alculo de observables como el condensado quiral y el va\-lor esperado del loop de Polyakov, y se ha hecho asimismo un tratamiento preliminar de las correcciones locales en el loop de Polyakov. Este determinante conduce a una rotura expl\'{\i}cita de la simetr\'{\i}a del centro, que es m\'as acentuada a temperaturas grandes. \'Este es el mecanismo por el cual el modelo quark acoplado con loop de Polyakov genera la transici\'on de fase de desconfinamiento. Un an\'alisis de los resultados muestra que ambas transiciones de fase (quiral y de desconfinamiento) se producen simult\'aneamente. En el tratamiento unquenched el enfriamiento de Polyakov persiste, aunque es menos efectivo que en el caso quenched. El c\'alculo del valor esperado del loop de Polyakov en representaci\'on adjunta es un ejemplo de que el tratamiento del modelo quark con integraci\'on en el grupo gauge de color es m\'as adecuado que el tratamiento de campo medio de ref.~\cite{Fukushima:2003fw}.

\chapter{Tensor Energ\'{\i}a-Impulso de Modelos de Quarks Quirales a bajas energ\'{\i}as}
\label{tensor_EM_MQQ}

El tensor energ\'{\i}a-impulso (TEI) juega un papel muy importante en teor\'{\i}a cu\'antica de campos, pues surge como una corriente de Noether del grupo de Poincar\'e. Es conservado en todas las teor\'{\i}as locales relativistas, incluso cuando no existen otras cargas conservadas. En QCD, el TEI da cuenta de la interacci\'on de los quarks y gluones con los gravitones. 

Desde un punto de vista fenomenol\'ogico, las colisiones profundamente
inel\'asticas proporcionan informaci\'on relevante sobre la fracci\'on de
momento que llevan los quarks y los gluones dentro de un hadr\'on a una escala
dada~\cite{Ji:1995sv}. Las determinaciones basadas en el intercambio de un
gravit\'on est\'an fuera de lugar debido a que la constante de gravitaci\'on
resulta peque\~n\'isima en comparaci\'on con los procesos d\'ebiles y
fuertes. El factor de forma gravitacional del pi\'on se puede usar para
determinar la anchura de desintegraci\'on de un bos\'on de Higgs ligero en dos
piones~\cite{Donoghue:1990xh}. En el pasado hubo algunos intentos de calcular
el TEI en el ret\'{\i}culo~\cite{Caracciolo:1989pt}, pero no se han encontrado resultados de inter\'es pr\'actico para los elementos de matriz entre estados hadr\'onicos con momentos diferentes.

En este cap\'{\i}tulo vamos a estudiar la estructura del TEI en varios modelos
de quarks quirales.\footnote{Consideraremos gravedad de Einstein. Esto quiere
  decir que haremos uso de la conexi\'on de Riemann, definida sin torsi\'on y
  preservando la m\'etrica. Una extensi\'on a gravedad con torsi\'on es
  posible~\cite{Hammond:2002rm}.} En concreto trataremos el Modelo Quark
Constituyente, el Modelo de Nambu--Jona-Lasinio (NJL)~\cite{njlmodel} y el
Modelo de Georgi-Manohar (GM)~\cite{Manohar:1983md}. El cap\'{\i}tulo est\'a
basado en la referencia~\cite{Megias:2005fj}.

\section{Tensor Energ\'{\i}a-Impulso}
\label{TEI}

El tensor energ\'{\i}a-impulso en cualquier teor\'{\i}a se puede calcular a\~nadiendo una m\'etrica externa~$g_{\mu\nu}(x)$ que se acople con los campos de materia de un modo completamente covariante. El TEI se obtiene de calcular la derivada funcional de la acci\'on con respecto a $g_{\mu\nu}(x)$, en torno a la m\'etrica plana $\eta_{\mu\nu}$,\footnote{Usaremos el convenio $\eta = \diag(1,-1,-1,-1)$.}
\begin{equation}
\frac{1}{2} \theta^{\mu \nu} (x) = \frac{\delta S}{\delta g_{\mu
\nu}(x)} 
\Big|_{g_{\mu \nu} = \eta_{\mu\nu} } 
\label{eq:theta_quark}
\end{equation} 
donde 
\begin{equation}
S = \int d^4 x \sqrt{-g} \;{\cal L}(x)\,.   
\end{equation}
A nivel cu\'antico el comportamiento a alta energ\'{\i}a de $\theta_{\mu\nu}$ se puede mejorar si se reali\-zan ciertas correcciones transversales convenientemente elegidas. Al hacer esto se pone de manifiesto una anomal\'{\i}a de la traza que relaciona $\theta_\mu^\mu$ con la divergencia de la corriente de dilataci\'on, lo cual se\~nala la rotura an\'omala de la invariancia de escala. Un valor esperado diferente de cero para $ \langle 0 | \theta_\mu^\mu |0 \rangle $ est\'a relacionado con la existencia de un condensado glu\'onico, que genera identidades de Ward de escala~\cite{Shifman:1988zk}.

En el desarrollo en potencias de los momentos externos de los campos que se considera en Teor\'{\i}a Quiral de Perturbaciones, los campos pseudoescalares $U$ y la m\'etrica $g_{\mu\nu}$ son orden ${\cal O}(p^0)$.
La estructura m\'as general de $\theta_{\mu\nu}$ hasta correcciones de orden cuatro, es~\cite{Donoghue:1991qv}
\begin{eqnarray}
\theta_{\mu \nu} = \theta_{\mu \nu}^{(0)}+\theta_{\mu \nu}^{(2)} +
\theta_{\mu \nu}^{(4)} + \cdots 
\end{eqnarray}
con
\begin{eqnarray}
\theta_{\mu \nu}^{(0)} &=& -\eta_{\mu \nu} {\cal L}^{(0)} , \\
\theta_{\mu \nu}^{(2)} &=& \frac{f_\pi^2}2 \langle D_\mu U^\dagger D_\nu U
\rangle - \eta_{\mu\nu} {\cal L}^{(2)} ,
\label{eq:en-mom} \\ 
\theta_{\mu \nu}^{(4)} &=& -  \eta_{\mu\nu}{\cal
L}^{(4)} + 2 L_4 \langle D_\mu U^\dagger D_\nu U \rangle \langle
\chi^\dagger U + U^\dagger \chi \rangle \nonumber \\ &+& L_5 \langle
D_\mu U^\dagger D_\nu U + D_\nu U^\dagger D_\mu U \rangle \langle
\chi^\dagger U + U^\dagger \chi \rangle \nonumber \\ &-& 2
L_{11}\left( \eta_{\mu \nu} \partial^2 - \partial_\mu \partial_\nu
\right) \langle D_\alpha U^\dagger D^\alpha U \rangle \nonumber \\ &-&
2 L_{13} \left(\eta_{\mu \nu} \partial^2 - \partial_\mu \partial_\nu
\right) \langle \chi^\dagger U + U^\dagger \chi \rangle \nonumber \\
&-& L_{12} \left( \eta_{\mu\alpha} \eta_{\nu \beta} \partial^2 +
\eta_{\mu\nu} \partial_\alpha \partial_\beta - \eta_{\mu \alpha}
\partial_\nu \partial_\beta - \eta_{\nu \alpha} \partial_\mu
\partial_\beta \right) \nonumber \langle D^\alpha
U^\dagger D^\beta U \rangle \,,
\end{eqnarray} 
donde  $\langle A \rangle = \tr \, A $ indica la traza en espacio de sabor. El desarrollo quiral del lagrangiano presenta una estructura del tipo~\cite{Donoghue:1991qv}
\begin{equation} 
{\cal L} = {\cal L}^{(0)}+{\cal L}^{(2,g)} +{\cal L}^{(2,R)}+{\cal L}^{(4,g)}
+{\cal L}^{(4,R)}+ \cdots \,, 
\label{eq:chl_flat}
\end{equation}
donde el super\'{\i}ndice $g$ indica contribuciones m\'etricas (acoplamiento m\'{\i}nimo con gravedad), y $R$ indica contribuciones que contienen el tensor de curvatura de Riemann (o sus contracciones). Las contribuciones m\'etricas se pueden obtener directamente del c\'alculo del lagrangiano quiral efectivo en espatio-tiempo plano. Sin embargo, los t\'erminos con $L_{11}-L_{13}$ son contribuciones genuinas de curvatura, pues no se pueden obtener del caso plano. Estos coeficientes de baja energ\'{\i}a surgen a nivel hadr\'onico debido a efectos cu\'anticos. 



\section{Acoplamiento de un Modelo Quark con Gravedad}
\label{acoplamiento_gravedad}

El acoplamiento de fermiones con gravedad es bien conocido~\cite{birrel}, pero no en el contexto de modelos de quarks quirales. En esta secci\'on haremos un estudio de este acoplamiento, de modo que no se introduzcan nuevos campos aparte de los del caso plano y la m\'etrica. Usaremos el formalismo de t\'etradas para espacio-tiempo curvo.\footnote{Para convenios, ver ref.~\cite{weinberg-gravitation}}

\subsection{Formalismo de t\'etradas}
\label{formalismo_tetradas}

Dado el tensor m\'etrico~$g^{\mu\nu}(x)$, introducimos una base local de vectores ortogonales (t\'etrada)
\begin{equation}
g^{\mu\nu}(x) = e^{\mu}_A(x) e^{\nu}_B(x) \eta^{AB}  \,.
\end{equation}
Las t\'etradas satisfacen ciertas relaciones de ortogonalidad
\begin{equation}
\delta^\mu_\nu = \eta^{AB} e^\mu_A e_{\nu B} = e^\mu_A e^A_\nu \,, 
\qquad \qquad \delta^A_B = g^{\mu \nu} e_\mu^A e_{\nu B} = e_\mu^A e_B^\mu .
\end{equation}
Bajo transformaciones generales de coordenadas $x^\mu \rightarrow x^{\prime \mu} (x)$ y de Lorentz $x^A \rightarrow \Lambda^A_B x^B$, las t\'etradas se transforman respectivamente como
\begin{equation}
e_\mu^A \to \frac{\partial x^\nu}{\partial x^{\prime \mu}}  e_\nu^A \,, \qquad 
e_\mu^A \to \Lambda^A_B (x) e_\mu^B \,.
\label{eq:tetrads}
\end{equation} 
Las t\'etradas transforman tensores de coordenadas en tensores de Lorentz (que se transforman de manera covariante bajo transformaciones de Lorentz locales), por ejemplo
\begin{eqnarray}
T^{AB} = e^A_\mu e^B_\nu T^{\mu \nu} . 
\end{eqnarray} 
Los tensores de Lorentz son invariantes bajo transformaciones de coordenadas $x^\mu \rightarrow x'^\mu$. Para un tensor general, por ejemplo $T_{\nu A}^\alpha$, los \'{\i}ndices griegos se transforman de manera covariante bajo transformaciones de coordenadas mientras que los latinos lo hacen bajo transformaciones de Lorentz, de modo que
\begin{eqnarray}
T_{\nu A}^\alpha \to \frac{\partial x^\mu}{\partial x'^\nu}
\frac{\partial x'^\alpha}{\partial x^\beta} \Lambda_A^B (x) T_{\mu
B}^\beta .
\end{eqnarray} 
La derivada covariante se define como
\begin{eqnarray}
d_\mu T_{\nu A}^\alpha &=& \partial_\mu T_{\nu A}^\alpha - \Gamma_{\nu
\mu}^\lambda T_{\lambda A}^\alpha + \Gamma_{\mu \lambda}^\alpha
T_{\nu A}^\lambda+ \omega_{AB\mu} T_{\nu}^{\alpha B}\,,
\end{eqnarray} 
donde la conexi\'on de Riemann viene dada por los s\'{\i}mbolos de Christoffel \begin{eqnarray}
\Gamma_{\lambda \mu}^\sigma = \frac12 g^{\nu \sigma} \left\{
\partial_\lambda g_{\mu\nu} + \partial_\mu g_{\lambda \nu} -
\partial_\nu g_{\mu \lambda} \right\} ,
\label{eq:christoffel}
\end{eqnarray} 
que son sim\'etricos en los \'{\i}ndices inferiores, $ \Gamma_{\lambda \mu}^\sigma = \Gamma_{\mu \lambda}^\sigma$ (no tiene torsi\'on). La derivada covariante $d_\mu$ se define con la conexi\'on adecuada actuando sobre cada \'{\i}ndice. Se tiene 
\begin{eqnarray}
d_\mu e_{\nu A} =  \partial_\mu e_{\nu A } - \Gamma_{\nu\mu}^\lambda e_{\lambda A} + \omega_{AB \mu } e_\nu^B =0  \,.
\label{eq:d_tetrada}
\end{eqnarray} 
Adem\'as, la condici\'on $d_\mu g^{\mu \nu} = 0 $, implica en particular   
\begin{eqnarray}
d_\mu \eta_{AB}= \omega_{AB\mu}  + \omega_{BA\mu} =0,  
\end{eqnarray} 
lo cual impone la restricci\'on de que la conexi\'on de esp\'{\i}n sea antisim\'etrica $ \omega_{AB\mu}=-\omega_{BA\mu}$. \'Esta viene dada por
\begin{eqnarray}
\omega_{AB\mu} = e_A^\nu \left[ \partial_\mu e_{\nu B} - \Gamma_{\nu
\mu}^\lambda e_{\lambda B} \right] \,.
\end{eqnarray} 
La derivada covariante $d_\mu$ act\'ua de manera diferente dependiendo del esp\'{\i}n de los campos correspondientes. Para un campo de esp\'{\i}n-$0$ U, esp\'{\i}n-$1/2$ $\Psi$, esp\'{\i}n-$1$ $A_\mu$ y esp\'{\i}n-$3/2$ $\Psi_\mu$, las propiedades de transformaci\'on son las siguientes
\begin{eqnarray}
U(x) &\to & U(x), \nonumber \\ \Psi (x) &\to& S(\Lambda(x)) \Psi (x), \\
A_\mu (x) &\to & \frac{\partial x^\nu}{\partial x'^\mu} A_\nu (x), \\
\Psi_\mu (x) &\to & \frac{\partial x^\nu}{\partial x'^\mu} S ( \Lambda
(x) ) \Psi_\nu (x).
\end{eqnarray} 
En el caso de transformaciones de Lorentz infinitesimales~$\Lambda_B^A = \delta_B^A + \epsilon_B^A$ con $\epsilon_{AB}=-\epsilon_{BA}$, se tiene~$S(\Lambda) = 1 -\frac{i}4 \sigma_{AB} \epsilon^{AB} $ donde $\sigma_{AB}=\frac{i}{2}[\gamma_A,\gamma_B]$. Para un campo escalar de esp\'{\i}n-0 se tiene la definici\'on est\'andar
\begin{eqnarray}
d_\mu U = \partial_\mu U \,.
\end{eqnarray} 
Para un vector (esp\'{\i}n-1), se tiene 
\begin{equation}
A_{\nu ;\mu}:=d_\mu A_\nu = \partial_\mu A_{\nu} - \Gamma_{\nu \mu}^\lambda A_\lambda \,,
\label{eq:dmuAnu}
\end{equation} 
que satisface adem\'as la propiedad\footnote{El tensor de curvatura de Riemann $R^\lambda_{\sigma \mu \nu}$ se define
\begin{eqnarray}
-R^\lambda_{\,\, \sigma \mu \nu} &=& \partial_\mu \Gamma^\lambda_{\nu \sigma}
- \partial_\nu \Gamma^\lambda_{\mu \sigma}+ \Gamma^\lambda_{\mu
\alpha} \Gamma^\alpha_{\nu \sigma} - \Gamma^\lambda_{\nu \alpha}
\Gamma^\alpha_{\mu \sigma}\,, \label{eq:riemann_tensor}
\end{eqnarray}
y sus contracciones permiten definir el tensor de Ricci~$R_{\mu \nu}$, y el de curvatura escalar~$R$
\begin{equation}
R_{\mu \nu} = R^\lambda_{\, \, \mu \lambda \nu } \,, 
\quad R = g^{\mu \nu} R_{\mu
\nu} \, .
\label{eq:ricci_curescalar}    
\end{equation}
Notar el signo opuesto de nuestra definici\'on para el tensor de Riemann en comparaci\'on con ref.~\cite{Donoghue:1991qv}. Aqu\'{\i} seguimos ref.~\cite{weinberg-gravitation}. 
}
\begin{eqnarray}
\left[ d_\mu , d_\nu \right] A_\alpha = R^\lambda_{\, \,  \alpha \mu \nu }
A_\lambda  \,.
\label{eq:dmudnuA} 
\end{eqnarray}
En el caso de fermiones de Dirac (esp\'{\i}n-$1/2$) la derivada covariante se define como
\begin{eqnarray}
d_\mu \Psi = \partial_\mu \Psi(x) - i \omega_\mu \Psi(x) \,,
\end{eqnarray} 
donde $\omega_\mu$ es la conexi\'on de Cartan de esp\'{\i}n,
\begin{eqnarray} 
\omega_\mu = \frac14 \sigma^{AB} \omega_{AB\mu}\,.
\end{eqnarray} 
Las matrices de Dirac $\gamma_A$ se encuentran en una representaci\'on fija independiente de $x$, y satisfacen las siguientes reglas de anticonmutaci\'on
\begin{eqnarray}
\gamma^A \gamma^B  + \gamma^B  \gamma^A  = 2
\eta^{AB}   .
\end{eqnarray}
Las matrices se pueden elegir
\begin{eqnarray}
\gamma_\mu (x) = \gamma_A e^A_\mu (x)
\end{eqnarray} 
y satisfacen 
\begin{eqnarray}
\gamma^\mu (x) \gamma^\nu (x) + \gamma^\nu (x) \gamma^\mu (x) = 2
g^{\mu \nu} (x)  .
\end{eqnarray}
La derivada covariante de una matriz de Dirac (independiente de $x$) es
\begin{eqnarray}
d_\mu \gamma_A = \partial_\mu \gamma_A - i \left[ \omega_\mu ,
\gamma_A \right] + \omega_{AB\mu} \gamma^B =0 .   \label{eq:d_gamma_cte}
\end{eqnarray} 
Teniedo en cuenta ec.~(\ref{eq:d_tetrada}) y (\ref{eq:d_gamma_cte}) se obtiene la siguiente identidad para las matrices de Dirac dependientes de~$x$
\begin{eqnarray}
d_\mu \gamma_\nu (x) =0 \,,  
\end{eqnarray} 
lo cual quiere decir que para el operador de Dirac libre, el orden de
colocaci\'on es irrelevante $\slashchar{d} \Psi =\gamma^\mu (x) d_\mu \Psi =d_\mu \gamma^\mu (x)
\Psi $.  Para un tensor de esp\'{\i}n-3/2
\begin{eqnarray}
\Psi_{\nu;\mu} := d_\mu \Psi_\nu = \partial_\mu \Psi_{\nu} - \Gamma_{\mu
\nu}^\lambda \Psi_\lambda - i \omega_\mu \Psi_\nu .
\end{eqnarray} 
Si aplicamos las definiciones anteriores a $d_\mu \Psi$ se obtienen las siguientes f\'ormulas, que ser\'an de utilidad
\begin{eqnarray}
\left[ d_\mu , d_\nu \right] \Psi &=& \frac{i}4 \sigma^{\alpha
\beta} R_{\alpha \beta \mu \nu} \Psi, \label{eq:dmu_dnu}\\ 
d^\mu d_\mu \Psi &=&
\frac{1}{\sqrt{-g}} \left\{ \left(\partial_\mu-i \omega_\mu \right)
\left[ \sqrt{-g} \,g^{\mu \nu} \left(\partial_\nu-i \omega_\nu \right)
\right] \Psi \right\}, \label{eq:dmu_dmu}
\end{eqnarray}
donde $ \sigma^{\alpha \beta } = e^\alpha_A e^\beta_B \sigma^{AB} $ es una matriz antisim\'etrica dependiente de $x$.

Los campos gauge pueden ser incluidos mediante la regla est\'andar de sustituci\'on m\'{\i}nima, lo cual da lugar a la derivada covariante de un fermi\'on
\begin{eqnarray}
\nabla_\mu \Psi = \left( d_\mu - i V_\mu \right) \Psi .
\end{eqnarray} 
Con esta notaci\'on, el operador de Dirac completo $i {\bf D}$ en presencia de campos externos de tipo vector, axial, escalar, pseudoescalar y gravitacionales se escribe\footnote{
La matriz pseudoescalar de Dirac en el caso curvo se define 
\begin{eqnarray}
\gamma_5 (x) = \frac{1}{4! \sqrt{-g}} \epsilon^{\mu\nu\alpha\beta}
\gamma_\mu (x) \gamma_\nu (x) \gamma_\alpha (x) \gamma_\beta (x) = \frac{1}{4!} \epsilon^{ABCD} \gamma_A \gamma_B
\gamma_C \gamma_D = \gamma_5 .
\end{eqnarray} 
}
\begin{eqnarray}
i {\bf D} &=& i\slashchar{d} - M U^5 - {\hat{m}_0} + \left(
\slashchar{v} + \slashchar{a} \gamma_5 - s - i \gamma_5 p \right) \,,
\label{eq:dirac_op}
\end{eqnarray} 
donde la barra indica
\begin{eqnarray}
\slashchar{V}  = \gamma^\mu (x) V_\mu (x) .
\end{eqnarray} 
$M$ es la masa constituyente de los quarks y hemos considerado la notaci\'on~$U^5 = U^{\gamma_5}$. La derivada covariante bajo transformaciones generales de coordenadas, de Lorentz, y quirales, act\'ua sobre los campos pseudoescalares (esp\'{\i}n-$0$), espinores de Dirac (esp\'{\i}n-$1/2$) y espinores de Rarita-Schwinger (esp\'{\i}n-$3/2$) de acuerdo con las f\'ormulas siguientes
\begin{eqnarray}
\nabla_\mu U &=& {\widehat D}_\mu U = \partial_\mu U - i [ v_\mu ,U] - i
\{ a_\mu ,U \}, \nonumber \\ 
\nabla_\mu \Psi &=& {\cal \widehat{D}}_\mu \Psi =
\partial_\mu \Psi - i ( \omega _\mu + v_\mu + \gamma_5 a_\mu ) \Psi,
\nonumber \\ 
\nabla_\mu \Psi_\nu &=& \partial_\mu \Psi_\nu - i(
\omega_\mu+v_\mu + \gamma_5 a_\mu ) \Psi_\nu - \Gamma_{\mu
\nu}^\lambda \Psi_\lambda \,,
\label{eq:nabla_campos}
\end{eqnarray}
y se corresponden con sustituir la derivada parcial por la derivada covariante~$\partial_\mu \to d_\mu $, dentro de la derivada covariante quiral~$D_\mu $. La notaci\'on $\widehat{D}_\mu$ significa la operaci\'on $[D_\mu,\;]$, preservando la quiralidad del objeto (ver ec.~(\ref{eq:der_cov_quiral_U})). Notar que con esta definici\'on, ni el objeto $ {\cal\widehat{D}}_\mu {\cal\widehat{D}}_\nu \Psi ( \neq \nabla_\mu \nabla_\nu \Psi )$ ni $ {\cal\widehat{D}}_\mu {\cal\widehat{D}}_\nu U $ son covariantes coordenados, ya que la segunda derivada no incluye la conexi\'on de Riemann~$\Gamma^\lambda_{\mu\nu}$.

\subsection{Operador de segundo orden} 
\label{operador_2_orden}

Cuando no existen fuentes gravitatorias, la contribuci\'on de paridad normal a la acci\'on efectiva se obtiene a partir del operador de segundo orden
\begin{eqnarray}
{\bf D}_5 {\bf D} &=& \left[ \slashchar{D}_L^2 + i {\cal
M}^\dagger 
\slashchar{D}_L - i \slashchar{D}_R {\cal M}^\dagger  + {\cal M}^\dagger {\cal
M} \right] P_R \nonumber \\  &+& 
\left[ \slashchar{D}_R^2 + i {\cal M}
\slashchar{D}_L - i \slashchar{D}_R {\cal M} + {\cal M} {\cal
M}^\dagger \right] P_L \,,
\label{eq:KG-flat} 
\end{eqnarray} 
donde ${\bf D}_5$ se define como en ec.~(\ref{eq:defD5}),
\begin{equation}
{\bf D}_5[s,p,v,a,U] = \gamma_5 {\bf D}[s,-p,v,-a,U^\dagger] \gamma_5 \,.
\end{equation}
${\bf D}_5$ corresponde a rotar ${\bf D}$ a espacio eucl\'{\i}deo, tomar su herm\'{\i}tico conjugado y volver a rotar a espacio de Minkowski. En la expresi\'on~(\ref{eq:KG-flat}), $P_{R,L} = \frac{1}{2}(1\pm \gamma_5)$, las derivadas covariantes quirales son
\begin{eqnarray}
D_\mu &=& \partial_\mu -i (v_\mu +\gamma_5 a_\mu) = D_\mu^R P_R + D_\mu^L P_L
\,,  \label{eq:Dmuquiral}\\
D^R_\mu &=& \partial_\mu -i( v_\mu + a_\mu) \,, \nonumber \\
D^L_\mu &=& \partial_\mu -i (v_\mu - a_\mu) \,, \nonumber
\end{eqnarray}
y el t\'ermino de masa
\begin{equation}
{\cal M} = M U^5 + (s + i \gamma_5 p) + \hat{m}_0 \,.
\end{equation}
Los campos gravitatorios se acoplan mediante covariantizaci\'on del operador
de Dirac, esto es con la sustituci\'on $\partial_\mu \rightarrow d_\mu =
\partial_\mu -\Gamma_{\mu\,\cdot}^{\,\cdot} -i\omega_\mu $ en
ec.~(\ref{eq:Dmuquiral}). Para fijar la notaci\'on, definimos en
ec.~(\ref{eq:nabla_campos}) la actuaci\'on de la derivada covariante quiral
sobre un espinor de Dirac~$\Psi$
\begin{equation}
{\cal\widehat{D}}_\mu = \partial_\mu -i (\omega_\mu + v_\mu + \gamma_5 a_\mu) \,.
\end{equation}
Teniendo en cuenta que, puesto que un espinor es un escalar en coordenadas, se tiene
\begin{equation}
{\cal\widehat{D}}_\mu \Psi = \nabla_\mu \Psi \,,
\end{equation} 
donde $\nabla_\mu = d_\mu -i(v_\mu + \gamma_5 a_\mu)$. Para el campo escalar en coordenadas $\slashchar{\nabla} \Psi $ se puede aplicar el mismo razonamiento, lo cual conduce a 
\begin{equation}
{\cal\widehat{D}}_\mu \slashchar{\nabla} \Psi = \nabla_\mu \slashchar{\nabla} \Psi  \,.
\end{equation} 
Esto significa que podemos considerar $ \slashchar{{\cal D}}_{L,R} =
\slashchar{\nabla}_{L,R} $ siempre y cuando act\'ue sobre campos espinoriales del siguiente modo  
\begin{eqnarray} 
{\bf D}_5 {\bf D} \Psi &=& \left[ \slashchar{\nabla}_L^2 + i
{\cal M} \slashchar{\nabla}_L - i \slashchar{\nabla}_R {\cal M}
+ {\cal M}^\dagger {\cal M} \right] P_R \Psi \nonumber \\ &+& \left[
\slashchar{\nabla}_R^2 + i {\cal M}^\dagger \slashchar{\nabla}_L
- i \slashchar{\nabla}_R {\cal M}^\dagger + {\cal M} {\cal
M}^\dagger \right] P_L \Psi \,.
\label{eq:KG-curved} 
\end{eqnarray} 
Si incluimos los campos gauge, se obtienen dos teor\'{\i}as tipo vector, una para campos left $V_\mu^L$ y otra para campos right $V_\mu^R$. Si suprimimos moment\'aneamente las etiquetas left y right, se tiene
\begin{eqnarray}
\slashchar{{\cal D}}^2 \Psi = \slashchar{{\nabla}}^2 \Psi = \left[
{\nabla}^\mu {\cal \nabla}_\mu - \frac12 \sigma^{\mu \nu} F_{\mu \nu}
+ \frac14 R \right] \Psi , 
\end{eqnarray} 
donde hemos hecho uso de la identidad
\begin{eqnarray}
\left[ \nabla_\mu , \nabla_\nu \right] \Psi &=& \left[ {\cal D}_\mu ,
{\cal D}_\nu \right] \Psi \nonumber \\ &=& \left[ D_\mu , D_\nu
\right] \Psi + \frac{i}4 \sigma^{\alpha \beta} R_{\alpha \beta \mu
\nu} \Psi \,. \label{eq:id_nabla2}
\end{eqnarray}
En la segunda igualdad de ec.~(\ref{eq:id_nabla2}) se ha hecho uso de ec.~(\ref{eq:dmu_dnu}). El laplaciano invariante coordenado y Lorentz para un espinor de Dirac viene dado por
\begin{eqnarray}
\nabla^\mu \nabla_\mu \Psi =  \frac{1}{\sqrt{-g}} {\cal D}_\mu \left( 
\sqrt{-g} g^{\mu \nu} {\cal D}_\nu \Psi \right) ,
\end{eqnarray} 
donde se ha aplicado ec.~(\ref{eq:dmu_dmu}). Con la notaci\'on quiral
de campos right y left, el operador de segundo orden se escribe
\begin{eqnarray}
{\bf D}_5 {\bf D} &=& \frac{1}{\sqrt{-g}} \left[ {\cal D}_\mu \left( 
\sqrt{-g} g^{\mu \nu} {\cal D}_\nu \right) \right] + {\cal V},
\label{eq:KG-heat} 
\end{eqnarray} 
con
\begin{eqnarray}
{\cal V} &=&  {\cal V}_R P_R + {\cal V}_L  P_L \label{eq:V_stcurvo} \\
{\cal V}_R &=& -\frac12 \sigma^{\mu \nu} {F}_{\mu \nu}^R  
+ \frac14 R - i \gamma^\mu {\nabla}_\mu  {\cal M} + {\cal
M}^\dagger {\cal M} ,\nonumber \\
{\cal V}_L &=& -\frac12 \sigma^{\mu \nu} {F}_{\mu \nu}^L  
+ \frac14 R - i \gamma^\mu {\nabla}_\mu  {\cal M}^\dagger + {\cal
M} {\cal M}^\dagger \,.  \nonumber
\end{eqnarray}

\section{Modelos de Quarks Quirales en presencia de \newline Gravedad}
\label{MQQ_ETC}

En esta secci\'on aprovecharemos los resultados obtenidos en sec.~\ref{acoplamiento_gravedad} y estudiaremos el acoplamiento con gravedad de dos modelos quirales concretos, que tienen en com\'un la incorporaci\'on de la rotura din\'amica de la simetr\'{\i}a quiral a nivel de un loop: el modelo de Nambu--Jona-Lasinio (NJL) y el modelo de Georgi-Manohar. 

En estos modelos, los quarks tienen una masa constituyente $M\sim 300\;\textrm{MeV}$. La principal diferencia entre ellos tiene que ver con la presencia o no de campos escalares din\'amicos $\overline{q}q$, respectivamente. Adem\'as, mientras que el modelo NJL genera de manera din\'amica la rotura espont\'anea de la simetr\'{\i}a quiral, el modelo GM comienza de por s\'{\i} en una fase de rotura de la simetr\'{\i}a quiral.

\subsection{Modelo de Nambu--Jona-Lasinio}
\label{MQQ_NJL}

El modelo de Nambu--Jona-Lasinio se introdujo en la secci\'on~\ref{sec:njl1}. La acci\'on del modelo en espacio-tiempo curvo de Minkowski con tensor m\'etrico~$g_{\mu\nu}(x)$ se escribe
\begin{equation}
S_{\NJL}=\int d^4 x \sqrt{-g} \,{\cal L}_{\NJL} \,,
\end{equation}
donde $g=\det(g_{\mu\nu})$ y el lagrangiano viene dado por
\begin{eqnarray}
{\cal L}_{\textrm{NJL}}&=& \overline{q}(i\slashchar\partial+\thru{\omega}-\hat{m}_0)q 
+\frac{1}{2a_s^2}\sum_{a=0}^{N_f^2-1}((\overline{q}\lambda_a q)^2
+(\overline{q}\lambda_a i\gamma_5 q)^2) \nonumber \\
&&\qquad -\frac{1}{2a_v^2}\sum_{a=0}^{N_f^2-1}((\overline{q}\lambda_a \gamma_\mu q)^2+
(\overline{q}\lambda_a \gamma_\mu \gamma_5 q)^2) \,.
\label{eq:lag_njl_grav}
\end{eqnarray}
La derivada $\partial_\mu -i\omega_\mu$ es covariante bajo transformaciones generales de coordenadas y bajo transformaciones de Lorentz, e incluye la conexi\'on de esp\'{\i}n
\begin{eqnarray}
\omega_\mu (x) = \frac{i}8 \left[ \gamma^\nu (x) ,
\gamma_{\nu;\mu} (x) \right] \,,
\end{eqnarray} 
donde la derivada covariante $\gamma_{\nu ; \mu}= d_\mu \gamma_\nu$ se define de la manera usual, ec.~(\ref{eq:dmuAnu}).
Haciendo uso del procedimiento est\'andar de bosonizaci\'on~\cite{eguchi}, como se vio en sec.~\ref{sec:njl1}, se introducen campos bos\'onicos din\'amicos internos auxiliares $(S,P,V,A)$, de modo que despu\'es de integrar formalmente los quarks se obtiene el funcional generador
\begin{equation}
Z_{\NJL}[g;s,p,v,a] = \int {\cal D}S {\cal D}P {\cal D}V {\cal D}A \, e^{i\Gamma_{\NJL}[g;\overline{S},\overline{P},\overline{V},\overline{A}]} \,,
\end{equation}
con ${\overline S}= s + S$, ${\overline P}= p + P$, ${\overline V}= v + V$, ${\overline A}= a + A$. La acci\'on efectiva es
\begin{equation}
\Gamma_{\NJL}[g;{\overline S},{\overline P},{\overline V},{\overline A}] = \Gamma_q[{\bf D}]+ \Gamma_m[g;S,P,V,A] \,,
\label{eq:ae_NJL_curv1}
\end{equation}
donde las contribuciones de los quarks a un loop y de los mesones a nivel \'arbol se escriben respectivamente
\begin{eqnarray}
\Gamma_q[{\bf D}] &=& -iN_c \Tr \log(i{\bf D}) \,, \label{eq:njlacefectqcurv} \\
\Gamma_m[g;S,P,V,A] &=& \int d^4 x \sqrt{-g}\,\left\{-\frac{a_s^2}{4}\tr(S^2+P^2)
+\frac{a_v^2}{4}\tr(V_\mu^2+A_\mu^2)
\right\} \,.
\label{eq:njlacefectcurv}
\end{eqnarray}
El operador de Dirac viene dado por
\begin{equation}
i {\bf D} = i\slashchar\partial + \thru{\omega} - \hat{m}_0 + \left(
\thruu{\overline V} + \thru{\overline A} \gamma_5 - {\overline S} - i\gamma_5 {\overline P} \right) \,.
\label{eq:dirac_op_curv} 
\end{equation} 
Para que la integral funcional en los campos bos\'onicos est\'e bien definida en espacio de Minkowski, es necesario usar la prescripci\'on $a_s^2 \to a_s^2 - i\epsilon$, $a_v^2 \to a_v^2 - i\epsilon$. 
La contribuci\'on $\gamma_5$-par de los quarks a la acci\'on efectiva puede ser regularizada mediante el esquema de Pauli-Villars
\begin{equation} 
\Gamma_q^+[{\bf D}] = -i{N_c \over 2} {\rm Tr} \sum c_i \log({\bf D}_5 {\bf D} +
\Lambda_i^2 +i\epsilon) \,.
\end{equation}
Para m\'as detalles, ver sec.~\ref{sec:njl1}.

\subsection{Modelo de Georgi-Manohar}
\label{MQQ_GM}

En presencia de gravedad, el lagrangiano del modelo de Georgi-Manohar~\cite{Manohar:1983md} se escribe 
\begin{equation} 
{\cal L}_{\rm GM} = \bar{q} \left( i\slashchar\partial +
\slashchar\omega - M U^5 - \hat{m}_0 + 
\frac{1}{2}(1-g_A) U^5 i\slashchar\partial U^5 \right) q 
=: \bar{q} \, i{\bf D} \,q \,,
\label{eq:GM} 
\end{equation}
donde $g_A$ es el acoplamiento axial de los quarks, que consideraremos diferente de uno, tal y como se sugiere en~\cite{Manohar:1983md}.
La acci\'on efectiva de este modelo es
\begin{equation}
\Gamma_{\rm GM} = -iN_c \Tr  \log (i{\bf D}) \,,
\label{eq:GM_ea}
\end{equation}
y por comparaci\'on directa con ec.~(\ref{eq:ae_NJL_curv1}) se puede ver que se corresponde con un modelo similar al NJL, sin t\'ermino de masa $\Gamma_m$ y con un operador de Dirac como ec.~(\ref{eq:dirac_op_curv}) con una elecci\'on espec\'{\i}fica de los campos din\'amicos de esp\'{\i}n 1
\begin{eqnarray} 
V_\mu &=& \frac{1}{4}(1- g_A) \left[ U^\dagger \partial_\mu U - \partial_\mu
U^\dagger U \right] \,,\label{eq:va_gm1} \\
A_\mu &=& \frac{1}{4}(1- g_A) \left[ U^\dagger \partial_\mu U + \partial_\mu
U^\dagger U \right] \,. 
\label{eq:va_gm2}
\end{eqnarray} 
En ec.~(\ref{eq:GM_ea}) implementaremos la misma regularizaci\'on de Pauli-Villars que en el modelo NJL.

\section{C\'alculo de la acci\'on efectiva}
\label{ae_gravedad}

En un desarrollo quiral de la acci\'on, la m\'etrica dependiente del espacio-tiempo es de orden cero y la derivada $\partial_\mu$ de orden uno. Esto implica en particular que $R^{\mu\nu\alpha\beta}$, $R^{\mu\nu}$, y $R$ son de orden 2. A nivel de un loop de quarks el desarrollo quiral se corresponde con un desarrollo en derivadas que debe de ser invariante bajo transformaciones gauge, de coordenadas y de Lorentz. Este desarrollo a baja energ\'{\i}a se puede obtener haciendo uso de la representaci\'on de tiempo propio del logaritmo
\begin{eqnarray}
\sum_i c_i {\rm Tr} \log \left( {\bf D}_5 {\bf D} + \Lambda_i^2
\right) = - {\rm Tr} \int_0^\infty \frac{d \tau}{\tau} e^{- i\tau {\bf D}_5 {\bf D} } \phi(\tau)   \,,
\end{eqnarray} 
donde $\phi(\tau) = \sum_i c_i e^{-\tau \Lambda_i^2}$. El operador que est\'a dentro del logaritmo es de tipo Klein-Gordon en espacio-tiempo curvo, y presenta cierta estructura espinorial, como se ve en ec.~(\ref{eq:KG-heat}). La forma de este operador es la adecuada para hacer un desarrollo del heat kernel en espacio-tiempo curvo. Para el elemento de matriz diagonal se tiene
\begin{eqnarray}
\langle x | e^{-i\tau {\bf D}_5 {\bf D} } | x \rangle &=& e^{ -
i\tau M^2 } \langle x | e^{-i\tau ( {\bf D}_5 {\bf D} -
M^2 ) } | x \rangle = \frac{i}{(4\pi i\tau )^2}
e^{-i\tau M^2 } \sum_{n=0}^\infty a_{n} (x) \left( i\tau \right)^n \,. \label{eq:des_hkkk_mqq}
\end{eqnarray}
Para el c\'alculo hasta ${\cal O}(p^4)$ es necesario llegar hasta $a_4$ en el desarrollo del heat kernel. Las contribuciones pueden separarse entre aquellas que son de espacio-tiempo plano, y las correspondientes a curvatura generadas por efectos cu\'anticos. Por el momento nos centraremos en el modelo NJL. Posteriormente particularizaremos las f\'ormulas para el modelo GM. Se obtiene lo siguiente~\cite{Megias:2004uj}
\begin{eqnarray}
a_0 &=& 1, \nonumber \\ 
a_1 &=& M^2-\overline{{\cal V}} + \frac16 R ,\nonumber \\ 
a_2 &=& \frac1{180} R_{\mu \nu \alpha \beta} R^{\mu \nu \alpha \beta}-
\frac1{180} R_{\mu \nu } R^{\mu \nu} + \frac{1}{12} \overline{\cal F}^{\mu \nu}
\overline{\cal F} _{\mu \nu} \nonumber \\ &+& \frac1{30} \overline\nabla^2 R - \frac16
\overline\nabla^2 \overline{\cal V}  + \frac12 \left[ M^2-\overline{\cal V} + \frac16 R
\right]^2  ,
\nonumber 
\end{eqnarray}
\begin{eqnarray}
a_3 &=& \frac16 \left[ M^2-\overline{\cal V} + \frac16 R
\right]^3  - \frac1{12} \overline\nabla^\mu \overline{\cal V} \,\overline\nabla_\mu
\overline{\cal V} + {\cal O} (p^6) ,\nonumber \\
a_4  &=& \frac1{24}\left[\overline{\cal V}-M^2\right]^4   + {\cal O} (p^6) \,.
\label{as_curvatura}
\end{eqnarray} 
La notaci\'on que estamos utilizando es $\overline{\cal F}_{\mu \nu} = i \left[ \overline{\cal D}_\mu , \overline{\cal D}_\nu \right ]$, 
$\overline\nabla^2 \overline{\cal V} = \overline\nabla^\mu \overline\nabla_\mu \overline{\cal V}$, donde
\begin{eqnarray}
\overline{\cal D}_\mu &=& \partial_\mu -i(\overline{V_\mu}+\gamma_5 \overline{A}_\mu) \,, \nonumber \\
\overline\nabla_\mu &=& d_\mu -i(\overline{V_\mu}+\gamma_5 \overline{A}_\mu) \,,
\end{eqnarray}
y $\overline{\cal V}$ viene dado por la misma expresi\'on (\ref{eq:V_stcurvo}), con la adici\'on de los campos bos\'onicos internos $(S,P,V,A)$. Las integrales que aparecen en la acci\'on son del tipo
\begin{equation}
\I_{2l} := M^{2l} \int_0^\infty \frac{d\tau}{\tau}\phi(\tau)(i\tau)^l
e^{-i\tau M^2} \,.
\end{equation} 
Los valores particulares que necesitamos en nuestro desarrollo son
\begin{eqnarray}
M^4 \I_{-4} &=& -\frac{1}{2}\sum_i
c_i(\Lambda_i^2+M^2)^2\log(\Lambda_i^2+M^2)
\,, \\ 
M^2\I_{-2} &=& \sum_i
c_i(\Lambda_i^2+M^2)\log(\Lambda_i^2+M^2) \,, \\ \I_0 &=& -\sum_i
c_i\log(\Lambda_i^2+M^2) \,, \\ \I_{2n} &=&\Gamma(n)\sum_i c_i
\left(\frac{M^2}{\Lambda_i^2+M^2}\right)^n \,, \qquad
\textrm{Re}(n)>0 \,.
\end{eqnarray}

Despu\'es del c\'alculo de las trazas de Dirac, el orden ${\cal O}(p^2)$ del
lagrangiano efectivo en el modelo NJL viene dado por
\begin{eqnarray}
{\cal L}^{(2)}_q  &=&{N_c\over(4\pi)^2} \Big\{ M^2 {\cal I}_0 \langle
\Nabla_\mu U^\dagger \Nabla^\mu U \rangle + 
2 M^3 {\cal
I}_{-2}\langle \overline{\it m}^\dagger U + U^\dagger \overline{\it m} \rangle
+ \frac{M^2}{6}{\cal I}_{-2} \langle R \rangle \Big\}, \nonumber 
\end{eqnarray} 
mientras que para el orden ${\cal O}(p^4)$ se tiene
\begin{eqnarray} 
{\cal L}^{(4)}_q  &=& {N_c\over (4\pi)^2} \Big\{  
- {1\over6} {\cal I}_0  \langle (\F_{\mu\nu}^R)^2 +
(\F_{\mu\nu}^L)^2 \rangle \nonumber + {\cal I}_0 \langle
\frac{7}{720} R_{\alpha \beta \mu \nu} R^{\alpha \beta \mu \nu} - 
\frac{1}{144} R^2 + \frac{1}{90} R_{\mu \nu} R^{\mu \nu} \rangle
\nonumber \\
&-&\frac{i }2 {\cal I}_2 \langle \F_{\mu\nu}^R \Nabla^\mu U^\dagger
\Nabla^\nu U + \F_{\mu\nu}^L \Nabla^\mu U \Nabla^\nu U^\dagger \rangle
\nonumber \\
&+& \frac{1}{12} {\cal I}_4 \langle (\Nabla_\mu U \Nabla_\nu U^\dagger
)^2 \rangle -\frac{1}{6} {\cal I}_4 \langle (\Nabla_\mu U \Nabla^\mu
U^\dagger )^2 \rangle \nonumber \\
&+& \frac{1}{6} {\cal I}_2 \langle \Nabla_\mu \Nabla_\mu U \Nabla^\nu \Nabla^\nu 
U^\dagger \rangle \nonumber \\ 
&+&  2 M^2{\cal I}_{-2} \langle \overline{\it m}^\dagger\overline{\it m} \rangle - M^2 {\cal I}_0   \langle ( \overline{\it m}^\dagger
U+U^\dagger \overline{\it m} )^2 \rangle  
\nonumber \\
&-& M {\cal I}_2 \langle \Nabla_\mu U^\dagger \Nabla^\mu U
(\overline{\it m}^\dagger U+U^\dagger \overline{\it m} ) \rangle \nonumber \\
&+& M {\cal I}_0 \langle \Nabla_\mu U^\dagger \Nabla^\mu \overline{\it m} +
\Nabla_\mu \overline{\it m}^\dagger \Nabla^\mu U \rangle \nonumber \\ 
&-& \frac{M}{6}{\cal I}_0 R\langle U^\dagger \overline{\it m}+
\overline{\it m}^\dagger U \rangle - \frac{1}{12} {\cal I}_2 R\,
\langle \Nabla_\mu U^\dagger \Nabla^\mu U \rangle \Big\} \,.
\label{eq:L4}
\end{eqnarray} 
En estas f\'ormulas $\langle \quad \rangle$ indica traza en espacio de sabor. La derivada covariante gauge y covariante Lorentz, y los tensores de fuerza que contienen los campos externos e internos (bosonizados) son
\begin{eqnarray} 
\Nabla_\mu U &=& \nabla_\mu U -i V_\mu^L U +iU  V_\mu^R, \\ 
\overline{F}_{\mu\nu}^r &=& 
\partial_\mu \overline{V}_\nu^r -\partial_\nu \overline{V}_\mu^r
-i [ \overline{V}_\mu^r , \overline{V}_\nu^r ], \nonumber 
\end{eqnarray} 
con $r=L, R$ y la combinaci\'on aditiva de esp\'{\i}n 0 
\begin{equation}
\overline{m} = (S+ i P- M U ) + \frac{1}{2B_0} \chi \,, \qquad
\chi = 2B_0(s +ip)\,.
\end{equation}
La constante de reescalamiento $B_0$ se elige de modo que~${\cal L}^{(2)}$ quede en la forma est\'andar de ec.~(\ref{eq:chl2}). Notar que ec.~(\ref{eq:L4}) no est\'a a\'un lista para poder ser comparada con el resultado de~\cite{gasser-leutwyler2,Donoghue:1991qv}. Para ello antes debemos eliminar todos los grados de libertad diferentes a los piones en la capa de masas. Procederemos en tres pasos: primero integraremos los grados de libertad vector y axial, despu\'es eliminaremos los campos escalares y finalmente haremos uso de las ecuaciones cl\'asicas de movimiento para los pseudoescalares. En el modelo de Georgi-Manohar \'unicamente ser\'a necesario considerar el \'ultimo paso.

\section{Ecuaciones de movimiento}
\label{eq:ec_mov_gravedad}

\subsection{Eliminaci\'on de los acoplamientos vector y axial}
\label{eq:elim_vector_axial}

En el modelo NJL, para eliminar los campos vector $V_\mu$ y axial $A_\mu$ en
la aproximaci\'on de campo medio es necesario minimizar el lagrangiano con
respecto a esos campos. Al orden que estamos considerando el desarrollo
quiral, ser\'a suficiente con tener en cuenta aquellos t\'erminos del
lagrangiano que contienen mesones vectoriales con dos \'{\i}ndices de Lorentz, esto es, el t\'ermino de masa y el orden dos que surge del determinante de los quarks
\begin{eqnarray}
{\cal L}^{(2)}_{A,V} &=&{N_c\over(4\pi)^2} M^2 {\cal I}_0 \langle
\Nabla_\mu U^\dagger \Nabla^\mu U \rangle + {a_v^2\over 4} \langle
V_\mu V^\mu + A_\mu A^\mu \rangle \,. \nonumber
\end{eqnarray}
Al minimizar, las ecuaciones de movimiento que se obtienen son similares a la elecci\'on concreta de los campos vector y axial en el modelo de Georgi-Manohar, ecs.~(\ref{eq:va_gm1})-(\ref{eq:va_gm2}),
\begin{eqnarray}
\V_\mu^R &=& v_\mu^R +
\frac{i}{2}(1-g_A) U^\dagger\nabla_\mu U \,, \qquad
\V_\mu^L = v_\mu^L + \frac{i}{2}(1-g_A) U \nabla_\mu
U^\dagger \,,
\end{eqnarray}
con $g_A = 1-2f_\pi^2/a_v^2$. Aplicando estas ecuaciones de movimiento se obtienen f\'acilmente las siguientes relaciones
\begin{eqnarray}
\F_{\mu\nu}^R &=& \frac{1}{2}(1+g_A)
F_{\mu\nu}^R+\frac{1}{2}(1-g_A)U^\dagger F_{\mu\nu}^L
U \nonumber \\
&&- \frac{i}{4}(1-g_A^2) \left(\nabla_\mu U^\dagger
\nabla_\nu U -\nabla_\nu U^\dagger \nabla_\mu U\right) \,,
\end{eqnarray}
\begin{eqnarray}
\F_{\mu\nu}^L &=& \frac{1}{2}(1-g_A) U F_{\mu\nu}^R
U^\dagger+\frac{1}{2}(1+g_A) F_{\mu\nu}^L   \nonumber \\
&&- \frac{i}{4}(1-g_A^2) \left(\nabla_\mu U
\nabla_\nu U^\dagger -\nabla_\nu U \nabla_\mu U^\dagger\right) \,, 
\\
\Nabla_\mu U &=& g_A \nabla_\mu U \, , \\ 
\Nabla^2 U &=& g_A \nabla^2 U + ig_A (1-g_A) U \nabla_\mu U^\dagger
\nabla^\mu U \,. \label{eq:umu}
\end{eqnarray}

\subsection{Eliminaci\'on de escalares}
\label{elim_escalares}

En el modelo NJL, la eliminaci\'on de los campos escalares se hace de manera similar a la de los campos vector y axial. Consideramos la rotaci\'on quiral
\begin{eqnarray}
S+ iP = \sqrt{U} \Sigma \sqrt{U} \,,   
\end{eqnarray} 
donde $\Sigma^\dagger = \Sigma$, y usando que $\Sigma=M+\Phi$, donde $\Phi$ es una fluctuaci\'on alrededor del valor del vac\'{\i}o, se tiene
\begin{eqnarray}
{\overline m} &=&  \sqrt{U} \Phi \sqrt{U} + \frac{1}{2B_0} \chi  \,.
\end{eqnarray}
 El t\'ermino de masa se escribe
\begin{eqnarray} 
{\cal L}_m &=& -{a_s^2\over 4} \langle M^2 + 2 M \Phi + \Phi^2 \rangle \,.
\end{eqnarray}
Haciendo uso de la ecuaci\'on del gap (\ref{eq:njlgap}), los t\'erminos lineales en $\Phi$ que no contienen campos externos se anulan. Como consecuencia, la parte del lagrangiano que contiene al campo escalar $\Phi$ es 
\begin{eqnarray}
 \cL_\Phi (x) &=& -\frac{N_c}{(4\pi)^2} \Bigg\langle 4 M^2 {\cal I}_0
   \Phi^2 + \frac{1}{3} M {\cal I}_0 R \Phi \nonumber \\ 
   &+& M {\cal I}_0
   \sqrt{U} \Phi \sqrt{U^\dagger } \left( U \Nabla^2 U^\dagger +
   \Nabla^2 U U^\dagger \right) \nonumber \\  
   &+& \frac{M^2}{B_0}(2 {\cal I}_0 - {\cal I}_{-2} ) \sqrt{U} \Phi \sqrt{U^\dagger }
   ( U \chi^\dagger + U^\dagger \chi ) \nonumber \\ 
   &+& \frac{M}{B_0} {\cal I}_2 \sqrt{U}
   \Phi \sqrt{U^\dagger } \Nabla_\mu U \Nabla^\mu U^\dagger \Bigg\rangle \,.
\end{eqnarray}
Minimizando respecto de $\Phi$, la ecuaci\'on cl\'asica de movimiento que se obtiene es 
 \begin{eqnarray}
\sqrt{U} \Phi \sqrt{U^\dagger } &=& -\frac{1}{24M} R + \frac{1}{4M} \left( 1- \frac{{\cal I}_2}{{\cal
I}_0} \right) \Nabla_\mu U \Nabla^\mu U^\dagger \nonumber \\ &-&
\frac{1}{4B_0}\left( 1 -\frac{{\cal I}_{-2}}{2 {\cal I}_0} \right) ( U \chi^\dagger + \chi U^\dagger) \,.
\end{eqnarray}
S\'olo queda sustituir esta ecuaci\'on dentro del lagrangiano $\cL_\Phi$ para obtener la contribuci\'on del lagrangiano efectivo proveniente de la integraci\'on de los campos escalares.

\subsection{Ecuaciones de movimiento cl\'asicas para pseudoescalares}
\label{ec_mov_pseudo}

Las ecuaciones de movimiento relevantes para el campo no linear $U$ se obtienen minimizando ${\cal L}^{(2)}$. Surgen una serie de relaciones que son v\'alidas incluso en presencia de curvatura
\begin{eqnarray} 
\langle \nabla^2 U^\dagger \nabla^2 U \rangle &=&  \langle \left(
\nabla_\mu U^\dagger \nabla^\mu U \right)^2 \rangle 
-\frac{1}{4} \langle \left(
\chi^\dagger U - U^\dagger \chi \right)^2 \rangle +\frac{1}{12} \langle 
\chi^\dagger U - U^\dagger \chi  \rangle^2 
\label{eq:id1}
\end{eqnarray} 
y
\begin{eqnarray} 
\langle \chi^\dagger \nabla^2 U + \nabla^2 U^\dagger \chi \rangle &=& 2 \langle
\chi^\dagger \chi \rangle - \frac{1}{2} \langle \left( \chi^\dagger U +
U^\dagger \chi \right)^2 \rangle \nonumber - \langle \left(
\chi^\dagger U + U^\dagger \chi \right) \nabla^\mu U^\dagger \nabla_\mu U
\rangle \nonumber \\ 
&&+ \frac{1}{6} \langle \chi^\dagger U +
U^\dagger \chi  \rangle^2 \,.
\label{eq:id2}
\end{eqnarray} 
En el caso del grupo U(3) de sabor, se tiene que $\Det \, U = e^{i\eta_0/f_\pi}$, que no es necesariamente igual a la identidad, y los dos \'ultimos t\'erminos $\langle \chi^\dagger U \pm U^\dagger \chi \rangle^2 $ en ecs.~(\ref{eq:id1}) y (\ref{eq:id2}) desaparecer\'an.\footnote{Existe otra identidad integral que nos va a resultar muy \'util
\begin{eqnarray}
&& \int d^4 x \sqrt{-g} \, \langle \nabla_\mu \nabla_\nu U^\dagger
\nabla^\mu \nabla^\nu U \rangle = \int d^4 x \sqrt{-g} 
\bigg[
\langle\nabla^2 U^\dagger \nabla^2 U \rangle 
+ i\langle F_{\mu\nu}^R \nabla^\mu U^\dagger \nabla^\nu U 
+F_{\mu\nu}^L \nabla^\mu U \nabla^\nu U^\dagger \rangle  \nonumber \\
&&\qquad\qquad\qquad\qquad -\langle F_{\mu\nu}^L U F^{\mu\nu}{}^R U^\dagger
\rangle
+ \frac{1}{2}\langle (F_{\mu\nu}^R)^2 +(F_{\mu\nu}^L)^2  \rangle
+R^{\mu \nu} \langle \nabla_\mu U^\dagger \nabla_\nu U
\rangle 
\bigg]\,.
\label{eq:ricci}
\end{eqnarray}
 En el \'ultimo t\'ermino aparece el tensor de Ricci~$R_{\mu\nu}$. Para llevar las f\'ormulas a la forma de Gasser-Leutwyler usamos la siguiente identidad, v\'alida en SU(3)
\begin{eqnarray}
\langle (\nabla_\mu U^\dagger \nabla_\nu U)^2 \rangle &=& - 2 \langle (
\nabla_\mu U^\dagger \nabla^\mu U )^2 \rangle + \langle \nabla_\mu
U^\dagger \nabla_\nu U \rangle^2 + \frac{1}{2} \langle \nabla_\mu
U^\dagger \nabla^\mu U \rangle^2 \,.
\label{eq:id_su(3)}
\end{eqnarray} 

}

\section{Coeficientes de Gasser-Leutwyler-Donoghue}
\label{Resultados_Gasser-Leutwyler-Donoghue}

En el desarrollo quiral del lagrangiano efectivo en la forma de Gasser-Leutwyler-Donoghue de ec.~(\ref{eq:chl_flat}), las contribuciones m\'etricas son
\begin{eqnarray} 
{\cal L}^{(2,g)}  &=& {f_\pi^2\over 4} \langle \nabla_\mu U^\dagger \nabla^\mu U
+(\chi^\dagger U + U^\dagger \chi) \rangle ,
\label{eq:chl2}
\end{eqnarray} 
y
\begin{eqnarray} 
{\cal L}^{(4,g)} &=& L_1 \langle \nabla_\mu U^\dagger \nabla^\mu U
  \rangle^2 + L_2 \langle \nabla_\mu U^\dagger \nabla_\nu U \rangle^2
  + L_3 \langle \left( \nabla_\mu U^\dagger \nabla^\mu
  U \right)^2\rangle \nonumber \\ &+& L_4 \langle \nabla_\mu U^\dagger
  \nabla^\mu U \rangle \langle \chi^\dagger U + U^\dagger \chi \rangle
  + L_5 \langle \nabla_\mu U^\dagger \nabla^\mu U (
  \chi^\dagger U + U^\dagger \chi) \rangle \nonumber \\
  &+& L_6
  \langle \chi^\dagger U + U^\dagger \chi \rangle^2 +
  L_7 \langle \chi^\dagger U - U^\dagger \chi \rangle^2 + L_8 \langle
  ( \chi^\dagger U)^2 + (U^\dagger \chi)^2 \rangle \nonumber \\ &-&
  iL_9 \langle F_{\mu\nu}^L \nabla^\mu U \nabla^\nu U^\dagger + F_{\mu\nu}^R
  \nabla^\mu U^\dagger \nabla^\nu U \rangle + L_{10} \langle
  F_{\mu\nu}^L U F^{\mu\nu}{}^R U^\dagger \rangle \nonumber \\ &+& H_1
  \langle (F_{\mu\nu}^R)^2 + (F_{\mu\nu}^L)^2 \rangle + H_2 \langle
  \chi^\dagger \chi \rangle \, .
\label{eq:chl4}
\end{eqnarray} 
Las contribuciones con curvatura del lagrangiano quiral se pueden escribir en la forma propuesta en ref.~\cite{Donoghue:1991qv}, y vienen dadas por
\begin{eqnarray}
{\cal L}^{(2,R)} &=& -H_0 R \,, \label{eq:chl2R} 
\end{eqnarray} 
y
\begin{eqnarray}
{\cal L}^{(4,R)} &=& -L_{11} R \langle \nabla_\mu U^\dagger \nabla^\mu
U \rangle -L_{12} R^{\mu \nu} \langle \nabla_\mu U^\dagger \nabla_\nu
U \rangle - L_{13} R \langle \chi^\dagger U + U^\dagger
\chi \rangle \nonumber \\
&+& H_3 R^2 + H_4 R_{\mu \nu} R^{\mu \nu} +
H_5 R_{\mu \nu \alpha \beta} R^{\mu \nu \alpha \beta} \,.
\label{eq:chl4R}
\end{eqnarray} 
Los t\'erminos de curvatura son un reflejo de la naturaleza compuesta de los campos pseudoescalares, pues en los modelos quirales que estamos considerando estos t\'erminos se corres\-ponden con el acoplamiento de los campos gravitatorios externos a nivel de quarks. Un valor no nulo de $H_0$ indica que existe una renormalizaci\'on fuerte finita de la constante gravitatoria de Newton~$G$, ya que el lagrangiano cl\'asico de Einstein es~${\cal L}=-R/(16 \pi G)$. 


Notar que la matriz pseudoescalar $U$ es un escalar bajo transformaciones de
Lorentz y de coordenadas. Por tanto, despu\'es (y s\'olo despu\'es) de haber aplicado las identidades (\ref{eq:id1})-(\ref{eq:id_su(3)}) se puede sustituir la derivada covariante en Lorentz y coordenadas por la derivada covariante $D_\mu$, esto es $\nabla_\mu U = \widehat{D}_\mu U$.

\subsection{Modelo de Georgi-Manohar}
\label{sec_gm_res}

Por simplicidad, comenzaremos mostrando los resultados de los coeficientes de Gasser-Leutwyler-Donoghue para el modelo de Georgi-Manohar, pues en este caso no existe contribuci\'on proveniente de campos escalares, esto es, de campos de esp\'{\i}n cero y paridad positiva, y la \'unica contribuci\'on procede del loop de quarks. Para este modelo, la cons\-tan\-te de desintegraci\'on d\'ebil del pi\'on es
\begin{equation}
f_\pi^2 = \frac{N_c}{4\pi^2}g_A^2 M^2 \I_0 \,.  \label{eq:f0GM}
\end{equation}
El factor de normalizaci\'on para el campo $\chi$ es
\begin{equation}
B_0 = \frac{M}{g_A^2}\frac{\I_{-2}}{\I_0} \,.
\end{equation}
Con
\begin{equation}
\rho \equiv \frac{M}{B_0} = M \frac{f_\pi^2}{|\langle \bar q q\rangle|}=g_A^2 \frac{\I_0}{\I_{-2}}
\end{equation}
el resultado que encontramos para los coeficientes de GLD es
\begin{eqnarray}
L_1 &=& \frac{N_c}{48(4\pi)^2}\Big[(1-g_A^2)^2 \I_0 + 4g_A^2(1-g_A^2) \I_2+2g_A^4 \I_4\Big] \,, \qquad
L_2 = 2L_1 \,,  \nonumber \\ 
L_3 &=& -\frac{N_c}{24(4\pi)^2}
\Big[3(1-g_A^2)^2\I_0 +8g_A^4\I_4 + 4g_A^2(3-4g_A^2) \I_2 \Big] \,, 
\qquad L_4 = 0 \,,  \nonumber 
\end{eqnarray}
\begin{eqnarray}
L_5
&=&\frac{N_c}{2(4\pi)^2}\rho g_A^2\left[\I_0- \I_2\right] \,, \qquad
L_6 =
0 \,, \qquad 
L_7 = -\frac{N_c}{24(4\pi)^2 N_f}\ga \left[6\rho\I_0-\ga \I_2\right] \,, \nonumber \\ 
L_8 &=&
-\frac{N_c}{24(4\pi)^2}\left[6\rho(\rho-g_A)\I_0+g_A^2 \I_2\right] \,,\qquad
L_9 =
\frac{N_c}{6(4\pi)^2}\left[(1-g_A^2)\I_0+2g_A^2 \I_2\right]\,,  \nonumber\\
L_{10} &=& -\frac{N_c}{6(4\pi)^2}\left[(1-g_A^2)\I_0+g_A^2
\I_2\right]\,, \qquad L_{11} = \frac{N_c}{12(4\pi)^2}g_A^2\I_2 \,, \nonumber\\ 
L_{12} &=& -\frac{N_c}{6(4\pi)^2}g_A^2\I_2 \,, \qquad
L_{13} = \frac{N_c}{12(4\pi)^2}\rho\I_0 = \frac{\rho}{48 M^2}
\frac{f_\pi^2}{g_A^2}\,, \nonumber \\
H_0 &=& -\frac{N_c N_f}{6(4\pi)^2}M^2\I_{-2} =
-\frac{N_f}{24}\frac{f_\pi^2}{\rho} \,, \qquad 
H_1 =
\frac{N_c}{12(4\pi)^2}\left[-(1+g_A^2)\I_0+g_A^2 \I_2\right] \,,\nonumber \\ 
H_2 &=& \frac{N_c}{12(4\pi)^2}\left[6\rho^2
  \I_{-2}-6\rho(\rho+g_A)\I_0+g_A^2\I_2\right] \,, \label{eq:DN_GM} \\
H_3 &=& -\frac{N_c N_f}{144(4\pi)^2}\I_0 = -\frac{N_f}{576
M^2}\frac{f_\pi^2}{g_A^2} \,, \qquad H_4 = \frac{N_cN_f}{90(4\pi)^2}\I_0  \,, \qquad H_5 = \frac{7N_c
N_f}{720(4\pi)^2}\I_0 \,. \nonumber 
\end{eqnarray}
Con los valores  $M=300 \,\textrm{MeV}$ y $g_A = 0.75$, el cutoff $\Lambda$ debe ajustarse para reproducir el valor emp\'{\i}rico $f_\pi = 93.2 \,\textrm{MeV}$. Esto conduce a
\begin{eqnarray}
&&\Lambda =  1470\,\textrm{MeV} \,,
 \quad B_0=4913\,\textrm{MeV} \,,
\nonumber \\ 
&&
{\cal I}_{-2} = 20.8 \,, \quad
{\cal I}_0 = 2.26  \,, \quad 
{\cal I}_2 = 0.922  \,, \quad 
{\cal I}_4 = 0.995    \,.
\end{eqnarray}
El modelo quark quiral constituyente (QC) se corresponde con la elecci\'on $g_A = 1$ en los coeficientes anteriores. Si se considera el mismo valor para $M$, para este modelo se tiene
\begin{eqnarray}
&&\Lambda =  828\,\textrm{MeV} \,, 
 \quad B_0=1299\,\textrm{MeV} \,,
\nonumber \\
&&
{\cal I}_{-2} = 5.50 \,, \quad
{\cal I}_0 = 1.27 \,, \quad
{\cal I}_2 = 0.781 \,, \quad
{\cal I}_4 = 0.963  \,.  
\label{eq:QC}
\end{eqnarray}
En la tabla~\ref{tab:table2} se muestran los valores num\'ericos de los coeficientes de GLD.


\subsection{Modelo de Nambu--Jona-Lasinio} 
\label{sec:njl_res}
Los coeficientes de GLD en este modelo tendr\'an dos contribuciones diferentes: una proveniente del loop de quarks e integraci\'on posterior de los campos de esp\'{\i}n 1, y otra proveniente de la integraci\'on de los campos de esp\'{\i}n 0. Para la primera contribuci\'on se tienen las mismas expresiones de ec.~(\ref{eq:DN_GM}). La constante de desintegraci\'on d\'ebil del pi\'on es
\begin{equation}
f_\pi^2 = \frac{N_c}{4\pi^2}g_A M^2 \I_0 \,. \label{eq:f0NJL}
\end{equation}
Notar que en este modelo $f_\pi^2$ tiene una potencia en $g_A$, mientras que en el modelo de GM la potencia es $g_A^2$, ec.~(\ref{eq:f0GM}). La diferencia se debe a la ausencia del t\'ermino de masa~${\cal L}_m$ en el modelo GM. Nuestra notaci\'on ser\'a la siguiente
\begin{equation}
B_0 = \frac{a_s^2 M}{2 f_\pi^2} = \frac{M}{g_A}\frac{\I_{-2}}{\I_0}\,, 
\qquad g_A = 1-2\frac{f_\pi^2}{a_v^2} \,.
\end{equation}
Con
\begin{equation}
 \rho \equiv \frac{M}{B_0}=g_A \frac{\I_0}{\I_{-2}} \,,  
\label{eq:const}
\end{equation}
las contribuciones de esp\'{\i}n $0^+$ son
\begin{eqnarray}
L^S_3 &=&
\frac{N_c}{4(4\pi)^2}\frac{g_A^4}{\I_0}\left[\I_0-\I_2\right]^2 \,, 
\qquad L^S_5 = \frac{N_c}{4(4\pi)^2}g_A^2(g_A-2\rho) \left[\I_0- \I_2\right]
\,, \nonumber \\ 
L^S_8 &=& \frac{N_c}{16(4\pi)^2}(g_A-2\rho)^2 \I_0 \,, \nonumber\\
L^S_{11} &=& \frac{N_c}{12(4\pi)^2}g_A^2\left[\I_0- \I_2\right] \,, 
\qquad L^S_{13} = \frac{N_c}{24(4\pi)^2}(g_A-2\rho)\I_0 \,, \nonumber\\ 
H^S_2 &=& 2L^S_8 \,, \qquad H^S_3 = \frac{N_c N_f}{144(4\pi)^2}\I_0 =
\frac{N_f}{576M^2}\frac{f_\pi^2}{g_A} \,. \label{GLD_S}
\end{eqnarray}
El resto de coeficientes $L^S_i$, $H^S_i$ son cero. La suma de las dos contribuciones (loop de quaks y escalares) dar\'a los coeficientes de GLD para este modelo. El resultado es el siguiente
\begin{eqnarray}
L_3 &=& -\frac{N_c}{24(4\pi)^2}
\Big[3(1-2g_A^2-g_A^4)\I_0 +8g_A^4\I_4 
+2g_A^2\left(2(3-g_A^2)-3g_A^2 \frac{\I_2}{\I_0}\right) \I_2 \Big] \,, 
\nonumber \\
L_5 &=&\frac{N_c}{4(4\pi)^2}g_A^3\left[\I_0- \I_2\right] \,, \qquad
L_8 =
\frac{N_c}{48(4\pi)^2}g_A^2\left[3\I_0-2\I_2\right] \,,\nonumber \\ 
L_{11} &=& \frac{N_c}{12(4\pi)^2}g_A^2\I_0 =
\frac{g_A f_\pi^2}{48M^2} \,, \qquad
L_{13} = \frac{N_c}{24(4\pi)^2}g_A\I_0 =\frac{f_\pi^2}{96M^2}\,, 
\nonumber \\
H_2 &=&
\frac{N_c}{24(4\pi)^2}\big[12\rho^2\I_{-2}+3g_A(g_A-8\rho)\I_0 + 2g_A^2 \I_2\big] \,,  \qquad H_3 = 0 \,. \label{eq:GLD_NJL}
\end{eqnarray}
El resto de coeficientes: $L_1$, $L_2$, $L_4$, $L_6$, $L_7$, $L_9$, $L_{10}$, $L_{12}$, $H_0$, $H_1$, $H_4$ y $H_5$; coinciden con los del modelo de GM (f\'ormulas (\ref{eq:DN_GM})). Notar, no obstante, que las expresiones de $f_\pi^2$ no coinciden en los dos modelos [ec.~(\ref{eq:f0GM}) y (\ref{eq:f0NJL})].

Este modelo reproduce la relaci\'on $L_3 = -6L_1$, siempre y cuando se desprecien los t\'erminos ${\cal O}(N_c g_A^4)$. Existen algunas diferencias con trabajos previos. Los valores $L_1$, $L_2$, $L_3$, $L_4$, $L_5$, $L_6$,
$L_9$, $L_{10}$, $H_1$ y $H_2$ coinciden con ref.~\cite{Bijnens:1992uz}. $L_8$ difiere en dos potencias de $g_A$ en el t\'ermino proporcional a $\I_2$. (Nuestros resultados reproducen los suyos para cada contribuci\'on por separado: contribuci\'on del loop de quarks y contribuci\'on de esp\'{\i}n cero.)

El valor de $L_7$ es diferente de cero, si se considera la condici\'on
$\textrm{Det}(U)=1$ debido a que estamos considerando la simetr\'{\i}a de sabor $\textrm{SU}(N_f)$. Tanto en ref.~\cite{Bijnens:1992uz} como en \cite{njlarriola} este t\'ermino no se obtiene, a pesar de que en estos trabajos se menciona expl\'{\i}citamente que consideran el grupo de sabor $\textrm{SU}(N_f)$. En el grupo $\textrm{U}(N_f)$ s\'{\i} se obtiene que $L_7=0$. 

Nuestros valores de $L_4$, $L_5$, $L_6$, $L_8$, $L_9$ y $L_{10}$
coinciden con los de \cite{njlarriola}. En esta referencia aparece un t\'ermino err\'oneo extra en $L_1$. $L_3$ se diferencia de ref.~\cite{njlarriola} en todos los factores excepto uno en $\I_4$. $H_1$ y $H_2$ no aparecen en esa referencia.

Los coeficientes $L_{11}$, $L_{12}$ y $L_{13}$, as\'{\i} como $H_{0,3-5}$, son
nuevos y constituyen el resultado principal de este cap\'{\i}tulo. $L_{11-13}$
fueron obtenidos tambi\'en hace alg\'un tiempo en un modelo quiral que incluye
bosonizaci\'on~\cite{Andrianov:1998fr}, y m\'as recientemente en el modelo
quark espectral~\cite{Megias:2004uj} (ver cap\'{\i}tulo~\ref{ae_quiral_SQM}).

Los valores num\'ericos de estos coeficientes, ec.~(\ref{eq:GLD_NJL}), aparecen en la tabla~\ref{tab:table2} para dos casos diferentes: el modelo NJL SU(3) generalizado, y el caso en que no se considera la integraci\'on de los campos de esp\'{\i}n 1, esto es~$g_A=1$. Para el primer caso se considera como valor razonable~$g_A=0.606$. Con $M=300\,\textrm{MeV}$, se tiene
\begin{eqnarray}
&&\Lambda = 1344 \, \textrm{MeV} \,,
 \quad B_0=4015\,\textrm{MeV} \,,
 \nonumber \\ 
&&
\I_{-2} = 17.0 \,, \quad
\I_0 = 2.10  \,, \quad
\I_2 = 0.907  \,, \quad
\I_4 = 0.993  \,.
\end{eqnarray}
Para el modelo NJL con $g_A=1$, los valores num\'ericos de $\Lambda$,
$B_0$, $\rho$ y $\I_{2n}$ son id\'enticos a los del modelo quark constituyente QC, ec.~(\ref{eq:QC}). Las LEC's en el modelo NJL con $g_A=1$ y en el QC se diferencian debido a la contribuci\'on de los escalares $L^S_{3,5,8,11,13}$ y
$H^S_{2,3}$, que no est\'an presentes en el caso QC.

\subsection{Resultados}
\label{resultados_coef_GLD}
\begin{table*}[tt]
\caption{\label{tab:table2} Constantes adimensionales de baja energ\'{\i}a y
  $H_0$ comparadas con otros modelos y con el valor que dan algunas
  referencias. Los valores mostrados para $L_{1-13},H_{1-5}$ deben ser
  multiplicados por $10^{-3}$. El valor de $H_0$ debe multiplicarse por $10^3\,\text{MeV}^2$. }
\begin{tabular} {lrrrrrrrrr}
\hline
\hline
  & TQP\footnotemark[1]  & NJL & NJL & QC & GM & SQM\footnotemark[2] & Large $N_c$\footnotemark[3] & Dual\footnotemark[2]\\  & & & ($g_A=1$) & & & (MDM)  & &  Large $N_c$\\
\hline
$L_1$  & 0.53 $\pm$ 0.25 & 0.77  & 0.76  & 0.76  & 0.78 & 0.79  & 0.9 &  0.79 \\
$L_2$ & 0.71 $\pm$ 0.27  & 1.54  & 1.52 & 1.52 & 1.56 &  1.58 & 1.8 &  1.58\\
$L_3$ & $-$2.72 $\pm$ 1.12  & $-$4.02  & $-$2.73 & $-$3.62  & $-$4.25 &  $-$3.17 & $-$4.3 & $-$3.17 \\
$L_4$ & 0  & 0 & 0 & 0  & 0 & 0 & 0 & 0 & \\
$L_5$ &  0.91 $\pm$ 0.15 & 1.26  & 2.32 & 1.08 & 0.44 & 2.0 $\pm$ 0.1 & 2.1 &  3.17\\
$L_6$ &  0 & 0 & 0 & 0  & 0 & 0  & 0 & 0 & \\
$L_7$ &  $-$0.32 $\pm$ 0.15 & $-$0.06  & $-$0.26  & $-$0.26  & $-$0.03  &  $-$0.07 $\pm$ 0.01 & $-$0.3 & \\
$L_8$ &  0.62 $\pm$ 0.20  & 0.65  & 0.89 & 0.46  & 0.04 & 0.08 $\pm$ 0.04  & 0.8 & 1.18 \\
$L_9$ &  5.93 $\pm$ 0.43  & 6.31 & 4.95 & 4.95 &  6.41 & 6.33 & 7.1 &  6.33 \\
$L_{10}$&  $-$4.40 $\pm$ 0.70\footnotemark[4]  & $-$5.25  & $-$2.47 & $-$2.47  &  $-$4.77 & $-$3.17 & $-$5.4 &  $-$4.75 \\
$L_{11}$&  1.85 $\pm$ 0.90\footnotemark[5]  & 1.22  & 2.01 & 1.24 & 0.82 & 1.58  & 1.6\footnotemark[5] & \\
$L_{12}$&  $-$2.7\footnotemark[5] &  $-$1.06   & $-$2.47 & $-$2.47 & $-$1.64  & $-3.17$ &  $-$2.7\footnotemark[5] &\\
$L_{13}$&  1.7 $\pm$ 0.80\footnotemark[5]  & 1.01 & 1.01  & 0.47 & 0.22 & 0.33 $\pm$ 0.01 & 1.1\footnotemark[5] & \\
$H_0 $   &   & $-$14.6 & $-$4.67  & $-$4.67 &  $-$17.7 &  1.09  &  & \\
$H_1$   &  &  $-$4.01  &  $-$2.78   & $-$2.78  &   $-$4.76  &   &   &  & \\
$H_2$   &    &  1.46   &  1.45  & 0.59 & 0.49 & $-$1.0 $\pm$ 0.2  & & & \\
$H_3$   &   &  0      & 0    & $-$0.50 &  $-$0.89 &   & &  &  \\
$H_4$   &   &  1.33   & 0.80     & 0.80 &  1.43  &   & &  & \\
$H_5$   &  &  1.16   & 0.70    &  0.70 & 1.25 &    & &  & \\
\hline
\hline
\end{tabular}
(1){ C\'alculo a dos loops de ref.~\cite{Amoros:2001cp}.} \\
(2){ Ref.~\cite{Megias:2004uj}, cap\'{\i}tulo~\ref{ae_quiral_SQM}. } \\
(3){ Ref.~\cite{Ecker:1988te}.} \\
(4){ Ref.~\cite{Bijnens:2002hp}.} \\
(5){ Ref.~\cite{Donoghue:1991qv}.}
\end{table*}
En la tabla~\ref{tab:table2} aparecen los resultados que hemos obtenido para los modelos de quarks quirales que se han tratado en este cap\'{\i}tulo: Quark Constituyente, Nambu--Jona-Lasinio con y sin mesones vectoriales, y Georgi-Manohar. Se ha incluido tambi\'en el resultado del c\'alculo en el modelo Quark Espectral del cap\'{\i}tulo~\ref{ae_quiral_SQM}. La primera columna  se corresponde con el c\'alculo de TQP a dos loops~\cite{Amoros:2001cp}. Se incluye tambi\'en el resultado obtenido en el modelo basado en $N_c$ grande con saturaci\'on por una \'unica resonancia~\cite{Ecker:1988te}. 

Los resultados para las constantes de baja energ\'{\i}a coinciden a grandes rasgos. Como regla, todos los modelos y ajustes dan el mismo signo para todos los coeficientes, con la excepci\'on de $H_0$ y $H_2$ en el modelo Quark Espectral. Para los coeficientes de Gasser-Leutwyler est\'andar $L_{1-10}$ el mejor acuerdo global con el c\'alculo de TQP a dos loops \cite{Amoros:2001cp} es el proporcionado por el modelo NJL con mesones vectoriales, para el que la chi cuadrada reducida es $\chi^2/DOF=2.2$, $(DOF=10)$, si bien los modelos QC y GM proporcionan resultados de calidad similar: $2.5$ y $3.6$ respectivamente.

Para los coeficientes nuevos no existen en la literatura valores ampliamente aceptados. El acuerdo m\'as cercano con las estimaciones de $N_c$ grande y saturaci\'on de resonancias de \cite{Donoghue:1991qv} para $L_{11-13}$ es el de NJL sin mesones vectoriales, para el que $\chi^2/DOF=0.29$, pero esto no es totalmente concluyente. Asimismo, es importante mencionar el notable acuerdo entre las predicciones del modelo Quark Espectral para estos tres coeficientes y aquellas provenientes del modelo quiral de bosonizaci\'on de ref.~\cite{Andrianov:1998fr}, para el que se obtiene
\begin{equation}
L_{11} = 1.58 \cdot 10^{-3} \,, \qquad
L_{12} = -3.2 \cdot 10^{-3} \,, \qquad
L_{13} = 0.3 \cdot 10^{-3} \,.
\end{equation}

\section{Conclusiones}
\label{conclusiones_gravedad}

En este cap\'{\i}tulo hemos calculado las constantes de baja energ\'{\i}a del tensor energ\'{\i}a-impulso en varios modelos de quarks quirales: Quark Constituyente, Nambu--Jona-Lasinio con y sin mesones vectoriales, y Georgi-Manohar. Algunas de estas constantes se obtienen directamente de los coeficientes est\'andar de Gasser-Leutwyler, mientras que otras, $L_{11-13}$ y $H_{0,3-5}$, son nuevas y proceden de operadores que no est\'an presentes en el lagrangiano quiral en espacio plano.

T\'ecnicamente, el mejor modo de proceder es considerar QCD en un espacio-tiempo curvo, ya que nos permite trabajar con el lagrangiano a bajas energ\'{\i}as, en lugar de su variaci\'on (el tensor energ\'{\i}a-impulso). Esto hace m\'as f\'acil tanto el c\'alculo como la imposici\'on de las restricciones debidas a las simetr\'{\i}as. El lagrangiano quiral en espacio-tiempo curvo contiene dos tipos de contribuciones. Por una parte, aquellas que surgen de un acoplamiento m\'{\i}nimo del lagrangiano en espacio plano con la m\'etrica, ${\cal L}^{(g)}$, y por otra aquellas contribuciones que contienen el tensor de curvatura de Riemann ${\cal L}^{(R)}$. En el esp\'{\i}ritu de no introducir nuevos campos diferentes a la m\'etrica, hemos considerado \'unicamente la gravedad de Einstein. En el caso de que se considerara torsi\'on o violaci\'on de la metricidad, en principio podr\'{\i}an aparecer nuevos t\'erminos. Al igual que ocurre con los acoplamientos gauge (por ejemplo, los momentos magn\'eticos), los t\'erminos gravitatorios ${\cal L}^{(R)}$ no pueden fijarse a partir de la covariancia general del lagrangiano quiral, y para obtenerlos es necesario acoplar directamente gravedad con los quarks y los gluones de QCD antes de integrar los campos y obtener el lagrangiano de bajas energ\'{\i}as.

Hemos calculado en estos modelos de quarks quirales las constantes de baja
energ\'{\i}a con un cierto grado de \'exito, y hemos aplicado la misma aproximaci\'on para los t\'erminos con curvatura ${\cal L}^{(R)}$. El acuerdo entre todos los modelos es razonable. Una comparaci\'on con los valores de TQP a dos loops~\cite{Amoros:2001cp} sugiere que NJL con mesones vectoriales es el que mejor funciona para los coeficientes est\'andar. Para los nuevos coeficientes $L_{11-13}$, el mejor acuerdo proviene de NJL sin mesones vectoriales, si bien el resultado no es concluyente.

\chapter{Modelo Quark Espectral y Acci\'on Efectiva Quiral}
\label{ae_quiral_SQM}

La estructura de QCD a bajas energ\'{\i}as en presencia de fuentes electrod\'ebiles y gravi\-ta\-cionales se describe muy bien mediante Teor\'{\i}a Quiral de Perturbaciones (TQP) \cite{gasser-leutwyler1, gasser-leutwyler2,Donoghue:1991qv}. En el sector mes\'onico, la rotura espont\'anea de la simetr\'{\i}a quiral es dominante a bajas energ\'{\i}as y el c\'alculo sistem\'atico de las correspondientes constantes de baja energ\'{\i}a (LEC's) ha sido lle\-vado a cabo recientemente hasta una precisi\'on de dos loops~\cite{Amoros:2001cp,Bijnens:2002hp} o mediante el uso de las ecuaciones de Roy~\cite{Ananthanarayan:2000ht}. Para los procesos fuertes y electrod\'ebiles que involucran mesones pseudoescalares, la mayor parte de las LEC's est\'an saturadas en t\'erminos de resonancias de intercambio~\cite{Ecker:1988te}, que pueden ser justificadas en el l\'{\i}mite de $N_c$ grande en una cierta aproximaci\'on de bajas energ\'{\i}as~\cite{Pich:2002xy}. En el caso de procesos gravitacionales se pueden aplicar las mismas ideas~\cite{Donoghue:1991qv}. Hoy en d\'{\i}a, TQP se usa como un test cualitativo y cuantitativo para cualquier modelo de la estructura de los hadrones a bajas energ\'{\i}as.

En este cap\'{\i}tulo nos proponemos analizar, en el contexto de TQP con
espacio-tiempo curvo, el modelo quark espectral propuesto recientemente en
ref.~\cite{spectralqm}. En primer lugar se mostrar\'a c\'omo calcular la
acci\'on efectiva de este modelo a un loop de quarks, y algunas de sus
propiedades. Posteriormente se har\'a un estudio de la parte an\'omala de la
acci\'on efectiva, con la obtenci\'on del t\'ermino est\'andar de
Wess-Zumino-Witten. Se ver\'a que la anomal\'{\i}a que se obtiene con este
modelo coincide con la anomal\'{\i}a de QCD. Se aplicar\'a el formalismo
desarrollado en el cap\'{\i}tulo~\ref{tensor_EM_MQQ} para el c\'alculo de la
contribuci\'on no an\'omala de la acci\'on efectiva, y se obtendr\'an las
expresiones correspondientes para los coeficientes de baja energ\'{\i}a
(LEC). Con el fin de considerar una realizaci\'on expl\'{\i}cita del modelo
espectral, se considerar\'a \'este dentro de un esquema de dominancia del
mes\'on vectorial, lo cual permitir\'a encontrar valores concretos para las
LEC's y comparar con resultados de otros modelos presentados en el
cap\'{\i}tulo~\ref{tensor_EM_MQQ}. Finalmente se comparar\'an las predicciones
del modelo espectral para estas constantes con las obtenidas en la
aproximaci\'on de una \'unica resonancia (SRA) en el l\'{\i}mite de $N_c$
grande~\cite{Donoghue:1991qv,Pich:2002xy}, lo cual conducir\'a a unas relaciones de dualidad entre los canales vector y escalar.

Este cap\'{\i}tulo est\'a basado en la referencia~\cite{Megias:2004uj}.

\section{Acci\'on Efectiva del Modelo Quark Espectral}
\label{SQM_sec1}

En la secci\'on~\ref{sec:mqe1} introdujimos el modelo quark espectral. La aproximaci\'on es similar en esp\'{\i}ritu al modelo de Efimov e Ivanov~\cite{Efimov:1988yd}, propuesto hace algunos a\~nos, y se basa en la introducci\'on formal de la representaci\'on de Lehmann generalizada para el propagador del quark. La acci\'on efectiva que obedece las identidades de Ward-Takahashi mediante la t\'ecnica de Delbourgo y West~\cite{gaugetechnique} corresponde en nuestro caso a una prescripci\'on de sustituci\'on m\'{\i}nima. Esto conduce a un determinante fermi\'onico de la forma\footnote{
Para un operador bilocal $A(x,x^\prime)$ (matrices en espacio de Dirac y de sabor) se tiene
\begin{eqnarray}
{\rm Tr} A = \int d^4 x
\sqrt{-g} \,{\rm tr} \langle A(x,x) \rangle \, , 
\end{eqnarray}
donde $\tr$ indica traza de Dirac y $\langle \;\rangle$ traza en espacio de sabor.
}
\begin{eqnarray}
\Gamma_{\rm SQM}[U,s,p,v,a,g] =-i N_c \int_{\cal C} d \omega \rho(\omega) {\rm Tr} \log
\left( i {\bf D} \right),
\label{eq:eff_ac} 
\end{eqnarray} 
donde el operador de Dirac viene dado por 
\begin{equation}
i {\bf D} = i\slashchar{d} - \omega U^5 - {\hat m_0} + \left(
\slashchar{v} + \slashchar{a} \gamma_5 - s - i \gamma_5 p \right) = iD-\omega U^5   \,.
\label{eq:dirac_opbiss} 
\end{equation} 
Estamos trabajando en espacio-tiempo curvo de Minkowski. La derivada $d_\mu$ es derivada covariante bajo transformaciones generales de coordenadas y transformaciones de Lorentz, e incluye la conexi\'on de esp\'{\i}n.
El tensor m\'etrico $g_{\mu\nu}$ es la fuente externa que representa el acoplamiento con un campo gravitatorio. La matriz $U^5=U^{\gamma_5}$ es la matriz de sabor que representa el octete de mesones pseudoescalares en la representaci\'on no lineal. Este operador de Dirac transforma de manera covariante bajo transformaciones quirales locales.\footnote{Para un estudio sobre el acoplamiento con gravedad de los modelos de quarks quirales, ver secciones~\ref{acoplamiento_gravedad} y \ref{MQQ_ETC}.} En lo sucesivo consideraremos el modelo con $N_f=3$.

Si se considera la matriz $U$ en el sector U(3) de sabor, la anomal\'{\i}a
U($1)_A$ se puede tener en cuenta a\~nadiendo el t\'ermino
habitual~\cite{chiralanomaly} 
\begin{eqnarray}
{\cal L}_A = - \frac{f_\pi^2}4 m_{\eta_1}^2 \left\{ \theta - \frac{ i}2
\left[ \log \det {\bar U} - \log \det {\bar U}^\dagger \right] \right\}^2 \,,
\label{eq:singlet} 
\end{eqnarray} 
donde $U = {\bar U} e^{i \eta_8 / (3f_\pi) } $, con $\det {\bar U} =
1$. Para $\theta=0$ este t\'ermino es invariante CP y SU$(N_f)_L \times $SU$(N_f)_R $.

La acci\'on efectiva del modelo tiene un aspecto similar a la del modelo NJL bosonizado (ver secci\'on~\ref{MQQ_ETC}). La principal diferencia tiene que ver con la interpretaci\'on del m\'etodo de regularizaci\'on. Por una parte, en los modelos NJL \'unicamente se puede regularizar sobre loops de quarks (l\'{\i}neas de quark cerradas). El hecho de que en el modelo quark espectral la ``regularizaci\'on'' de Lehmann se produzca sobre l\'{\i}neas de quark abiertas tiene importantes consecuencias en cuanto a la consistencia de los c\'alculos a energ\'{\i}as altas tanto en una interpretaci\'on puramente hadr\'onica como part\'onica. 

Dado que el contorno de integraci\'on para la variable espectral~$\omega$ es en general complejo, resulta complicado pasar a espacio eucl\'{\i}deo y separar la acci\'on en una parte real y otra imaginaria. En lugar de espacio eucl\'{\i}deo, podemos considerar el espacio de Minkowski e introducir, como hicimos en sec.~\ref{operador_2_orden}, el operador auxiliar
\begin{eqnarray}
-i {\bf D}_5 &=& \gamma_5 \left( i \slashchar{d} -\omega {U^5}^\dagger -\hat{m}_0 +
\slashchar{v} - \gamma_5 \slashchar{a} -
s + i\gamma_5 p \right) \gamma_5 \,.
\end{eqnarray} 
De este modo, la acci\'on efectiva con paridad normal se escribe
\begin{equation}
\Gamma_{\rm SQM}^+ = -\frac{i}2 N_c \int_{\cal C} d \omega \rho(\omega) {\rm Tr} \log
\left( {\bf D}_5 {\bf D} \right) \,.
\label{eq:np}
\end{equation}

\section{Anomal\'{\i}as Quirales}
\label{chiral_anomalies}

Una de las ventajas m\'as importantes de la regularizaci\'on espectral es que conduce a observables hadr\'onicos finitos e independientes de la escala, lo cual es un requerimiento b\'asico de todo procedimiento de regularizaci\'on. No obstante, esto no significa o implica necesariamente que la acci\'on efectiva total en presencia de campos externos sea finita, ya que incluso en el caso de que los campos pi\'onicos sean cero, $U=1$, existen procesos no hadr\'onicos. En realidad, ocurre que la renormalizaci\'on de la funci\'on de onda del fot\'on es proporcional a $\rho_0^\prime$ \cite{spectralqm}, de modo que depende de la escala~$\mu$ y por tanto diverge en ciertos esquemas de regularizaci\'on (por ejemplo, en regularizaci\'on dimensional). Esta dependencia en escala surge tambi\'en en otros t\'erminos no hadr\'onicos de la acci\'on efectiva. 

En \cite{spectralqm} se encuentra que las desintegraciones~$\pi^0 \to 2
\gamma $ y $ \gamma \to 3 \pi $ se muestran de acuerdo con los valores correctos que se esperan de la anomal\'{\i}a quiral de QCD. Con ayuda de la acci\'on efectiva, ec.~(\ref{eq:eff_ac}), vamos a ver en esta secci\'on que esto es cierto tambi\'en para todos los procesos an\'omalos. En primer lugar calcularemos la anomal\'{\i}a quiral, y mostraremos que en presencia de campos externos la anomal\'{\i}a no depende del campo pi\'onico~$U$, y por tanto coincide con la anomal\'{\i}a en QCD debido a las condiciones espectrales $\rho_1=\rho_2=\rho_3=\rho_4=0$. Despu\'es veremos c\'omo surge en este contexto el t\'ermino est\'andar de Wess-Zumino-Witten~\cite{Wess:yu,Witten:tw}.

\subsection{C\'alculo de la anomal\'{\i}a quiral}
\label{calculo_anomalia_quiral}

Bajo transformaciones quirales locales (vector y axial) el operador de Dirac se transforma 
\begin{eqnarray} 
 {\bf D} \to e^{+i\epsilon_V (x) -i\epsilon_A (x) \gamma_5 } \,{\bf D}\,
e^{-i\epsilon_V (x)-i\epsilon_A (x) \gamma_5 },
\end{eqnarray} 
con
\begin{eqnarray} 
\epsilon_V (x) = \sum_a  \epsilon_V^a (x) \lambda_a \,, \qquad
\epsilon_A (x) = \sum_a  \epsilon_A^a (x) \lambda_a \,.
\end{eqnarray} 
Infinitesimalmente, la transformaci\'on es
\begin{eqnarray}
\delta {\bf D} = i [ \epsilon_V , {\bf D}] - i \{ \epsilon_A \gamma_5, {\bf D} \} \,.
\end{eqnarray}
Si consideramos una transformaci\'on quiral en la acci\'on efectiva, ec.~(\ref{eq:eff_ac}), sin ninguna regularizaci\'on adicional, se tiene
\begin{eqnarray}
\delta S = - i N_c {\rm Tr} \int_{\cal C} d \omega \rho (\omega) \left[ \delta {\bf D}{\bf D}^{-1} \right] \,. 
\end{eqnarray} 
Teniendo en cuenta la propiedad c\'{\i}clica de la traza, se obtiene s\'olo una contribuci\'on procedente de la variaci\'on axial
\begin{eqnarray} 
\delta_A S &\equiv &
{\cal A}_A = \int d^4 x \,{\rm tr} \int_{\cal C} d \omega \rho(\omega) \langle 2 i \epsilon_A \gamma_5
\rangle  = \rho_0 \int d^4 x \,{\rm tr} \langle 2 i \epsilon_A \gamma_5 \rangle \,, 
\end{eqnarray}  
un resultado que es ambiguo incluso en presencia de regularizaci\'on
espectral, debido a la traza dimensional infinita~\cite{fujikawa}. Para evitar la ambig\"uedad es necesario introducir una regularizaci\'on extra. Como es bien sabido, no existe una regularizaci\'on que preserve la simetr\'{\i}a quiral, de modo que la anomal\'{\i}a es generada. 

El c\'alculo se puede hacer con m\'etodos est\'andares. Una regularizaci\'on conveniente es la regularizaci\'on $\zeta$~\cite{detdiracop}, que permite calcular directamente la anomal\'{\i}a a partir del propio operador de Dirac (no su cuadrado), y no precisa de ninguna redefinici\'on de la matriz~$\gamma_5$. Esto conduce a
\begin{eqnarray} 
\delta_A S \equiv {\cal A}_A &=& {\rm Tr} \int_{\cal C} d \omega \rho(\omega) \left( 2 i
\epsilon_A \gamma_5 \left[ i {\bf D} \right]^0 \right)  \nonumber \\
&=& \int d^4 x \, {\rm tr} \int_{\cal C} d \omega \rho(\omega) \langle 2 i
\epsilon_A (x) \gamma_5 \langle x | {\bf D}^0 | x
\rangle \rangle \,,
\end{eqnarray}  
donde la potencia cero del operador de Dirac se entiende como una continuaci\'on anal\'{\i}tica que puede escribirse en t\'erminos de coeficientes de Seeley-DeWitt para operadores de Dirac~\cite{detdiracop}:
\begin{eqnarray}
\langle x | {\bf D}^0 | x \rangle &=& \frac1{(4\pi)^2} \Big\{ {1\over 2}{\bf
               D}^4 + {1\over 3}({\bf D}^2\Gamma_\mu^2+\Gamma_\mu {\bf D}^2\Gamma_\mu 
               +\Gamma_\mu^2{\bf D}^2) \nonumber \\
               &&\qquad\qquad\qquad\qquad\qquad + {1\over 6}
               \left(\Gamma_\mu^2\Gamma_\nu^2 + (\Gamma_\mu\Gamma_\nu)^2 +
               \Gamma_\mu\Gamma_\nu^2\Gamma_\mu \right) \Big\} \,,
\end{eqnarray} 
donde $\Gamma_\mu = {1\over 2}\{\gamma_\mu,{\bf D}\}$. La combinaci\'on~$\{\gamma_\mu , {\bf D}\}$ es un operador multiplicativo, de modo que equivale a una funci\'on. El resultado para acoplamientos generales en cuatro dimensiones ha sido obtenido de~\cite{detdiracop}. Una inspecci\'on directa muestra que, puesto que la dependencia en $\omega$ viene dada por~$i {\bf D} = i D - \omega U^5$, el resultado se puede escribir como la suma de un t\'ermino independiente de~$\omega$ m\'as un polinomio en $\omega$
\begin{eqnarray} 
{\cal A}_A &=& \int_{\cal C} d \omega \rho( \omega ) \left( {\cal A}_A [ s,p,v,a ] + 
{\cal A}_A [ s,p,v,a,\omega ,U] \right) 
=\rho_0 {\cal A}_A [ s,p,v,a ] \,,
\label{eq:qcd_anom} 
\end{eqnarray}
donde el t\'ermino polin\'omico dependiente de~$\omega$ se anula, por las
condiciones espectrales (los momentos positivos son cero). Esto muestra que la
anomal\'{\i}a del modelo quark espectral coincide con la anomal\'{\i}a de QCD
despu\'es de introducir una regularizaci\'on adicional conveniente,
independientemente de los detalles de la funci\'on espectral. Esto es un punto
importante, ya que si la acci\'on efectiva~$ \Gamma [U,s,p,v,a] $ en
ec.~(\ref{eq:eff_ac}) fuera finita e invariante quiral, aparentemente no habr\'{\i}a raz\'on para la existencia de anomal\'{\i}as.

\subsection{T\'ermino de Wess-Zumino-Witten}
\label{sec:WZWterm}

Mostraremos aqu\'{\i} d\'onde y c\'omo surgen estas divergencias. Por simplicidad, conside\-remos el l\'{\i}mite quiral~$\hat{m}_0=0$, los campos externos los haremos cero y trabajaremos en espacio-tiempo plano, de modo que~$ i {\bf D} = i \slashchar{\partial} $. Conseguiremos una representaci\'on conveniente si introducimos el campo
\begin{eqnarray} 
U_t^5 = e^{ { i} t \sqrt{2} \gamma_5 \Phi /f_\pi} \,,
\end{eqnarray} 
que permite interpolar entre el vac\'{\i}o $U_{t=0}^5= {\bf 1}$, y la matriz completa~$U_{t=1}^5 = U^5$. Podemos escribir la siguiente identidad trivial para la acci\'on efectiva con sustracci\'on del vac\'{\i}o:
\begin{eqnarray}
\Gamma_{\rm SQM}[U,s,p,v,a] - \Gamma_{\rm SQM}[{\bf 1},s,p,v,a] &=& - i N_c \int_0^1 dt
\frac{d}{dt} \int_{\cal C} d \omega \rho(\omega) {\rm Tr} \log \left( i D -
  \omega U^5_t \right) \nonumber \\ &=& i N_c\int_0^1 dt \int_{\cal C} d\omega \rho(\omega) {\rm Tr}\left[ \omega \frac{d U_t^5}{dt} \frac1{
i D - \omega U^5_t } \right] \,. \nonumber
\label{eq:GammaU}
\end{eqnarray} 
Puesto que estamos interesados en procesos con paridad anormal, es suficiente con identificar los t\'erminos que contienen el tensor de Levi-Civit\`a~$\epsilon_{\mu\nu\alpha\beta}$, que por invariancia Lorentz precisan de al menos cuatro derivadas. Teniendo en cuenta el hecho de que las derivadas act\'uan sobre su derecha, se tiene
\begin{eqnarray}
\Gamma_{\rm SQM}^{- (4)} &=& - i N_c \int_0^1 dt \int_{\cal C} d \omega \rho(\omega)
\int d^4 x \int \frac{d^4 k}{(2 \pi)^4 } \frac1{[k^2- \omega^2]^5} \nonumber \\ &\times & \Tr\left\{ -\omega \gamma_5
U_t^\dagger \frac{d U_t }{dt} \omega \left[ \omega U_t^\dagger i \slashchar{\partial} U_t \right]^4 \right\},
\label{eq:s_ab}
\end{eqnarray}  
donde el super\'{\i}ndice $(4)$ indica ${\cal O}(p^4)$. Tras el c\'alculo de
las trazas e integrales, finalmente se obtiene
\begin{eqnarray}
\Gamma_{\rm SQM}^{- (4)} &=& \rho_0 \frac{N_c}{48\pi^2} \int_0^1 dt \int d^4 x
\,\epsilon_{\mu\nu\alpha\beta} \,\langle U_t^\dagger
\frac{d U_t }{dt} U_t^\dagger \partial^\mu U_t U_t^\dagger
\partial^\nu U_t U_t^\dagger \partial^\alpha U_t U_t^\dagger
\partial^\beta U_t \rangle , \nonumber 
\label{eq:wzw} 
\end{eqnarray} 
que coincide con el t\'ermino de Wess-Zumino-Witten
(WZW)~\cite{Wess:yu,Witten:tw}, si usamos que $\rho_0=1$. Los campos externos
pueden ser incluidos mediante el uso de ec.~(\ref{eq:s_ab}), lo cual genera el
t\'ermino de WZW en la forma de Bardeen. En realidad, la
diferencia~$\Gamma_{\rm SQM}[U,s,p,v,a]-\Gamma_{\rm SQM}[{\bf 1},s,p,v,a]$ es finita y preserva invariancia gauge, pero rompe la simetr\'{\i}a quiral lo cual genera la anomal\'{\i}a de ec.~(\ref{eq:qcd_anom}).


\section{Desarrollo quiral de la acci\'on efectiva}
\label{des_quiral_ae}

A partir de la acci\'on de ec.~(\ref{eq:eff_ac}) podemos calcular el desarrollo en derivadas en el contexto de espacio-tiempo curvo~(para los detalles, ver la secci\'on~\ref{ae_gravedad}). Teniendo en cuenta la f\'ormula del desarrollo del heat kernel, ec.~(\ref{eq:des_hkkk_mqq}), los coeficientes que se obtienen son los mismos que se obtuvieron en el modelo NJL, ec.~(\ref{as_curvatura}), con la salvedad de considerar la sustituci\'on~$M \to \omega$, y el hecho de que en el modelo espectral no se introducen campos internos auxiliares para bosonizar (los s\'{\i}mbolos no tienen barra: ${\cal V}$, $\nabla_\mu$, ${\cal F}_{\mu\nu}$). Despu\'es de usar las condiciones espectrales~$\rho_n=0,\; n>0$, la contribuci\'on de paridad normal para la acci\'on efectiva se escribe
\begin{eqnarray}
-\frac{i}2 {\rm Tr} \log {\bf D}_5 {\bf D} &=& -\frac12
 \frac{N_c}{(4\pi^ 2)} \int d^4 x \sqrt{-g} \int_{\cal C} d\omega \rho(\omega)
 \nonumber \\ &\times& {\rm tr} \langle -\frac12 \omega^4 \log \omega^2 a_0 +
 \omega^2 \log \omega^2 a_1 -\log (\omega^2/\mu^2) a_2 +
 \frac1{\omega^2} a_3 + \frac1{\omega^4} a_4 + \cdots \rangle \nonumber \\
&=& \int d^4 x \sqrt{-g} \left( {\cal L}^{(0)} + {\cal L}^{(2)} + {\cal L}^{(4)} + \cdots
 \right) \,. \label{eq:inp_ae_sqm}
\end{eqnarray} 
Despu\'es del c\'alculo de las trazas de Dirac, para el orden ${\cal O}(p^2)$ del lagrangiano efectivo se tiene
\begin{eqnarray}
{\cal L}^{(2)} &=&{N_c\over(4\pi)^2} \int_{\cal C} \rho(\omega) \Big\{ -
\omega^2 \log \omega^2 \langle \nabla_\mu U^\dagger \nabla^\mu U
\rangle \nonumber \\ &+& 2 \omega^3 \log \omega^2 \langle {\it
m}^\dagger U + U^\dagger {\it m} \rangle + \omega^2 \log \omega^2
\frac1{12} \langle R \rangle \Big\} \,, \label{eq:np_ae_2_sqm}
\end{eqnarray} 
y para el orden ${\cal O}(p^4)$
\begin{eqnarray} 
{\cal L}^{(4)} &=& {N_c\over (4\pi)^2} \int_{\cal C} \rho(\omega) \Big\{
\nonumber \\
&+& {1\over6} \log \omega^2 \langle (F_{\mu\nu}^R)^2 +
(F_{\mu\nu}^L)^2 \rangle - \log \omega^2 \langle
\frac7{720} R^{\alpha \beta \mu \nu} R_{\alpha \beta \mu \nu} -
\frac{1}{144} R^2 + \frac1{90} R^{\mu \nu} R_{\mu \nu} \rangle
\nonumber \\
&-&\frac{i}3 \langle F_{\mu\nu}^R \nabla_\mu U^\dagger \nabla_\nu U +
F_{\mu\nu}^L \nabla_\mu U \nabla_\nu U^\dagger \rangle 
+ \frac1{12} \langle (\nabla_\mu U \nabla_\nu U^\dagger )^2 \rangle
-\frac16 \langle (\nabla_\mu U \nabla^\mu U^\dagger )^2 \rangle
\nonumber \\
&+& \frac16 \langle \nabla^\mu \nabla^\nu U \nabla_\mu \nabla_\nu 
U^\dagger \rangle- \frac16 \langle F_{\mu\nu}^L U F_{\mu\nu}^R
U^\dagger \rangle \nonumber \\ 
&+&  \log \omega^2 \omega^2 \left( 2 \langle {\it m}^\dagger{\it m} \rangle
+ \langle ( {\it m}^\dagger
U+U^\dagger {\it m} )^2 \rangle \right) 
\nonumber \\
&-& {1\over2} \omega \langle \nabla_\mu U^\dagger \nabla^\mu U ({\it m}^\dagger
U+U^\dagger {\it m} ) \rangle 
-\log \omega^2 \omega  \langle \nabla_\mu U^\dagger \nabla^\mu {\it m} +
\nabla_\mu {\it m}^\dagger \nabla^\mu U \rangle \nonumber \\ 
&-&\omega \log \omega^2 \frac16 R \langle U^\dagger {\it m}+ {\it
m}^\dagger U \rangle + \frac1{12} R\, \nabla_\mu U^\dagger \nabla^\mu U
\rangle  \Big\} \,. \label{eq:np_ae_4_sqm} 
\end{eqnarray} 
En estas f\'ormulas $m \equiv s+ip=\chi/2B_0$. Notar que los momentos que aparecen hasta este orden son~ $\rho_0 =1 $, $\rho_1=0$ y $\rho_2=0$, as\'{\i} como los momentos logar\'{\i}tmicos~$ \rho^\prime_0 $,
$\rho^\prime_1 $ y $\rho^\prime_2 $. Tras aplicar las ecuaciones de movimiento
cl\'asicas del campo~$U$, ecs.~(\ref{eq:id1})-(\ref{eq:id2}), la identidad
integral de ec.~(\ref{eq:ricci}) y la identidad v\'alida en SU(3),
ec.~(\ref{eq:id_su(3)}), se llega a la forma est\'andar del lagrangiano dada
por ecs.~(\ref{eq:chl2})-(\ref{eq:chl4}) para las contribuciones m\'etricas y
ecs.~(\ref{eq:chl2R})-(\ref{eq:chl4R}) para las contribuciones con
curvatura. Los valores que se obtienen para la constante de desintegraci\'on
d\'ebil del pi\'on y el condensado de quarks en el l\'{\i}mite quiral son
\begin{eqnarray}
f_\pi^2 &=& - \frac{ 4 N_c}{(4\pi)^2} \rho_2' \,, \\ 
f_\pi^2 B_0 &=& -
\langle \bar q q  \rangle =
\frac{4N_c}{(4\pi)^2} \rho_3'\,, \label{eq:fB0_qq_sqm}
\end{eqnarray} 
y los coeficientes LEC's se escriben
\begin{eqnarray}
L_3 &=& -2 L_2 = -4 L_1 = -\frac{N_c}{(4 \pi)^2} \frac{\rho_0}6,  \qquad
L_4 = L_6=0\,, \nonumber \\ L_5 &=& -\frac{N_c}{(4\pi)^2} \frac{\rho_1'}{2 B_0}\,,  \qquad L_7 = \frac{ N_c }{(4 \pi)^2} \frac1{2 N_f}\left(  \frac{\rho_1'}{2B_0}
+ \frac{\rho_0}{12} \right) \,, \nonumber \\
L_8 &=& \frac{N_c}{(4\pi)^2} \left[ \frac{\rho_2'}{ 4 B_0^2 } - 
\frac{\rho_1'}{4 B_0} - \frac{\rho_0}{24} \right] \,, \qquad L_9 = -2 L_{10} =
\frac{ N_c }{(4 \pi)^2} \frac{\rho_0}3 \,, \nonumber \\ 
L_{12} &=& - 2 L_{11} = -\frac{N_c}{(4\pi)^2} \frac{\rho_0}6 \,, \qquad L_{13}
= -\frac{N_c}{(4\pi)^2} \frac{\rho_1^\prime}{12 B_0} = \frac16 L_5\,,
\label{eq:LEC_sqm_list} \\ 
H_0 &=& -\frac{f_\pi^2}4 \frac{N_f}{6}\,, \qquad H_1 = \frac{ N_c }{(4 \pi)^2} \frac{\rho^\prime_0}6 \,, \qquad H_2 = \frac{ N_c }{(4 \pi)^2} \left( \frac{\rho^\prime_2}{B_0^2} +  
\frac{\rho^\prime_1}{2 B_0} + \frac{\rho_0}{12} \right) \,, \nonumber\\
H_3 &=& \frac{N_c}{(4\pi)^2} N_f \frac{\rho_0'}{144}\,, \qquad H_4 = -\frac{N_c}{(4\pi)^2} N_f \frac{\rho_0'}{90}\,, \qquad H_5
= -\frac{N_c}{(4\pi)^2} N_f \frac{7\rho_0'}{720}\,. \nonumber 
\label{eq:lec's_g}
\end{eqnarray} 
El valor para $L_7$ se corresponde con el modelo SU(3) de sabor. Para el modelo U(3), se obtiene del c\'alculo que $L_7=0$, pero entonces el t\'ermino de ec.~(\ref{eq:singlet}) deber\'{\i}a ser a\~nadido, de modo que el valor de~$L_7$ se modificar\'{\i}a.

Como vemos, los coeficientes $L_1, L_2,L_3, L_4,L_6 , L_9 , L_{10} $ son
n\'umeros puros, y coinciden con los que se esperan en el l\'{\i}mite en que
la regularizaci\'on se elimina~\cite{Espriu:1989ff}. Esto tiene que ver con el
car\'acter adimensional de las LEC's, y que involucran por tanto el momento
cero $\rho_0=1$. El hecho de que~$H_1$ sea proporcional a~$\rho_0^\prime$ se
corresponde con una funci\'on de onda del campo gauge dependiente de la escala, o divergente. Quiere esto decir que la parte finita de~$H_1$ depende del esquema de regularizaci\'on.

A partir de los valores de $f_\pi^2 = 93.2\,{\rm MeV}$ y $L_5= 2.1\cdot 10^{-3}$~\cite{Ecker:1988te}, se obtiene
\begin{eqnarray}
L_7&=&-\frac{L_5}{2N_f}+\frac{N_c}{384 \pi^2 N_f} \simeq -0.09 \cdot
10^{-3}, \nonumber \\
 L_8&=&\frac{L_5}{2}-\frac{N_c}{384
\pi^2}-\frac{f_\pi^2}{16B_0^2} \simeq 0.13 \cdot 10^{-3}, \nonumber \\
H_2&=&-L_5+\frac{N_c}{192 \pi^2}-\frac{f_\pi^2}{4B_0^2} \simeq -1.02 \cdot
10^{-3} .
\end{eqnarray}

En cuanto a las contribuciones con curvatura, el valor no nulo de $H_0$
conduce a una corre\-cci\'on fuerte para la constante gravitatoria de
Newton~$G$. Esta correcci\'on es proporcional al cociente entre la escala
hadr\'onica y la escala de Planck $2\pi N_f f_\pi^2 G/3$, lo cual es num\'ericamente despreciable.

\section{Resultados para el Modelo de Dominancia Vectorial}
\label{MDM_results}

Hasta ahora todas nuestras consideraciones han sido hechas para una funci\'on espectral general sujeta a una serie de propiedades que deben cumplir sus momentos y momentos logar\'{\i}tmicos. Es deseable construir una forma expl\'{\i}cita para esta funci\'on pues esto conducir\'a a importantes consecuencias fenomenol\'ogicas del modelo. Con este fin, en ref.~\cite{spectralqm} se adopta la siguiente expresi\'on para el factor de forma del pi\'on
\begin{equation}
F_V(t)=\frac{M_V^2}{M_V^2+t} \,,
\label{eq:mqeFv}
\end{equation}
donde $M_V$ indica la masa del mes\'on $\rho$. Esta forma corresponde al {\it
  esquema de dominancia del mes\'on vectorial}, que reproduce muy bien los
datos experimentalres recientes \cite{volmer}. La expresi\'on del factor de forma del pi\'on que se deriva del modelo espectral depende de los momentos pares y negativos de $\rho(\omega)$. Por comparaci\'on con (\ref{eq:mqeFv}) se llega a la siguiente identificaci\'on \cite{spectralqm}
\begin{equation}
\rho_{2-2n} 
= \frac{2^{2n+3}\pi^{3/2}f_\pi^2}{N_c M_V^{2n}}
\frac{n\Gamma(n+3/2)}{\Gamma(n+1)} \,,
\qquad
n = 1, 2, 3, \ldots
\label{eq:mqempar}
\end{equation}
La condici\'on $\rho_0=1$ conduce a 
\begin{equation}
f_\pi^2 = \frac{N_c M_V^2}{24\pi^2} \,,
\label{eq:fpi_mrho}
\end{equation} 
que es una relaci\'on que se obtiene a menudo en los modelos de quarks quirales cuando se considera este esquema de dominancia. Esto proporciona una estimaci\'on razonable de la masa del mes\'on $\rho$,  $M_V = 826\,\textrm{MeV}$ para $f_\pi = 93\, \textrm{MeV}$, y $M_V=764\,\text{MeV}$ para $f_\pi = 86 \, \text{MeV}$ en el l\'{\i}mite quiral. 

Notar que si en (\ref{eq:mqempar}) hici\'eramos una prolongaci\'on an\'alitica en el \'{\i}ndice $n$, obtendr\'{\i}amos para los momentos positivos $\rho_{2n}=0$, $n= 2, 3, \ldots$ debido a que la funci\'on $\Gamma(n)$ presenta singularidades en enteros no positivos. Los momentos logar\'{\i}tmicos de $\rho(\omega)$ se pueden eva\-luar f\'acilmente mediante prolongaci\'on anal\'{\i}tica de los momentos $\rho_n$ en el plano complejo de $n$ \cite{spectralqm},
\begin{equation}
\rho_n^\prime = \int_{\cal C} d\omega\,\log(\omega^2)\omega^n\rho(\omega)
= 2 \frac{d}{dz}\int_{\cal C} d\omega\,\omega^z \rho(\omega)\Big|_{z=n}
= 2\frac{d}{dz}\rho_z\Big|_{z=n} \,,
\end{equation}
que conduce a
\begin{equation}
\rho_{2n}^\prime = \left(-\frac{M_V^2}{4}\right)^n
\frac{\Gamma(n)\Gamma(\frac{5}{2}-n)}{\Gamma(5/2)}
\,, \qquad
n = 1, 2, 3,\ldots  
\end{equation}
Los momentos contienen toda la informaci\'on necesaria para c\'alculos
pr\'acticos, sin embargo resulta interesante escribir una f\'ormula
expl\'{\i}cita para la funci\'on espectral. El problema matem\'atico consiste
en invertir la f\'ormula $\rho_{2n}=\int_{\cal C} d\omega\, \omega^{2n}\rho_V(\omega)$, con los momentos dados por (\ref{eq:mqempar}). La soluci\'on del problema conduce a \cite{spectralqm}
\begin{equation}
\rho_V(\omega)=\frac{1}{2\pi i}\frac{1}{\omega}
\frac{1}{(1-4\omega^2/M_V^2)^{d_V}} \,,
\label{eq:rov}
\end{equation}
con $d_V=5/2$. Esta funci\'on presenta un polo simple en el origen, y cortes de rama que empiezan en $\omega =\pm M_V/2$. 

La funci\'on espectral vector, $\rho_V$, corresponde a la parte par de la funci\'on~$\rho$: $\rho_V(\omega) = \left(\rho(\omega) + \rho(-\omega)\right)/2$. Para la parte impar, que denominaremos funci\'on espectral escalar, $\rho_S(\omega) = \left(\rho(\omega) - \rho(-\omega)\right)/2$, debe suponerse una cierta forma funcional que sea adecuada, que satisfaga las condiciones espectrales impares $\rho_{2n+1}=0$, $n \ge 0$, y reproduzca el valor del momento logar\'{\i}tmico~$\rho^\prime_3= -4\pi^2\langle \bar{q}q\rangle/N_c$, (ec.~(\ref{eq:fB0_qq_sqm})). En ref.~\cite{spectralqm} se sugiere una forma an\'aloga a ec.~(\ref{eq:rov}),
\begin{equation}
\rho_S (\omega) = 
\frac{1}{2\pi i} \frac{16(d_S-1)(d_S-2)\rho^\prime_3} {M_S^4
(1-4\omega^2/M_S^2)^{d_S}} \,. \label{rhos}
\end{equation} 
Los datos del ret\'{\i}culo para la masa constituyente de los quarks favorece el valor $d_S=5/2$~\cite{spectralqm}.

En el modelo de dominancia vectorial (MDM), el propagador del quark de ec.~(\ref{eq:prop_q_sqm_re}) se escribe
\begin{eqnarray}
S({p}) = \int_{\cal C} d \omega \frac{ \rho_V (\omega) \slashchar{p} +
\rho_S(\omega)\omega }{ p^2 - \omega^2 } = \frac{Z(p^2) }
{\slashchar{p} -M(p^2)}\,,
\label{eq:spec1} 
\end{eqnarray} 
donde el contorno de integraci\'on ${\cal C}$ consta de dos partes. La primera
comienza en~$+\infty-i0$~siguiendo el eje real positivo, rodea el polo
$+M_V/2$ haciendo una media circunferencia en el sentido de las agujas del
reloj, y vuelve a $+\infty+i0$ siguiendo el mismo eje real positivo. La
segunda parte del contorno comienza en $-\infty+i0$ y sigue el eje real
negativo hasta el polo $-M_V/2$, lo rodea en sentido de las agujas del reloj,
y vuelve a $-\infty-i0$ siguiendo el mismo eje real negativo. Estas dos
secciones est\'an conectadas en el infinito con semic\'{\i}rculos. Este
contorno de integraci\'on es el que se usa para~$\rho_V$. Para $\rho_S$ se
considera el mismo contorno~${\cal C}$, salvo que los polos est\'an en~$\pm M_S/2$.

En este modelo se obtienen los siguientes valores para los momentos logar\'{\i}tmicos
\begin{eqnarray}
{\rho^\prime_1}^{\rm MD} &=& \frac{ 8 \pi^2 \langle \bar q q \rangle
}{ N_c M_S^2} = - \frac{5 M_Q M_S^2 }{6 M_V^2}\,, \nonumber \\
{\rho^\prime_2}^{\rm MD} &=& - \frac{4 \pi^2 f_\pi^2 }{N_c} = - \frac{M_V^2}{6} \,,\\ 
{\rho^\prime_3}^{\rm MD} &=& - \frac{ 4 \pi^2 \langle \bar q q \rangle
}{N_c}= \frac{5 M_Q M_S^4}{12 M_V^2} \nonumber \,,
\label{eq:md}
\end{eqnarray} 
donde $M_Q$ es la masa constituyente de los quarks, que viene dada por~\cite{spectralqm}
\begin{eqnarray}
M_Q \equiv M(0) = - \frac{48 M_V^2 \pi^2 \langle \bar q q \rangle }{5 N_c M_S^4 }\,.
\end{eqnarray} 
Haciendo uso de estos valores se tiene
\begin{eqnarray}
L_5 &=& \frac{N_c}{96 \pi^2} \frac{M_V^2} {M_S^2} \,,\label{eq:mdd}\\ L_7 &=&
\frac{N_c}{32 \pi^2 N_f} \left( \frac1{12} -\frac{M_V^2} {6 M_S^2}
\right), \\ L_8 &=& \frac{N_c}{16 \pi^2}\left( - \frac{M_V^{10}}{150
M_Q^2 M_S^8} + \frac{M_V^2}{12 M_S^2} - \frac1{24} \right) .
\end{eqnarray} 
En la tabla~\ref{tab:table2} del cap\'{\i}tulo~\ref{tensor_EM_MQQ} se muestran lo resultados correspondientes al modelo quark espectral en su realizaci\'on MDM para las constantes~$L_{5,7,8}$, as\'{\i} como las predicciones para~$L_{1,2,3,4,6,9,10}$, que son comunes al esquema de~\cite{Espriu:1989ff}.  Adem\'as aparecen los coeficientes $L_{11-13}$, correspondientes a las contribuciones con curvatura del lagrangiano quiral. Estos valores num\'ericos se han obtenido considerando $M_V=770\,\text{MeV}$, $M_S=970(21)\,\text{MeV}$ y $M_Q=303(24)\,\text{MeV}$.~\footnote{Para una discusi\'on sobre estos resultados y su comparaci\'on con otros modelos, ver secci\'on~\ref{resultados_coef_GLD}. Estos valores de $M_S$ y $M_Q$ se han obtenido en ref.~\cite{spectralqm} a partir de un ajuste con el modelo espectral de los datos para la masa constituyente de los quarks obtenidos en el ret\'{\i}culo~\cite{Bowman:2002bm}.}

Para el modelo espectral en su versi\'on SU(2) de sabor, en ausencia de
correcciones de loops mes\'onicos, se tiene\footnote{Hacemos uso de las
  relaciones dadas en ref.~\cite{gasser-leutwyler2} para pasar de la forma del
  lagrangiano quiral en SU(3) a la forma en SU(2). Estas relaciones son~$\bar l_1
= 192 \pi^2 ( 2 L_1 + L_3) $, $\bar l_2 = 192 \pi^2 L_2 $, $\bar l_3 = 256 \pi^2 ( 2 L_4 + L_5 - 4 L_6 - 2 L_8 ) $, $\bar l_4 = 64 \pi^2 ( 2
L_4 + L_5) $, $\bar l_5 = - 192 \pi^2 L_{10}$, $ \bar l_6 = 192 \pi^2
L_9 $, $\bar l_{11} = 192 \pi^2 L_{11}$ , $\bar l_{13} = 256 \pi^2 l_{13}$. La constante $l_{12}$ no est\'a renormalizada por el loop pi\'onico.}
\begin{eqnarray} 
\bar l_1 &=& - \bar l_2 = - \frac12 \bar l_5 = - \frac14 \bar l_6 =
-N_c\,, \\ 
\bar l_3 &=& \frac{4 N_c }3 + \frac{16 N_c M_V^{10}}{75 M_Q^2 M_S^8}\,, \\
\bar l_4 &=& \frac{ 2 N_c M_V^2  }{3 M_S^2}\,. 
\end{eqnarray} 
Los radios cuadr\'aticos medios vector y escalar del pi\'on vienen dados por~\cite{gasser-leutwyler1} 
\begin{equation}
\langle r^2 \rangle_V = \frac1{16\pi^2 f_\pi^2 } \bar l_6 =\frac{6}{M_V^2}\,,\qquad
\langle r^2 \rangle_S = \frac{3}{8 \pi^2 f_\pi^2} \bar l_4 =\frac{6}{M_S^2}\,.
\label{eq:radii} 
\end{equation} 
Las componentes escalar (esp\'{\i}n-0) y tensorial (esp\'{\i}n-2) de los factores de forma gravitacionales ($\theta_0$ y $\theta_2$ respectivamente)~\cite{Donoghue:1991qv}, producen el mismo radio cuadr\'atico medio
\begin{eqnarray}
\langle r^2 \rangle_{G,0} &=&  \langle r^2 \rangle_{G,2} =
\frac{N_c}{48\pi^2 f_\pi^2 } \,, 
\end{eqnarray}
independientemente de la realizaci\'on particular del modelo espectral. Si saturamos los factores de forma con mesones escalares y tensoriales~$f_0$ y $f_2$, para sus masas se tiene
\begin{eqnarray}
M_{f_0}= M_{f_2} = 4 \pi f_\pi \sqrt{3/N_c} = 1105-1168 \,{\rm MeV} \,,  
\end{eqnarray} 
dependiendo de si se toma~$f_\pi=88$ o $93$ MeV, respectivamente. El valor experimental para el mes\'on tensorial m\'as ligero es~$M_{f_2}^{\rm exp}= 1270 \,{\rm MeV} $. Tal y como se discute en~\cite{Donoghue:1991qv}, el factor de forma~$\theta_0$ (correspondiente a la traza del tensor energ\'{\i}a-impulso) se acopla con mesones escalares, mientras que~$\theta_2$ (correspondiente a la parte de~$\theta_{\mu\nu}$ sin traza) se acopla con mesones tensoriales~(esp\'{\i}n-2).

Hay que decir que el mes\'on escalar de masa~$M_{f_0}$, que domina el tensor energ\'{\i}a-impulso, no necesariamente coincide con el mes\'on escalar de masa~$M_S$, que domina el factor de forma escalar. En realidad se tiene~$M_{f_0}=\sqrt{2}M_V$, mientras que~$M_S$ es una magnitud libre. Esto surge de manera natural en la aproximaci\'on espectral, donde el factor de forma escalar~$F_S$ en el l\'{\i}mite quiral involucra los momentos impares, mientras que~$\theta_0$ involucra los pares. En particular, los radios cuadr\'aticos medios son proporcionales a~$\rho_1^\prime$ y $\rho_0$, respectivamente.

\section{L\'{\i}mite de $N_c$ grande y Dualidad}
\label{large_Nc_duality}

En virtud del hecho de que nuestro resultado se ha obtenido en la
aproximaci\'on de un loop de quarks,\footnote{El modelo espectral no se ha
  desarrollado m\'as all\'a de un loop.} no podemos esperar que el modelo d\'e
mejores resultados para las LEC's que la contribuci\'on de orden m\'as bajo en
un contaje en~$N_c$, el cual est\'a formado por un n\'umero infinito de
intercambios de resonancias~\cite{Pich:2002xy}. Por otra parte, el c\'alculo de estas contribuciones en~$N_c$ grande requiere el uso de suposiciones adicionales, tales como la convergencia de una serie infinita de estados y, por otra parte, una estimaci\'on de las contribuciones de las resonancias m\'as altas. En la pr\'actica, se puede trabajar en la aproximaci\'on de una \'unica resonancia (SRA), lo cual conduce a una reducci\'on de los par\'ametros~\cite{Donoghue:1991qv,Pich:2002xy}:
\begin{eqnarray} 
2 L_1^{\rm SRA} &=& L_2^{\rm SRA} = \frac14 L_9^{\rm SRA} = - \frac13
L_{10}^{\rm SRA} = \frac{f_\pi^2}{8 M_V^2}\,,  \label{eq:SRA1}\\ 
L_5^{\rm SRA} &=&
\frac{8}{3} L_8^{\rm SRA} = \frac{f_\pi^2}{4 M_S^2}\,, \\ 
L_3^{\rm SRA} &=& -
3 L_2^{\rm SRA} + \frac12 L_5^{\rm SRA}\,,  \label{eq:SRAmitad} \\ 
2 L_{13}^{\rm SRA} &=& 3 L_{11}^{\rm SRA} + L_{12}^{\rm SRA} =
\frac{f_\pi^2}{4 M_{f_0}^2}\,, \label{eq:SRA13}\\ 
L_{12}^{\rm SRA} &=& -\frac{f_\pi^2}{2 M_{f_2}^2}\,, 
\label{eq:SRA2}
\end{eqnarray} 
donde~$f_\pi$, $M_V$ y $M_S$ indican las contribuciones de orden m\'as bajo en $N_c$ para estas magnitudes. En la obtenci\'on de estas f\'ormulas para~$L_1-L_{10}$, se han ajustado las contribuciones de los mesones pseudoescalares y axiales con objeto de reproducir las reglas de suma quirales para las funciones de correlaci\'on de dos puntos VV-AA y SS-PP, adem\'as de exigir un comportamiento convergente a altas energ\'{\i}as para los factores de forma hadr\'onicos.\footnote{En particular, $M_P/M_S=M_A/M_V=\sqrt{2}$, donde~$M_P$ es la masa del pi\'on excitado.} Obviamente, el imponer m\'as ligaduras a cortas distancias implica el uso de m\'as resonancias. 

Los valores de~$L_{11,12,13}$ se han obtenido del intercambio de una \'unica
resonancia escalar y tensorial~\cite{Donoghue:1991qv}. Por una parte, es
necesario considerar un mes\'on tensorial con objeto de proporcionar un valor
no nulo para~$L_{12}$, y por otra parte, los mesones tensoriales contribuyen
tambi\'en a otras LEC's~\cite{Toublan:1995bk}, lo cual no est\'a tenido en
cuenta en ecs.~(\ref{eq:SRA1})-(\ref{eq:SRA2}). Por tanto, con objeto de
simplificar la discusi\'on, en lo que sigue nos restringiremos a los
acoplamientos no gravitacionales~$L_1-L_{10}$. Notar que, si bien el poder
predictivo es grande, se consigue en t\'erminos de dos razones
adi\-men\-sio\-nales~$f_\pi/M_V$ y $f_\pi/M_S$. Obviamente, en el l\'{\i}mite
quiral se espera que tanto $M_V$ como $M_S$ escalen como~$f_\pi$. Por tanto,
con objeto de preservar las reglas de contaje en $N_c$ grande, se deber\'{\i}a
tener que
\begin{equation}
M_V=c_V f_\pi/\sqrt{N_c}  \,, \qquad
M_S=c_S f_\pi/\sqrt{N_c} \,, 
\end{equation}
donde~$c_V$ y $c_S$ son coeficientes independientes de~$N_c$. El hecho sorprendente es que en el modelo quark espectral, las constantes de baja energ\'{\i}a dependen de las razonas adimensio\-nales~$\rho_1^\prime/B_0$ y $\rho_2^\prime/B_0^2$. En vista de esto, resulta tentador calcular los momentos logar\'{\i}tmicos espectrales a partir de las reglas de $N_c$ grande, de un modo que sea modelo-independiente. En primer lugar vemos que las razones~$L_1 : L_2 : L_9$ en el modelo quark espectral coinciden con las de SRA. Los valores de~$L_5$ y $L_8$ pueden ser usados para determinar~$\rho_1^\prime$ y $\rho_2^\prime$ respectivamente, de modo que se tiene
\begin{eqnarray}
{\rho^\prime_1}^{\rm SRA} &=&  \frac{ 8 \pi^2 \langle \bar q q \rangle }{ N_c
M_S^2}, \label{eq:rho_sra1}\\
{\rho^\prime_2}^{\rm SRA} &=& - \frac{4 \pi^2 f_\pi^2 }{N_c} = -
\frac{M_V^2}{6} ,
\label{eq:rho_sra2} 
\end{eqnarray}
lo cual est\'a de acuerdo con ecs.~(\ref{eq:mdd}) y (\ref{eq:fpi_mrho}). Esto no es sorprendente, pues la f\'{\i}sica de SRA y del modelo quark espectral en su versi\'on MDM es similar. La \'unica diferencia es que de ecs.~(\ref{eq:rho_sra1})-(\ref{eq:rho_sra2}) no se puede deducir el valor de la masa constituyente de los quarks~$M_Q=M(0)$, que viene dada por el cociente~$M_Q=\rho_{-1}/\rho_{-2}$ (ecs.~(\ref{eq:ab_sqm})-(\ref{eq:mz_sqm})). Para determinar~$M_Q$ ser\'{\i}a necesario calcular los t\'erminos de~${\cal O}(p^6)$ en el lagrangiano quiral y comparar con SRA en el l\'{\i}mite~$N_c$ grande.

Por otra parte, no es posible hacer compatibles $L_8$ o $L_{10}$. El desacuerdo con los corres\-pon\-dientes valores en~$N_c$ grande se debe a que el modelo espectral viola la regla de suma SS-PP y la segunda regla de Weinberg VV-AA. Esta violaci\'on tambi\'en ocurre en otros modelos de quarks~\cite{Peris:1998nj,Bijnens:2003rc} (no ocurre en los modelos no locales; ver~\cite{WBcoim,Dorokhov:2003kf}). En efecto, en el modelo no existe intercambio de mes\'on axial en~$L_{10}$ ($1/4$ de la contribuci\'on total) ni de mes\'on pseudoescalar en~$L_8$ ($1/4$ de la contribuci\'on total). Por otra parte, para el valor de~$f_\pi$ que se obtiene de ec.~(\ref{eq:fpi_mrho}), las constantes $L_1$, $L_2$, $L_4$, $L_5$, $L_6$, $L_9$ reproducen las identidades en $N_c$ grande que aparecen en~\cite{Ecker:1988te}. Este acuerdo se puede ver en la tabla~\ref{tab:table2} si se considera un factor de correcci\'on~ $ 24\pi^2 f_\pi^2 / N_c M_V^2 = 1.15$. Se podr\'{\i}a forzar que $L_3$ coincidiera con la estimaci\'on de $N_c$ grande tomando~$M_V=M_S$. Esto concuerda con la observaci\'on en la aproximaci\'on unitaria quiral de ref.~\cite{Pelaez:2003dy}, de que en el l\'{\i}mite de~$N_c$ grande, los mesones escalar y vector son degenerados.\footnote{Para~$N_c=3,10,20,40$, en ref.~\cite{Pelaez:2003dy} se obtiene~$M_S /M_V = 0.58,0.84,0.96, 0.98 $, respectivamente, con $M_S$ y $M_V$ las partes reales de los polos en la segunda hoja de Riemann.} Por tanto, el intentar compatibilizar el l\'{\i}mite de $N_c$ grande en la SRA con el modelo quark espectral produce una degeneraci\'on de los mesones escalar y vector. Esta degeneraci\'on fue sugerida en~\cite{Gilman:1967qs} en el contexto de reglas de suma superconvergentes y han sido interpretadas m\'as recientemente en base a simetr\'{\i}as que se restablecen~\cite{Weinberg:xn}.

Parece claro que cualquier modificaci\'on en el modelo quark espectral afectar\'a \'unicamente a~$L_8$ y $L_{10}$. Si se considera~$M_S=M_V=2 \pi f_\pi \sqrt{6/N_c} $ para $N_c$ grande en la aproximaci\'on SRA, se obtienen las siguientes relaciones de dualidad
\begin{equation} 
2 L_1 = L_2 = - \frac12 L_3 =\frac12 L_5 = \frac23 L_8= \frac14 L_9 =
- \frac13 L_{10} = \frac{N_c}{192\pi^2} \,.
\label{eq:L_dual} 
\end{equation} 
Esto conduce a las relaciones de dualidad para las masas
\begin{eqnarray}
M_A = M_P = \sqrt{2} M_V = \sqrt{2} M_S = 4\pi \sqrt{\frac{3}{N_c}} f_\pi \, .   
\end{eqnarray} 
La nueva relaci\'on~$M_A=M_P$ concuerda con el valor experimental dentro del error del~$30\%$ que se espera de considerar el l\'{\i}mite $N_c$ grande. Haciendo uso de ec.~(\ref{eq:radii}) se obtiene
\begin{equation}
\langle r^2 \rangle^{1/2}_S = \langle r^2 \rangle^{1/2}_V = \frac{\sqrt{N_c}}{2\pi f_\pi} \,.
\end{equation}  
Estas relaciones est\'an sujetas a correcciones en $m_\pi$ y en \'ordenes m\'as altos en~$N_c$. Num\'ericamente se tiene
\begin{equation}
\langle r^2 \rangle^{1/2}_S = \langle r^2 \rangle^{1/2}_V = 0.58-0.62~{\rm fm} \,,
\end{equation}
dependiendo de si se toma $f_\pi=88$ o $93$ MeV. El valor del radio escalar es pr\'oximo al que se obtiene de TQP hasta dos loops~\cite{Colangelo:2001df}, $0.78\,{\rm fm}$.

En el caso SU(2), el modelo de dualidad con $N_c$ grande conduce a
\begin{eqnarray}
- \bar l_1 =  \bar l_2 = \frac32 \bar l_3 = \frac32 \bar l_4  =
  \frac13 \bar l_5 = \frac14 \bar l_6 = N_c \,.
\end{eqnarray} 
Los valores recientes obtenidos a partir del an\'alisis de la colisi\'on~$\pi\pi$ a nivel de dos loops~\cite{Colangelo:2001df} y de factores de forma vector y escalar~\cite{Bijnens:1998fm} a dos loops son
\begin{eqnarray} 
\bar l_1 &=& -0.4 \pm 0.6 \, , \qquad 
\bar l_2 = 6.0 \pm 1.3  \, , \qquad 
\bar l_3 = 2.9 \pm 2.4,  \nonumber \\ 
\bar l_4 &=& 4.4 \pm 0.2  \, , \qquad \quad
\bar l_5 = 13.0 \pm 1.0 \, , \quad
\bar l_6 = 16.0 \pm 1.0  \,.
\end{eqnarray} 
Los coeficientes~$\bar l$ son  m\'as susceptibles de poder compararse con TQP ya que los loops quirales generan un cambio constante $c=\log (\mu^2/m^2)$, que es el mismo para todos ellos. Por tanto, tiene sentido comparar diferencias donde los logaritmos se cancelan.
\begin{eqnarray} 
\bar l_2 -\bar l_1 &=& 2 N_c \quad ( {\rm Exp.}\, 6.4 \pm 1.4 )\,,
\nonumber \\ \bar l_3 -\bar l_1 &=& \frac{5N_c}3 \quad ( {\rm Exp.}\,
3.3 \pm 2.5 )\,, \nonumber \\ \bar l_4 -\bar l_1 &=& \frac{5N_c}3 \quad
( {\rm Exp.}\, 4.8 \pm 0.6 )\,, \\ \bar l_5 -\bar l_1 &=& 4 N_c \quad (
{\rm Exp.}\, 13.4 \pm 1.2 )\,, \nonumber \\ \bar l_6 -\bar l_1 &=& 5 N_c
\quad ( {\rm Exp.}\, 16.4 \pm 1.2 )\,. \nonumber 
\end{eqnarray}
El acuerdo es excelente dentro de las incertidumbres, y esto sugiere una precisi\'on del orden de~$1/N_c^2$ en lugar de la que cabr\'{\i}a esperar {\it a priori}~$1/N_c$.

El cambio constante de los loops pi\'onicos se produce con una escala~$\mu=513 \pm 200\,{\rm MeV}$, lo cual es comparable con la masa del mes\'on~$\rho$. Considerando las ecs.~(\ref{eq:SRA1})-(\ref{eq:SRAmitad}) corres\-pondientes a SRA, con los valores f\'{\i}sicos~$f_\pi=93.2\,\textrm{MeV}$, $M_S=1000\;\textrm{MeV}$ y $M_V=770\,\textrm{MeV}$, tal y como se hace en~\cite{Pich:2002xy}, se tiene
\begin{equation}
\bar l_2 - \bar l_1 = 8.3\,, \quad \bar
l_3 - \bar l_1 = 6.2\,, \quad \bar l_4 - \bar l_1 = 6.2\,, 
\quad \bar l_5 - \bar l_1 = 15.2\,, \quad \bar l_6 - \bar l_1= 18.7\,. 
\end{equation}
Se podr\'{\i}an obtener unos valores m\'as razonables considerando~$M_S=600\,{\rm MeV}$, pero entonces la relaci\'on SRA, $M_P=\sqrt{2}M_S$, predicir\'{\i}a un valor demasiado peque\~no para la masa del estado pi\'onico excitado.

Esta discusi\'on favorece fenomenol\'ogicamente las relaciones de dualidad~ec.~(\ref{eq:L_dual}) frente a las relaciones de SRA, ecs.~(\ref{eq:SRA1})-(\ref{eq:SRAmitad}), con par\'ametros f\'{\i}sicos.

\section{Conclusiones}
\label{conclusiones_sqm_etc}

En este cap\'{\i}tulo se ha estudiado el desarrollo quiral en el modelo quark
espectral pro\-pues\-to recientemente, en presencia de fuerzas externas
electrod\'ebiles y gravitatorias. El modelo est\'a basado en una
representaci\'on de Lehmann para el propagador del quark con una funci\'on
espectral no convencional, que es en general una funci\'on compleja con cortes
de rama. Se ha escrito la acci\'on efectiva que reproduce las identidades de Ward-Takahashi, y gracias a una serie infinita de condiciones espectrales hemos obtenido la contribuci\'on an\'omala quiral a la acci\'on. Esta contribuci\'on aparece convenientemente normalizada sin necesidad de eliminar la regularizaci\'on. Adem\'as, la contribuci\'on no an\'omala se puede escribir en t\'erminos de 13 constantes de baja energ\'{\i}a. Los valores num\'ericos muestran un acuerdo razonable con los esperados fenomenol\'ogicamente, si bien existen algunas dis\-cre\-pan\-cias para~$L_8$ y $L_{10}$. \'Estas se podr\'{\i}an explicar de manera natural como fallos del modelo a la hora de reproducir las condiciones quirales a cortas distancias, y sugiere que \'este necesita ser mejorado. Por otra parte, si se intenta comparar las LEC's no-gravitacionales restantes con las predicciones de $N_c$ grande en la aproximaci\'on de una \'unica resonancia, tiene lugar una nueva reducci\'on de par\'ametros. En particular, el mejor acuerdo se encuentra para el caso de mesones escalar y vector degenerados.

Se han estimado las LEC's gravitatorias $L_{11}$, $L_{12}$ y $L_{13}$ en el contexto de los modelos de quarks quirales. Estas constantes dependen de las propiedades de curvatura de la m\'etrica en espacio-tiempo curvo. Este c\'alculo permite la determinaci\'on de algunos elementos de matriz del tensor energ\'{\i}a-impulso. Nuestro an\'alisis sugiere que el acoplamiento del mes\'on escalar con el condensado de quarks~$m_0 \overline{q}q$, y el mes\'on escalar acoplado con la traza del tensor energ\'{\i}a-impulso~$\theta_\mu^\mu$, no coinciden necesariamente. Estos dos operadores se comportan de manera diferente bajo simetr\'{\i}a quiral, ya que~$m_0 \overline{q}q$ se anula en el l\'{\i}mite quiral mientras que~$\theta_\mu^\mu$ no lo hace. Esto se materializa en el modelo quark espectral en el hecho de que estos dos mesones escalares dependen de momentos espectrales impares y pares, respectivamente. Por otra parte, se obtiene~$M_{f_0}=M_{f_2}=
\sqrt{2} M_V = \sqrt{2} M_S = 4\pi \sqrt{3/N_c} f_\pi $, que constituye un
resultado muy razonable si tenemos en cuenta la aproximaci\'on de un loop de
quarks en que estamos trabajando. Se han discutido otras relaciones de
dualidad quark-mes\'on, lo cual ha permitido una determinaci\'on bastante
precisa de las LEC's ya conocidas, y se muestran de acuerdo con los valores
conocidos a dos loops dentro de los errores experimentales.

\chapter{Conclusiones}
\label{Conclusiones}

\section{Resumen y Conclusiones}
\label{resumen_conclusiones}

En esta tesis se ha hecho un estudio detallado de algunos efectos de temperatura y de curvatura en QCD y en algunos modelos de quarks quirales. Las conclusiones y logros m\'as significativos de este trabajo han sido los siguientes:
\begin{itemize}
\item Se ha construido un desarrollo del heat kernel invariante gauge orden por orden a temperatura finita, dentro del formalismo de tiempo imaginario, para espacio-tiempo plano. Se ha considerado un tratamiento general v\'alido en cualquier gauge, y en pre\-sen\-cia de campos escalares que pueden ser no abelianos y no est\'aticos. Para preservar la invariancia gauge a temperatura finita se ha hecho uso del loop de Polyakov, y se ha llegado hasta orden 6 en un contaje en dimensiones de masa.

\item Se ha aplicado el desarrollo del heat kernel para el c\'alculo de la acci\'on efectiva de QCD a un loop, incluyendo fermiones sin masa, en la regi\'on de temperaturas grandes. Se ha considerado un loop de Polyakov no est\'atico. Se ha estudiado la invariancia gauge del resultado, y en concreto la rotura expl\'{\i}cita de la simetr\'{\i}a del centro por efecto de los fermiones.
  
\item Se ha obtenido la acci\'on de la teor\'{\i}a efectiva dimensionalmente
  reducida de QCD, v\'alida en el r\'egimen de temperaturas grandes. Esto ha
  permitido obtener nuevos t\'erminos de orden 6 no calculados en la
  literatura, tanto en el sector fermi\'onico como en el glu\'onico.

\item Se ha propuesto un modelo fenomenol\'ogico que permite describir con
  gran \'exito los datos del ret\'{\i}culo tanto para el loop de Polyakov
  renormalizado como para la energ\'{\i}a libre de un quark pesado, en el
  r\'egimen de temperaturas inmediatamente por encima de la transici\'on de
  fase. Este modelo da cuenta de contribuciones no perturbativas provenientes
  de condensados glu\'onicos, y se ha obtenido una predicci\'on para el valor
  del condensado glu\'onico de dimensi\'on 2 en el r\'egimen de temperaturas
  considerado, $T_c \le T \le 6T_c$. El resultado se muestra de acuerdo con otras predicciones existentes tanto a temperatura cero como a temperatura finita.  

\item Se ha estudiado la analog\'{\i}a existente entre el loop de Polyakov y el potencial quark-antiquark a temperatura cero. Esto ha permitido encontrar una relaci\'on entre el condensado glu\'onico de dimensi\'on 2 y la tensi\'on de la cuerda.

\item Se ha introducido el loop de Polyakov de color en los modelos de quarks quirales a nivel de un loop de quarks, siguiendo un esquema de acoplamiento m\'{\i}nimo, y hemos visto que esto permite resolver algunas inconsistencias que presentaban estos modelos en su tratamiento est\'andar a temperatura finita. En concreto, la integraci\'on sobre el grupo gauge da lugar a una conservaci\'on de trialidad, y el contaje en~$N_c$ se muestra de acuerdo con las predicciones de Teor\'{\i}a Quiral de Perturbaciones.

\item Se ha calculado el lagrangiano efectivo quiral a temperatura finita de los modelos Nambu--Jona-Lasinio y Quark Espectral a nivel de un loop de quarks y a nivel \'arbol para los mesones, en la aproximaci\'on quenched, y se ha obtenido una predicci\'on para las constantes de baja energ\'{\i}a de Teor\'{\i}a Quiral de Perturbaciones. 

\item Se han analizado algunas correcciones de orden mayor para los modelos de
  quarks quirales acoplados con el loop de Polyakov. En concreto correcciones
  glu\'onicas, locales, y las provenientes de ir m\'as all\'a de un loop de quarks. Se ha encontrado que los efectos t\'ermicos est\'an exponencialmente suprimidos a temperaturas peque\~nas, y vienen domi\-nados por loops mes\'onicos. Adem\'as, se ha analizado la influencia del determinante fermi\'onico sobre algunos observables como el condensado de quarks y el valor esperado del loop de Polyakov, y se han estudiado sus implicaciones sobre las transiciones de fase quiral y de desconfinamiento de color.

\item  Se ha estudiado el acoplamiento de los modelos de quarks quirales con gravedad, y se ha analizado la correspondiente estructura del tensor energ\'{\i}a-impulso a bajas energ\'{\i}as para cuatro modelos concretos: Quark Constituyente, Georgi-Manohar, Nambu--Jona-Lasinio y Quark Espectral. Se ha obtenido una predicci\'on para los coeficientes de baja energ\'{\i}a correspondientes a los t\'erminos no m\'etricos con contribuciones de curvatura.

\item Se ha obtenido la contribuci\'on an\'omala quiral a la acci\'on efectiva en el modelo quark espectral. Despu\'es de introducir una regularizaci\'on conveniente, el resultado no depende de los detalles de la funci\'on espectral, de modo que coincide con la anomal\'{\i}a de QCD.

\item Se han comparado los resultados del modelo quark espectral para las constantes quirales de baja energ\'{\i}a, con las predicciones de~$N_c$ grande en la aproximaci\'on de una \'unica resonancia. El mejor acuerdo se encuentra para el caso de mesones escalar y vector degenerados, dando lugar a unas relaciones de dualidad quark-mes\'on, que han permitido una determinaci\'on precisa de las constantes de baja energ\'{\i}a conocidas.

\end{itemize}

%
%
%





\newpage
\section{Anexo de art\'{\i}culos publicados}
\label{anexo_articulos_publicados}

Esta tesis est\'a basada en las siguientes publicaciones.

\begin{enumerate}

\item { \bf Revistas internacionales: }

\begin{itemize}

\bibitem{xxMegias:2002vr}
  E.~Meg\'{\i}as, E.~Ruiz Arriola and L.~L.~Salcedo,
  \emph{The Polyakov loop and the heat kernel expansion at finite temperature},
  \emph{Phys. Lett.} {\bf B563}, 173-178 (2003), \newline
  [{\tt arXiv:hep-th/0212237}].

\bibitem{xxMegias:2003ui}
E.~Meg{\'\i}as, E.~Ruiz~Arriola, and L.~L. Salcedo,
\emph{Thermal heat kernel expansion and the one-loop effective action of  QCD at finite temperature},
\newblock \emph{Phys. Rev.} {\bf D69}, 116003 (2004),
 [{\tt arXiv:hep-ph/0312133}].

\bibitem{xxMegias:2004uj}
  E.~Meg\'{\i}as, E.~Ruiz Arriola, L.~L.~Salcedo and W.~Broniowski,
\emph{Low energy chiral Lagrangian from the spectral  quark
  model},
\newblock \emph{Phys. Rev.} {\bf D70}, 034031 (2004),
  [{\tt arXiv:hep-ph/0403139}].

\bibitem{xxMegias:2005fj}
  E.~Meg\'{\i}as, E.~Ruiz Arriola and L.~L.~Salcedo,
\emph{Energy momentum tensor of chiral quark models at low energies},
\newblock  \emph{Phys. Rev.} {\bf D72}, 014001 (2005), \newline
  [{\tt arXiv:hep-ph/0504271}].

\bibitem{xxMegias:2005ve}
  E.~Meg\'{\i}as, E.~Ruiz Arriola and L.~L.~Salcedo,
\emph{Dimension two condensates and the Polyakov loop above the deconfinement phase transition},
\newblock  \emph{JHEP} {\bf 0601}, 073 (2006),
  [{\tt arXiv:hep-ph/0505215}].

\bibitem{xxMegias:2007pq}
  E.~Meg\'{\i}as, E.~Ruiz Arriola and L.~L.~Salcedo,
\emph{Power corrections in the quark-antiquark potential at finite temperature},
\newblock  \emph{Phys. Rev.}  {\bf D75}, 105019 (2007), \newline
  [{\tt arXiv:hep-ph/0702055}].

\bibitem{xxMegias:2004hj}
  E.~Meg\'{\i}as, E.~Ruiz Arriola and L.~L.~Salcedo,
\emph{Polyakov loop in chiral quark models at finite temperature},
\newblock   \emph{Phys. Rev.}  {\bf D74}, 065005 (2006), \newline
  [{\tt arXiv:hep-ph/0412308}].


\bibitem{xxMegias:2006bn}
  E.~Meg\'{\i}as, E.~Ruiz Arriola and L.~L.~Salcedo,
\emph{Chiral Lagrangian at finite temperature from the Polyakov-chiral quark
  model},
\newblock \emph{Phys. Rev.}  {\bf D74}, 114014 (2006), \newline
  [{\tt arXiv:hep-ph/0607338}].

\end{itemize}

\vspace{0.5cm}

\item {\bf Actas de congresos:}
 
\begin{itemize}

\bibitem{xxMegias:2004bj}
  E.~Meg\'{\i}as,
\emph{One-loop effective action of QCD at high temperature using the heat  kernel method}. Actas de 9th Hadron Physics and 8th Relativistic Aspects of Nuclear Physics (HADRON-RANP 2004). 
\newblock \emph{AIP Conf. Proc.}  {\bf 739}, 443-445 (2005),
  [{\tt arXiv:hep-ph/0407052}].

\bibitem{xxMegias:2004kc}
  E.~Meg\'{\i}as, E.~Ruiz Arriola and L.~L.~Salcedo,
\emph{Polyakov loop at finite temperature in chiral quark models}.
Actas de la conferencia Mini-Workshop on Quark Dynamics: Bled 2004.
\newblock \emph{Bled Workshops in Physics}, Vol. 5, No. 1, P\'ag. 1-6  (2004),
  [{\tt arXiv:hep-ph/0410053}].

\bibitem{xxMegias:2004gy}
  E.~Meg\'{\i}as, E.~Ruiz Arriola and L.~L.~Salcedo,
\emph{Chiral lagrangians at finite temperature and the Polyakov loop}.
Actas de 6th International Conference on Quark Confinement and the Hadron Spectrum. 
\newblock  \emph{AIP Conf. Proc.}  {\bf 756}, 436-438 (2005), \newline
 [{\tt arXiv:hep-ph/0411293}].

\bibitem{xxMegias:2005pe} E.~Meg\'{\i}as, E.~Ruiz~Arriola and L.~L.~Salcedo,
  \emph{Non-perturbative contribution to the Polyakov loop above the
    deconfinement phase transition}. Actas de 18th International Conference on
  Ultra-Relativistic Nucleus-Nucleus Collisions: Quark Matter 2005 (QM 2005).
  \emph{Romanian Reports in Physics}, Vol. 58, No. 1, P\'ag. 81-85 (2006),
  \newblock [{\tt arXiv:hep-ph/0510114}].

\bibitem{xxMegias:2005qf}
  E.~Meg\'{\i}as, E.~Ruiz~Arriola and L.~L.~Salcedo,
\emph{Polyakov loop at low and high temperatures}. Actas de 29th Johns Hopkins Workshop in Theoretical Physics.
\newblock  \emph{JHEP Proceedings of Science}, PoS(JHW2005)025, (2006), 
 [{\tt arXiv:hep-ph/0511353}].

\bibitem{xxMegias:2006df}
  E.~Meg\'{\i}as, E.~Ruiz Arriola and L.~L.~Salcedo,
\emph{The quantum and local Polyakov loop in chiral quark models at finite
  temperature}. Actas de 7th International Conference on Quark Confinement and the Hadron Spectrum.
\newblock \emph{AIP Conf. Proc.}  {\bf 892}, 444-447 (2007), 
  [{\tt arXiv:hep-ph/0610095}].

\bibitem{xxMegias:2006ke}
  E.~Meg\'{\i}as, E.~Ruiz Arriola and L.~L.~Salcedo,
\emph{Dimension-2 condensates and Polyakov chiral quark models}. Actas de 4rd International Conference on Quarks and Nuclear Physics.
\newblock \emph{The European Physical Journal}  {\bf A31}, 553-556 (2007), 
  [{\tt arXiv:hep-ph/0610163}].

\end{itemize}

\end{enumerate}

\appendix

\chapter{Transformaciones Gauge}
\label{app:gauge}
En este ap\'endice explicaremos qu\'e se entiende por transformaci\'on gauge y discutiremos ciertas propiedades que cumple una transformaci\'on gauge a temperatura finita. Estu\-dia\-remos la rotura de la simetr\'{\i}a del centro del grupo gauge al considerar una teor\'{\i}a con fermiones. Vamos a seguir en parte la referencia \cite{NPB}.

\section{Definiciones}
Consideremos un operador $f(M,D_\mu)$ construido con $M$ y $D_\mu$ en sentido algebraico. Una configuraci\'on gauge transformada $(M^U,A_\mu^U)$ es una de la forma
\begin{eqnarray}
M^U(x) &=& U^{-1}(x)M(x)U(x) \, , \nonumber  \\
A_\mu^U(x) &=& U^{-1}(x)\partial_\mu U(x) + U^{-1}(x)A_\mu(x)U(x) \, ,  
\end{eqnarray}
donde la transformaci\'on gauge $U(x)$ es una funci\'on que toma valores sobre matrices en el espacio interno. Esta transformaci\'on corresponde a una transformaci\'on de semejanza de $D_\mu$ de la forma $D_\mu^U=\partial_\mu + A_\mu^U(x)=U^{-1}(x)D_\mu U(x)$, donde $U(x)$ se considera que es un operador multiplicativo en el espacio de Hilbert $\cal H$ de las funciones de onda. Debido a que $f(M,D_\mu)$ est\'a construido con $M$, $D_\mu$ y c-n\'umeros, se sigue que $f(M,D_\mu)$ tambi\'en se transforma bajo una transformaci\'on de semejanza
\begin{equation}
f(M^U,D_\mu^U) = U^{-1}f(M,D_\mu)U  \,  .
\end{equation}
$U(x)$ pertenece a cierto grupo gauge G y el campo gauge $A_\mu(x)$ es un elemento del \'algebra de Lie de G. La clase de matrices $M(x)$ debe ser cerrada bajo transformaciones gauge. $U(x)$ debe ser una funci\'on continua del espacio-tiempo y a temperatura finita ha de ser peri\'odica (salvo una posible fase global) como funci\'on de $x_0$. Notar que una transformaci\'on gauge deja invariante el espectro de $f(M,D_\mu)$, por tratarse de una transformaci\'on de semejanza.

\section{Gauges estacionarios}

En c\'alculos expl\'{\i}citos suele ser usual fijar el gauge a trav\'es de la
condici\'on $\partial_0 A_0=0$, que no implica p\'erdida de generalidad ya que
este gauge siempre existe.\footnote{Este gauge es conocido en la literatura como 'gauge de Polyakov', aunque nosotros nos referiremos a \'el tambi\'en como gauge estacionario.} Esto quiere decir que para cada configuraci\'on existe una transformaci\'on gauge que la lleva a la configuraci\'on estacionaria. Una vez fijado este gauge, queda a\'un cierta libertad. Cuando se trabaja en el gauge estacionario, para comprobar la invariancia gauge es necesario encontrar el resto de transformaciones compatibles con este gauge y ver que todas ellas producen el mismo resultado. A continuaci\'on vamos a determinar cu\'al es la transformaci\'on gauge m\'as general de este tipo.

Sean $A_\mu$ y $B_\mu$ dos configuraciones estacionarias y sea $U$ una transformaci\'on gauge que transforma $A_\mu$ en $B_\mu$. Esto quiere decir
\begin{equation}
B_0(\vec{x}) = U^{-1}(x)\partial_0 U(x) + U^{-1}(x)A_0(\vec{x})U(x) \,.
\end{equation}
Notar que el primer t\'ermino cambia la magnitud de $A_0$ y el segundo simplemente lo rota en el espacio interno. Podemos simplificar esta ecuaci\'on si hacemos uso de la variable auxiliar $V(x)= \exp(x_0 A_0(\vec{x})) U(x)$, con lo que queda
\begin{equation}
B_0(\vec{x}) = V^{-1}(x)\partial_0 V(x)\,.
\label{eq:apptg1}
\end{equation}
La soluci\'on m\'as general de (\ref{eq:apptg1}) va a estar formada por una transformaci\'on gauge arbitraria independiente del tiempo y por una transformaci\'on cuya dependencia temporal sea lineal\footnote{Una dependencia no lineal dar\'{\i}a lugar a una contribuci\'on temporal en $B_0$.}
\begin{equation}
V(x)=U_0(\vec{x})e^{x_0 B_0(\vec{x})} \,. 
\end{equation}
Un modo conveniente de escribir la transformaci\'on es haciendo uso del cambio de variable
\begin{equation}
B_0(\vec{x}) = U_0^{-1}(\vec{x})(A_0(\vec{x})+\Lambda(\vec{x}))U_0(\vec{x})\,,
\end{equation}
con lo cual finalmente queda
\begin{equation}
U(x)= e^{-x_0 A_0(\vec{x})}e^{x_0(A_0(\vec{x})+\Lambda(\vec{x}))}U_0(\vec{x})\,.
\end{equation}
Ahora debemos imponer la condici\'on de que $U(x)$ es funci\'on peri\'odica de $x_0$, salvo una posible fase global
\begin{equation}
U(x_0+\beta,\vec{x})=e^{i\alpha}U(x_0,\vec{x})\,.
\label{eq:apptg2}
\end{equation}
Aqu\'{\i} $\alpha$ es una fase global escalar multiplicada por la matriz identidad. Esto conduce a la restricci\'on
\begin{equation}
e^{\beta(A_0(\vec{x})+\Lambda(\vec{x} ))}=e^{i\alpha} e^{\beta A_0(\vec{x})} \,,
\label{eq:apptg3}
\end{equation}
lo cual va a producir una discretizaci\'on en la parte temporal de la transformaci\'on gauge. De (\ref{eq:apptg3}) se deduce que $A_0(\vec{x})$ y $\Lambda(\vec{x})$ deben conmutar con $\exp(\beta A_0(\vec{x}))$. Si el espectro de la matriz unitaria $\exp(\beta A_0(\vec{x}))$ es no degenerado, \'esta puede ser diagonalizada en una base que es esencialmente \'unica e independiente de $\vec{x}$. En este caso $A_0(\vec{x})$ y $\Lambda(\vec{x})$ deben ser diagonales en la misma base y por tanto van a conmutar entre s\'{\i}. Esto da lugar a que la condici\'on sobre $\Lambda$ sea
\begin{equation}
e^{\beta \Lambda(\vec{x})}=e^{i\alpha} \,, \qquad
[A_0(\vec{x}),\Lambda(\vec{x})]=0 \,.
\end{equation}
La primera condici\'on conduce a que los valores propios de $\Lambda(\vec{x})$ sean de la forma $\chi_j = i(\alpha +2\pi n_j)/\beta,\; n_j \in \mathbb{Z}$. Notar que por continuidad estos enteros deben ser independientes de $\vec{x}$. Finalmente la transformaci\'on gauge queda 
\begin{equation}
U(x)=e^{x_0 \Lambda(\vec{x})}U_0(\vec{x}) \,,
\end{equation}
expresi\'on v\'alida cuando el espectro de $\exp(\beta A_0(\vec{x}))$ es no degenerado.

\section{Particularizaci\'on al grupo gauge SU($N_c$)}

\subsection{Simetr\'{\i}a del centro del grupo gauge}
Consideremos espec\'{\i}ficamente el grupo gauge SU($N_c$). En la ecuaci\'on (\ref{eq:apptg2}), tomando en cada miembro el determinante y teniendo en cuenta que $\Det(U)=1$, obtenemos que los valores permitidos de $\alpha$ son cuando $\Det[\exp(i\alpha)]=1$, esto es $\alpha=2\pi n/N_c,\; n \in \mathbb{Z}$. Puesto que solamente est\'an permitidos valores discretos para $\alpha$, esto implica que la matriz $\Lambda$ debe ser independiente de $\vec{x}$, por continuidad. Como ejemplo, en SU(2) los valores propios de $\Lambda$ son de la forma $\chi_j=i\pi n_j/\beta,\; n_j \in \mathbb{Z}$. Para SU($N_c$), con $N_c>2$, es siempre posible elegir una representaci\'on fundamental en la cual todos los generadores diagonales excepto uno tengan al menos un valor propio cero [por ejemplo, las matrices de Gell-Mann $\lambda_3$ y $\lambda_8$ para SU(3)]. La transformaci\'on $U$ se escribir\'a
\begin{equation}
U(x) = \exp(x_0 \lambda_a \Lambda^a) U_0(\vec{x})\,, 
\end{equation}
donde $\lambda_a/2i$ son los generadores diagonales del grupo. Los t\'erminos $\Lambda^a$ correspondientes a cada uno de los generadores con un valor propio cero deben ser de la forma $\Lambda^a =i2\pi n_a/\beta$. El otro generador $\lambda_{N_c^2-1}$ viene dado por
\begin{equation}
\lambda_{N_c^2-1}=\diag(1,1,\cdots,1-N_c)\rho \,,
\label{eq:p1}
\end{equation}
donde $\rho$ es un factor de normalizaci\'on. En este caso $\Lambda^{N_c^2-1}=i2\pi n/(N_c\beta)$ dan lugar a transformaciones gauge permitidas. Esto quiere decir que adem\'as de la simetr\'{\i}a gauge SU($N_c$), exite una simetr\'{\i}a extra global ${\mathbb Z}(N_c)$, que es el centro del grupo gauge. Esta simetr\'{\i}a es generada por la acci\'on de transformaciones gauge locales que son peri\'odicas en la variable temporal, salvo un elemento arbitrario de ${\mathbb Z}(N_c)$,
\begin{equation} 
U(x_0+\beta,\vec{x})= z\, U(x_0,\vec{x})\,, \qquad z=e^{i2\pi n/N_c} \,,
\end{equation}
m\'odulo transformaciones gauge locales estrictamente peri\'odicas.

\subsection{Rotura expl\'{\i}cita de la simetr\'{\i}a del centro}

La situaci\'on cambia si hay fermiones en la teor\'{\i}a. Puesto que los fermiones transforman como $\psi \rightarrow U \psi$, no hay factores $U^{-1}$ que cancelen la fase global. Por tanto, con objeto de que las condiciones de contorno temporales para fermiones queden inalteradas bajo transformaciones gauge, s\'olo est\'an permitidas transformaciones que satisfagan~(\ref{eq:apptg2}) con $\alpha=0$. Esto quiere decir que los fermiones rompen la simetr\'{\i}a del centro del grupo gauge que est\'a presente en todas las teor\'{\i}as gauge puras. En consecuencia, la forma m\'as general de $\Lambda^a$ para una teor\'{\i}a SU($N_c$) con fermiones es $\Lambda^a=i2\pi n_a/\beta$. 

La rotura de la simetr\'{\i}a del centro del grupo gauge se manifiesta en que algunos de los m\'{\i}nimos absolutos degenerados del potencial efectivo de la teor\'{\i}a gauge pura dejan de serlo cuando la teor\'{\i}a incluye fermiones. No obstante, es posible probar que estos m\'{\i}nimos seguir\'an siendo puntos estacionarios del potencial efectivo completo con fermiones. En el gauge de Polyakov $A_0$ es independiente del tiempo y diagonal. Una matriz diagonal arbitraria de su($N_c$) se puede escribir siempre como una combinaci\'on lineal de matrices que tengan al menos un cero en la diagonal y la matriz $\lambda_{N_c^2-1}$ dada en (\ref{eq:p1}). \'Unicamente esta \'ultima matriz pondr\'a de manifiesto el m\'{\i}nimo que estamos buscando, por lo comentado anteriormente. El potencial efectivo de QCD que calculamos en el cap\'{\i}tulo~\ref{QCD_efective_action} se puede escribir como
\begin{equation}
{\cal L}_{0,q}(x) = -\frac{(2\pi)^2}{3\beta^4}N_f \tr B_4\left(\half+\overline\nu\right) \,, 
\qquad
\Omega(x)=e^{i2\pi\overline\nu} \,,
\quad
-\frac{1}{2}<\overline\nu<\frac{1}{2} \,,
\label{eq:p2}
\end{equation}
para el sector fermi\'onico y
\begin{equation}
{\cal L}_{0,g}(x) = \frac{2\pi^2}{3\beta^4} \widehat{\tr} B_4\left(\widehat\nu\right) \,, 
\qquad
\widehat\Omega(x)=e^{i2\pi\widehat\nu} \,,
\quad
0<\widehat\nu<1
\label{eq:p3}
\end{equation}
para el sector glu\'onico. $\tr$ es traza en la representaci\'on fundamental del grupo gauge y $\widehat{\tr}$ es en la representaci\'on adjunta. Los valores propios del loop de Polyakov en la representaci\'on fundamental son $\omega_A=\exp(i2\pi\nu_A),\; A=1,\dots,N_c$, y en la representaci\'on adjunta $\omega_{A\dot{A}}=\exp(i2\pi(\nu_A-\nu_{\dot{A}})),\; A,\dot{A}=1,\dots,N_c$. Si hacemos uso de la representaci\'on en serie de los polinomios de Bernoulli \cite{tablas}
\begin{equation}
B_{2\ell}(x)=\frac{(-1)^{\ell-1}2(2\ell)!}{(2\pi)^{2\ell}}\sum_{n=1}^\infty
\frac{\cos(2\pi n x)}{n^{2\ell}}\,,
\qquad
0 \le x \le 1 \,,
\quad
n=1, 2, \ldots 
\end{equation}
y nos limitamos a considerar el potencial efectivo para $\lambda_{N_c^2-1}$ obtenemos
\begin{eqnarray}
{\cal L}_{0,q}(x) &=& \frac{4N_f}{\pi^2\beta^4}
\sum_{n=1}^\infty \frac{(-1)^n}{n^4}\left\{
(N_c-1)\cos(2\pi n \rho)+\cos((N_c-1)2\pi n \rho)
\right\}\,, \\
{\cal L}_{0,g}(x) &=& -\frac{2}{\pi^2\beta^4}
\sum_{n=1}^\infty \frac{1}{n^4}\left\{
2(N_c-1)\cos(2\pi n N_c \rho) + (N_c-1)^2
\right\} \,.
\end{eqnarray}
Los m\'{\i}nimos de ${\cal L}_{0,g}$ se encuentran en $\rho=m/N_c$, con $m$ entero. Si diferenciamos el lagrangiano ${\cal L}_{0,q}$ respecto a $\rho$ se puede comprobar que estos m\'{\i}nimos se corresponden exactamente con puntos estacionarios (m\'{\i}nimos o m\'aximos) de la parte fermi\'onica. En consecuencia, el potencial efectivo total siempre va a tener puntos estacionarios en $\rho=m/N_c$.

\chapter{Integrales en tiempo propio con regularizaci\'on dimensional}
\label{app:integralesQCD}

Para obtener el lagrangiano efectivo de QCD quiral a un loop del cap\'{\i}tulo~\ref{QCD_efective_action} hemos necesitado calcular las trazas en espacio interno y las integrales en $\tau$. En este ap\'endice calcularemos la expresi\'on gen\'erica de la siguiente integral regulada dimensionalmente
\begin{equation}
I_{\ell,n}^\pm(\nu) = \int_0^\infty \frac{d\tau}{\tau} 
(4\pi\mu^2\tau)^\epsilon \tau^\ell \varphi_n^\pm(e^{i2\pi\nu})
\,, \quad 
\nu,\ell,\epsilon \in \mathbb{R}
\,, \quad
n=0,1,2,\ldots
\end{equation}
Las funciones $\varphi_n$ las definimos en su momento como
\begin{equation}
\varphi_n^\pm(\Omega;\tau/\beta^2) = 
\frac{(4\pi\tau)^{1/2}}{\beta} \sum_{p_0^\pm} \tau^{n/2}Q^n e^{\tau Q^2}
\,, \quad
Q = i p_0^\pm - \frac{1}{\beta}\log(\Omega)\,,
\end{equation}
donde en la versi\'on bos\'onica sumamos sobre las frecuencias de Matsubara $p_0^+ = 2\pi n/\beta$, y en la versi\'on fermi\'onica sobre $p_0^-=2\pi (n+\half)/\beta$. Centr\'emonos por el momento en la versi\'on bos\'onica de la funci\'on $\varphi_n$. Vamos a tener
\begin{equation}
I_{\ell,n}^+(\nu) = (4\pi\mu^2)^\epsilon 
\frac{\sqrt{4\pi}}{\beta} \left(\frac{2\pi i}{\beta}\right)^n \sum_{k\in \mathbb{Z}}(k-\nu)^n
\int_0^\infty d\tau \tau^{\ell+\epsilon+(n-1)/2} e^{-(\frac{2\pi}{\beta})^2(k-\nu)^2\tau} 
\,, \quad \nu \not\in \mathbb{Z} \,.     
\end{equation}
Debido a la sumatoria en $k \in \mathbb{Z}$, la funci\'on es peri\'odica en $\nu$ con per\'{\i}odo 1. El caso $\nu \in \mathbb{Z}$ ser\'a discutido m\'as tarde. La integral sobre $\tau$ se calcula y se obtiene
\begin{equation}
I_{\ell,n}^+(\nu) = i^n (4\pi\mu^2)^\epsilon
\left(\frac{\beta}{2\pi}\right)^{2(\ell+\epsilon)}
\frac{\Gamma(\ell+\epsilon+(n+1)/2)}{\Gamma(\half)}
\sum_{k\in\mathbb{Z}}\frac{(k-\nu)^n}{|k-\nu|^n}\frac{1}{|k-\nu|^{2(\ell+\epsilon)+1}} \,.
\end{equation}
Definamos $\nu = k_0 + \widehat\nu$, donde $0<\widehat\nu<1$ y $k_0 \in
\mathbb{Z}$. La suma sobre $k$ la podemos dividir en una suma para $k \le k_0$ y otra para $k > k_0$
\begin{eqnarray}
I_{\ell,n}^+(\nu) &=& i^n (4\pi\mu^2)^\epsilon
\left(\frac{\beta}{2\pi}\right)^{2(\ell+\epsilon)}
\frac{\Gamma(\ell+\epsilon+(n+1)/2)}{\Gamma(\half)} \nonumber \\
&&\qquad \times \left( \sum_{k \le k_0} \frac{(-1)^n}{(k_0+\widehat\nu-k)^{2(\ell+\epsilon)+1}}
+\sum_{k>k_0}\frac{1}{(k-k_0-\widehat\nu)^{2(\ell+\epsilon)+1}}
\right) \,.
\label{eq:appsum}
\end{eqnarray}
Si hacemos uso de la funci\'on $\zeta$ de Riemann generalizada \cite{tablas}
\begin{equation}
\zeta(z,q)=\sum_{n=0}^\infty \frac{1}{(n+q)^z} 
\, \qquad
[{\textrm Re}\,z>1,\;q\ne 0,-1,-2,\ldots] \,,
\end{equation}
llegamos a la siguiente expresi\'on
\begin{eqnarray}
I_{\ell,n}^+(\nu) &=& (4\pi)^\epsilon
\left(\frac{\mu\beta}{2\pi}\right)^{2\epsilon}
\left(\frac{\beta}{2\pi}\right)^{2\ell}
\frac{\Gamma(\ell+\epsilon+(n+1)/2)}{\Gamma(\half)} \nonumber \\
&&\qquad\qquad\times\bigg[(-i)^n \zeta(1+2\ell+2\epsilon,\widehat\nu)
+i^n \zeta(1+2\ell+2\epsilon,1-\widehat\nu)\bigg] \,,
\end{eqnarray}
donde $\widehat\nu = \nu \;({\textrm{mod}} \; 1)$, $0<\widehat\nu<1$. 

Las versiones bos\'onica y fermi\'onica de las funciones $\varphi_n$ est\'an relacionadas por $\Omega \rightarrow -\Omega$, esto es $\varphi_n^+(\Omega)=\varphi_n^-(-\Omega)$. Por tanto $I_{\ell,n}^-$ se puede obtener a partir de las integrales $I_{\ell,n}^+$ con el cambio $\nu \rightarrow \nu+\half$,
\begin{eqnarray}
I_{\ell,n}^-(\nu) &=& (4\pi)^\epsilon
\left(\frac{\mu\beta}{2\pi}\right)^{2\epsilon}
\left(\frac{\beta}{2\pi}\right)^{2\ell}
\frac{\Gamma(\ell+\epsilon+(n+1)/2)}{\Gamma(\half)} \nonumber \\
&&\qquad\qquad\times\bigg[(-i)^n \zeta(1+2\ell+2\epsilon,\half+\overline\nu)
+i^n \zeta(1+2\ell+2\epsilon,\half-\overline\nu)\bigg] \,,
\end{eqnarray}
donde $\overline\nu = (\nu+\half)\;({\textrm{mod}} \; 1)-\half$, $-\half<\overline\nu<\half$.
Notar que
\begin{equation}
I_{\ell,2n}^\pm(\nu) = (-1)^n 
\frac{\Gamma(\ell+\epsilon+n+\half)}{\Gamma(\ell+\epsilon+\half)}
I_{\ell,0}^\pm(\nu)
\,, \qquad
I_{\ell,2n+1}^\pm(\nu) = (-1)^n
\frac{\Gamma(\ell+\epsilon+n+1)}{\Gamma(\ell+\epsilon+1)}
I_{\ell,1}^\pm(\nu) \,.
\end{equation}
Estas funciones son peri\'odicas en $\nu$ y bajo paridad se comportan
\begin{equation}
I_{\ell,n}^\pm(\nu) = (-1)^n I_{\ell,n}^\pm(-\nu) \,.
\end{equation}

En el problema de la reducci\'on dimensional de la teor\'{\i}a de Yang-Mills \'unicamente se suma sobre fluctuaciones cu\'anticas no est\'aticas $(n \ne 0)$. Con objeto de preservar las propiedades de periodicidad y paridad de las funciones $I_{\ell,n}^+$, definimos las integrales bos\'onicas sin el modo est\'atico eliminando la frecuencia $k=k_0$ cuando $\widehat\nu<\half$ y la frecuencia $k=k_0+1$ cuando $\widehat\nu>\half$. Haciendo esto en (\ref{eq:appsum})  se obtiene
\begin{eqnarray}
I_{\ell,n}^{\prime +}(\nu) &=& (4\pi)^\epsilon
\left(\frac{\mu\beta}{2\pi}\right)^{2\epsilon}
\left(\frac{\beta}{2\pi}\right)^{2\ell}
\frac{\Gamma(\ell+\epsilon+(n+1)/2)}{\Gamma(\half)} \\
&& \qquad
\times \left\{
\begin{matrix}
 (-i)^n \zeta(1+2\ell+2\epsilon,1+\widehat\nu)
 +i^n \zeta(1+2\ell+2\epsilon,1-\widehat\nu) 
\,,\quad  &&0 \le \widehat\nu < \half \,,  
\cr
(-i)^n \zeta(1+2\ell+2\epsilon,\widehat\nu)
 +i^n \zeta(1+2\ell+2\epsilon,2-\widehat\nu)
\,,\quad &&\half < \widehat\nu \le 1 \,.
\end{matrix}
\right. \nonumber
\end{eqnarray}
Estas funciones son finitas, incluso para valores enteros de $\widehat\nu$.

Consideremos ahora $\nu \in \mathbb{Z}$. En este caso el modo est\'atico $p_0^+=0$ de las integrales $I_{\ell,n}^+(\nu)$ con $n \ne 0$ no contribuye. Este modo va a contribuir solamente en $I_{\ell,0}^+$ dando origen a divergencias infrarrojas o ultravioletas. En regularizaci\'on dimensional la integral  $I_{\ell,0}^+(\nu)|_{p_0=0}$ con $\nu \in \mathbb{Z}$ se define como cero ya que no tiene una escala natural. Esto conduce a la siguiente prescripci\'on
\begin{eqnarray}
I_{\ell,n}^+(\nu) &=& I_{\ell,n}^{\prime +} 
= (4\pi)^\epsilon
\left(\frac{\mu\beta}{2\pi}\right)^{2\epsilon}
\left(\frac{\beta}{2\pi}\right)^{2\ell}
\frac{\Gamma(\ell+\epsilon+(n+1)/2)}{\Gamma(\half)} \nonumber \\
&&\qquad\qquad\qquad\times\left\{
\begin{matrix}
2(-1)^{n/2} \zeta(1+2\ell+2\epsilon) 
\,,&&(n\;\textrm{par})  
\cr
0
\,,&&(n\;\textrm{impar})
\end{matrix}
\right.
\nu \in \mathbb{Z} \,.
\end{eqnarray}

\chapter{Lagrangiano Efectivo de QCD en SU(2)}
\label{app:resultadosSU(2)}

En este ap\'endice presentaremos el lagrangiano efectivo de QCD quiral a un loop a temperatura alta calculado en el cap\'{\i}tulo~\ref{QCD_efective_action}, para SU(2) en el sector de quarks y en el sector glu\'onico, incluyendo todos los t\'erminos hasta dimensi\'on de masa 6. Los resultados vienen dados en el esquema $\MS$, y hemos considerado expl\'{\i}citamente un cutoff infrarrojo. Las convenciones son las que aparecen en la secci\'on~\ref{resultados_su2}.

\begin{eqnarray}
\cL_{\text{\'arbol}}(x) &=& \frac{1}{4 g^2(\mu)}\,\vec F_{\mu\nu}^2 \,, \\ \nonumber \\
\cL_{0,g}(x) &=& 
\frac{\pi^2 T^4}{3} 
\left(-\frac{1}{5}+4\hnu^2 (1-\hnu)^2\right)
\,,  \\ \nonumber\\
 \cL_{2,g}(x) &=& -\frac{11}{96 \pi^2}\left[\frac{1}{11}+2\log\left(\frac{\mu}{4\pi T}\right)-\psi(\hnu)-\psi(1-\hnu)\right] 
\vec F_{\mu\nu \parallel}^2
\nonumber \\
&&- \frac{11}{96 \pi^2}\left[\frac{\pi T}{m}+\frac{1}{11}+2\log\left(\frac{\mu}{4\pi T}\right)+\gamma_E-\frac{1}{2}\psi(\hnu)-\frac{1}{2}\psi(1-\hnu)\right] \vec F_{\mu\nu \perp}^2
\nonumber \\
&&+\frac{1}{24\pi^2}\vec{E}_i^2 -\frac{1}{48\pi^2}\left(\frac{\pi T}{m}\right)\vec{E}_{i \perp}^2 \,, \\  \nonumber\\
 \cL_{3,g}(x) &=&\frac{61}{2160\pi^2}\left(\frac{1}{4\pi T}\right)^2\left[8\left(\frac{\pi T}{m}\right)^3+2\zeta(3)-\psi^{\prime\prime}(\widehat\nu)-\psi^{\prime\prime}(1-\widehat\nu)\right](\vec{F}_{\mu\nu} \times \vec{F}_{\nu\alpha})\cdot \vec{F}_{\alpha\mu} \nonumber \\
&&-\frac{1}{48\pi^2}\left(\frac{1}{4\pi T}\right)^2\left[\psi^{\prime\prime}(\widehat\nu)+\psi^{\prime\prime}(1-\widehat\nu)\right]\vec{F}_{\lambda\mu\nu \parallel}^2 \nonumber \\
&&+\frac{1}{96\pi^2}\left(\frac{1}{4\pi T}\right)^2\left[16\left(\frac{\pi T}{m}\right)^3+4\zeta(3)-\psi^{\prime\prime}(\widehat\nu)-\psi^{\prime\prime}(1-\widehat\nu)\right]\vec{F}_{\lambda\mu\nu \perp}^2 \nonumber 
\end{eqnarray}
\begin{eqnarray}
&&+\frac{1}{480\pi^2}\left(\frac{1}{4\pi T}\right)^2\left[\psi^{\prime\prime}(\widehat\nu)+\psi^{\prime\prime}(1-\widehat\nu)\right]\vec{F}_{\mu\mu\nu \parallel}^2 \nonumber \\
&&-\frac{1}{960\pi^2}\left(\frac{1}{4\pi T}\right)^2\left[16\left(\frac{\pi T}{m}\right)^3+4\zeta(3)-\psi^{\prime\prime}(\widehat\nu)-\psi^{\prime\prime}(1-\widehat\nu)\right]\vec{F}_{\mu\mu\nu \perp}^2
\nonumber \\
&&-\frac{3}{80\pi^2}\left(\frac{1}{4\pi T}\right)^2\left[\psi^{\prime\prime}(\widehat\nu)+\psi^{\prime\prime}(1-\widehat\nu)\right]\vec{F}_{0\mu\nu \parallel}^2\nonumber \\
&&+\frac{3}{160\pi^2}\left(\frac{1}{4\pi T}\right)^2\left[-8\left(\frac{\pi T}{m}\right)^3+4\zeta(3)-\psi^{\prime\prime}(\widehat\nu)-\psi^{\prime\prime}(1-\widehat\nu)\right]\vec{F}_{0\mu\nu \perp}^2  \nonumber \\ 
&&-\frac{1}{10\pi^2}\left(\frac{1}{4\pi T}\right)^2\left(\frac{\pi T}{m}\right)^3 \vec{E}_{0i \perp}^2  \nonumber \\  
&&+\frac{1}{240\pi^2}\left(\frac{1}{4\pi T}\right)^2\left[\psi^{\prime\prime}(\widehat\nu)+\psi^{\prime\prime}(1-\widehat\nu)\right]\vec{E}_{ii \parallel}^2 \nonumber \\
&&-\frac{1}{480\pi^2}\left(\frac{1}{4\pi T}\right)^2\left[-8\left(\frac{\pi T}{m}\right)^3+4\zeta(3)-\psi^{\prime\prime}(\widehat\nu)-\psi^{\prime\prime}(1-\widehat\nu)\right]\vec{E}_{ii \perp}^2  \nonumber \\
\noindent &&+\frac{1}{240\pi^2}\left(\frac{1}{4\pi T}\right)^2\left[\psi^{\prime\prime}(\widehat\nu)+\psi^{\prime\prime}(1-\widehat\nu)\right]\epsilon_{ijk}(\vec{E}_i\times \vec{E}_j)\cdot \vec{B}_k \\
\noindent &&+\frac{1}{240\pi^2}\left(\frac{1}{4\pi T}\right)^2\left[8\left(\frac{\pi T}{m}\right)^3-4\zeta(3)-\psi^{\prime\prime}(\widehat\nu)-\psi^{\prime\prime}(1-\widehat\nu)\right]\epsilon_{ijk}(\vec{E}_{i \perp}\times\vec{E}_{j \perp})\cdot \vec{B}_{k \parallel} \,, \nonumber
\end{eqnarray}

\begin{eqnarray}
\cL_{0,q}(x) &=& \frac{2}{3} \pi^2  T^4 N_f \left(\frac{2}{15}-\frac{1}{4}(1-4 \bnu^2)^2\right)
\,, \\ \nonumber \\ 
\cL_{2,q}(x) &=&
\frac{N_f}{96 \pi^2}\left[2\log\left(\frac{\mu}{4\pi T}\right)
-\psi(\half+\bnu)-\psi(\half-\bnu)\right] \vec{F}_{\mu\nu}^2 
-\frac{N_f }{48 \pi^2}\vec{E}_i^2
\,, \\  \nonumber  \\ 
\cL_{3,q}(x) &=& \frac{N_f}{960\pi^2}\left(\frac{1}{4\pi T}\right)^2
\left[\psi^{\prime\prime}(\half +\bnu)+\psi^{\prime\prime}(\half-\bnu)\right]
      \\
&&
\times 
\Big(
\frac{16}{3}(\vec{F}_{\mu\nu} \times \vec{F}_{\nu\alpha}) \cdot \vec F_{\alpha\mu}+\frac{5}{2}\vec{F}_{\lambda\mu\nu}^2
-\vec{F}_{\mu\mu\nu}^2
-2\epsilon_{ijk}(\vec{E}_i\times\vec{E}_j) \cdot \vec{B}_k
+3\vec{F}_{0\mu\nu}^2
-2\vec{E}_{ii}^2
\Big)
\,.  \nonumber
\end{eqnarray}

$\vec{a}\times\vec{b}$ es el producto vectorial de $\vec{a}$ y $\vec{b}$, esto es
\begin{equation}
(\vec{a}\times\vec{b})_i = \epsilon_{ijk} a_j b_k \,.
\end{equation}
Como vemos, las contribuciones de los quarks no distinguen entre componentes
paralelas y perpendiculares. Esto se debe a que en SU(2) una funci\'on par en
$\overline\nu$ en la representaci\'on fundamental es necesariamente un
c-n\'umero. Puesto que todas las funciones $\varphi_n(\Omega)$ involucradas en
los t\'erminos de dimensi\'on 6 son pares $[\varphi_n(\Omega) +
\varphi_n(\Omega^{-1})= c \cdot \mathbf{1}_{2\times 2}]$, la dependencia en
$\overline\nu$ de las ecs.~(\ref{eq:L2qpp}) y (\ref{eq:L3qpp}) sale fuera de
la traza, de modo que $A_0$ no ser\'a una direcci\'on privilegiada en espacio de color. Este propiedad no se cumple en la representaci\'on adjunta (sector glu\'onico), ni tampoco en otros grupos SU($N_c$) (por ejemplo, ec.~(\ref{eq:45a})).

Las divergencias infrarrojas est\'an sujetas a que $\nu$ sea entero, de modo que no existen en el sector fermi\'onico, y se cancelan en las contribuciones glu\'onicas que \'unicamente involucran componentes paralelas.

\chapter{Lagrangiano Efectivo del Modelo Quark Quiral acoplado con el loop de Polyakov}
\label{sec:hhkk} 

En este ap\'endice se explicar\'a en detalle el c\'alculo del lagrangiano quiral efectivo a temperatura finita presentado en la secci\'on~\ref{lagrangiano_quiral_Tfinita}. El c\'alculo se divide en tres partes. En primer lugar se construir\'a el operador de Klein-Gordon a partir del operador de Dirac y su adjunto para la parte real de la acci\'on efectiva. Haciendo uso de la representaci\'on de Schwinger de tiempo propio, deberemos calcular el heat kernel para este operador. Para ello haremos uso de la t\'ecnica desarrollada en el cap\'{\i}tulo~\ref{heat_kernel}. Calcularemos las trazas en los grados de libertad internos (en nuestro caso, sabor). Finalmente, haremos uso de las ecuaciones de movimiento con objeto de tener en cuenta el hecho de que los campos pi\'onicos est\'an en la capa de masas.

\section{Operador de Klein-Gordon efectivo} 
\label{op_KG_ef}

El operador de Dirac que aparece en el determinante fermi\'onico se comporta de ma\-ne\-ra covariante bajo transformaciones quirales. Esto implica que, en principio, habr\'{\i}a que considerar tanto el acoplamiento vector como el axial. Conseguiremos una gran simplificaci\'on en nuestro tratamiento si hacemos uso de los convenios de ref.~\cite{normalpc, abnormalpc}, donde se muestra que es suficiente con llevar a cabo el c\'alculo en el caso de un acoplamiento vector, y posteriormente reconstruir el resultado quiral total de un modo conveniente.

Consideremos el siguiente operador de Dirac con un acoplamiento tipo vector
\begin{equation}
{\mathbf D} = \thruu{D}+h, \qquad h=m+z \,,
\end{equation}
donde $h$ incluye el campo del pi\'on $m$, que es orden ${\cal O}(p^0)$, y el t\'ermino de masa $z$ que rompe expl\'{\i}citamente la simetr\'{\i}a quiral, y que tomamos ${\cal O}(p^2)$. Nuestra notaci\'on es la siguiente
\begin{eqnarray}
h_{LR} &=& MU +\frac{1}{2B_0^*}\chi \,, \nonumber \\
h_{RL} &=& MU^\dagger +\frac{1}{2B_0^*}\chi^\dagger \,.
\end{eqnarray}
La parte real de la acci\'on efectiva es, formalmente
\begin{equation}  
\Gamma_q^+[v,h] =-\frac{1}{2}\Tr\log({\mathbf D}^\dagger {\mathbf D})=: \int_0^\beta dx_0
\int d^3 x \,{\cal L}^*_q(x)\,,
\end{equation}
donde el operador de Klein-Gordon relevante viene dado por
\begin{eqnarray}
{\mathbf D}^\dagger {\mathbf D} &=&
-D_\mu^2-\frac{1}{2}\sigma_{\mu\nu}F_{\mu\nu}-\gamma_\mu{\D}_\mu h +
m^2 + \overline{h}^2 \,, \nonumber \\ \overline{h}^2 &=& 
h^2-m^2=\{m,z\}+z^2\,.
\end{eqnarray}
El problema radica en hacer un desarrollo en derivadas covariantes para la acci\'on efectiva. Podemos identificar el operador de masa
$M(x)=-\frac{1}{2}\sigma_{\mu\nu}F_{\mu\nu}-\gamma_\mu {\widehat
D}_\mu h + \overline{h}^2$. Haciendo uso de la representaci\'on de Schwinger de tiempo propio, el lagrangiano efectivo en espacio eucl\'{\i}deo se puede escribir como
\begin{eqnarray}
{\cal L}^*_q &=& \frac{1}{2}\int_{0}^\infty
\frac{d\tau}{\tau}\phi(\tau)\,\Tr \,e^{-\tau D^\dagger D} = \frac{1}{2}\int_{0}^\infty \frac{d\tau}{\tau}\phi(\tau)\frac{e^{-\tau
M^2}}{(4\pi\tau)^2}\sum_n \tau^n \tr \,b_n^T \,.
\label{eq:lagran}
\end{eqnarray}
En esta representaci\'on haremos uso de la regularizaci\'on de Pauli-Villars~\cite{njlarriola}
\begin{eqnarray}
\phi(\tau)=\sum_i c_i e^{-\tau\Lambda_i^2} \,.
\label{eq:pv}
\end{eqnarray}
Hasta ${\cal O} (p^4) $ obtenemos las siguientes contribuciones para los coeficientes de Seeley-DeWitt t\'ermicos, despu\'es de haber tomado la traza de Dirac
\begin{eqnarray}
\hb_0 &=& 4 \varphi_0(\Omega) \,,\nonumber \\
\hb_{1/2} &=& 0 \,, \nonumber \\
\hb_1 &=& -4 \varphi_0(\Omega){\overline h}^2 =
-4\varphi_0(\Omega)\big(\{m,z\}+z^2\big)\,, \nonumber \\
\hb_{3/2} &=& 0 \,, \nonumber \\
\hb_2 &=&
2\varphi_0(\Omega)\left((h_\mu)^2+{\overline
h}^4-\frac{1}{3}F_{\mu\nu}^2\right) -\frac{2}{3}{\overline\varphi}_2
E_i^2 \nonumber \\ 
&=&
2\varphi_0(\Omega)\left((m_\mu)^2+\{m_\mu,z_\mu\}+\{m,z\}\{m,z\}-\frac{1}{3}F_{\mu\nu}^2\right)-\frac{2}{3}{\overline\varphi}_2(\Omega)E_i^2
+{\cal O}(p^6) \,, \nonumber \\
\hb_{5/2}&=&
-\frac{2}{3}\varphi_1\{E_i,(\overline h^2)_i\} = -\frac{2}{3}\varphi_1
\{E_i,\D_i\{m,z\}\}= {\cal O}(p^5) \,, \nonumber \\
\hb_3 &=&-\frac{2}{3}\varphi_0(\Omega)\bigg(
m_\mu\{m_\mu,\{m,z\}\}+\{m,z\}m_\mu m_\mu+ \{F_{\mu\nu},m_\mu
m_\nu\}-m_\mu F_{\mu\nu} m_\nu  \nonumber \\
&& +\frac{1}{2}(m_{\mu\nu})^2 \bigg) 
+\frac{1}{3}{\overline\varphi}_2(m_{0\mu})^2 + {\cal O}(p^5)\,, \nonumber 
\end{eqnarray}
\begin{eqnarray}
\hb_{7/2} &=& {\cal O}(p^5) \,, \nonumber \\
\hb_4 &=&
\frac{1}{6}\varphi_0(\Omega)(m_\mu m_\mu m_\nu m_\nu +m_\mu m_\nu
m_\nu m_\mu -m_\mu m_\nu m_\mu m_\nu) + {\cal O}(p^5)\,.
\end{eqnarray}

\section{Trazas de sabor e identidades \'utiles}  
\label{trazas_sabor}

Para $N_f=3$ sabores se tiene la siguiente identidad de $\textrm{SU}(3)$
\begin{equation}
\tr(ABAB)=-2\tr(A^2 B^2)+\frac{1}{2}\tr(A^2)\tr(B^2)+(\tr(AB))^2 \,,
\end{equation}
donde $A$ y $B$ son matrices herm\'{\i}ticas $3 \times 3$ de traza cero. De aqu\'{\i} se tiene
\begin{equation}
\tr_f(m_{\mu}m_{\nu}m_{\mu}m_{\nu}) = -2\tr_f((m_{\mu})^2(m_{\nu})^2)+\frac{1}{2}\tr_f((m_{\mu})^2)\tr_f((m_{\nu})^2)+(\tr_f(m_{\mu}m_{\nu}))^2\,, \label{eq:trf1}
\end{equation}
\begin{equation}
\tr_f(m_{0}m_{\mu}m_{0}m_{\mu}) = -2\tr_f((m_{0})^2(m_{\mu})^2)+\frac{1}{2}\tr_f((m_{0})^2)\tr_f((m_{\mu})^2)+(\tr_f(m_{0}m_{\mu}))^2 \,.
\label{eq:trf2}
\end{equation}
Otras identidades \'utiles son
\begin{equation}
\tr_f((m_{\mu\nu})^2) = 
\tr_f((m_{\mu\mu})^2)-2\tr_f(F_{\mu\nu}m_\mu m_\nu)
+\tr_f(mF_{\mu\nu}mF_{\mu\nu})
-M^2\tr_f(F_{\mu\nu}^2)  \,,
\label{eq:trf3}
\end{equation}
\begin{equation}
\!\!\!\!\!\!\!\!\!\!\!\!\!\!\!\!\!\!\!\!\!\!\!\!\!\!\!\!\!\!\!\!\!\!\!\!\!\!\!\!\!\!\!\!\!\!\!\!\!\!\!\!\!
\tr_f((m_{0\mu})^2) = 
\tr_f(m_{00}m_{\mu\mu})-2\tr_f(E_i [m_0,m_i])
-2\tr_f(E_{0i}m m_i) \,,
\label{eq:trf4}
\end{equation}
donde hemos hecho uso de la propiedad $X_{\mu\nu}=X_{\nu\mu}+[F_{\mu\nu},X]$. Podemos aplicar las ecuaciones de movimiento, ec.~(\ref{eq:ecm}), para obtener
\begin{eqnarray}
\tr_f(m_\mu z_\mu) &=&
\frac{1}{2B^*_0M^2}\tr_f(m_\mu m_\mu m x)
-\frac{1}{4B^*_0M}\tr_f(m x m x)
+\frac{M}{4B^*_0}\tr_f(x^2) \nonumber \\
&&+\frac{1}{8MN_fB^*_0}\tr_f([m,x])\tr_f([m,x])
\,,
\label{eq:trf5} \\
\tr_f(m_{\mu\mu}m_{\nu\nu}) &=&
\frac{1}{M^2}\tr_f(m_\mu m_\mu m_\nu m_\nu)
-\frac{1}{2} \tr_f(m x m x) 
+\frac{M^2}{2}\tr_f(x^2) \nonumber \\
&&+\frac{1}{4N_f}\tr_f([m,x])\tr_f([m,x]) \,, 
\label{eq:trf6}\\
\tr_f(m_{00}m_{\mu\mu}) &=&
\frac{1}{M^2}\tr_f(m_0 m_0 m_\mu m_\mu)
-M\tr_f(m_{00}x) 
-\frac{1}{M}\tr_f(m_0 m_0 m x)  \nonumber \\
&&+\frac{1}{2MN_f}\tr_f(m_{00}m)\tr_f([m,x]) \,.
\label{eq:trf7}
\end{eqnarray}
donde se han introducido los campos normalizados $x=2B^*_0 z$. La notaci\'on es
la siguiente: 
\begin{equation}
x_{LR} = \chi  \,, \qquad x_{RL} = \chi^\dagger \,.
\end{equation}
Haciendo uso de (\ref{eq:trf1})-(\ref{eq:trf7}) podemos calcular la traza en espacio de sabor de los coeficientes de Seeley-DeWitt. Esto conduce a  
\begin{eqnarray}
\tr_f \hb_0 &=& 4N_f \varphi_0(\Omega) \,, \nonumber \\ \tr_f
\hb_1 &=& -\varphi_0(\Omega)\left(\frac{4}{B^*_0}\tr_f(m x)
+\frac{1}{B^{*2}_0}\tr_f(x^2)\right) \,, \nonumber \\ 
\tr_f \hb_2 &=& 2\varphi_0(\Omega)\tr_f(m_\mu m_\mu) +\frac{2}{B^*_0
M^2}\varphi_0(\Omega)\tr_f(m_\mu m_\mu mx) \nonumber \\
&& +\frac{1}{B^*_0}\left(\frac{1}{B^*_0}-\frac{1}{M}\right)\varphi_0(\Omega)\tr_f(mx
mx)+\frac{M}{B^*_0}\left(\frac{M}{B^*_0}+1\right)\varphi_0(\Omega)\tr_f(x^2) \nonumber \\
&&-\frac{2}{3}\varphi_0(\Omega)\tr_f(F_{\mu\nu}^2)
-\frac{2}{3}\overline\varphi_2(\Omega)\tr_f(E_i^2) +\frac{1}{2MN_f
B^*_0}\varphi_0(\Omega)\tr_f([m,x])\tr_f([m,x]) \,, \nonumber \\
\tr_f 
\hb_3 &=& -\frac{4}{3}\varphi_0(\Omega)\tr_f(F_{\mu\nu}m_\mu
m_\nu) -\frac{1}{3}\varphi_0(\Omega)\tr_f(mF_{\mu\nu}mF_{\mu\nu})
+\frac{1}{3}M^2\varphi_0(\Omega)\tr_f(F_{\mu\nu}) \nonumber \\
&&-\frac{1}{6}M^2\varphi_0(\Omega) \tr_f(x^2) 
+\frac{1}{6}\varphi_0(\Omega)\tr_f(mx mx)
-\frac{2}{B^*_0}\varphi_0(\Omega)\tr_f(m_\mu m_\mu mx) \nonumber \\
&&-\frac{1}{3M}\overline\varphi_2(\Omega)\tr_f(m_0m_0mx)
-\frac{M}{3}\overline\varphi_2(\Omega)\tr_f(m_{00}x) 
-\frac{2}{3}\overline\varphi_2(\Omega)\tr_f(E_i[m_0,m_i]) \nonumber \\
&&-\frac{2}{3}\overline\varphi_2(\Omega)\tr_f(E_{0i}mm_i)
-\frac{1}{3M^2}\varphi_0(\Omega)\tr_f(m_{\mu}m_{\mu}m_{\nu}m_{\nu}) \nonumber \\&&+\frac{1}{3M^2}\overline\varphi_2(\Omega)\tr_f(m_0m_0m_{\mu}m_{\mu})
-\frac{1}{12N_f}\varphi_0(\Omega)\tr_f([m,x])\tr_f([m,x]) \nonumber \\
&&+\frac{1}{6MN_f}\overline\varphi_2(\Omega)\tr_f(m_{00}m)\tr_f([m,x])
) \,, \nonumber \\ 
\tr_f \hb_4 &=&
-\frac{1}{12}\varphi_0(\Omega)\tr_f(m_\mu m_\mu)\tr_f(m_\nu m_\nu)
-\frac{1}{6}\varphi_0(\Omega)\tr_f(m_\mu m_\nu)\tr_f(m_\mu m_\nu) \nonumber \\
&&+\frac{2}{3}\varphi_0(\Omega)\tr_f(m_\mu m_\mu m_\nu m_\nu) \,.
\label{eq:csdw}
\end{eqnarray}

\section{Integrales en tiempo propio}
\label{integrales_TP}

Las integrales en tiempo propio b\'asicas que definimos son
\begin{eqnarray}
\J_{l}(\Lambda,M,\nu) &:=& \int_0^\infty \frac{d\tau}{\tau}
\phi(\tau)\tau^l e^{-\tau M^2} \varphi_0(\Omega) \,,
\\
\overline\J_{l}(\Lambda,M,\nu) &:=& \int_0^\infty \frac{d\tau}{\tau}
\phi(\tau)\tau^l e^{-\tau M^2} \overline\varphi_2(\Omega) \,,
\end{eqnarray}
donde $\Omega=e^{i2\pi\nu}$ es una matriz SU($N_c$) en espacio de color. Haciendo uso de la f\'ormula de Poisson para la sumatoria, podemos escribir $\varphi_0$ y $\overline\varphi_2$ del siguiente modo
\begin{eqnarray}
\varphi_0(\Omega) &=& \sum_{n\in\mathbb Z}
e^{-\frac{n^2\beta^2}{4\tau}}(-\Omega)^n \,, 
\end{eqnarray}
\begin{eqnarray}
\overline\varphi_2(\Omega) &=& \frac{\beta^2}{2\tau}\sum_{n\in\mathbb
Z}n^2 e^{-\frac{n^2\beta^2}{4\tau}}(-\Omega)^n \,.
\end{eqnarray}
La contribuci\'on de temperatura cero viene dada por el t\'ermino $n=0$, y
para \'el es necesario aplicar una regularizaci\'on (aqu\'{\i} usamos Pauli-Villars). En los t\'erminos $n \ne 0$ la regulari\-zaci\'on puede ser eliminada, pues el ba\~no t\'ermico act\'ua de por s\'{\i} como un regulador ultravioleta. Esta aproximaci\'on est\'a justificada a temperaturas suficientemente peque\~nas $T \ll \Lambda_{\rm PV}$. T\'{\i}picamente $\Lambda_{\rm PV} \approx 1\,{\rm GeV}$ de modo que incluso para $T\approx M \approx 300 \,{\rm MeV}$ la aproximaci\'on es v\'alida. El c\'alculo de las integrales conduce a  
\begin{eqnarray}
\J_{l}(\Lambda,M,\nu) &=&\mathbf{1}_{N_c\times N_c}\Gamma(l)\sum_i
c_i(\Lambda_i^2+M^2)^{-l} \label{eq:Jl1} \\ &&\qquad + 2\left(
\frac{\beta}{2M}\right)^l\sum_{n=1}^\infty n^l K_l(n\beta M)
((-\Omega)^n + (-\Omega)^{-n}) \,, \quad \textrm{Re}(l)>0 \,,
\nonumber \\ 
\J_{0}(\Lambda,M,\nu) &=& -\mathbf{1}_{N_c\times
N_c}\sum_i c_i \log(\Lambda_i^2+M^2) \nonumber \\
&&\qquad + 2\sum_{n=1}^\infty K_0(n\beta
M) ((-\Omega)^n + (-\Omega)^{-n}) \,, \label{eq:Jl2} \\
\J_{-1}(\Lambda,M,\nu) &=& \mathbf{1}_{N_c\times N_c}\sum_i c_i
(\Lambda_i^2+M^2)\log(\Lambda_i^2+M^2) \nonumber \\
&&\qquad +\frac{4M}{\beta}\sum_{n=1}^\infty \frac{K_1(n\beta M)}{n}
((-\Omega)^n + (-\Omega)^{-n}) \,, \label{eq:Jl3}  \\
\J_{-2}(\Lambda,M,\nu) &=& -\mathbf{1}_{N_c\times
N_c}\,\frac{1}{2}\sum_i c_i (\Lambda_i^2+M^2)^2\log(\Lambda_i^2+M^2) \nonumber \\ 
&& \qquad + 8  \left(\frac{M}{\beta}\right)^2 \sum_{n=1}^\infty \frac{K_2(n\beta M)}{n^2}((-\Omega)^n + (-\Omega)^{-n}) \,, \label{eq:Jl4} \\
\overline\J_{l}(\Lambda,M,\nu) &=&
\frac{\beta^{l+1}}{(2M)^{l-1}}\sum_{n=1}^\infty n^{l+1} K_{l-1}(n\beta
M) ((-\Omega)^n + (-\Omega)^{-n}) \,, \quad l \in {\mathbb R} 
\,. \label{eq:Jl5}
\end{eqnarray}

\section{Ecuaciones cl\'asicas de movimiento}
\label{ec_clas_mov}

A orden ${\cal O}(p^2)$ el lagrangiano quiral se escribe 
\begin{eqnarray}
{\cal L}_q^{* (2)} &=&\int_0^\infty \frac{d\tau}{\tau}\phi(\tau)
\frac{e^{-\tau M^2}}{(4\pi)^2} \tr_c \varphi_0(\Omega)
\left[\tr_f(m_\mu m_\mu)-\frac{4}{\tau}\tr_f(m z)\right]
\nonumber \\
&=& \frac{1}{(4\pi)^2}\bigg(
\tr_c {\cal J}_{0}(\Lambda,M,\nu) \tr_f(m_\mu m_\mu)
-4 \tr_c {\cal J}_{-1}(\Lambda,M,\nu) \tr_f(mz)
\bigg) 
\nonumber 
\end{eqnarray}
\begin{eqnarray}
&=& \frac{M^2}{(4\pi)^2}\tr_c{\cal J}_{0}(\Lambda,M,\nu)
\left(
\tr_f(\D_\mu U^\dagger \D_\mu U)
-\frac{2}{M}\frac{\tr_c {\cal J}_{-1}(\Lambda,M,\nu)}{\tr_c {\cal J}_{0}(\Lambda,M,\nu)}
\tr_f(z_{RL} U + z_{LR} U^\dagger)
\right)
\nonumber \\
&=& \frac{M^2}{(4\pi)^2}\tr_c{\cal J}_{0}(\Lambda,M,\nu)
\tr_f\left(\D_\mu U^\dagger \D_\mu U
-(\chi^\dagger U + \chi U^\dagger)
\right)  \,,
\end{eqnarray}
donde la normalizaci\'on del campo $\chi$ viene dada por el factor 
\begin{equation}
\chi = 2B^*_0 \,z_{LR}\,, \qquad \chi^\dagger = 2B^*_0 \,z_{RL} \,, \qquad B^*_0= \frac{1}{M}\frac{\tr_c{\cal J}_{-1}(\Lambda,M,\nu)}{\tr_c{\cal J}_{0}(\Lambda,M,\nu)} \,.
\end{equation}
y
\begin{equation}
\frac{\f^2}{4}=\frac{M^2}{(4\pi)^2}\tr_c{\cal J}_{0}(\Lambda,M,\nu) \,.
\end{equation}
Si minimizamos la acci\'on a este orden, se obtienen las ecuaciones de movimiento de Euler-Lagrange
\begin{equation}
m_{\mu\mu}m +m_\mu m_\mu -\frac{M}{2}[m,x]+\frac{M}{2N_f}\tr_f([m,x])=0 \,.
\label{eq:ecm}
\end{equation}
El \'ultimo t\'ermino en ec.~(\ref{eq:ecm}) viene de imponer la condici\'on~$\Det(U)=1$, pues estamos considerando un grupo de sabor~SU($N_f$).

\section{Lagrangiano Efectivo} 
\label{lagrang_efect}

El lagrangiano efectivo se puede escribir como
\begin{equation}
{\cal L}^*_q = {\cal L}_q^{* (0)} + {\cal L}_q^{* (2)} + {\cal L}_q^{* (4)} +
\cdots \,.
\end{equation}
Haciendo uso de la expresi\'on del lagrangiano en ec.~(\ref{eq:lagran}), los coeficientes de Seeley-DeWitt de ec.~(\ref{eq:csdw}) y despu\'es de calcular la integral en tiempo propio con regularizaci\'on de Pauli-Villars, se obtiene
\begin{eqnarray}
\cL_q^{* (0)} &=& \frac{2N_f}{(4\pi)^2}\tr_c\J_{-2}(\Lambda,M,\nu) \,,
\\
\cL_q^{* (2)} &=& \frac{\f^2}{4}\tr_f\left(
\D_\mu U^\dagger\D_\mu U -(\chi^\dagger U +\chi U^\dagger)
\right) \,,
\nonumber \\
\cL_q^{* (4)} &=& -L^*_1\tr_f(u_\mu u_\mu)\tr_f(u_\nu u_\nu)
-L^*_2\tr_f(u_\mu u_\nu)\tr_f(u_\mu u_\nu)
-L^*_3\tr_f(u_\mu u_\mu u_\nu u_\nu ) \nonumber \\
&&-\overline{L}^*_3 \tr_f(u_0u_0u_\mu u_\mu)
+2L^*_4\tr_f(u_\mu u_\mu)\tr_f(x u)
+2L^*_5\tr_f(u_\mu u_\mu ux)   \nonumber \\
&&+2\overline L^*_5\tr_f(u_0u_0ux)
+2\overline L_5^{* \prime} \tr_f(u_{00}x)
-2(L^*_6+L^*_7)\tr_f(ux)\tr_f(ux)   \nonumber \\
&&-2(L^*_6-L^*_7)\tr_f(ux)\tr_f(x u)
+2\overline L^{* \prime} \tr_f(u_{00}u) \tr_f([u,x])
-2L^*_8\tr_f(ux ux)     \nonumber \\
&&-2L^*_9\tr_f(F_{\mu\nu}u_\mu u_\nu)
-2\overline L^*_9 \tr_f(E_i[u_0,u_i])
-2\overline L_9^{* \prime} \tr_f(E_{0i}uu_i)
\nonumber \\
&&+L^*_{10}\tr_f(uF_{\mu\nu}uF_{\mu\nu})
+2H^*_1\tr_f(F_{\mu\nu}^2)
+2{\overline H}^*_1 \tr_f(E_i^2)
-H^*_2\tr_f(x^2) \,,
\label{eq:lag4}
\end{eqnarray}
donde se ha usado la notaci\'on~$m=Mu$. Los coeficientes que aparecen en ec.~(\ref{eq:lag4}) se han escrito de manera que se correspondan con la convenci\'on de Gasser-Leutwyler.

\end{document}